## *Projecte de fi de carrera*

Disseny d'un prototipus de xarxa MESH sense fils multiràdio i multicanal sobre OLSR modificat amb canal de senyalització dedicat

**Autor: Miguel Ángel Jaume Otero**
**Director: Josep Paradells Aspas**

Barcelona, febrer 2008

# PREFACI I AGRAÏMENTS

Aquest projecte[1] ha estat desenvolupat en el marc de les comunicacions mallades sense fils i ha estat proposat pel Grup de Xarxes Sense Fils del Departament d'Enginyeria Telemàtica de la Universitat Politècnica de Catalunya.

A aquest projecte es realitzarà el disseny, la implementació i l'avaluació d'un prototipus multiràdio i multicanal amb la utilització de *hardware* específic pel muntatge del prototipus. Amb el desenvolupament d'aquest treball s'aportaran mètodes que permetin millorar el rendiment global de les xarxes *mesh*.

*Abans de passar al contingut del projecte he d'agrair l'ajuda que he rebut per part de moltes persones i que han contribuït en la realització d'aquest projecte. Per això he de donar les gràcies al meu director de projecte Josep Paradells per orientar-me al treball, a en Josep Lluís Ferrer per ajudar-me en fases complicades del projecte i en Miquel Catalán per les seves indicacions sempre útils i oportunes.*

*També agrair a tota la gent que ha estat al meu costat durant aquests mesos de feina i que han contribuït en la realització d'aquest projecte més del que s'imaginen amb el seu suport. I, especialment, agrair la paciència i comprensió de la meva família, ja que són ells els que més han cregut en jo en tot aquest camí tan llarg que m'ha dut fins aquí.*

---

[1] Tot el document ha estat escrit en català amb algunes paraules i temps verbals dialectals, pròpies de la forma de parlar mallorquina. Aquesta aclaració es fa per advertir de, tot i que han tractat de ser evitades, l'aparició de possibles paraules específiques que no són incorrectes gramaticalment però que ho poden semblar pel seu poc ús a l'estàndard.

# ÍNDEX

# ÍNDEX DE FIGURES

# ÍNDEX DE TAULES

# 1 INTRODUCCIÓ

## 1.1 CONTEXTE DEL PROJECTE

Des del seu origen, les **xarxes de telecomunicacions** han sofert una gran evolució degut al gran desenvolupament de la tecnologia. Inicialment les xarxes de comunicacions que donaven accés als serveis eren cablades i era impensable utilitzar el medi sense fils per proporcionar serveis similars als que s'oferien en les xarxes cablades degut a l'hostilitat d'aquest medi.

Però els avenços tecnològics, l'abaratiment dels costos d'aquestes tecnologies i l'aparició d'estàndards consistents van fer que les xarxes sense fils es convertissin en opcions a considerar, ja que proporcionen una major flexibilitat en la connexió, podent accedir als serveis independentment de la localització. Això es veu en especial en l'àmbit de les comunicacions mòbils que, amb l'aparició de nous estàndards, cada vegada inclouen serveis més variats, de millor qualitat i amb costos més assequibles. Els estàndards més populars han estat **GSM** (2ª generació de telefonia mòbil), **GPRS** (generació 2,5), **UMTS** (3ª generació). Aquest desenvolupament tecnològic, acompanyat d'una reducció de costos en els nous terminals, unes tarifes més ajustades i uns terminals més petits i manejables han fet que la telefonia clàssica fixa quedés relegada a un segon terme.

Però no només en el camp de la telefonia mòbil es veuen aquests progressos. Altres tecnologies sense fils han experimentat un increment notable, com és el cas de la tecnologia *wifi*. Aquesta tecnologia es basa en estàndards desenvolupats pel **Institute of Electrical and Electronic Engineers (IEEE)** i permet proporcionar accés sense fils als usuaris permetent la configuració de xarxes sense fils. La disponibilitat i l'abaratiment de la tecnologia han estat clau per l'expansió de tecnologia *wifi*.

Una mostra de l'augment de l'ús de la tecnologia *wifi* es veu en la següent gràfica (figura 1.1) [1] on es pot veure l'evolució del nombre d'usuaris *wifi* a Espanya i del mercat econòmic (resultat *100€) que aquestes unitats generen.

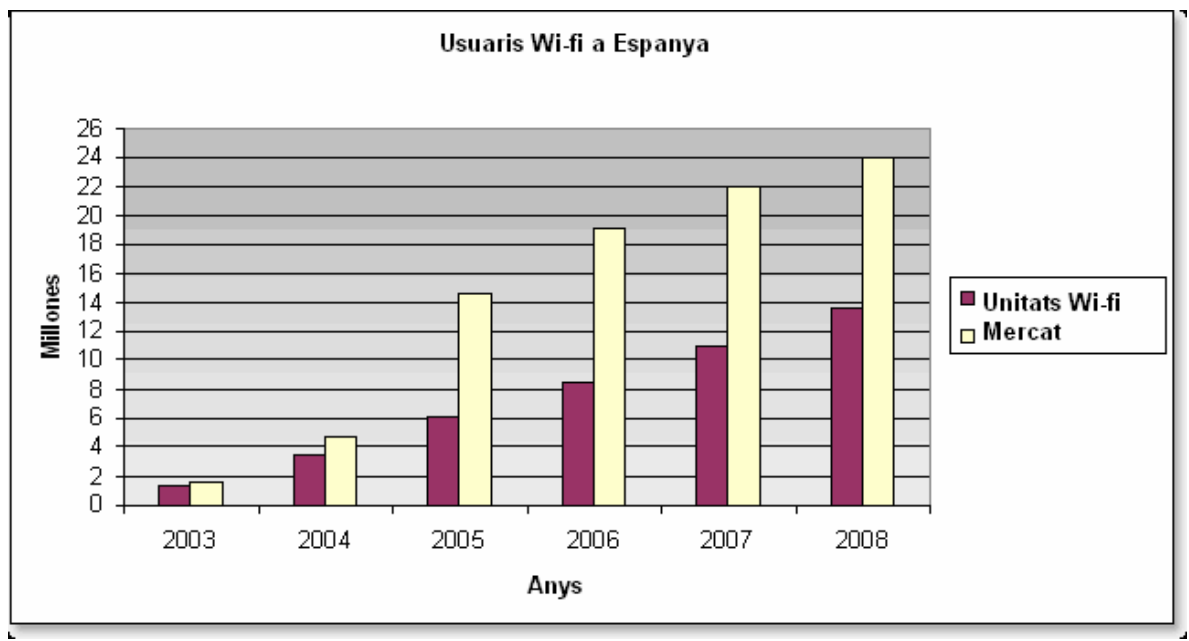

**Figura 1.1: Evolució usuaris Wi-Fi**

A més, la instal·lació de punts d'accés a llocs freqüentats massivament pel públic com les administracions públiques, hospitals, centres comercials o aeroports, ha permès proporcionar **connectivitat** a tots els usuaris baix aquests punts d'accés. D'aquesta manera, qualsevol usuari amb la tecnologia necessària baix la cobertura d'aquests punts d'accés pot tenir connectivitat a la xarxa Internet o a altres serveis que s'ofereixin des d'aquests *hotspots.*

La creació de nous **estàndards** van en el camí de proporcionar una connectivitat cada vegada més global i més extensa. És el cas d'estandarditzacions com la que proposa el WiMax Forum amb **WiMax** [2]. **WiMax** o 802.16 tracta de donar cobertura a una àrea més extensa, com podria ser el domini que abraça tota una ciutat. Seria proporcionar accés als serveis des de qualsevol punt de la localitat baix la cobertura de WiMax.

També s'han definit uns altres tipus de xarxes sense fils i que seran el focus d'actuació d'aquest projecte i són les **xarxes *mesh*** o **xarxes mallades sense fils**. Aquestes xarxes són aquelles que es constitueixen amb la col·locació intel·ligent de diferents punts d'accés, que donen cobertura a una certa àrea geogràfica determinada i a on els punts d'accés conformen una xarxa mallada amb rutes establertes per protocols d'encaminament. D'aquesta forma, s'aconsegueix augmentar la cobertura inicial amb la simple agregació de nodes.

Les **xarxes** *mesh* funcionen com a sistemes distribuïts i estan conformades de nodes amb poca potència de transmissió amb un cost i consum de recursos menor que els dels clàssics punts d'accés relatats abans. Normalment funcionen amb sistemes encastats i es tracten de xarxes autoconfigurables que es poden fer servir a situacions on la xarxa no existeix. Al ser autoconfigurables, és idònia per a la implantació d'aquests punts d'accés en llocs com els d'una flota de vehicles d'emergències com de ambulàncies, bombers…

Els avantatges dels nodes de la **xarxa** *mesh* també es converteixen en desavantatges en temes de qualitat de servei, seguretat, estabilitat i rendiment. Degut a això, diverses línies d'investigació han intentat millorar aquestes mancances, una de les quals serà la que englobarà a aquest projecte.

## 1.2  MOTIVACIÓ DEL PROJECTE

La motivació bàsica del projecte serà el de desenvolupar una de les línies d'investigació obertes per les **xarxes** *mesh*. En concret, seria la de contribuir a tenir un augment en el **rendiment** d'aquests tipus de xarxes mitjançant l'aplicació de diverses tècniques.

S'ha vist que les xarxes mallades sense fils poden estar constituïdes per un gran nombre de punts d'accés. Les freqüències d'aquest medi seran sempre les mateixes i seran aquelles en les que les diferents versions d'**IEEE 802.11** ens permeten funcionar, essent sempre un espectre limitat.

Per comunicar-se, aquests nodes necessiten tenir la mateixa freqüència sintonitzada per tal de poder realitzar l'intercanvi d'informació. Però els recursos d'una mateixa freqüència a un espai concret són limitats, reduint el rendiment en les comunicacions quan més nodes hi ha sintonitzats a la mateixa freqüència. I això és un dels problemes més greus amb els que ens podem trobar, ja que s'està limitant enormement la capacitat de les estacions que composen la xarxa *mesh*.

Els casos de xarxes amb una topologia lineal compostes per vàries estacions són un bon exemple. Hi ha estudis que ens indiquen quina és la capacitat teòrica màxima assolible [28] en el cas d'utilitzar un sol canal, però a la realitat es veu que després d'incrementar en nombre de nodes la xarxa de topologia lineal, aquesta capacitat assolida quedarà molt enfora de les prestacions indicades a l'estudi, disminuint a l'hora que es van afegint nodes.

Per aquesta causa, aquest projecte estarà destinat en solucionar o millorar aquests problemes derivats de la utilització de freqüències, ideant un conjunt de tècniques que millorin el **rendiment global** de la xarxa *mesh*. Aquestes tècniques consistiran en la utilització de més canals en les comunicacions entre els nodes de la xarxa.

# 1.3  OBJECTIUS DEL PROJECTE

Després d'haver vist quina és la motivació principal del projecte es passarà ara a enumerar quins són els principals **objectius** derivats d'aquest treball, resumint que és el que s'espera aconseguir finalment, i a quines conclusions s'espera arribar amb l'obertura de possibles línies futures d'actuació amb la possibilitat de millorar el conjunt de solucions que es pugui arribar a presentar en aquest treball.

Sabent que l'objectiu fonamental serà el d'incrementar el **rendiment** final d'una xarxa *mesh* composta per un seguit d'estacions que comparteixin la mateixa freqüència, desglossarem aquesta idea principal en altres objectius secundaris que s'esperen assolir en la realització del treball. Tots aquests objectius són els següents:

- Profunditzar en el **concepte** i **funcionament** de les **xarxes** *mesh*, identificant les limitacions en el seu ús en quan a rendiment a l'hora de realitzar transmissions, especialment quan augmenta el nombre d'estacions amb els que s'opera.
- Plantejament d'una **solució global** per augmentar el rendiment de la situació amb l'ús de nodes que presentin més d'una interfície per node i que, per tant, permetin dedicar-les a diferents afers.
- Introducció d'un **canal de senyalització dedicat**, completament separat de les dades sense interferir en la transmissió de dades útils.

- Implementació d'un **protocol multicanal** que ens permeti arribar a totes les estacions del sistema emprant més d'una freqüència per arribar a les destinacions. El protocol haurà de gestionar la informació rebuda de l'entorn i aplicar una tècnica amb la que seleccionar quina serà la millor freqüència per a la transmissió en cadascun dels enllaços, per tal de poder maximitzar el rendiment de la **xarxa** *mesh*.
- Muntatge d'un **prototipus** dissenyat amb les funcionalitats programades als punts anteriors i, després, execució del sistema a diferents escenaris, valorant posteriorment els resultats obtinguts.
- Extreure unes **conclusions** després d'haver calculat una sèrie de resultats en situacions diferents, valorar les millores que s'hagin pogut derivar dels resultats i, finalment, comentar quines són les possibles **futures línies de treball** en les que posteriors investigacions poden centrar el seu treball.

Es pot veure que un dels objectius és, precisament, obtenir una solució pràctica a sobre d'una plataforma *hardware*. D'aquesta manera els resultats obtinguts s'aproximaran molt més a la problemàtica real del tema en qüestió que amb unes simulacions que, tot i ser correctes, mai podran reproduir les condicions reals d'un entorn ni els problemes que sorgeixen al realitzar el muntatge i instal·lació de les propostes dissenyades.

La **metodologia** emprada per arribar als objectius descrits és pot observar a la figura 1.2. Tractarà primer d'una primera fase de **cerca** de documentació i d'altres propostes realitzades. Quan es té recopilada suficient informació es passarà a pensar en les possibles **alternatives de disseny** per arribar als objectius plantejats, passant seguidament a realitzar el **disseny del prototipus** que s'hagi seleccionat com el més factible i amb més possibilitats de què ens produeixi millors resultats. Una vegada dissenyat el prototipus, es fa la **implementació pràctica** del disseny seleccionat amb una de les eines que es té a disposició. Finalment, es realitzen una sèrie de **proves** dels que s'obtenen una sèrie de resultats i que han de passar un procés de **validació**. Si aquests resultats no són els esperats, es modificaran tant l'etapa del disseny com la d'implementació del disseny, depenent de la profunditat del canvi que sigui necessari realitzar. Aquestes revisions es faran tantes vegades com faci falta fins arribar a tenir uns resultats plenament satisfactoris.

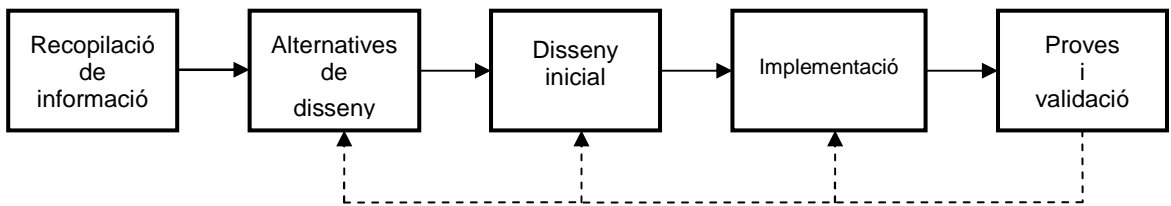

**Figura 1.2: Procés d'implementació**

## 1.4 ESTRUCTURA DEL DOCUMENT

El present document està estructurat seguint la metodologia de treball de la figura 1.2, ja que es segueixen els passos de recopilació d'informació, estudi d'alternatives, disseny, implementació i d'avaluació de resultats. A continuació es mostra com s'ha agrupat la informació als diferents capítols.

- A l'**apartat 2**, es fa una introducció de l'evolució que han patit les xarxes d'àrea local sense fils fins que s'ha arribat a les **xarxes *mesh***. Després s'expliquen els conceptes bàsics de les xarxes mallades sense fils, descrivint quines característiques tenen, quins avantatges i punts crítics poden tenir i quines possibles aplicacions poden tenir.

- El **capítol 3** és una descripció dels diferents **algorismes d'encaminament** compatibles amb les xarxes *mesh* i s'expliquen d'una manera molt especial els protocols OLSR i AODV ja que seran els que es podran fer servir més fàcilment al nostre prototipus.

- Per acabar amb els apartats destinats a la recopilació d'informació, en el **capítol 4** es fa un estudi de les diferents solucions que s'han plantejat anteriorment per poder emprar l'espectre **multicanal** per augmentar rendiment. També s'introdueixen les causes per les que són interessants la introducció d'aquest tipus de solucions, justificant la necessitat d'investigacions com aquesta.

- L'**apartat 5** està dedicat al **disseny** i a la **implementació** de la solució del prototipus. Primerament es fa un plantejament per solucionar el problema, basant-se en alguns treballs que s'havien exposat a l'apartat 4, passant després a veure com algunes de les propostes que s'havien pensat eren irrealitzables. Seguidament, es plasmen les diferents **alternatives de disseny** que han estat proposades en algun moment del procés desenvolupat, fins que s'arriba a la solució definitiva, que és descrita amb detall a la darrera part del capítol.

- Després d'haver explicat la part de disseny i implementació, al **capítol 6** es recullen els escenaris de proves del disseny implementat, les dades que s'han obtingut quan s'han fet proves damunt d'aquests escenaris, acabant amb unes **valoracions** dels resultats calculats.

- L'**apartat 7** correspon al de les **conclusions finals** que s'han extret del desenvolupament del treball, de les fites aconseguides en la realització del projecte, dels resultats obtinguts en l'execució del prototipus i de l'obertura de noves línies de treball que permetin perfeccionar el que s'ha implementat en aquest projecte.

- Finalment, a la darrera part del document es presenten una sèrie de **annexes** on apareixeran aspectes del disseny que no són necessaris per a la comprensió del projecte, però que han estat utilitzats o realitzats en el procés de desenvolupament del projecte i a on la seva aportació pot resultar aclaridora en algunes parts del document. A aquests annexes apareix informació respecte al codi font, on es comentaran la funció que fan els arxius implementats i quins arxius han estat modificats al cas d'emprar codi ja programat. També es recollirà informació damunt les especificacions tècniques dels *meshcube*, el llistat de canals ortogonals, els missatges dels protocols d'encaminament, alguns *scripts* creats i per quina causa es van crear, informació sobre el software emprat per les proves així com d'altres paràmetres que s'han emprat per configurar el prototipus.

# 2 XARXES MESH

## 2.1 INTRODUCCIÓ A LES WLAN

Amb el pas del temps, les xarxes de telecomunicacions han anat evolucionant molt ràpidament tant en concepte com en les prestacions que ens poden arribar a oferir. Inicialment, totes les xarxes de comunicacions d'àmbit local (**LAN**) eren cablades: xarxes de tipus Ethernet, on els equips s'interconnectaven entre ells directament o mitjançant altres elements capaços de commutar dades, com el *switch*, xarxes Token Ring…

Les primeres xarxes cablades tenien una capacitat inicial molt limitada però, a mesura que la tecnologia relacionada amb el món de les telecomunicacions ha anat evolucionant, aquestes xarxes cada vegada han augmentat més la seva capacitat i han possibilitat l'aparició de nous serveis i noves aplicacions.

D'aquesta manera, els usuaris poden accedir a una variada gama de serveis a una velocitat destacable. El problema que ens trobem és que, al ser xarxes cablades, la mobilitat i comoditat de l'usuari no són les millors i, per tant, només podríem accedir a aquests recursos des d'unes localitzacions i situacions molt concretes.

Per tal d'aconseguir millorar aquesta mobilitat i poder tenir una xarxa més flexible, apareixen les xarxes locals sense fils (**WLAN**) que tracten d'aprofitar l'espectre ràdio. El problema principal que presenten aquestes xarxes és el de superar l'hostilitat del medi ràdio degut a les interferències que creen altres aparells, les atenuacions pels obstacles que ens trobem i altres tipus de condicions atmosfèriques o ambientals que poden afectar a les connexions.

Per tractar de superar aquests inconvenients, s'han realitzat una sèrie d'estàndards i protocols per tractar d'aprofitar el millor possible tot l'espectre que disposem. De fet, els organismes reguladors van assignar unes bandes de freqüències a on es podrien realitzar les comunicacions *wifi*. Aquestes es corresponen a les bandes de 2,4 GHz i de 5

GHz. A la taula A.1 de l'annex A.1 es pot veure la distribució general de la banda de freqüències [3] i, ressaltat en negreta, les freqüències que corresponen a les xarxes d'aquest tipus. S'han dut a terme algunes estandarditzacions per alguns dels organismes reguladors (**ETSI**, **IEEE**), però les que són d'ús comú actualment són les que van ser estandarditzades per l'**IEEE**. L'organisme regulador va proposar una sèrie d'estàndards que són els que la indústria de telecomunicacions ha acollit massivament. Aquests estàndards són: **IEEE 802.11a** (que actua a la banda de 5 GHz i ofereix fins a 54 Mbps), **IEEE 802.11b** (que actua a la banda de 2,4 GHz i dóna fins a 11 Mbps) i **IEEE 802.11g** (que actua a la banda de 2,4 GHz i arriba fins a 54 Mbps, gràcies a la tècnica **OFDM**). Apart d'aquestes tres versions, que són les més emprades als dispositius actuals, s'han implementat o s'estan desenvolupant noves versions de la IEEE 802.11 que afegeixen funcionalitats a les versions més utilitzades i que permeten assolir majors prestacions.

Apart d'aquestes versions d'**IEEE 802.11** [4], existeixen unes altres de més innovadores que afegeixen algunes particularitats a les versions ja plenament desenvolupades. Alguns exemples són:

- **IEEE 802.11h:** estàndard a 5 GHz suplementari per complir amb la normativa europea. Agrega característiques com la selecció dinàmica i la potència de transmissió variable.

- **IEEE 802.11c:** versió que especifica mètodes per a la commutació *wifi* per connectar diferents tipus de xarxes mitjançant xarxes sense fils.

- **IEEE 802.11d**: és com un controlador que té en compte les diferències regionals en quant a les diferents freqüències disponibles depenent del lloc en què estem a aquell moment.

- **IEEE 802.11e:** defineix la qualitat de servei dins les comunicacions sense fils. Així, es pode ajustar les xarxes per aplicacions multimèdia i de Voice Over Internet Protocol (VoIP), molt sensibles als retards.

- **IEEE 802.11f:** especificació que funciona baix l'estàndard IEEE 802.11g i que s'aplica a la interconnexió entre punts d'accés de distints fabricants, permetent el *roaming*.

- **IEEE 802.11i:** és el que reuneix les especificacions de seguretat. Inclou autentificació amb Extensible Authentication Protocol (EAP), Remote Authenticated Dial-In User Service (RADIUS) i encriptació amb l'algorisme Advanced Encryption Standard (AES).

- **IEEE 802.11j:** amb canals addicionals per l'espectre de freqüències japonès.

- **IEEE 802.11k:** s'afegeix la funcionalitat d'intercanvi de capacitats entre punts d'accés i clients.
- **IEEE 802.11n:** estàndard amb el que es pretén arribar a 100 Mbps a una xarxa sense fils.
- **IEEE 802.11s:** proposada pel grup Wi-Mesh, defineix la capa física i d'enllaç per a xarxes en malla. La proposta s'enfocarà per múltiples dimensions: la capa MAC, l'encaminament, desenvolupament de la seguretat i interconnexió entre punts d'accés de diferents fabricants, implementant una versió compatible per tots ells.

L'estàndard **IEEE 802.11** ha permès popularitzar les xarxes sense fils per la seva comoditat i facilitat en l'ús. Tanmateix, no s'ha d'oblidar que tot i les millores aconseguides, les xarxes sense fils no arriben al nivell de qualitat en l'intercanvi de dades al que es pot arribar amb una xarxa cablada. Així, dins les **WLAN** podem distingir uns factors positius envers d'uns altres de negatius fruit de la utilització de dites xarxes [5].

Els avantatges de la utilització de les WLAN es resumeixen en:
- **Mobilitat.** Es pot accedir als recursos de la xarxa mentre el node està en moviment o des d'un àmbit diferent a l'habitual (aeroports, cafès amb accés *wifi*...).
- **Flexibilitat.** El no tenir un medi cablejat permetrà configurar els llocs de treball d'una manera molt més flexible sense haver de realitzar grans canvis. També permet que el nombre d'usuaris canviï sense afectar el comportament general de la xarxa.
- **Costos.** Són molt menors que a una xarxa cablada, especialment a l'hora de desplegar la xarxa inicialment. El no haver d'instal·lar medi cablejat permet un estalvi en instal·lacions i en obres viàries.

En quant als desavantatges, s'han de tenir en compte quins són al moment de desplegar la xarxa sense fils, per tal de contrastar si realment estem davant d'una opció vàlida. Alguns d'ells són:
- **Seguretat.** És molt més senzill interceptar una comunicació sense fils que una que transcorre per un medi amb cables. Per tant, es requeriran mecanismes de seguretat específics per aquests casos que no sempre estan plenament desenvolupats.
- **Rang.** L'abast d'una xarxa *wifi* és bastant limitat, especialment en situacions de velocitats elevades, molt més sensibles a les interferències externes.

- **Capacitat.** Encara que amb els nous estàndards el creixement de la capacitat ha estat notable, la capacitat que es pot assolir amb una xarxa sense fils és sensiblement inferior a la capacitat que proporciona una xarxa fixa de tipus Ethernet. A més, el medi aeri és més limitat i els recursos són més escassos, perjudicant als casos on hi ha més connexions simultànies.

- **Fiabilitat.** El medi aeri és hostil per a les comunicacions. Al no haver cap tipus de protecció material (com apantallaments), totes les ones que circulen pel medi poden distorsionar les nostres comunicacions. Qualsevol tipus d'aparell domèstic, obstacle o condicions ambientals adverses poden afectar greument els nostres intents d'accedir als recursos.

Per poder superar aquests desavantatges, s'estan fent nombrosos estudis de diferents configuracions de xarxes sense fils i es treballa en la implementació de nous protocols que permetin mitigar els inconvenients derivats de l'ús de les **WLAN**. I una de les modalitats en què es treballa són les xarxes mallades sense fils (**WMN**).

## 2.2  ENTORN MULTICANAL

Com s'ha vist a l'apartat anterior, l'estàndard **IEEE 802.11** té diferents versions, les quals poden funcionar damunt diferents freqüències depenent de la naturalesa de l'especificació que estem utilitzant. L'**IEEE** defineix dues possibles bandes de freqüències de funcionament: la banda de **2,4 GHz** i la banda de **5 GHz** [6]. Ambdues són bandes de freqüència que ja havien estat assignades a altres usos, però que la seva utilització es pot compatibilitzar amb la utilització de les xarxes WLAN. Així, la banda de **2,4 GHz** coincideix amb la banda ISM, utilitzada per alguns aparells mèdics i d'investigació i que després ha estat utilitzada com de lliure ús, mentre que la banda de **5 GHz** correspon a la banda anomenada UNII per la franja de 5,15 GHz a 5,35 GHz i que, inicialment, estava pensada com a banda de freqüències de lliure ús i una altra banda ISM que es mou a la franja entre 5,725 GHz i 5,875 GHz amb les mateixes aplicacions que l'altra banda ISM. De totes formes, la majoria d'aparells que no són *wifi* emeten en la banda de 2,4 GHz.

A la banda de **2,4 GHz** disposem d'un ample de banda total de 83 MHz (fins a 2,483 GHz). Aquesta banda de freqüències ens permet tenir fins a 14 canals diferents en aquest espectre. Aquests canals tenen una amplària mitjana de 22 MHz, que és, pràcticament, una tercera part de la freqüència disponible. Això provocarà que hagi interferències entre canals veïns degut al solapament de les seves bandes laterals, limitant la quantitat de recursos disponibles per aquesta banda.

En canvi, a la banda de **5 GHz** es disposen de fins a 300 MHz repartits en dues bandes, la que abasta des de 5,15 GHz fins a 5,35 GHz, i 100 MHz més compresos entre 5,75 GHz i 5,85 GHz. En aquest cas, tenim fins a 52 canals dins aquest rang de freqüències que també interfereixen entre ells però que, al ser un espectre de freqüències més ampli, limita en menor nombre la capacitat màxima que podem arribar a assolir, ja que es tindran més canals no solapats a l'espectre.

 A nivell teòric, es pot arribar a tenir velocitats de fins a **54 Mbps** a la banda de **5 GHz**, mentre que la màxima assolible per la banda de **2,4 GHz** és de **11 Mbps**, ja que les freqüències més altes donen lloc a capacitats més elevades encara que pateixen més distorsions amb la distància i cauen més ràpidament que les freqüències més baixes.

Per tant, el que veiem, és que els conjunts *wireless* poden funcionar a dues bandes diferents, que depenent de la banda podem arribar a tenir unes característiques o unes altres que ens poden servir segons l'entorn on la xarxa es pugui conformar i que, tant en un cas com en l'altre, tenim la possibilitat d'escollir diferents freqüències pel seu funcionament o, el que ve a ser el mateix, podrem escollir el millor canal pel funcionament de les nostres connexions; és a dir, estem davant del que es diu un entorn multicanal.

## 2.3  TIPUS DE WLAN

Les xarxes sense fils d'àrea local es poden classificar en dos tipus depenent de la interacció que realitzin entre ells els diferents elements que conformen la xarxa [5].  A l'hora de desplegar la xarxa es tindran en compte alguns factors de la zona que la xarxa ha de cobrir i els elements que disposem per dissenyar un tipus o un altre. Les xarxes **WLAN** les classifiquem en dos tipus: amb **infraestructura** i **ad-hoc**.

### 2.3.1   WLAN amb infraestructura

Aquest és el cas on hi ha dos tipus diferenciats d'estacions que conformen la xarxa: les estacions-client i els punts d'accés. Les estacions són les que proporcionen als clients l'accés als recursos de la xarxa. Són simples transmissors de dades, no tenen cap tipus d'intel·ligència sobre la xarxa i, per tant, no prenen cap tipus de decisió sobre la commutació o encaminament de paquets. En canvi, els punts d'accés són els que donen cobertura a totes les estacions i són les que tenen tota la intel·ligència de la xarxa, encarregant-se de la commutació de paquets, d'accés a altres xarxes, del control dels nodes estació...

Un exemple clàssic d'una **WLAN amb infraestructura** el podem veure a la fig. 2.1.

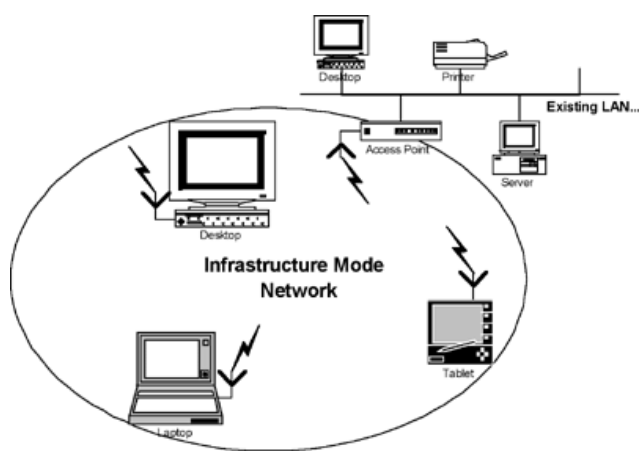

**Figura 2.1: WLAN amb infraestructura**

Els avantatges d'aquests tipus de xarxa és que són molt simples, ja que haurà una estació que concentrarà la intel·ligència de la xarxa sense necessitat que les altres la incorporin. Per contra, és un tipus de xarxa que li manca flexibilitat ja que la presa de decisions només es podrà realitzar des d'aquell punt concret. A més, en cas d'error del punt d'accés, tota la xarxa cauria a darrera ja que és el que proporciona tots els serveis.

### 2.3.2   WLAN ad-hoc

A les xarxes ad-hoc es pretén que totes les estacions tinguin una certa autonomia i que, per tant, estiguin capacitades per prendre decisions respecte a commutació i encaminament de paquets. A les xarxes **ad-hoc** no es requerirà un punt d'accés per donar cobertura, tal com succeïa a les que tenien infraestructura, ja que totes les estacions tenen la facultat de relacionar-se entre elles sense el concurs d'una estació diferent a elles. A la figura 2.2 es pot veure un exemple gràfic del que es comenta.

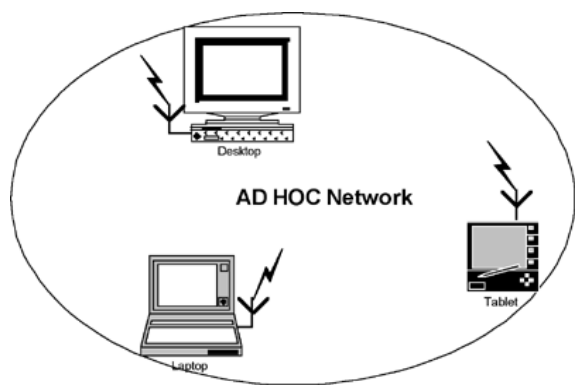

**Figura 2.2: WLAN ad-hoc**

Les xarxes **ad-hoc** presenten una major flexibilitat que les anteriors. Tenen totes les funcionalitats per encaminar la informació cap a una destinació, així com totes les capacitats per associar-se i conformar les seves pròpies xarxes locals, relacionant-se amb altres estacions del mateix tipus. A més poden incloure funcions que suportin la mobilitat dels terminals, podent interactuar mentre la corresponent estació està en moviment. Com a inconvenient, cal ressaltar que tots els elements requeriran eines d'una major complexitat que incorporin totes aquestes funcionalitats.

Com es veurà a continuació, gran part dels nous avenços en temes de xarxes sense fils estan relacionats amb les xarxes ad-hoc, ja que són les que més fàcilment ens poden proporcionar els objectius de connexió on sigui a qualsevol moment.

En aquest aspecte, la **Internet Engineering Task Force (IETF)** ha creat un grup de treball encarregat de desenvolupar noves idees que permetin perfeccionar el funcionament de les xarxes **ad-hoc**. Aquest grup és el nomenat **Mobile Ad-hoc Network (MANET) group** que s'encarrega de crear solucions que permetin encaminar els paquets dins una xarxa ad-hoc amb més d'un *hop*. L'encaminament és difícil de resoldre ja que aquestes xarxes són molt flexibles i hi ha molts de canvis en el nombre d'estacions que conformen la zona de cobertura, així com en les seves posicions, fruit de la seva mobilitat.

## 2.4  XARXES MESH

### 2.4.1  Introducció a les xarxes mesh

Amb el propòsit de crear xarxes que tinguin un abast important es van crear un altre tipus de xarxes, les **xarxes mallades sense fils** o **xarxes *mesh*** (**WMN**) [7]. Com hem vist abans, les xarxes amb infraestructura proporcionen una certa cobertura dins un àrea molt limitada, i una vegada que es surt d'aquesta zona, es perd la connexió. Xarxes d'aquests tipus pot haver en gran nombre a moltes i diverses localitzacions, gairebé sempre de forma aïllada unes de les altres.

Per superar aquest aïllament, es va proposar des d'àmbits militars noves solucions a les ja existents i que permetessin connectar diverses cel·les aïllades d'una forma fiable i guanyant en flexibilitat. D'aquí sorgeix la idea de les **Wireless Mesh Network (WMN)**, xarxes que permetrien aquestes interconnexions. Una vegada que el cost tecnològic ha baixat a causa de què els materials s'han abaratit força i com ha succeït en altres múltiples ocasions, el què és d'àmbit militar passa a ser utilitzat per la societat civil.

Aquesta solució consisteix en crear un sistema de punts d'accés que es comuniquen entre ells mitjançant una xarxa mallada sense fils donant cobertura a una determinada zona. D'aquesta manera, un element que està a dins del rang d'acció d'un d'aquests punts d'accés podrà connectar-se a un altre element que estigui baix el paraigües d'un altre punt d'accés que, al mateix temps, es comunica amb una ruta amb el punt d'accés previ. A la figura 2.3 es pot veure un exemple de **xarxa *mesh***.

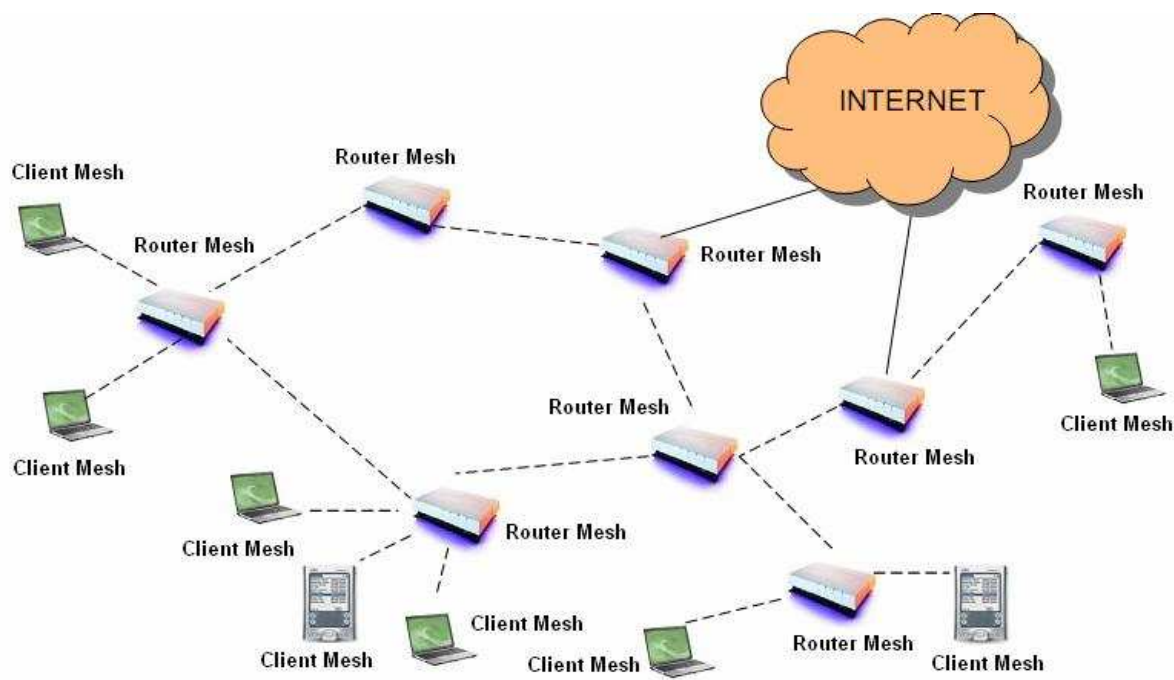

**Figura 2.3: Exemple de xarxa mesh**

A la il·lustració podem veure els diferents element que composen l'arquitectura d'una xarxa mesh: els **clients** *mesh* i els **routers** *mesh*.

- Els **clients mesh** tenen les funcions per poder interactuar amb altres estacions i routers de la mateixa xarxa.
- Els **routers mesh** formen un *backbone* sense fils amb topologia mallada per on es transmeten les comunicacions entre els **clients mesh**. Els routers constitueixen una estructura multihop sense fils i dinàmica que incorporen pel seu funcionament algorismes d'encaminament automàtic i connexions **ad-hoc**. Per tant, incorporen funcions d'enrutament i funcions de gateway cap a altres tipus de xarxes.

## 2.4.2   Arquitectura de les xarxes mesh

Depenent de les funcions que tinguin incorporades els nodes que conformen les **xarxes** *mesh*, podem fer una classificació en tres tipus d'arquitectura de xarxa ben diferenciats [7].

- **WMN amb infraestructura.** Correspon al cas on els **routers _mesh_** conformen un _backbone_ que permet que els clients es connectin amb ells, podent accedir a informació aliena a la seva pròpia xarxa gràcies a aquesta infraestructura. Aquests routers mesh es comuniquen entre ells mitjançant enllaços sense fils i possibiliten la integració de diferents tipus de xarxes en un espai molt més ampli. Per aquest fet, són les WMN més comuns (figura 2.4).

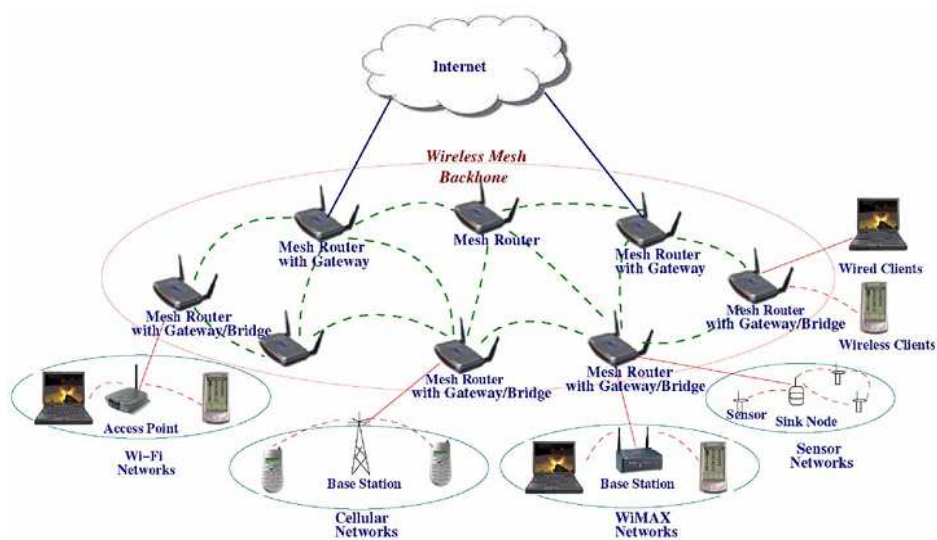

**Figura 2.4: WMN amb infraestructura**

- **WMNs client.** A aquest cas, tots els nodes incorporen totes les funcionalitats de enrutament i de configuració per tal de tenir comunicació punt a punt amb un altre node dels que constitueix la xarxa. Les estacions són autònomes i realitzen el que s'anomena _client meshing_. És una arquitectura amb manca d'infraestructura (figura 2.5) i que estarà molt limitada en espai.

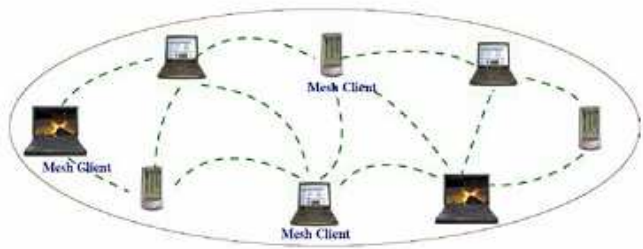

**Figura 2.5: WMN client**

- **WMNs híbrides.** Tracta de combinar el _client meshing_ amb una certa infraestructura dins la mateixa arquitectura. Hi ha un _backbone_ de routers mesh que interactua amb les xarxes de clients mesh i que proporciona connectivitat amb altres tipus de xarxes com Internet (figura 2.6).

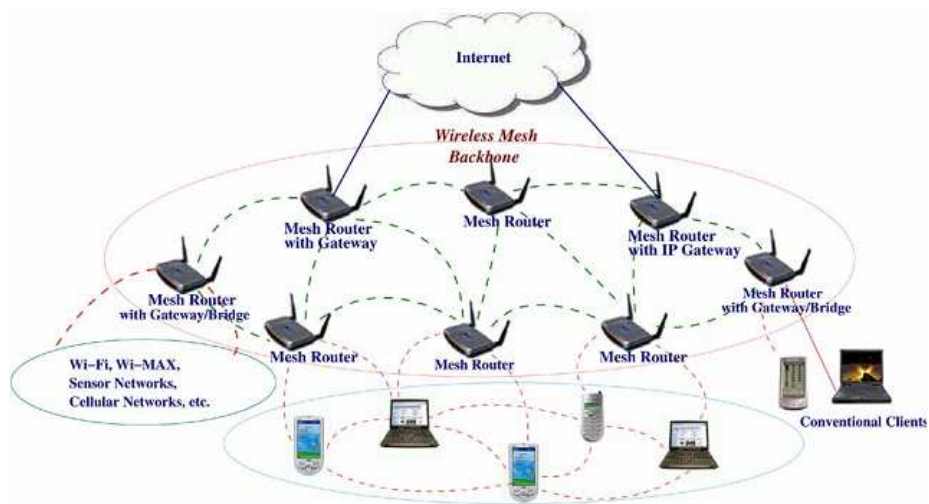

**Figura 2.6: WMN híbrida**

## 2.4.3   Característiques de les WMN

Les WMN presenten una sèrie de propietats destacables que es resumeixen en les següents característiques [7].

- **Estructura multihop sense fils.** Volem que el sistema tingui el major abast possible i per fer-ho, es col·loquen múltiples mesh routers que cobreixin aquelles zones on hagi clients mesh. Els routers es comunicaran mitjançant enllaços sense fils creant la xarxa mallada i poden estar separats uns quants salts de l'estació router *mesh* a la que es vol accedir, sense tenir la necessitat de què hagi línia de visibilitat entre ells.
- **Capacitat de autoconfiguració, autorecuperació i connexions ad-hoc.** Els routers *mesh* estableixen i mantenen la connectivitat entre ells. Per això el cost de desplegament és baix comparat amb altres tecnologies.
- **Múltiples tipus d'accés a la xarxa.** Les xarxes *mesh* suporten l'accés troncal a Internet com les connexions punt a punt. A més, també es suporta la integració de les WMN amb altres tipus de xarxes i l'aprovisionament de serveis als usuaris finals d'aquestes xarxes.
- **Mobilitat depenent el tipus de nodes.** Un dels objectius de les xarxes *mesh* és aconseguir mobilitat dels usuaris finals. Així, mentre els routers *mesh* tindran una mobilitat mínima, els clients *mesh* la poden tenir molt elevada.

- **Potència limitada depenent dels nodes mesh de la xarxa.** Depenent de l'accés a la potència que puguin tenir els nodes de la xarxa mesh, aquest pot ser un factor limitant. S'ha de tenir en compte que estem parlant de nodes que poden ser mòbils i que, per tant, no sempre tindran accés a les fonts d'energia. Per aquest fet, les WMN procuren optimitzar el consum de potència dels seus nodes.

- **Compatibilitat i interoperabilitat amb les tecnologies sense fils actuals.** Les WMN han de ser compatibles amb les diferents tecnologies sense fils existents com la IEEE 802.11, WiMax, xarxes de comunicacions mòbils...

Les característiques són les comunes a les xarxes ad-hoc ja que suporten les funcionalitats d'aquests tipus de xarxes (mobilitat, potència limitada), però a més, comprenen capacitats addicionals a les xarxes ad-hoc (infraestructura de tipus *backbone*, pot haver múltiples ràdios...). Per això es pot dir que les **WMN** formen un grup específic dins les xarxes **ad-hoc**.

## 2.4.4  Avantatges i punts crítics de les WMN

Les WMN tenen els avantatges que proporcionen les xarxes ad-hoc sumant-li els que provenen de les funcionalitats pròpies de les xarxes mesh. Els avantatges es poden concretar de forma resumida als següents punts [8]:

- **Cobertura més àmplia.** La disposició dels routers mesh i el seu mode de comunicació fa que les zones que cobreixen les WMN siguin més àmplies que amb altres xarxes sense fils.

- **Eficiència espectral excel·lent i major capacitat.** Les diferents àrees de cobertura que abraça la WMN poden reutilitzar els mateixos canals de freqüència i, per tant, resultar lliures d'interferències, permeten un augment de la eficiència espectral i, conseqüentment, de la capacitat total del sistema.

- **Integració de diferents xarxes.** Permet accedir a diferents tipus de xarxes de forma transparent gràcies a l'estructura de routers mesh, que incorporen les funcionalitats per compatibilitzar les diferents tecnologies de xarxa.

- **Costos de desplegament baixos.** Dins l'estructura mallada de la WMN no hi ha estacions base, només nodes a una localització molt concreta que són els que conformen la malla. A més, al ser un sistema sense fils, el cost de desplegar ja és menys costós que amb xarxes d'una altra natura.

- **Complexitat baixa.** Els routers *mesh* seran d'una menor complexitat que altres tipus d'elements com estacions base.

- **Flexibilitat i suporta mobilitat.** Com totes les estructures sense fils, les WMN seran flexibles ja que els canvis de localització o de nombre d'àrees de cobertura es podran dur a terme sense grans problemes. A més es suporta la mobilitat de les estacions client que actuen dins les zones de cobertura.

Les WMN podran proporcionar aquests avantatges sempre que es superin una sèrie de punts crítics [7] que es donen a l'hora de dissenyar una xarxa d'aquest tipus. Aquests punts crítics a tenir en compte són:

- **Tècniques de ràdio.** Per incrementar la capacitat i flexibilitat de la xarxa s'han fet servir algunes tècniques d'aquest tipus. Una de les que es fa servir és la d'emprar més d'una interfície ràdio per la comunicació. Altres tècniques es refereixen a les antenes direccionals, a les antenes intel·ligents o a l'aprofitament de l'entorn multicanal.

- **Escalabilitat.** A l'augmentar la grandària de les xarxes, la capacitat total per node s'anirà degradant progressivament tal i com augmenti el nombre d'estacions. Per això, els protocols s'hauran de fer escalables per mantenir les prestacions de la xarxa encara que el nombre de nodes augmenti.

- **Connectivitat *mesh*.** Molts dels avantatges que ens donen les WMN són deguts a la connectivitat de la malla. Per mantenir això es necessiten mecanismes de control de la topologia i d'autoconfiguració de la xarxa.

- **Facilitat en l'ús.** Els protocols s'han de dissenyar de tal manera que la xarxa sigui el més autònoma possible. Per això s'haurien de tenir protocols de gestió de xarxa eficients que permetin configurar els paràmetres de la WMN.

- **Compatibilitat i interoperabilitat.** Per donar servei al major nombre de clients possible s'han de tenir en compte la gran quantitat de xarxes diferents que podem arribar a tenir. Els routers *mesh* han de ser capaços d'integrar xarxes heterogènies d'una forma transparent.

- **Banda ampla i QoS.** S'haurà de donar servei a xarxes ben diferenciades i heterogènies i que, per tant, tindran requeriments de QoS ben diferents.

- **Seguretat.** Encara que han estat proposades moltes solucions de seguretat per WLAN, aquestes no són totalment aplicables a les WMN. La dificultat resideix en què no pot haver una autoritat central de seguretat ja que estem parlant d'un sistema totalment distribuit.

## 2.4.5 Aplicacions de les WMN

Els avenços que les WMN proporcionen poden ser aprofitats en molts d'àmbits diversos. La flexibilitat de la xarxa, la zona de cobertura, la mobilitat i altres factors poden ser interessants per moltes situacions. Algunes de les aplicacions on WMN poden ser de gran ajuda són les que es relaten a continuació [7]:

- **Ús domèstic.** La WMN té una zona de cobertura que abraça tota la casa. D'aquesta manera es podrà connectar a la xarxa tots els diferents components repartits per la vivenda, permetent que es puguin controlar de forma centralitzada.
- **Comunitats sense fils.** Per donar accés a una comunitat de veïns, les WMN poden ser una bona opció. Cadascun dels veïns tindrà accés a un router *mesh*, que el permetrà accedir als recursos de la xarxa global.

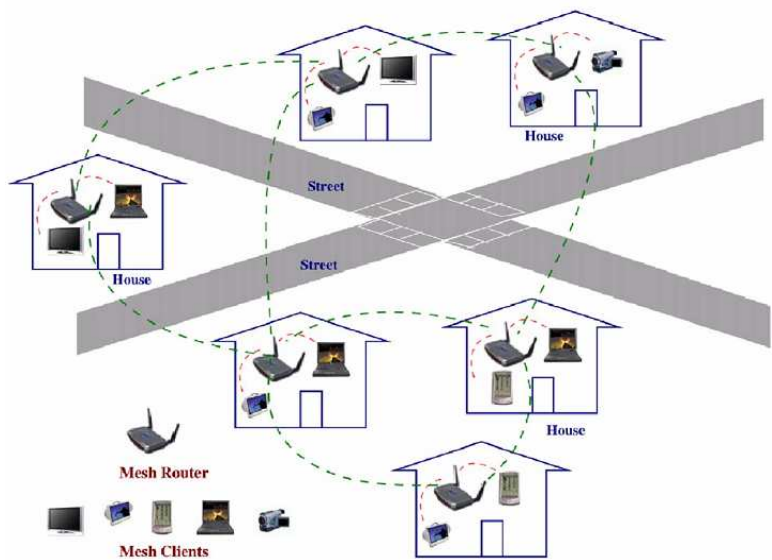

**Figura 2.7: Comunitat sense fils**

- **Xarxes corporatives.** Pot ser útil per empreses que ocupen una oficina, tot un edifici o, fins i tot, per un conjunt d'edificis contigus ja que abaratiria els costos de la

instal·lació d'una xarxa cablada de tipus Ethernet. També seria aplicable als casos d'hospitals, campus universitaris, edificis d'administracions públiques...

- **Xarxes d'àrea metropolitana.** Les WMN poden ser de gran ajuda en xarxes d'una major extensió que les anteriors que s'han exposat, connectant diverses comunitats, empreses, cases...

- **Sistemes de transport.** També poden ser de gran utilitat per comunicar i donar servei als sistemes de transport massiu com els autobusos, els trens o el metro. A cadascun d'aquests vehicles hauria un dispositiu que donaria cobertura i que es comunicaria amb els altres router *mesh*, tenint accés a les altres xarxes que proporcionen informació. Fins i tot es poden fer servir pels serveis d'emergències, ja sigui per una flota de bombers, de automòbils de policia o per vehicles militars.

- **Edificis "intel·ligents".** La WMN pot servir per controlar els diferents automatismes dels edificis tal com són la llum dels corredors, els ascensors, l'aire condicionat i altres mecanismes que poden ser monitoritzats i controlats.

- **Xarxes d'emergència.** Pot haver situacions inesperades com podrien ser fenòmens metereològics adversos (inundacions, huracans...), fenòmens naturals destructius (terratrèmols, volcans, *tsunamis*...) o altres tipus de catàstrofes (incendis, accidents, atemptats...) que requereixen d'una gestió ràpida que una xarxa *mesh* pot proporcionar degut a l'autoconfiguració dels nodes que conformen la xarxa.

- **Xarxes temporals.** Per cobrir les necessitats d'events temporals com fires, events esportius o congressos, les xarxes *mesh* són una opció còmoda ja que els routers *mesh* podrien conformar una xarxa independentment de la seva localització en la que, després, els assistents podrien accedir als recursos oferts i es podrien comunicar amb les estacions connectades a aquesta xarxa.

Un clar exemple d'una **xarxa *mesh*** que inclou una flota de vehicles és el que es veu a la il·lustració (figura 2.8). Els edificis i les estructures tenen els routers mesh que formen el *backbone* de la xarxa, mentre que les flotes de vehicles (ambulàncies, cotxes de policies, bombers...) són els clients mesh que accedeixen al **router *mesh*** que li dóna cobertura en aquell moment i que li permet comunicar-se amb les seves institucions matrius (hospitals seus d'emergències). D'aquesta forma s'aconsegueix flexibilitat i major rapidesa als serveis d'emergència.

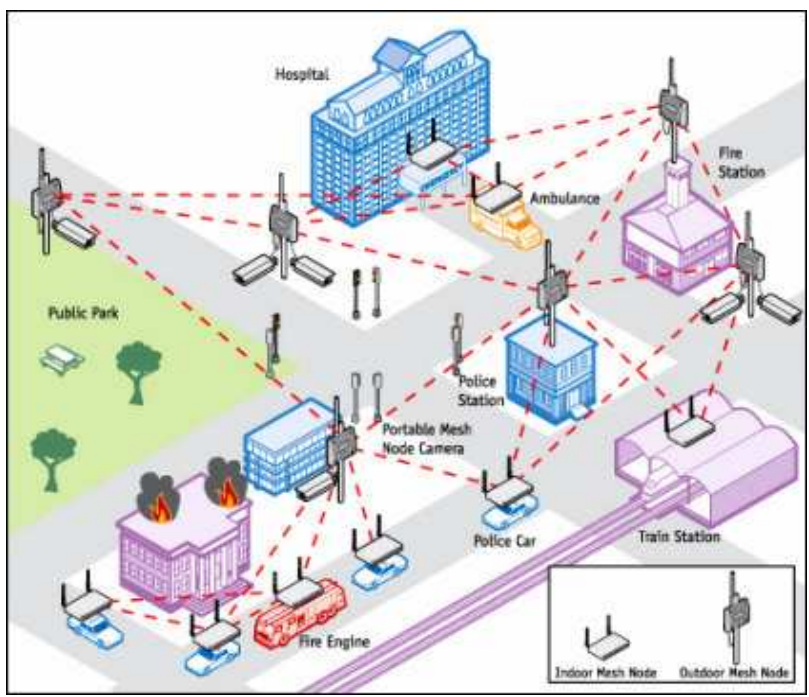

**Figura 2.8: Aplicacions WMN**

# 3 PROTOCOLS D'ENCAMINAMENT

A l'apartat anterior, mentre s'exposaven les particularitats d'una **Wireless Mesh Network (WMN)**, s'ha ressaltat com a punt crític d'aquests tipus de xarxes els protocols d'encaminament que s'havien d'utilitzar per tal de configurar les rutes cap a les destinacions. S'ha de tenir en compte que els nodes que conformen la **WMN**, moltes vegades, no són estacions amb grans bateries ni amb gran capacitat de càlcul. A més, molts d'aquests nodes exploten la seva mobilitat per aprofitar les seves qualitats i, per aquest fet, s'hauran d'utilitzar protocols d'encaminament específics per xarxes sense fils d'aquest tipus, però emprant la mínima quantitat de prestacions possible.

Per això, organismes com l'**IETF**, a instàncies del **MANET** group, han proposat per a la seva implementació una sèrie de protocols més adients per a xarxes **ad-hoc**. S'han proposat protocols d'encaminament diferenciats, que aprofiten diverses tècniques pel descobriment de rutes. Tenint en compte els tipus de tècniques, es pot fer una classificació dels protocols en diferents famílies, que tindran en comú gran part del seu funcionament. Les famílies de protocols d'encaminament per xarxes **ad-hoc** més importants són: els protocols **proactius**, els protocols **reactius**, els protocols **híbrids**, els protocols **basats en qualitat de servei (QoS)** i altres tipus de protocols com són els basats en el consum de potència.

Com s'ha dit anteriorment, les **WMN** són un subconjunt dins les xarxes **ad-hoc** i, per tant, incorporaran un d'aquests protocols per configurar les rutes més òptimes que les rutines hagin calculat.

Després d'haver presentat els diferents tipus d'algorismes d'encaminament i haver comentat les seves característiques clau, s'entrarà més en detall en els dos algorismes d'encaminament més estesos i que són compatibles amb el sistema que es dissenyarà i amb el material que s'utilitzarà. Els protocols són el **Ad-hoc On-Demand Distance Vector (AODV)** i el **Optimized Link State Routing (OLSR)**.

## 3.1  TIPUS D'ALGORISMES D'ENCAMINAMENT

Els diferents tipus d'algorismes d'encaminament es basen en distints conceptes a l'hora de buscar les rutes òptimes cap a tots els nodes. A continuació es farà un repàs de les tècniques més utilitzades.

### 3.1.1  Protocols proactius

Els protocols **proactius** [9] són aquells que han de mantenir una taula actualitzada de les rutes cap a les altres destinacions a tots els nodes de la xarxa. La informació continguda a les taules dels nodes ens indicarà quin és el millor camí en termes de rendiment per arribar a la destinació.

Quan una estació ha d'enviar una informació cap a l'altre extrem de la comunicació, primer es consulta la direcció destí a la taula d'encaminament. Si la direcció destí es troba a la taula s'envien les dades pel camí que s'indica. En cas contrari, el node ha d'activar el procés de descobriment de veïns. Això es pot fer, enviant un missatge HELLO cap a totes les estacions veïnes que, al mateix temps, tornaran a fer un altre *broadcast* fins a arribar a la direcció destí desitjada. Quan l'estació a la que va dirigit el missatge rep el HELLO, aquesta contesta de forma *unicast* cap a l'origen de la comunicació. Al rebre la confirmació del HELLO, el node origen actualitza la seva pròpia taula. En molts de protocols, els nodes intermedis poden agafar la informació que circula per la xarxa per anar configurant les seves pròpies taules.

Un altre aspecte important d'aquests protocols és l'actualització de les taules dels nodes. Es sap que les xarxes que estem tractant poden canviar molt ràpidament, tant en nombre de nodes com en les posicions dels respectius nodes. Per això, serà necessari que hagi una actualització periòdica de les taules cada un cert període de temps. El procés que es realitza és similar al que es fa per descobrir nous nodes de la xarxa. Si hi ha canvis als nodes, la informació s'expandeix per tots els nodes que conformen la xarxa. D'aquesta forma tots els nodes s'assabenten dels canvis que hagi pogut haver a la topologia de la xarxa.

També es pot donar el cas de què una entrada de la taula ja no existeixi com a tal. En aquest cas, quan es fa l'actualització, s'eliminaria l'entrada de la taula d'encaminament.

Hi ha uns quants exemples de protocols **proactius** molt significatius: l'**Optimized Link State Routing (OLSR)** [10], el **Destination-Sequenced Distance Vector (DSDV)** [11], el **Wireless Routing Prootocol (WRP)** [12] i el **Topology Dissemination Based on Reverse Path Forwarding (TBRPF)** [13]. Tots tenen el mateix comportament descrit anteriorment, però cadascun d'ells té les seves peculiaritats. L'**OLSR** basa el seu comportament en els **MPR**, el **DSDV** fa un càlcul de les rutes seguint l'algorisme Bellman-Ford, al **WRP** cada node manté quatre taules diferent per a la comunicació (taula de distàncies, taula d'enrutament, taula de cost de cada enllaç i taula de retransmissió de missatges i al **TBRPF** hi ha una sèrie de canvis als missatges d'actualització respecte als HELLO.

Els **avantatges** que proporcionen els protocols **proactius** vénen en gran part de la conservació de les rutes durant un espai de temps relativament llarg. Al tenir actualizacions periòdiques, a la majoria dels casos el node destí apareixerà a la taula del node. Això voldrà dir que la informació s'enviarà amb l'únic retard d'haver de consultar a la taula d'encaminament. Un altre avantatge és el d'actualització de les taules, que permet tenir coneixement de la topologia de la xarxa.

A la banda contrària destacarien com a **desavantatges** la càrrega que suporta la xarxa degut als HELLO i a la quantitat d'informació que tenen les taules. A més, pot haver un lleuger malbaratament de recursos degut a rutes que encara apareixen i que ja no siguin vàlides i que, encara, no han estat actualitzades.

Els protocols **proactius** són més efectius per xarxes amb usuaris amb menys mobilitat i que requereixin enviar una elevada quantitat d'informació.

## 3.1.2  Protocols reactius

Els protocols **reactius** [9] serien el cas contrari als protocols **proactius**. Aquests protocols no mantenen cap tipus de taula als nodes de la xarxa. De fet es qualifiquen habitualment com protocols d'encaminament **sota demanda**. Això vol dir que, només es crearà una ruta quan hagi demanda d'enviar informació a aquell moment, és a dir, l'arbre de rutes només es configurarà quan es necessiti per transmetre cap a la destinació.

Aquesta filosofia implicarà que sempre que es vulgui transmetre s'haurà de fer un descobriment dels veïns ja que no es tindrà emmagatzemada cap informació sobre les destinacions. El descobriment es farà de manera similar a com es feia als protocols **proactius**, és a dir, enviant un *broadcast* cap al destí i, quan el destí rep el missatge de petició enviat, aquest respon amb un missatge *unicast* que confirma la ruta que s'ha d'emprar.

El manteniment de les rutes es realitza d'una manera molt distinta. Aquí, les rutes només es mantenen mentre el node emissor tingui dades per transmetre. Una vegada el node deixa de transmetre, la ruta és eliminada per aquest mateix node. En aquest cas no es guarda cap tipus de taula amb les rutes a utilitzar per accedir als nodes de la xarxa. De totes formes, encara que no es guarden les taules, les dades queden durant un curt periode de temps a la caché dels nodes.

A l'igual que amb els proactius, han sortit diversos protocols **reactius** que presenten peculiaritats dins el seu funcionament com a família. Alguns d'ells són: l'**Ad-hoc On-Demand Distance Vector (AODV)** [22], el **Dynamic Source Routing (DSR)** [14] o el **Dynamic MANET On-Demand (DYMO)** [15]. Una de les poques diferències que presenten **AODV** i **DSR**, és que **DSR** inclou tot el camí a la capçalera dels paquets, mentre que **AODV** només inclou el següent salt. **DYMO** és un protocol gairebé experimental l'objectiu del qual és malbaratar el menor nombre de recursos possible.

Els protocols **reactius** volen un sistema molt senzill i que sobrecarregui poc el sistema. Els protocols només gastaran aquells recursos que necessitin a cada moment. A més es procura que els nodes utilitzin la menor quantitat de recursos possible amb l'eliminació de les taules als nodes.

Per contra, cada vegada que s'han de transmetre dades s'haurà de fer un nou descobriment de veïns i crear les rutes corresponents. Això, inevitablement, crea una latència més elevada que al cas dels protocols proactius.

La utilització dels protocols **reactius** sembla més adient per xarxes amb poc tràfic de xaxa, per xarxes on hagi una gran mobilitat i on, per tant, la topologia es modifiqui contínuament o per xarxes on els nodes tinguin una capacitat de processament mínima.

### 3.1.3  Protocols híbrids

Entre les solucions proactives i reactives que s'han exposat, s'ha proposat un altre tipus de solució intermèdia que recull propietats de les dues tendències. El protocol **híbrid** més reconegut és el **Zone Routing Protocol (ZRP)** [16] i mescla el comportament **proactiu** i **reactiu**.

El que fa aquest protocol és funcionar de manera proactiva dins un determinat radi R, que ve a ser la distància que hi ha entre el node que inicia la comunicació i l'extrem de la topologia (que estarà a R salts de l'origen), mentre que actuarà de forma reactiva en els nodes que queden fora d'aquest radi. Aquesta suposició es realitza perquè s'entén que els nodes pròxims tindran més tràfic i canviaran menys que els nodes externs.

Amb aquests protocols guanyem en robustesa i es pot guanyar en rendiment, sempre que la xarxa es comporti enviant més tràfic als nodes interiors que als exteriors. Al mateix temps, és una família de protocols complexa i que pot presentar un rendiment molt pobre quan hi ha canvis de topologia als nodes interiors o quan s'envia més tràfic als nodes fora el radi que als interiors.

### 3.1.4  Protocols basats en QoS

Les futures xarxes sense fils transportaran diverses aplicacions multimèdia de veu, vídeo i dades. Aquestes aplicacions són molt sensibles als retards variables ja que funcionen en temps real i requeriran uns termes d'ample de banda i de retard determinats. Per aquesta raó, faran falta que els protocols d'encaminament també tinguin en compte la **qualitat de servei (QoS)** a l'hora de crear els arbres de rutes.

La provisió de **QoS** en aquestes xarxes és una tasca complicada ja que, a més de la **QoS** a proporcionar, s'ha de tenir amb compte els canvis de topologia propis dels nodes sense fils i la compartició que es realitza del medi.

Per tant, el concepte dels protocols canvia una mica ja que ara haurem de suportar la **QoS** als nodes trobant una ruta que satisfaci els requeriments de les aplicacions. Dos dels protocols que s'han proposat per dur a terme aquesta idea de proveir **QoS** a aquestes xarxes són: el **Quality of Service aware Routing Based on Bandwidth Estimation Routing Protocol (QoSRBonBERP)** [18] i el **Quality of Service aware Source-Initiated Routing Protocol (QoSSIRP)** [17].

El primer d'ells és semblant en funcionament a l'**AODV**, però afegint als seus missatges HELLO uns camps d'ample de banda consumit i una marca temporal per discriminar el moment en què s'ha realitzat la mesura (quan més recent més vàlid).

L'aspecte clau del protocol és el càlcul de l'**ample de banda consumit**, que és el factor que proporciona la qualitat de servei al protocol. Per a l'execució d'aplicacions multimèdia necessitarem uns millors paràmetres en quant a ample de banda. Pel càlcul d'aquest ample de banda, l'estimació es realitzarà calculant l'ús d'ample de banda que fa el propi node i enviant la seva dada cap a la destinació. Aquest procés es pot generalitzar a tots els veïns. Quan el node destí rep el HELLO, determina l'ample de banda disponible tenint en compte els amples de banda dels seus veïns. L'ample de banda residual, serà extret d'aquell que sigui el *coll de botella* a aquella ruta.

El segon dels protocols afegeix **control de qualitat** a les diferents fases del protocol d'encaminament: descobriment de rutes, selecció de rutes i manteniment de rutes. També és un protocol sota demanda, com l'anterior.

El descobriment de les rutes es realitza enviant missatges específics QRREQ que portaran una sèrie de mètriques que seran les utilitzades per enviar la informació sobre el node a totes les destinacions. Aquestes mètriques són la latència, l'ample de banda disponible, el nivell de senyal i la potència de la bateria del element. La destinació, quan rep el QRREQ, envia un QRREP *unicast* cap a l'origen.

L'elecció de les rutes es farà segons el criteri de mètrica que es vulgui emprar. Aquell que tingui un valor més ajustat a les necessitats de l'aplicació serà el que es seleccionarà. En cas de no haver cap paràmetre prioritari, es donarà major importància a la potència de les bateries.

Aquests protocols poden ser més utilitzats en un futur quan ja estigui més consolidat el tema de la **QoS** a les **WMN** i a les xarxes **ad-hoc**, en general.

### 3.1.5 Altres tipus de protocols d'encaminament

Apart dels protocols ja comentats, hi ha un grapat més que també es poden utilitzar i que utilitzen tècniques diferents a les mencionades. Alguns d'aquestes famílies de protocols són:

- **Geogràfics.** Són protocols que configuren el seu arbre de rutes depenent de la informació geogràfica de l'entorn. Un exemple de protocol geogràfic és el **Geographic Circuit Routing Protocol (GCRP)** [19].
- **De potència.** Tenen en compte com a factor limitant la bateria disponible de cadascun dels nodes. Amb aquesta informació es configura la mètrica que servirà per aconseguir les rutes òptimes [20].
- **Multicast.** Els protocols multicast que s'utilitzen a les xarxes cablades no valen per la fragilitat dels arbres de rutes a les xarxes ad-hoc, per això s'han creat específics per aquestes xarxes com: **Ad-hoc Multicast Route (AMRoute)** [21], **On Demand Multicast Routing Protocol (ODMRP)** [21] i **Ad-hoc Multicast Routing protocol using Increasing id-numberS (AMRIS)** [21].

## 3.2 AD-HOC ON-DEMAND DISTANCE VECTOR (AODV)

Com hem vist, **AODV** [22] és un protocol **reactiu** i recull totes les propietats que caracteritzen a un protocol d'aquest tipus, és a dir, crea rutes només quan s'ha de transmetre amb l'avantatge de què no inunda de missatges de control el medi. Per contra, l'haver de crear noves rutes cada vegada que es transmet implica una latència més elevada que amb els proactius, excepte en els casos que s'hagin de fer poques transmissions que al tenir menys *overhead* envia més ràpidament. Tots els detalls de l'**AODV** es poden veure al **RFC3561** [22] i va ser elaborat per Charles E. Perkins [23] de la University of California Santa Barbara (UCSB) [24]. A continuació es resumiran els detalls de funcionament del protocol.

## 3.2.1  Creació de les rutes

Cada vegada que un dels nodes vulgui iniciar una transmissió, el node que té implantat l'**AODV**, haurà de realitzar el procés de descobriment de rutes per configurar les rutes òptimes.

El procés de descobriment s'inicia amb la transmissió per part del node emissor d'un missatge Route Request (RREQ) (figura A.1) cap a tots els seus nodes veïns. Aquest missatge RREQ du la direcció IP de la destinació a la que va dirigit el missatge, la direcció IP d'origen, els nombres de seqüència de l'origen i del destí, el compte del número de salts i una sèrie de flags, com el de destinació que diu que només contesti a la petició el destí del missatge. El número de salts és limitat i és el que marca l'extensió que pot arribar a tenir la xarxa.

Quan els nodes veïns reben el RREQ, primer determinen si la petició va destinada a ells. Si no és així, tornen a fer un broadcast del RREQ al seus propis veïns, i així es repetirà el procés successivament fins que arribi al destí desitjat pel creador del missatge RREQ. A **AODV** els nodes intermedis juguen un paper molt important ja que es suposa que tenen una certa capacitat de processament i que guarden certa informació sobre el node que els ha enviat aquest RREQ. Com que els missatges RREQ no guarden totes les direccions dels nodes per les que els paquets van passant, els mateixos nodes van canviant el camp de l'adreça d'origen pel seu, de manera que així, el següent node pugui tenir emmagatzemada l'adreça del que li ha enviat el RREQ.

Cada node també guarda el nombre de seqüència que li arriba del RREQ. Quan es rep un missatge i es veu que el nombre de seqüència és menor o igual que el que surt a l'entrada del node, el missatge serà descartat. Mentre que, si és major, es copiarà el nombre de seqüència i es seguirà amb el procés normal de l'algorisme.

Al rebre el RREQ el node destinatari, aquest genera un Route Reply (RREP) (figura A.2) que s'enviarà de forma unicast cap a l'origen. Com que tots els nodes intermedis han guardat l'adreça originària dels missatges, el camp de destí del RREP s'anirà omplint amb l'adreça de la taula del respectiu node. Així el missatge anirà avançant de forma inversa cap a l'origen per la ruta que més ràpidament havia arribat al destí i, una vegada arriba al iniciador de la transmissió, aquella ruta queda configurada com l'òptima per comunicar ambdós extrems. Els RREP també tenen un número de seqüència que permeten visionar si els missatges són recents o si, en canvi, són antics. Si tinguessin un número de seqüència inferior al que té a la taula el node corresponent, el RREP seria descartat immediatament.

## 3.2.2 Manteniment de les rutes

Seguint la tendència dels protocols **reactius** en general, el manteniment de les rutes només es manté mentre es necessiti, és a dir, la ruta existirà mentre hagi transmissions en ella. Si una ruta, creada per a la transmissió d'una determinada informació, deixa de ser utilitzada, el sistema optarà per eliminar aquesta ruta, alliberant els conseqüents recursos. L'alliberació de la ruta es durà a terme després d'esgotar un cert interval de temps que és determinat pel mateix **AODV**. És una forma de tenir un major aprofitament dels recursos disponibles. Aquesta eliminació de la ruta es realitzarà encara que no hagi canvis de topologia a la xarxa.

Un altre tipus de missatges que contribueix al manteniment de les rutes són els missatges Route Error (RERR) (figura A.3). Els nodes que conformen la ruta activa, van monitoritzant el que succeeix per la xarxa amb la fi de mantenir aquestes rutes i detectar possibles pèrdues de paquets. La generació d'un missatge de tipus RERR pot ser deguda a una pèrdua de paquets al transmetre'ls cap al següent *hop*, quan es rep un paquet amb una destinació desconeguda pel node o quan es rep un RERR per part d'un veí d'alguna de les rutes que són actives. Quan s'envia el RERR, es fa amb el propòsit de què tots els nodes estiguin assabentats dels enllaços que han sofert problemes i que, per tant, ja no són accessibles des de la ruta que abans havia estat activa. El RERR durà indicat a un dels seus camps a quines destinacions ja no es pot accedir.

## 3.3   OPTIMIZED LINK STATE ROUTING (OLSR)

L'**OLSR** [10] és un protocol proactiu que va ser proposat per Clausen i Jacquet. Com és norma als protocols proactius, **OLSR** és un algorisme de encaminament basat en les taules i en la topologia de la xarxa. Funcionarà millor amb xarxes que generen i transmeten una quantitat de tràfic destacable.

D'**OLSR** s'han fet algunes implementacions i la que es detallarà aquí és una de les que té millors prestacions i opcions, a més de ser compatible amb el hardware que s'utilitzarà per la realització del projecte. Aquesta és la implementació anomenada *unik* [25] on A. Tonnesen ha desenvolupat el **OLSR Daemon (OLSRD)**. El seu treball es pot veure reflectit a la seva *Master Thesis* [26]. Als següents apartats es detallarà el funcionament bàsic d'OLSR.

### 3.3.1   Implementació de l'OLSRD

La implementació d'**OLSRD** es pot trobar a la web [25]. El codi font està escrit en C i es pot utilitzar a gairebé a qualsevol plataforma compatible amb Linux, Windows o MacOS. Les característiques principals de la implementació són les que es relaten a continuació:

- **Modularitat.** El codi ha estat implementat de la forma més modular possible. Això ha facilitat que es puguin realitzar canvis sobre certs aspectes, realitzant modificacions directes sobre una part del codi sense afectar a la resta del codi. Després es comentarà una funcionalitat que incorpora **OLSRD**, els *plugins*. Són mòduls afegits al programa general i que li afegeixen una funcionalitat a aquest.
- **Estructures de dades consistents.** S'utilitzen els mateixos tipus de dades a totes les estructures, possibilitant la consistència de la informació que s'està manejant.
- **Transparent a versions IP.** El programa suporta que el sistema funcioni tant amb adreces IP versió 4 de 32 bits com amb adreces IP versió 6 de 128 bits. Fins i tot es suporta els sistemes que comparteixen adreces IP de les dues versions.
- **Codi estructurat.** S'ha procurat fer un codi el més estructurat i llegible possible, dins la complexitat que té implícita l'eina dissenyada.
- **Plataforma independent del codi.** La part del codi que depengui de la plataforma ha d'anar separada de la resta del codi, aplicant el disseny modular que s'ha comentat al primer punt. D'aquesta manera, les implementacions per altres plataformes es simplifiquen degut a que queden completament identificades.

A la figura 3.1 es poden veure quines són les entitats bàsiques que conformen OLSR.

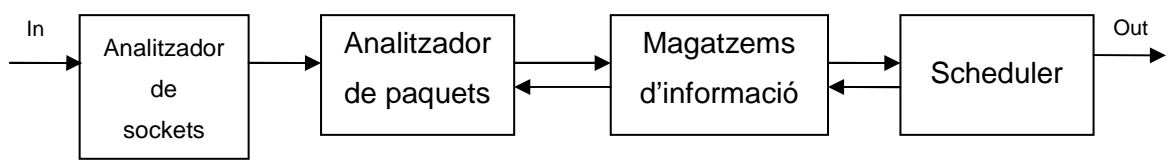

**Figura 3.1: Diagrama de blocs d'OLSR**

- **Analitzador de sockets.** L'analitzador revisa tota la informació que li entra i depenent de l'entrada que tingui crida a una funció de socket relacionada amb aquest tràfic entrant.

- **Analitzador de paquets.** Aquest element rep tot el tràfic OLSR entrant i pot tractar el tràfic de tres maneres diferents: descartant el paquet si el que rebem és invàlid, fer un determinat processament dels paquets o, simplement, passar-lo cap a altres entitats.

- **Magatzems d'informació.** Aquí es guarden totes les taules d'informació que el protocol utilitza pel seu funcionament. Es guarda tota la informació de l'estat de la xarxa en aquell moment i que el tràfic OLSR pot consultar segons les necessitats que tingui. Alguns exemples de magatzems són els següents:

  o *Informació de múltiples interfícies*: Recull un conjunt de nodes que funcionen amb més d'una interfície.

  o *Estat dels enllaços*: Taula mantinguda per calcular l'estat dels enllaços caps els veïns.

  o *Conjunt de veïns*: Es registren tots els veïns que estan a només un salt del node originari. Aquesta taula s'actualitza dinàmicament amb la informació de l'estat dels enllaços.

  o *Conjunt de veïns a 2 salts*: Conté els nodes que estan a 2 salts del node local.

  o *Taula de MPR*: Apareixen els MPR (s'expliquen al subapartat 3.3.2.4) de la xarxa actual.

  o *Taula de MPR seleccionat*: Apareixen tots els nodes que tenen un Multi Point Relay (MPR) determinat seleccionat.

  o *Informació de la topologia*: Conté la informació sobre l'estat dels enllaços a tot el domini de l'encaminament.

o   *Taula de duplicats*: Té informació de missatges processat i enviats recentment.

- **Scheduler.** És el que du el control dels intervals de temps del programa. Així, el *scheduler* ens indica quan s'ha de transmetre segons el tipus de missatges de control o com s'ha de conformar el tràfic de sortida.

## 3.3.2  Funcionament del nucli d'OLSRD

En aquest apartat es descriu quines són les funcionalitats bàsiques de **OLSRD** i com es realitza el procés de creació i manteniment de rutes.

### 3.3.2.1  Missatges d'OLSRD

Abans de començar a descriure el procés en sí, presentarem el format de missatges que els nodes s'intercanviaran a aquesta implementació. Els missatges **OLSR** s'envien inserits dins del camp de missatge d'un paquet UDP i es transmeten pel port 698. El missatge **OLSR** genèric que s'utilitza és el que es veu a la figura A.4 i la informació damunt els camps de la figura es pot trobar a l'annex A.3.

Es tenen tres tipus de missatges definits que tindran unes particularitat sobre el missatge OLSR genèric. Els diferents missatges possibles són:

- **Missatge HELLO.** Són transmesos a tots els veïns i s'utilitzen per realitzar el sondeig d'enllaços i per fer el càlcul dels MPR.
- **Missatge TC.** Són els missatges de control de topologia, que s'encarreguen de monitoritzar l'estat dels enllaços.
- **Missatges MID.** Aquests missatges es transmeten quan els nodes tenen més d'una interfície. Els missatges llisten totes les adreces IP que tinguin els nodes.

### 3.3.2.2  Sondeig dels enllaços

Abans de començar a realitzar qualsevol tipus d'acció, es consulta la taula d'enllaços del node. Per saber quins són els enllaços actius per a la comunicació s'envien missatges HELLO cap a les destinacions des de totes les interfícies. D'aquesta manera es van obtenint els enllaços disponibles per ser utilitzats en les futures comunicacions.

### 3.3.2.3  Detecció dels veïns

El següent pas és determinar els veïns, per tenir les taules convenientment actualitzades, i com es pot accedir a ells. El procés és força simple: primer el node que vol fer la detecció, envia un missatge HELLO buit cap a tots els seus veïns. Quan el node de destí rep el HELLO, afegeix el node origen a la seva taula com a destinació asimètrica. Aquest receptor també envia aquest HELLO cap a l'originari que, al rebre el HELLO que ell mateix havia transmès, declara el destí com a veí simètric a la taula de veïns. Seguidament, l'origen transmet un altre HELLO incloent-hi l'adreça del destí, que al rebre-ho canviarà l'estat del node d'asimètric a simètric.

L'estat de la taula de veïns depèn de l'estat dels enllaços i està sotmès als seus canvis. Un veí només podrà ser simètric si existeix un enllaç que comuniqui amb el veí. En el cas de què una entrada a aquest enllaç sigui eliminada, també serà eliminada l'entrada del veí ja configurada sempre que no hagi un altre enllaç que es comuniqui amb dit veí.

També actualitzem les taules de dos salts amb els nodes que es poden accedir des dels veïns simètrics d'un salt. Així, tots els HELLO rebuts pel node local des d'un dels veïns simètrics amb direccions d'una estació diferent a les que ja estan registrades, és afegida a la taula de dos salts del node local.

### 3.3.2.4  Selecció de Multi Point Relay (MPR)

El **MPR** és un dels conceptes nous que s'introdueix a **OLSR**. Com s'ha vist, el procés de detecció de veïns i de sondeig de enllaços fan un *broadcast* de missatges HELLO per tota la xarxa fins que es creen les rutes. Hi ha mecanismes de control per no realitzar tantes retransmissions com és el nombre de seqüència. Si un node rep un paquet amb un nombre de seqüència més baix o igual que el que ja havíem rebut anteriorment, aquest paquet no es retransmet. Amb la **inundació** aconseguim arribar a les rutes òptimes, però a preu d'omplir tota la xarxa de missatges HELLO.

El concepte de **MPR** intenta superar el problema de la **inundació**. Això es fa seleccionant una sèrie de nodes veïns simètrics al node local i que permeti arribar a qualsevol dels nodes a dos salts que apareixen a les taules. D'aquesta manera es limita en gran mesura el nombre de rutes utilitzades per la transmissió de missatges a sobre de la xarxa. Un exemple del que es descriu es pot veure gràficament a les següents figures 3.2 i 3.3.

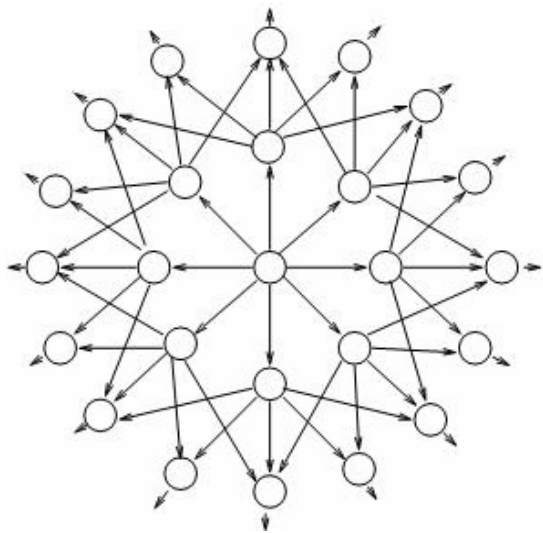

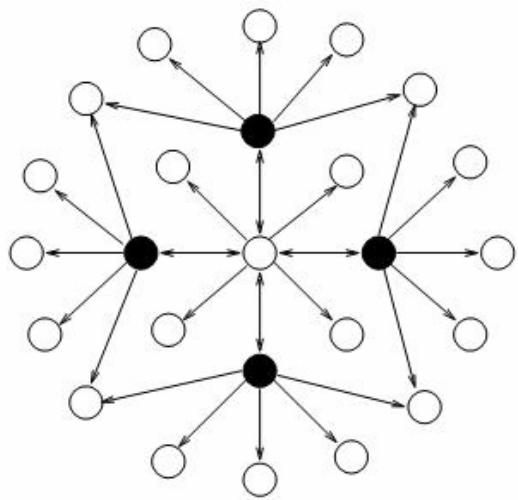

**Figura 3.2: Topologia sense MPR**        **Figura 3.3: Topologia amb MPR**

La selecció del **MPR** es realitzarà marcant uns determinats nodes que compliran el requisits per ser **MPR**, és a dir, nodes a un *hop* que ens permetran arribar a tots els nodes a dos salts sense accedir més d'una vegada a cap d'aquests nodes. Una vegada decidit el **MPR** es fa saber a tots els veïns per tal de configurar les rutes òptimes. Els nodes ho marquen omplint als missatges de HELLO el camp de tipus de veí com a MPR_NEIGH. Les adreces dels nodes **MPR** es guarden a unes taules específiques on es guarden tots els MPR del node local. Quan es rep un HELLO amb el camp MPR_NEIGH, la taula es va actualitzant.

### 3.3.2.5  Missatges de control de topologia

Amb la difusió d'aquests missatges de control de topologia, es vol tenir un control de l'estat dels diferents enllaços, així com de possibles canvis de topologia. De fet són generats quan es detecta un canvi a les taules de **MPR**. Aquests missatges, per no confondre'ls amb altres més antics porten un camp anomenat Advertised Neighbor Sequence Number (ANSN) i que és el nombre de seqüència del node publicitat.

### 3.3.2.6  Càlcul de les rutes

Pel càlcul de rutes s'ha proposat un algorisme senzill que es basa en el descobriment dels camins més curts. Primer s'afegeixen tots els nodes que estan només a un salt i amb els quals es té un enllaç simètric i després s'afegeix tots aquells que estan a dos salts i amb els que es hi ha també enllaç simètric. Després s'afegeixen tots els altres nodes que estan al conjunt de Topology Control (TC) i que estan a més de dos salts. Serà amb els missatges TC, difosos pels **MPR**, que descobrirem les rutes. Primer s'aniran afegint els nodes que estan a tres salts, després els que estan a quatre salts i així successivament fins que s'arriba a configurar totalment l'arbre de rutes òptimes de la xarxa.

### 3.3.2.7  Múltiples interfícies

L'eina **OLSRD** suporta la creació de rutes damunt nodes que tenen més d'una interfície. Cada interfície tindrà una adreça única per la que es diferencia de les altres. Per confeccionar les rutes amb nodes amb múltiples interfícies, s'envien missatges Multiple Interface Declaration (MID) entre ells, missatges que contenen un llistat de les adreces de xarxa del node i que corresponen a cadascuna de les interfícies que composen aquest node.

## 3.3.3  Funcions auxiliars d'OLSRD

Amb el propòsit de complementar el funcionament general d'**OLSR** es van implementar noves característiques que augmenten les prestacions de l'algorisme.

### 3.3.3.1  Interfícies no OLSR

Un dels problemes que podria haver a una xarxa d'estacions mòbils és que un dels nodes o conjunt de nodes que constitueixen l'entorn, no és compatible en cap de les seves interfícies amb **OLSR**. Per oferir connectivitat al node o subxarxa incompatible, **OLSRD** ofereix un tipus de missatges anomenats Host and Network Association (HNA). El HNA contindrà un camp amb l'adreça IP de la xarxa i un altre amb la màscara de la subxarxa. D'aquesta manera es poden incloure nodes aliens a l'entorn i que no suporten **OLSR**.

### 3.3.3.2  Notificacions de capa d'enllaç

Encara que **OLSR** és independent de la plataforma *hardware* fent-se la detecció i manteniment dels enllaços des de capes relativament altes, pot donar-se el cas que es requereixin missatges de capes més baixes, com quan es produeix una ruptura d'un dels enllaços. Amb les notificacions provinents de la capa d'enllaç, s'aconsegueix una prompta detecció del problema, afegint robustesa al sistema.

### 3.3.3.3  Histèresi

El concepte d'histèresi ens serveix d'ajuda per prevenir canvis de rutes en moments de baixa qualitat a l'enllaç per algun motiu concret. Així, els enllaços són més resistents enfront els intervals de tràfics o a algunes pèrdues de connexió.

La estratègia que es segueix en la histèresi es basa en dues regles: la regla d'estabilitat i la regla d'inestabilitat. La primera s'aplica cada vegada que es rep correctament un paquet **OLSR** i es tradueix en un augment del paràmetre de qualitat de l'enllaç, mentre que la segona s'utilitza sempre que un dels paquets que circula per la xarxa es perd i resulta en pèrdua de qualitat d'enllaç.

Aquesta qualitat d'enllaç és un nivell que serà el que determinarà si una ruta és simètrica o asimètrica, o si s'ha d'eliminar. Si aquest valor és superior al UPPER_THRESHOLD, un enllaç passa de convertir-se d'asimètric a simètric. El cas contrari es dóna quan el valor és inferior al LOWER_THRESHOLD.

Tots els paràmetres de llindars i d'escalat d'histèresi amb el que es fan les operacions són configurables des d'**OLSRD**.

### 3.3.3.4  Redundància dels Topology Control (TC)

Es basa en un paràmetre configurable de l'OLSRD anomenat TC_REDUNDANCY i pot servir per fer més entenedora la topologia de la xarxa. Aquest paràmetre pot assolir tres valors: 0, on la difusió dels missatges TC es realitza només als MPR; 1, la difusió pot realitzar-se des dels MPR i els selectors de MPR; 2, on tots els enllaços són simètrics i els missatges TC s'envien per tota la xarxa.

### 3.3.3.5  Redundància Multi Point Relay (MPR)

Quan s'explicava el concepte de MPR, hem vist que tots els nodes a dos salts només podien ser accedits des d'un MPR. Així es guanya en optimització de recursos, però també es pot perdre en robustesa, ja que si suposem el cas en què un dels enllaços es perdi, resultarà que ja no es podrà accedir a aquell node en concret. Per això, **OLSRD** permet configurar un paràmetre anomenat MPR_COVERAGE on es permet que els nodes a dos salts siguin accessibles des de *n* **MPR** diferents. Un MPR_COVERAGE podria ser interessant en escenaris on hi ha una elevada mobilitat dels nodes.

## 3.3.4  Plugins

Els *plugins* són una de les característiques més destacables de l'**OLSRD**. Quan s'ha introduït el codi, s'ha parlat de què es va voler implementar el protocol de la forma més modular possible perquè després, en un futur, es poguessin aplicar ampliacions i millores damunt del codi que s'havia dissenyat.

Els *plugins* són una d'aquestes millores al codi original. És una entitat que actua com a mòdul diferencial de la resta i que vendria a ser el que és una llibreria dinàmica (dll), és dir, un codi que s'integraria amb l'estructura d'**OLSRD** (figura 3.4). El mateix algorisme dissenyat per *unik* proporciona una interfície que permet configurar aquest mòdul complementari de forma lliure i independent de la resta del codi d'**OLSRD**.

Exemples de *plugins* ja implementats són el *plugin* de potència, on es creen estructures que tenen en compte el consum de bateries dels nodes i, per tant, configurar rutes segons aquesta informació i el plugin de connexió a Internet dinàmica, que permet anar afegint i eliminant rutes.

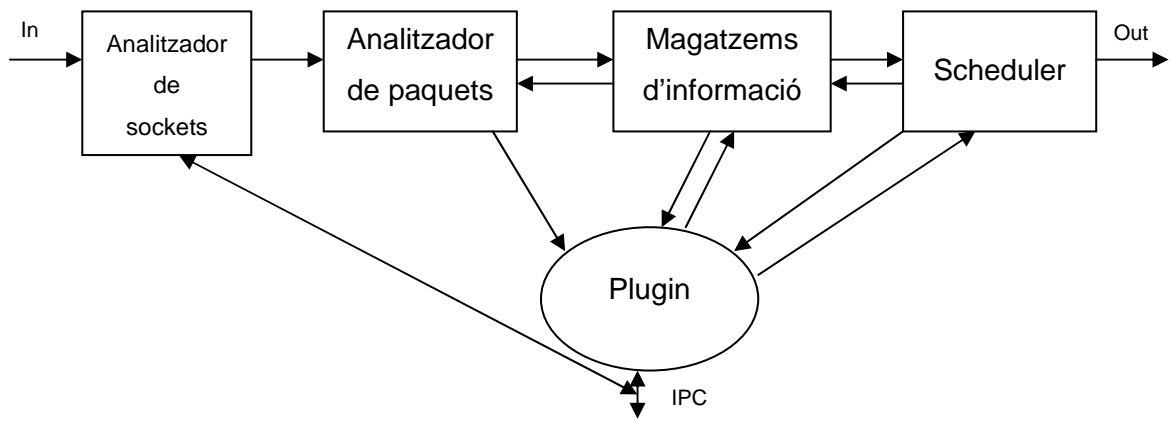

**Figura 3.4: Plugin OLSR**

# 4 PROTOCOLS MULTICANAL

Tant les **xarxes *mesh*** com els **protocols d'encaminament**, especialment **OLSR**, seran bàsics per explicar el procés que s'ha seguit per l'eina dissenyada.

A continuació, s'exposaran els altres conceptes que serviran de base en la realització del projecte: els **protocols multicanal**. A dins aquest apartat es tractarà d'explicar com és possible tenir protocols multicanal, quina necessitat i quins avantatges podem tenir fruit de la utilització d'aquests protocols i quines varietats de protocols multicanal s'han proposat fins a l'actualitat. Tot aquest estudi sobre l'estat de l'art dels algorismes multicanal ens servirà per conèixer quines són les millors opcions que hi ha a l'hora d'aplicar aquestes solucions a les Wireless Mesh Networks (WMN), ens facilitarà la comprensió de la solució adoptada a l'hora de realitzar el disseny i, a més, podrà ser utilitzada com a *background* per a la implementació de l'objectiu del treball.

## 4.1 INTRODUCCIÓ A L'ESPECTRE MULTICANAL

Al capítol 2 ja s'ha introduït l'existència de l'espectre multicanal. Als documents del CNAF [3] es veu com els espais assignats al funcionament de la xarxes sense fils i, com a conseqüència a les WMN, són les bandes de **2,4 GHz** i de **5 GHz**. Aquestes bandes es divideixen en una sèrie de canals d'una grandària de 5 MHz, que és l'espai que serà assignat a cadascuna de les comunicacions que es facin a la xarxa. D'aquesta manera, a la banda de 2,4 GHz es tenen fins a 14 canals possibles per a la seva utilització, mentre que a la banda de 5 GHz es tenen fins a 52 canals possibles.

Aquesta canalització, teòricament, implica que al primer conjunt de freqüències podrien haver fins a 14 comunicacions simultànies i fins a 52 simultànies al cas d'utilitzar les freqüències més elevades. El problema és que, a la pràctica, totes les comunicacions ocupen més espai que els 5 MHz que serien el espai d'un canal, envaint l'espai freqüencial dels canals contigus i, per tant, distorsionant la informació dels nodes que utilitzen aquests canals. Una mostra del que s'han comentat es pot veure a la il·lustració 4.1. Les bandes laterals de la comunicació sempre afectaran als canals adjacents. Segons algunes proves experimentals que es poden veure a la web [27], la comunicació d'un canal afecta a l'espai freqüencial que abasta entre la nostra freqüència central i la freqüència de 3 canals més cap a la dreta, i la freqüència central i la que es troba 3 canals més cap a l'esquerra, ja que l'amplada de banda és de 22 MHz a 2,4 GHz i de 20 MHz a 5 GHz. Exemplificant-ho un poc, una comunicació que opera sobre el canal 6 de la freqüència afectaria a totes les comunicacions que estiguessin operant entre el canal 3 i el canal 9.

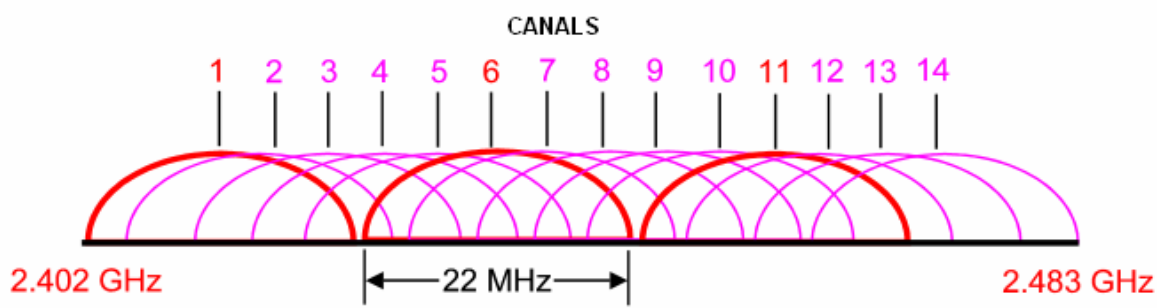

**Figura 4.1: Canals 2,4 GHz**

Això voldrà dir que l'espai freqüencial quedarà molt limitat per aquestes interferències cocanal essent aprofitada tan sols una tercera part de l'espai inicialment ofert, ja que només es podran utilitzar simultàniament aquells canals que siguin totalment ortogonals entre ells i que no solapin la seva informació. Per aquesta raó, a la banda de **2,4 GHz** només tindrem la possibilitat de tenir **3 comunicacions simultànies**, mentre que a la de **5 GHz** podrà ser de fins a **13 connexions simultànies**.

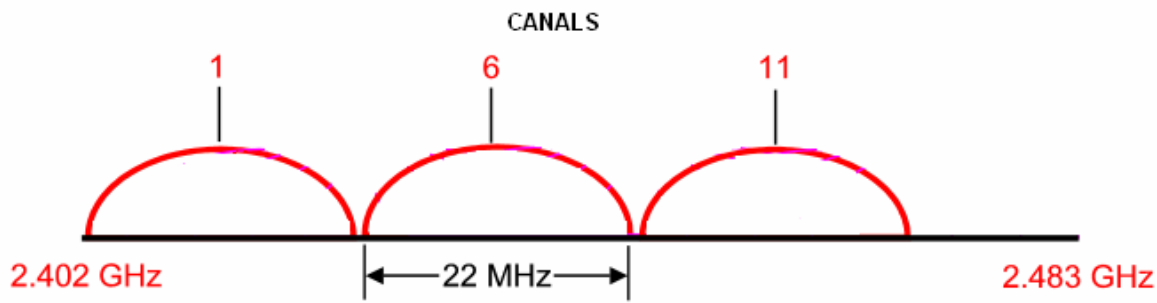

**Figura 4.2: Canals ortogonals 2,4 GHz**

## 4.2 NECESSITAT D'UN PROTOCOL MULTICANAL

A l'apartat anterior, s'ha observat que les comunicacions a diferents canals causen interferències al canals contigus. Coneixent aquestes interferències que es produeixen al sistema, es podria pensar que la millor solució i la més simple seria la d'emprar els mateixos canals de freqüència per a totes les comunicacions i per a tots els nodes, tal com es pot veure als diagrames de la figura 4.3.

Però aquesta solució, encara que simple i sense requeriments extra de processament, no és la millor possible i, especialment, quan es vol expandir la xarxa a un nombre determinat de nodes que tenen un cabdal de tràfic important. Això és degut a que si tots els nodes estan compartint la mateixa freqüència a tot l'entorn, els recursos oferts a aquella freqüència dins el rang de cobertura de l'estació seran compartits per cada vegada un major nombre de connexions, resultant en que a major nombre de nodes, menor capacitat serà assignada a cada usuari. Per fer-ho més comprensible, es pot exemplificar la situació d'una manera semblant a la que es va il·lustrant a les següents figures.

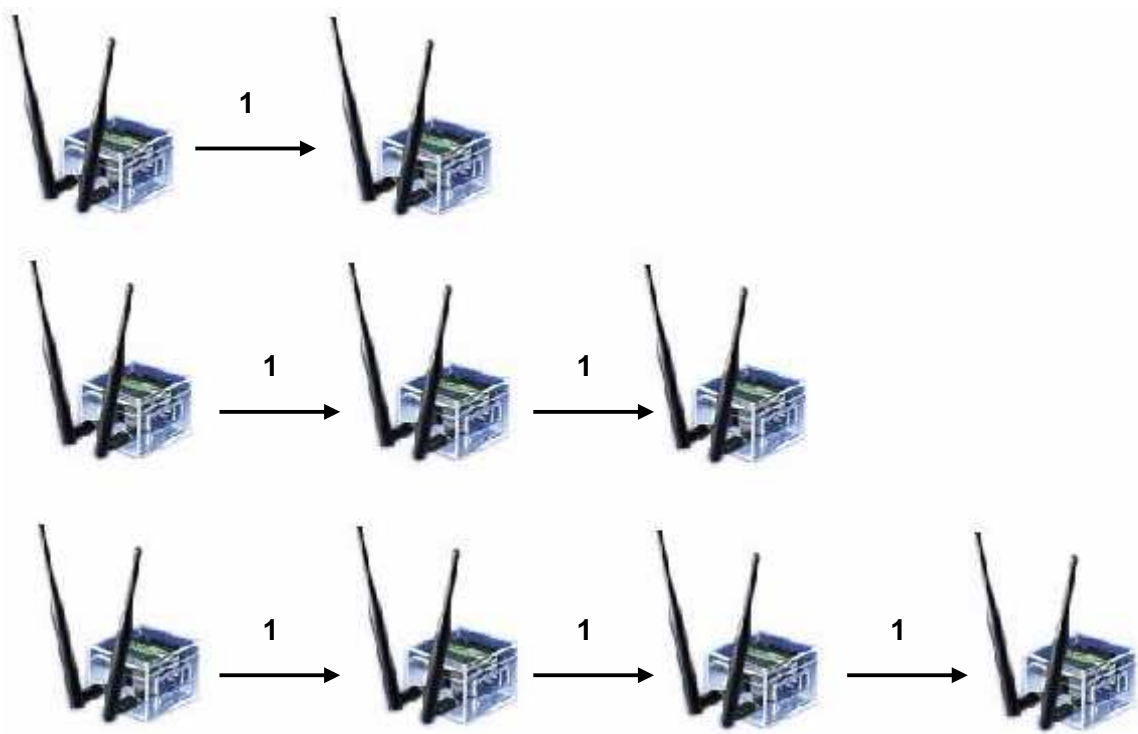

**Figura 4.3: Necessitat del multicanal**

Si tenim només dos nodes (figura 4.3) dins el rang de cobertura que es comuniquen entre ells utilitzant un canal per a la connexió, s'utilitzaran tots els recursos que disposem a la xarxa, però si tenim casos com el segon de la figura 4.3 o el tercer de la figura 4.3 on tots els nodes es comuniquen entre ells simultàniament, provocarà que, al cas de tres nodes, la mateixa capacitat que es tenia, s'hagi de dividir entre les dues comunicacions. Igualment, al cas que s'inclou amb quatre nodes, a l'haver tres connexions dins el mateix rang de cobertura, aquests recursos totals encara es repartiran més, pertocant a cadascun dels nodes un flux de tràfic menor. Les raons de per què es divideix el tràfic i en quina quantitat es tractarà més endavant amb més detall al apartat 4.3.

Si afegint tants pocs nodes i tantes poques connexions sobre un mateix canal es produeix tal minva de recursos, imaginem el que podria succeir amb una xarxa relativament densa, amb molt de tràfic entre les estacions i baix un mateix rang de cobertura, els recursos que se li assignarien a cadascun dels nodes serien mínims en comparació dels que es podrien arribar a assolir tenint en compte la velocitat de transmissió dels nodes.

Per solucionar aquest problema, es poden aplicar una sèrie de mesures que permeten aprofitar millor l'espectre radioelèctric en el seu conjunt i, per tant, obtenir una major quantitat de recursos per cada enllaç. Aquest és l'objectiu dels **protocols multicanal**.

Hi ha moltes línies d'implementació de protocols multicanal que apunten cap a tècniques diferenciades, però totes comparteixen l'objectiu d'aprofitar les freqüències que disposem. Les línies diferenciades són degudes a que la implementació pràctica d'aquests protocols presenta una sèrie de problemes difícils de resoldre en molts de casos:

- Els nodes que conformen una **xarxa *mesh*** no sempre tindran una gran capacitat de processament ni possibilitat d'accedir a fonts d'energia degut a la seva mobilitat. Per aquest fet, els protocols o tècniques multicanal haurien de ser el més simple possible.
- Si la solució és complexa, algunes estacions poden excedir un temps raonable degut a la seva pobra capacitat de processament.
- Tots els nodes hauran de ser capaços de sintonitzar els canals independentment dels altres per fer compatible el sistema.

- S'ha de tenir en compte el problema de la interferència dels canals i seleccionar per a les transmissions de dades canals ortogonals que no perjudiquin ni comparteixin els mateixos recursos que altres transmissions.

- A l'hora de realitzar una connexió punt a punt, per transmetre i per rebre s'ha de tenir la mateixa freqüència a les estacions. En cas contrari, no és possible la transmissió de dades. El problema s'eixampla quan s'escalen alguns models a xarxes majors amb més tràfic. Els que són nodes centrals, possiblement escoltaran a vàries estacions simultàniament i si s'utilitzen diversos canals s'haurà d'estar commutant contínuament de freqüència amb la possibilitat de convertir-se en un coll de botella ja que els altres nodes hauran d'esperar a que el node central (destí de la transmissió) sintonitzi el mateix canal que ells.

- A més, el procés de sintonitzar el canal i d'associació amb l'estació veïna no és immediat i no sempre té per què ser ràpid. Si aquest temps d'associació és massa gran, la suposada millora que havíem de notar amb la introducció de la solució multicanal, quedarà totalment eclipsada.

Tenint en compte tot això, i amb l'afany de superar els diferents entrebancs que s'han presentat, s'han anat elaborant i proposant una sèrie de tècniques i de **protocols multicanal** tractant d'obtenir millores respecte als entorns de canal únic.

## 4.3  CAPACITAT A L'ENTORN MULTICANAL

Dins aquest apartat es farà un estudi de les capacitats que podem tenir a una WMN, donant-li una gran importància al paràmetre de capacitat, ja que serà a sobre del que més es treballarà a aquest document. La capacitat màxima teòrica dels entorns sense fils generals, les fites màximes que es poden assolir a un espai en canal únic amb varis nodes i l'impacte que té el multicanal a sobre del sistema són alguns dels aspectes que es tracten als següents punts.

### 4.3.1  Capacitat teòrica màxima

Abans d'entrar en les fites que es poden arribar a assolir a entorns multicanal ben distribuïts i de quina capacitat podem tenir arribar a tenir amb les nostres tècniques multicanal, cal tenir en compte altres factors respecte a la velocitat de transmissió a la que es pot arribar al medi sense fils [28].

Els aparells que funcionen a l'espectre multicanal ofereixen **capacitat teòrica màxima (TMT)** de fins a **11 Mbps** en el cas de què les estacions emprin **IEEE 802.11b** i de fins a **54 Mbps** al cas de **IEEE 802.11a**. Però això sempre que es transmetin les dades netament sense cap tipus de capçaleres ni cues. Tal com succeeix al medi cablat, les transmissions per medi *wireless* segueixen un model de capes on cadascuna d'elles va afegint una capçalera diferent que influirà a la transmissió útil de les dades. A l'afegir aquestes dades haurà una disminució en quant a velocitat de transmissió de les dades. Tanmateix, aquests no són els únics factors que afecten comunicacions ja que, si fos així, es podrien arribar a velocitats de fins a 10 Mbps, semblants a les obtingudes als medis cablats. La capacitat que tindríem per damunt de la capa MAC seria com el que indica a 1.

$$TMT_{APP} = \frac{\beta}{\alpha + \beta} \cdot TMT_{802.11}(bps)$$

**Equació 1: Capacitat màxima teòrica**

Els paràmetres α i β representen la longitud de les capçaleres per damunt de la capa MAC i la longitud del datagrama respectivament. Operant aquesta relació que apareix amb el **Theoretical Maximum Throughput (TMT)** que prové de la MAC, obtindrem el flux de dades de la nostra aplicació.

Però al medi *wireless*, a les capes física i MAC, s'utilitzen uns procediments específics de l'estàndard, diferents als utilitzats a medis com Ethernet i que influiran de manera determinant en el rendiment de les transmissions de dades. Així, s'hauran de tenir en compte quines capçaleres s'estan utilitzant per la transmissió (preàmbuls, cues, espai entre trames...), així com els mecanismes que s'utilitzen per establir comunicació amb els veïns. A l'estar tractant amb un medi amb un entorn hostil, aquests mecanismes són més complexes que a altres situacions i, per tant, afectaran més a la velocitat de transferència dels nodes. Per aquesta raó, **IEEE 802.11** defineix que els ACK de les comunicacions siguin sempre a 1 Mbps, ja que prevaldrà la fiabilitat, assegurant l'arribada del paquet. A més a casos com el de RTS/CTS, s'hauran de tenir en compte els temps assignats pels intervals de RTS, de CTS, de DIFS i d'ACK. Apart, també afectaran al global de les comunicacions factors com la longitud de la finestra de contenció. Per tant el **TMT$_{MAC}$** es calcularà com es diu a continuació.

$$TMT_{MAC} = \frac{MSDU_{SIZE}}{delaySDU}$$

**Equació 2: Capacitat màxima teòrica per MAC**

$$delaySDU = (T_{DIFS} + T_{SIFS} + T_{BO} + T_{RTS} + T_{ACK} + T_{DATA}) \cdot 10^{-6} s$$

**Equació 3: Delay SDU**

Al calcular tots aquests factors, és lògic pensar que s'estarà per davall de 11 Mbps. De fet, als experiments realitzats per diferents tecnologies de transmissió *wireless*, s'ha arribat a assegurar que la velocitat real màxima rondarà els valors de **7 i 8 Mbps**, metre que la versió IEEE 802.11a, la velocitat real serà un poc superior a **40 Mbps** i no els 54 Mbps de les especificacions d'IEEE 802.11a.

## 4.3.2   Capacitat de les WMN

S'han fet nombrosos estudis matemàtics sobre la capacitat que tenen les xarxes sense fils. Un dels més utilitzats i que serveix com a referència a moltes propostes de treballs relacionats amb capacitats és l'article de Gupta i Kumar [29].

A l'article es postula que, després d'una sèrie de càlculs i suposicions respecte al sistema que tenim, s'arriba a la conclusió de què, per una xarxa on els nodes estiguin aleatòriament col·locats i en un medi sense interferències, el marge superior de *throughput* que es pot assignar a cadascun dels nodes és el següent:

$$\lambda(n) = \Theta\left(\frac{W}{\sqrt{n \cdot \log n}}\right)$$

**Equació 4: Fita de Gupta i Kumar**

Aquesta fita que proposen és veritat sempre que els nodes estiguin situats a un espai tridimensional. Al cas de què estiguem parlant d'un àrea superficial de dos dimensions, el **throughput** assignat a cada node no podrà superar la fita següent:

$$\lambda(n) = \Theta\left(\frac{W}{\sqrt{n}}\right)$$

**Equació 5: Fita Gupta i Kumar per xarxes aleatòries en dos dimensions**

Això pel cas de les xarxes aleatòries en un entorn bidimensional. En canvi, si es fa una sel·lecció del posicionament dels nodes, la fita canviaria:

$$\lambda(n) = \Theta\left(\frac{W}{\sqrt[\alpha]{n}}\right)$$

**Equació 6: Fita de Gupta i Kumar per xarxes arbitràries**

La fita donaria un **throughput** per node més elevat ja que l'atenuació α sempre serà superior a 2 degut a que tindríem en compte el posicionament de les estacions.

De totes formes, les fites proposades són el marge superior de la capacitat que es pot arribar a tenir. A la gran majoria dels casos, la capacitat que tindrem serà inferior a la que s'indica a les fites. Així tenim que als articles de Jun *et al.* [30] i Blake *et al.* [31] s'indica com evoluciona la capacitat a una WMN a mesura que se li van afegint nodes. De fet a una topologia en línia com la que es veu a la figura 4.3 queda demostrat que la capacitat de la xarxa es va degradant a un ritme de $O\left(\frac{1}{n}\right)$ on *n* és el nombre de salts totals necessaris per arribar a la destinació. A una topologia en línia de fins a 2 salts, la nostra capacitat es veuria, doncs, reduïda aproximadament a la meitat. Cal dir que hi ha estudis [32] que indiquen que, fins i tot, la degradació podria ser més elevada i que també dependria d'altres certs factors com de tipus *hardware*, de configuració de la xarxa, condicions de l'entorn *wireless*... Una mostra de l'evolució que tindríem afegint nodes a la xarxa i seguint amb la degradació proposta de $O\left(\frac{1}{n}\right)$ seria semblant a la que es veu a la gràfica de la figura 4.4.

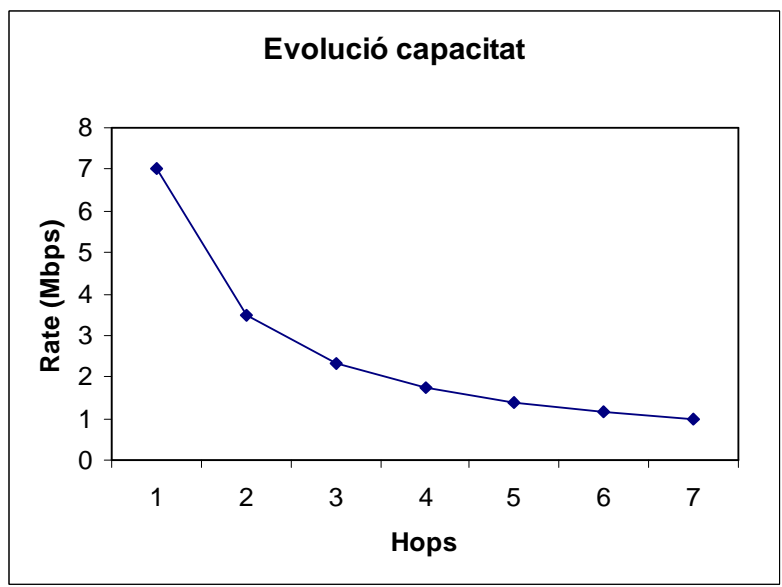

**Figura 4.4: Evolució capacitat de sistemes d'únic canal**

La utilització de múltiples canals ortogonals ens permetrà superar aquestes reduccions de capacitat que marca la gràfica anterior aconseguint arribar a fites properes a les màximes assolibles.

Segons les consideracions que es realitzen als articles de Gupta i Kumar [29] i la que apareix als de Kyasanur i Vaidya [33], s'arriba a la conclusió de què **la capacitat d'un entorn multicanal és la mateixa que la d'un entorn d'únic canal**, sempre que cada interfície tingui un canal dedicat. A l'article de Kyasanur i Vaidya [33], es tenen en compte els casos on tenim vàries interfícies que seleccionen entre un ventall de *c* canals. Degut a això es mesuren nous límits (per alt i per baix) de capacitat on apareixen els paràmetres *m* **interfícies i *c* canals**, quantificant l'impacte que té per una xarxa el afegir noves interfícies ràdio i que tinguin capacitat per canviar de canal.

A l'estudi es determina quina és la pèrdua de capacitat que es té enfront la fita màxima de Gupta i Kurmar [29] tenint en compte la relació $\left(\dfrac{c}{m}\right)$. Si aquesta relació és 1, es podrà tenir la situació òptima de capacitat però a mesura que aquest valor va augmentant (habitualment tindrem més canals que interfícies), la capacitat anirà decreixent progressivament d'acord amb unes fites calculades a l'estudi. Una de les dades importants que s'extreu de l'estudi és la de que per una **WMN aleatòria**, la capacitat pot arribar a ser l'òptima sempre que $\left(\dfrac{c}{m}\right) \leq \log n$ on *n* és el número de salts de la xarxa.

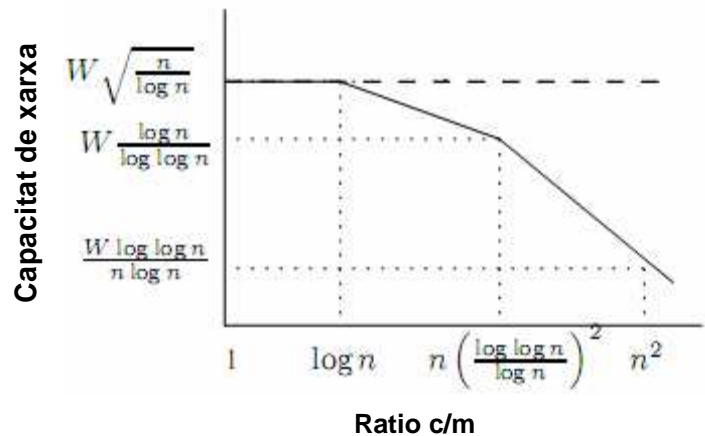

**Figura 4.5: Evolució capacitat a multicanal i multi interfície**

A l'estudi també apareixen nous conceptes que són decisius per calcular la capacitat que podem tenir com el temps de canvi d'interfícies i canals, i que també afectaran negativament al rendiment general de la WMN en qüestió. Per tant, encara que teòricament amb una interfície per canal tenim assegurada la fita màxima de capacitat, el retard produït pels canvis d'interfície faran perdre rendiment total.

### 4.3.3  Factors limitants a una WMN

Les fites calculades a l'anterior subapartat són unes cotes rarament assolibles a la majoria de sistemes normals. De fet, a un entorn multicanal existeixen una gran quantitat de factors tècnics i pràctics que, finalment, determinen la **capacitat real** d'una xarxa WMN. Són una sèrie de **factors limitants** que impedeixen a la gran majoria dels casos arribar a tenir la capacitat màxima i que només apareixen quan les solucions proposades són dutes a terme de forma pràctica.

Un dels factors limitants són produïts per la inclusió de més d'una interfície ràdio dins un mateix node [34]. El tenir més d'un transmissor a un node provoca una pèrdua de rendiment en la comunicació. Aquesta afectació dependrà de la naturalesa de les targetes sense fils i, per tant, del fabricant en qüestió. Aquest problema es pot solucionar en part posant-hi un **apantallament** adient a cadascuna de les targetes instal·lades.

Una altra forma de solucionar-ho és aplicar a les interfícies ràdio uns **filtres** més selectius que eliminin les bandes laterals que produeixen aquestes interferències que degraden les prestacions del sistema.

Però sorgeixen més problemes en el cas de què tinguem escenaris amb més d'un *hop*, i a on als nodes intermedis es realitzi un canvi de canal a l'hora de realitzar una transmissió. A aquests casos sempre comptarem amb més d'una interfície per tal de poder realitzar aquests canvis de canal. Segons Robinson *et. al* [34] i Ramachandran *et al.* [35], la proximitat entre antenes a un mateix node provoca **acoblament** entre elles i que, per tant, la capacitat del sistema quedi minvada. Això és producte de la pèrdua d'ortogonalitat dels canals, encara que estiguin sintonitzades les interfícies a canals ben diferenciats.

Aquesta interferència anterior també es produeix entre antenes sintonitzades a diferents ràdios [36] (per exemple, IEEE 802.11a i IEEE 802.11b). Encara que les dos interfícies siguin totalment ortogonals, es produeix un descens en el *throughput*.

A més, l'increment de les tecnologies *wifi* i d'altres aparelles que funcionen en la banda ISM a tots els àmbits de la vida domèstica i industrial ha fet que hagi una gran quantitat d'aparells interactuant per l'espectre ràdio. Tot i que aquests aparells actuïn a distintes freqüències que la WMN pertinent, la **coexistència** de múltiples ràdios a un mateix entorn (Bluetooth, WiMax, UMTS...) serà un factor més de pèrdues a la capacitat total de l'enllaç [37], produint-se fenòmens com intermodulacions, bloquejos a la recepció del senyal i aparició de soroll no desitjat.

Altres autors com Cheng *et al.* [75] exposen limitacions majors a les comunicacions a l'espectre de 5 GHz que al de 2,4 GHz. El disseny dels components no és prou bo per filtrar les interferències laterals dels canals (***leakage***) produint interferències als canals adjacents. A la pràctica es veurà que al nostre cas, les interferències són elevades afectant a molts canals adjacents, teòricament ortogonals. Es donen vàries solucions per minimitzar aquests efectes negatius: major **separació** de les antenes, **orientació** de les antenes, **apantallament** o implementació de **filtres** més acurats.

## 4.4   MÈTRIQUES DE LES WMN

Un dels conceptes fonamentals de qualsevol protocol d'encaminament és la **mètrica** que s'empra per calcular les distàncies o els pesos que hi ha entre els diferents nodes que conformen la xarxa. La forma més habitual de calcular aquestes mètriques és comptant els salts de distància entre les estacions. Per calcular la millor ruta, es selecciona aquella que té un menor nombre de salts entre la estació origen i la de destí [38]. Aquesta mètrica no té en compte ni l'ample de banda dels enllaços, ni el tràfic existent, ni la memòria del sistema. Per això, per xarxes sense fils de tipus WMN hi ha noves mètriques pel càlcul de noves rutes com són **Expected Transmission Count (ETX)** [38] i **Weighted Cumulative Expected Transmission Time (WCETT)** [39].

### 4.4.1   Expected Transmission Count (ETX)

L'**ETX** és una mètrica que té en compte les pèrdues de paquet a l'hora de transmetre dades. És, per tant, una tècnica diferent a l'encaminament basat en salts, ja que aquí es calcula l'**ETX** a cada enllaç i es selecciona la ruta que tingui un menor **ETX** a la seva transmissió.

El càlcul d'**ETX** es basarà en la mesura de paquets que es perden a una transmissió. Es suposa que perquè la informació enviada sigui la correcta, els paquets enviats hauran de ser correctes, així com els paquets de reconeixement que s'envien com a resposta de la transmissió dels paquets d'informació. D'aquesta manera la probabilitat total d'error vendrà a ser la següent:

$$p = 1 - (1 - p_f) \cdot (1 - p_r)$$

**Equació 7: Probabilitat d'error**

on $p_f$ és la probabilitat *forward*, és a dir, la probabilitat de què un paquet de dades que s'envia sigui erroni, mentre que $p_r$ és la probabilitat *reverse*, que ve a ser la probabilitat de què el paquet de reconeixement contingui errors.

La capa MAC d'**IEEE 802.11** retransmetrà les dades totes les vegades que la capa MAC ho permeti, fins que hagi una transmissió que sigui totalment correcta. Així, pot passar que es transmetin *k-1* paquets de forma errònia fins que s'envia un de correcte. D'això es pot extreure l'esperança, que vendria a ser el nombre mig de transmissions que necessitem per a què es transmeti un paquet correctament. A continuació, es poden observar les expressions analítiques a la que es refereix el text anterior i que culmina en el càlcul d'aquesta esperança i que és, en definitiva, el valor d'**ETX** que estem calculant.

$$s(k) = p^{k-1} \cdot (1 - p)$$

**Equació 8: Càlcul de s(k)**

$$ETX = \sum_{k=1}^{\infty} k \cdot s(k) = \frac{1}{1 - p}$$

**Equació 9: Càlcul d'ETX**

Aquest valor d'**ETX** serà el que marcarà la qualitat dels enllaços. Quan més pròxim estigui a 1 aquest valor, voldrà dir que menor nombre de retransmissions s'hauran de fer i que, per tant, més segura i fiable és la xarxa per la que circulen els paquets.

La mètrica **ETX** presenta uns resultats millors que la de salts clàssica buscant el camí més curt cap a la ruta. **ETX** és utilitzat a molts protocols com per exemple a OLSR. Tanmateix, s'ha d'advertir que **ETX** és una mètrica que està basada en les pèrdues dels enllaços i no en els seus amples de banda. Això provocaria que, en cas d'haver dues interfícies amb diferents capacitat l'una de l'altra, la seleccionada no seria la que ens assegura major taxa de transmissió sinó que seria aquella on les pèrdues de l'enllaç fossin menors. A més, ETX, tampoc té en compte que en una topologia de més d'un *hop* es pugui donar el cas de tenir diferents canals seleccionats a tota la topologia. Per tant, es tracta d'una mètrica que dóna camins subòptims.

## 4.4.2  Weighted Cumulative Expected Transmission Time

Per superar alguns dels entrebancs que presenta als protocols multicanal **ETX**, es va proposar la mètrica **WCETT** [39]. Aquesta mètrica és, llavors, utilitzada per protocols com **MR-LQSR** [39]. La mètrica **WCETT** es basa en el terme mig d'un paràmetre anomenat Expected Transmission Time (ETT) i que té en compte tant la grandària del paquet que s'està transmetent, com l'amplada de banda que disposem.

L'ETT el definim com el càlcul d'**ETX** tenint en compte l'ample de banda de l'entorn. Per calcular el paràmetre ETT necessitarem saber la longitud del paquet que estem transmetent, l'ample de banda que disposem a l'enllaç i el **ETX**, tal com ha estat calculat a l'equació 9. D'aquesta manera, ETT seria calculat tal com es veu a continuació:

$$ETT = ETX \cdot \frac{S}{B}$$

**Equació 10: Càlcul d'ETT**

On S és la longitud del paquet que transmetem i B és l'ample de banda del medi.

L'expressió de l'equació 10 ens permet solucionar un dels problemes que es comentaven respecte d'ETX, l'ample de banda. A aquest càlcul, apart de la fiabilitat empírica del medi, també comptaran força les prestacions ofertes pel medi. A més, **WCETT**, recull l'impacte sobre la mètrica que tenen el que hagin diferents canals en la mateixa topologia. **WCETT** recollirà un major valor quan el mateix canal estigui repetit als diferents enllaços de l'estructura.

El **WCETT** farà una ponderació dels ETT a cadascun dels diferents enllaços que estaran a diferents freqüències. El **WCETT** es pot entendre de dues maneres diferents segons el punt de vista que estem buscant. Així primerament, es pot entendre com la suma de ETT's d'una mateixa ruta. Es podria expressar com el següent sumatori

$$WCETT = \sum_{i=1}^{n} ETT_i$$

**Equació 11: Càlcul de WCETT**

Però aquí tenim un problema similar al que ens trobàvem a l'anterior mètrica descrita, ja que no es tenen en compte els diferents canals que es poden seleccionar a la topologia. Per això, per calcular **WCETT** s'afegeix un paràmetre que opera amb el canal amb el que s'està actuant en aquell moment. De fet, es va incrementant el paràmetre sempre que hagi un canal repetit al llarg de la mateixa topologia.

$$X_j = \sum_{Hop\_i\_canal\_j} ETT_i \quad \text{on } 1 \le j \le k$$

**Equació 12: Segon terme de MR-LQSR**

Les j representen els diferents canals, mentre que i representa el salt sobre el que s'està mirant en aquell moment. Per calcular el *throughput* total del sistema, primer es detecta el canal que representa el coll de botella al sistema i que serà aquell que tingui una suma més alta d'ETT dins de la xarxa. Així, doncs, el calcular **WCETT** seria tal com indica l'expressió 13.

$$WCETT = \max_{1 \le j \le k} X_j$$

**Equació 13: Possible càlcul WCETT**

D'aquesta manera aconseguim tenir en compte els diferents canals a la topologia. Però aquesta tampoc serà la definició completa de **WCETT** ja que aquesta definició implica que només tinguem en compte el camí que presenta el coll d'ampolla quan es vol tenir present les possibles modificacions que presenti el sistema. Per tant, una opció pel càlcul de WCETT seria ponderar les dues expressions que s'han calculat, la 12 i la 13. L'expressió ponderada podria quedar tal com es veu a continuació:

$$WCETT = (1-\beta) \cdot \sum_{i=1}^{n} ETT_i + \beta \cdot \max_{1 \le j \le k} X_j$$

**Equació 14: Càlcul de WCETT**

El paràmetre β serà un valor entre 0 i 1 que serà el que realitzarà la ponderació corresponent donant importància a un o a un altre terme. Aquesta expressió es pot interpretar com un balanceig entre el comportament del sistema en global i el comportament en local.  Així el primer terme reflecteix els recursos que s'han gastat, mentre que el segon és aquell que reflecteix aquells salts que causen un major impacte global al sistema.

Hi ha un segon punt de vista per interpretar aquesta expressió i és el que es refereix al balanceig entre **latència** i ***throughput***. Mentre que el primer terme representa la latència del camí en qüestió, el segon, que mesura l'impacte que tenen els colls de botella a un *hop* en concret, representaria la mesura del *throughput* del sistema.

Amb el paràmetre configurable β li donem més pes a un o altre factor depenent de les condicions del medi en què funcionem amb aquesta mètrica. Amb valors baixos de β donem menys importància a la diversitat dels canals, contràriament al que es fa quan aquests valors són més elevats.

## 4.5  TÈCNIQUES MULTICANAL

Abans de començar a explicar quin ha estat el procés que s'ha seguit pel disseny, primer es farà una introducció de l'estat de l'art de les tècniques multicanal, descrivint quines particularitats es fan servir a cadascun dels casos i quines són les característiques més interessants que podrien ser aprofitables en el desenvolupament de la nostra idea.

Aquestes diferents tècniques s'han aplicat tenint en compte distints punts de vista i, en aquesta memòria, s'han classificat atenent a uns criteris similars que complien cadascuna d'aquestes solucions.

Les diferents tècniques es poden agrupar seguint criteris molt diversos. Es poden classificar en tipus d'assignació per interfícies, en tècniques d'única interfície o de múltiples interfícies, per estratègies de selecció de canals, diferenciades en principis d'operació...

S'ha optat per fer la classificació dels diferents grups protocols multicanal estudiats seguint el tipus d'**assignació de canals** a les respectives interfícies [40].

Segons aquest criteri, distingim els següents tipus d'**assignació de canals**:

- **Assignació estàtica de canals.** Seran totes aquelles estratègies a on un mateix canal és assignat a una mateixa interfície durant un període de temps molt llarg. Seran tècniques que encaixaran millor en sistemes on el retard causat pel canvi de canal sigui elevat. A més, no requeriran de cap tipus de coordinació especial a l'hora d'efectuar les comunicacions. Dins l'**assignació estàtica** es podran distingir encara dos subgrups més:

  o *Assignació comuna de canal.* S'assignen un mateix conjunt de canals a les interfícies de tots els nodes. D'aquesta forma el sistema s'assegura connectivitat com si es tractés d'un sistema de canal únic.

  o *Assignació variable de canal.* També es fixa l'assignació dels canals però aquí es pot assignar un conjunt diferent de canals als nodes de la xarxa. Això permet utilitzar un major nombre de canals però, alhora, provoca que les rutes cap als diferents nodes siguin més llargues, amb risc de què hagi particions a la xarxa.

- **Assignació dinàmica de canals.** En aquest cas, les interfícies poden seleccionar qualsevol dels canals, canviant freqüentment de canal sempre que ho requereixin les condicions de l'entorn. Com que per a què hagi una comunicació entre els dos nodes aquests han de compartir canal, el que es requerirà serà un **mecanisme elaborat de coordinació**, que decidirà en quin moment s'ha de realitzar els canvis de canal. Si aquest mecanisme està ben implementat, es tindrà un sistema a on es pot gestionar una gran quantitat de canals amb poques interfícies.

- **Assignació híbrida de canals.** Són unes tècniques que combinen les característiques dels dos grups comentats i que, per tant, funcionaran amb algunes interfícies que tindran assignats el canal un període llarg de temps amb altres que aniran seleccionant els diferents canals a cada moment. D'aquesta manera, les tècniques d'assignació híbrida permeten simplificar de manera important els mecanismes de coordinació necessaris per poder realitzar les comunicacions.

## 4.5.1  Assignació estàtica

### 4.5.1.1  Tècnica de colorejat grafs

Una de les possibilitats que es presenten per fer una assignació de canals a tot un entorn és utilitzar una **tècnica de colorejat de grafs** [41]. Així, es defineix la WMN com un graf on cadascuna de les estacions és un dels vèrtexs del graf. Dos vèrtexs que es comuniquen entre ells són el que anomenem enllaç de comunicació, mentre que aquells que no es comuniquen directament però que provoquen una **interferència cocanal** a algun d'aquests vèrtexs s'anomena enllaç interferent.

Per confeccionar el graf i omplir cadascun dels enllaços amb el valor d'un canal diferent es poden fer servir dues tècniques diferenciades: el **colorejat fort** o el **colorejat basat en el grau d'interferència**.

El **colorejat fort** (figura 4.6) no té en compte el grau d'interferència i per configurar s'assignen els canals depenent de la distància. Així només es podrà assignar el mateix "color" a nodes que estiguin almenys a més de dos nodes de distància ja que es suposa que no haurà problemes amb les interferències, mentre que el colorejat basat en el grau d'interferència té en compte la interferència que el canal causa a un node veí actuant en conseqüència a l'hora de configurar les rutes.

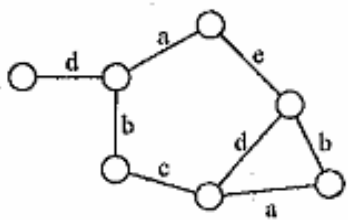

**Figura 4.6: Tècnica de grafs**

A la figura 4.6 es veu un exemple de **colorejat fort**. Al gràfic, cada lletra representa un canal distint que només es repeteix a una distància de com a mínim dos nodes.

### 4.5.1.2   Multi-Radio Unification Protocol (MUP)

**MUP** [42] es presenta com un protocol d'encaminament que suporta múltiples interfícies i múltiples canals. Per la implementació del protocol s'aprofita la estructura de capes de comunicació, i **MUP** s'instal·larà a la capa d'enllaç de dades, sense necessitat de modificar les capes de xarxa, transport i aplicació. També actuarà independentment de les NIC que actuen per davall de la capa d'enllaç. La raó d'utilitzar la capa 2 és deguda a que per tractar amb canals i interfícies serà molt més simple funcionar des de la capa 2 que des d'una capa de nivell superior. Un reflex de l'arquitectura de **MUP** es pot veure a la figura 4.7.

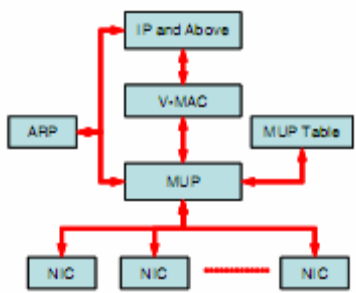

**Figura 4.7: Arquitectura MUP**

Com es veu al diagrama anterior, **MUP** té una taula que es va actualitzant amb les dades dels veïns que interactuen amb el node en qüestió. Dins aquesta taula hi ha dades que serveixen per identificar totalment el veí (taula 4.1) amb informacions respecte a les qualitats dels enllaços, així com dels canals que es poden utilitzar, i que serà el que ens permetrà configurar les rutes òptimes. La V-MAC que apareix al diagrama ve a ser un identificador general de tot el node i que permet diferenciar els nodes de les interfícies.

| Camp | Descripció (per cada veí) |
|---|---|
| Veí | Adreça IP del veí |
| Status | Indica si el veí suporta MUP |
| Llista MAC | Adreces MAC d'aquest veí |
| Llista de qualitats | Qualitats del canal associats amb el veí |
| Canal | Canal preferit per la comunicació |
| Temps de selecció | Darrera decisió sobre selecció de canal feta |
| Temps de paquet | Darrera vegada que un paquet ha estat enviat o rebut des del veí |
| Llista de temps de sondes | Llista dels temps pels missatges no reconeguts |

**Taula 4.1: Camps de MUP**

Per tant, com gran part dels protocols d'encaminament, **MUP** té implementat un procés de descobriment dels veïns que hi ha a la xarxa. Al cas de **MUP** el que primerament es fa és veure si els nodes suporten el protocol. Això es fa mitjançant l'enviament *broadcast* de paquets ARP, que les destinacions contestaran amb l'adreça MAC de cadascuna de les seves interfícies. Aquesta contestació no voldrà dir que aquella destinació suporti MUP, per això l'emissor original del missatge envia un missatge Carrier Sense (CS) cap als destins, que aquests contestaran amb un CS-ACK si realment tenen implantat el protocol MUP. Si no hi ha resposta al CS enviat, es proven de fer una sèrie de retransmissions que una vegada esgotades vendran a dir que aquell node en concret no suporta el protocol **MUP**. Si això succeeix, l'entrada a la taula **MUP** serà eliminada.

Per calcular la qualitat de cadascuna d'aquestes entrades, s'utilitza la tècnica del Round Trip Time (RTT) a l'enllaç. L'aplicació de la tècnica consisteix en l'enviament de paquets sonda periòdics cap a la destinació que quan arriben al destí són contestats íntegrament. D'aquesta forma podem mesurar el RTT quan el paquet arribi de nou a l'origen de la comunicació.

La forma de mesurar la qualitat és a partir de l'expressió 15. Com els missatges sonda que s'han comentat abans són enviats periòdicament, el valor de RTT pot ser diferent a cada prova. El que s'aconsegueix amb l'expressió és aplicar una certa memòria del que hi havia anteriorment per confeccionar el nou valor, que s'anomenarà **Smoothed Round Trip Time (SRTT)**. Així, ajustant el paràmetre $\alpha$ s'aconsegueix, sempre que es vulgui, que hagi canvis molt menys bruscos, evitant canvis de canal innecessaris.

$$SRTT = \alpha \cdot RTT_{new} + (1 - \alpha) \cdot SRTT$$

**Equació 15: Càlcul de SRTT**

L'elecció del canal vendrà donada per aquella entrada a la taula que tingui el valor de **SRTT** més baix. Quan les condicions del medi que conforma la xarxa canviïn, això també quedarà reflectit al valor de **SRTT** i pot provocar canvis de canal. El problema principal és que el canvi de canal representarà una latència extra quan es realitza aquest procés. Per això, en moltes ocasions, convindrà tenir un paràmetre $\alpha$ que mesuri els canvis d'una forma molt més suau.

### 4.5.1.3  Multi-Radio Link Quality Source Routing (MR-LQSR)

El protocol **MR-LQSR** és un protocol que suporta l'encaminament a múltiples canals i a múltiples interfícies. A entorns multicanal, criteris com l'encaminament pel camí més curt no són els més adients. **MR-LQSR** proporciona mecanismes suficients per tenir en compte les particularitats d'un medi multicanal.

**MR-LQSR** [39] és una combinació del protocol LQSR i de la nova mètrica desenvolupada: WCETT. El LQSR és un protocol basat en l'estat de l'enllaç derivat del protocol Dynamic Source Routing (DSR) [14]. El protocol MR-LQSR té quatre components diferenciades.

- La **primera component** engloba tots els mecanismes necessaris per realitzar els descobriments de veïns del node.
- La **segona component** assigna als enllaços que es tenen amb els veïns detectats anteriorment uns pesos.
- Un **tercer punt** que conté tots els processos a realitzar per a la propagació de la informació del node a les altres estacions de la xarxa.
- El **darrer component** és aquell que comprova els enllaços cap a una determinada estació per seleccionar el millor camí per un destí en concret, és a dir, atenent a uns pesos es configura l'arbre de rutes òptim.

Pel descobriment de veïns es realitza un procediment gairebé igual que el que realitza DSR. Primerament, s'envien Route Request (RREQ) cap a tots els veïns de la xarxa. Quan els nodes reben aquests RREQ comproven si són els destinataris d'aquests missatges i si ho són generen una resposta unicast cap a l'origen. En cas contrari, es reenvia el missatge de forma broadcast novament.

El tercer component de **MR-LQSR** també és similar a com es realitza el protocol DSR. Cada cert temps, **MR-LQSR** envia informació als seus nodes perquè d'aquesta manera es puguin anar configurant les millors rutes.

Tant el segon com el darrer component estan relacionats amb la mètrica utilitzada pel protocol d'encaminament. **MR-LQSR** i que és la **Weighted Cumulative Expected Transmission Time (WCETT)**. La mètrica és la que ens permet aplicar pesos als diferents enllaços entre els nodes. Aquests enllaços poden estar funcionant sobre diferents canals i, d'aquí podrem extreure la possibilitat de configurar la ruta segons les mètriques dels diferents canals.

Una vegada que els pesos han estat assignats per **WCETT**, es tindran les mètriques que permetran configurar les rutes. Aquells enllaços que tinguin les millors mètriques seran els seleccionats per poder transmetre. Cada cert temps, la informació de **MR-LQSR** s'anirà modificant per tal de garantir un millor funcionament del sistema.

### 4.5.1.4  Load-Aware Channel Assignment and Routing

És una tècnica a on es tracta de crear un algorisme el més **equitatiu** possible entre els nodes i que, a més, està destinada a fer-se servir en entorns multicanal i de múltiples interfícies. Els autors del protocol proposen una nova metodologia d'assignació de canal centralitzada a cadascuna de les diferents interfícies i un algorisme d'encaminament per adreçar els paquets distribuint la càrrega de la millor manera possible [43].

Abans de realitzar qualsevol tipus d'assignació o operació d'encaminament, primerament es realitza una **estimació inicial de la càrrega** als enllaços. Aquestes dades seran les que ens serviran per determinar com s'haurà de fer aquesta assignació de canals. Per fer-ho es poden utilitzar vàries expressions per calcular la capacitat estimada. La més ajustada és calculada a Gopalan *et al.*  [44] i s'expressa de la següent manera:

$$\phi_l = \sum_{s,d} \frac{P_l(s,d)}{P(s,d)} \cdot B(s,d)$$

**Equació 16: Càrrega dels enllaços**

$\phi$ representa la càrrega total estimada per un enllaç en concret, mentre que P simbolitza el nombre de camins acceptables entre un parell de nodes i $P_l$ és el nombre de camins acceptables que travessen un enllaç. El paràmetre B és la càrrega estimada en un model de tràfic teòric imposat inicialment (tal com es veu a la figura 4.9).

 Una vegada tenim aquesta càrrega estimada per l'enllaç, la meta és **assignar els canals** a les interfícies on l'ample de banda disponible a aquestes interfícies és, com a mínim, igual que la càrrega de tràfic estimada. Per assignar aquests canals, es poden donar diferents casos ja que assumim que es tenen fins *q* interfícies per node, i depenent de l'ocupació de les interfícies, el protocol actuarà de forma diferent.

Si els nodes que es comuniquen per l'enllaç tenen interfícies lliures, assignaran aquell canal que minimitzi el grau d'interferència. Si és només un dels nodes el que té interfícies lliures, s'assigna un dels canals que té sintonitzat a una de les interfícies el node sense interfícies lliures, essent el seleccionat aquell que minimitza el grau d'interferència. Finalment, en el cas de què tots dos nodes tinguin totes les interfícies assignades, agafaran un canal comú entre ambdós que minimitzi la interferència. En cas de què no existís cap canal comú, els nodes canviaran cap al possible millor canal comú que puguin tenir en quant a rendiment. El següent diagrama exemplifica el que es vol relatar (figura 4.8).

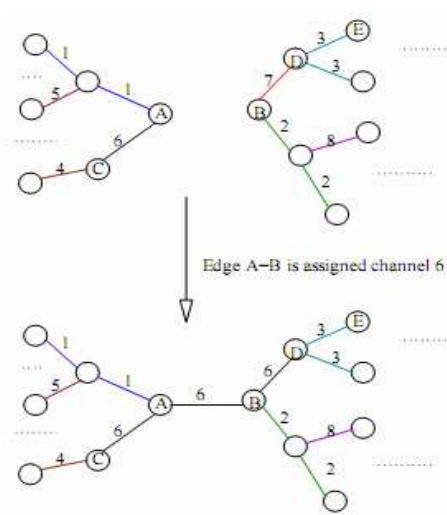

**Figura 4.8: Encaminament per balanceig de la càrrega**

Quan es té l'assignació de canal hi ha un mètode que avaluarà l'efectivitat d'aquesta configuració de canal. Per fer això es calcula la nova capacitat del canal estimada després de fer l'assignació. L'expressió representativa seria la següent:

$$bw_i = \frac{\phi_i}{\sum\limits_{j \in Intf(i)} \phi_i} \cdot C$$

**Equació 17: Ample de banda**

on tenim la càrrega estimada *i* enfront del conjunt de càrregues estimades que estan dins la zona d'interferència, mentre que C és la capacitat del canal ràdio.

Quan tenim totes les dades d'estimació de capacitat, és quan passem a la part d'encaminament de l'algorisme. L'algorisme pot ser de dos tipus diferents: **encaminament pel camí més curt** (com Bellman-Ford) i de **balanceig de càrrega** que tractarà de distribuir la càrrega entre tots els enllaços d'una manera similar.

El procés descrit anteriorment seria la **fase d'exploració** i aquesta fase es repetiria fins trobar les millors assignacions de canal i els millors arbres de rutes possibles. Quan s'arriba a aquest punt, s'entra en una **fase de convergència** on es realitzen les mateixes funcions que a la fase d'exploració, però tenint en compte que l'algorisme d'encaminament només encaminarà aquells fluxes que no han estat conformats, és a dir, aquells que no han trobat un camí amb el suficient ample de banda per satisfer les seves demandes de tràfic. Aquesta fase es repeteix fins que s'arriba a un estat de **convergència**. Quan s'aconsegueix aquest estat, l'algorisme acaba la seva execució. A la figura 4.9 es pot veure quin és el procés complet que s'ha descrit.

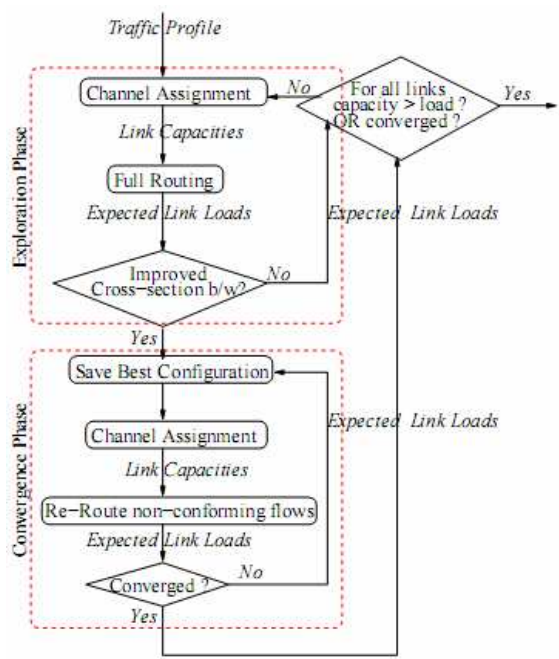

**Figura 4.9: Procés de la tècnica de balanceig de càrrega**

### 4.5.1.5  MobiMESH

**MobiMESH** [45] és una arquitectura dissenyada per tal d'aprofitar l'espectre multicanal i suportar mobilitat dels diferents nodes. A diferència d'una gran part de les propostes que s'han estudiat, **MobiMESH** presenta una solució que s'ha dut a terme a la pràctica, amb resultats més manejables que els que presenten les simulacions, ja que es tenen en compte factors que les simulacions no utilitzen.

Els principals objectius amb els que es dissenya **MobiMESH** són els següents:

- **Connectivitat.** La xarxa ha d'aportar connectivitat a tots els usuaris, ja siguin usuaris ad-hoc o clients WLAN estàndards.
- **Transparència.** La xarxa hauria d'oferir la mateixa interfície que si fos una xarxa WLAN corrent.
- **Mobilitat.** Els clients haurien de ser capaços de transitar tranquil·lament per tota l'àrea de cobertura sense perdre cap tipus de funcionalitat de les què ofereix la xarxa. MobiMESH suporta *handover*, localització de clients i altres procediments relacionats amb la mobilitat.
- **Integració.** L'arquitectura **MobiMESH** està dissenyada per ser fàcilment integrable amb altres xarxes heterogènies.

**MobiMESH** ha estat dissenyada com una xarxa mallada híbrida ja que està formada per dues subxarxes ben diferenciades: una **xarxa troncal** i una **xarxa d'accés**.

La **xarxa troncal** és la responsable de realitzar les funcions d'**encaminament**, de **mobilitat** i d'**integració** amb altres xarxes. Tots els components que formen aquesta subxarxa tenen capacitat per realitzar l'enrutament, col·laborant amb les diferents tasques d'encaminament. Per la configuració de rutes, els nodes utilitzen el protocol d'encaminament proactiu Optimized Link State Routing (OLSR) [10]. L'OLSR és de gran utilitat ja que amb els missatges HNA es permet traçar rutes fins a altres xarxes externes. A més, aquests nodes que conformen aquesta secció són els que proporcionen connectivitat als clients WLAN i tenen altres components bàsics pel bon funcionament del sistema com el servidor DHCP, les bases de dades per suportar la mobilitat i un o més *gateways* que són els que realitzaran la integració amb altres xarxes.

La **xarxa d'accés** és aquella infraestructura on els clients es connecten als punts d'accés proveint **connectivitat** a la xarxa troncal. Aquesta xarxa estarà dissenyada de tal manera que sembli una xarxa WLAN normal sense la necessitat d'haver d'instal·lar *software* especialitzat.

Tant la xarxa troncal com la d'accés empren la mateixa tecnologia i això podria conduir a tenir una elevada interferència, minvant la capacitat del sistema. Per solucionar-ho, es sintonitzen les dues xarxes a canals diferents (1 i 11, per exemple), minimitzant l'efecte de les interferències.

El component que connecta les dues subxarxes és l'**Access Router (AR)** i actua simultàniament com a punt d'accés i com a router *mesh*. Els AR estaran equipats almenys amb dues interfícies ràdio per poder realitzar aquesta connexió de les dues subxarxes. A la figura 4.10 es pot veure el que s'ha comentat respecte a les subxarxes. Aquest component tindrà les seves interfícies sintonitzades als dos canals en què operen les dues subxarxes.

En quan a l'organització IP, les dues subxarxes tindran direccions IP diferents (10.0.0.0 al *backbone* i 192.168.1.0 a la xarxa d'accés). L'AR tindrà dos IP diferents per comunicar-se simultàniament amb les dues xarxes.

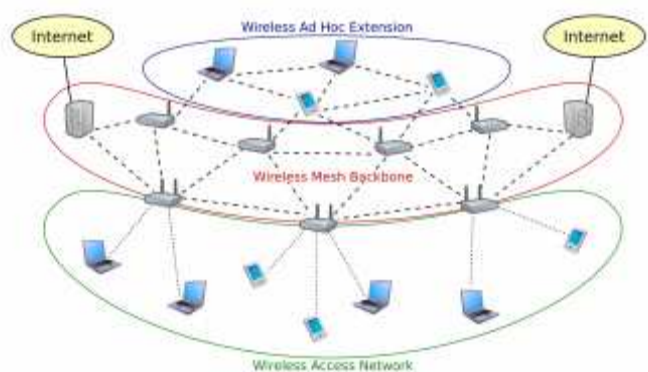

**Figura 4.10: Xarxa mesh estesa**

Com s'ha dit, **MobiMESH** suporta la mobilitat dins la xarxa de tots els seus clients i això, voldrà dir que s'ha de suportar la mobilitat d'aquests clients a la xarxa d'accés. La mobilitat dels clients serà gestionada de diferent manera depenent de si aquest terminal és un client *mesh* o és un client WLAN estàndard. Al primer cas, la mobilitat serà suportada pels propis terminals ja que aquests tenen implantats el protocol d'encaminament ad-hoc corresponent que fa automàtiques les funcions de mobilitat. Pels altres clients, la mobilitat serà més complicada de garantir ja que estem parlant de mecanismes de mobilitat de nivell de capa MAC, mentre que la mobilitat a la *mesh* es fa a sobre el nivell de capa de xarxa. Per tant, la capa MAC s'encarregarà de realitzar les associacions a la estació base corresponent que li proporciona cobertura i, després comunicarà a la capa de xarxa dels canvis soferts a nivell de capa MAC. Això es fa per informar als mecanismes d'encaminament de la capa de xarxa de la nova localització del client WLAN, informació que permetrà que els paquets es puguin encaminar a aquestes estacions i quin AR serà el més propici per arribar fins a elles.

Per realitzar la integració amb altres xarxes, s'empren AR especials que tindran una de les seves interfícies connectada a l'altra xarxa possibilitant la comunicació amb aquestes xarxes externes al sistema.

**MobiMESH** ha estat implementada físicament  en una plataforma on havien fins quatre AR i ha estat provat, resultat en un cert augment de *throughput*.

### 4.5.1.6  Planificació de freqüències

Per augmentar la capacitat total de la WMN, alguns autors han proposat tècniques basades en la **planificació de freqüències** per augmentar la capacitat de la xarxa. A l'article de Huang *et al.* [46], es dissenya un espai freqüencial compost per diversos anells concèntrics on cadascun d'aquests anells funcionarà a un canal diferent a l'altre.

L'estructura de la planificació consistirà amb un *gateway* central, que tindrà accés a serveis com Internet, i que es comunicarà amb la resta de nodes intermedis de la xarxa. Cadascun d'aquests nodes intermedis estarà emplaçat a un anell concèntric diferent que està sintonitzat a una freqüència diferent.

Els nodes intermedis es comuniquen amb el *gateway* principal mitjançant *i* salts, depenent de la distància a la que estigui l'anell en qüestió del *gateway* central. Al quart anell, s'hauran de realitzar quatre salts per accedir a l'element central de l'estructura. Cada node intermedi està constituït per dues interfícies ràdio sintonitzades a les freqüències dels anells adjacents. D'aquesta forma, els nodes podran anar intercanviant dades al llarg de tota l'estructura, comunicant-se sempre amb els nodes adjacents.

Aquesta **planificació freqüencial** de l'espai permet reduir considerablement el domini de la col·lisió, ja que es limitarà a tots aquells usuaris que vulguin accedir als mateixos serveis des del mateix anell, al mateix instant de temps.

### 4.5.1.7 Addició intel·ligent d'interfícies

Altres tècniques advoquen per afegir interfícies ràdio per augmentar la capacitat de la xarxa. Però, a diferència de la majoria de solucions, no s'afegeixen interfícies de manera uniforme a tots els nodes sinó a uns quants i de manera selectiva.

La particularitat de la proposta realitzada per Aoun *et al.* [47] resideix en seleccionar aquells nodes que són *colls de botella* i que seran aquells als que se li **afegiran les interfícies ràdio necessàries** per evitar el major nombre de col·lisions a aquests nodes concrets. D'aquesta manera s'aconsegueixen optimitzar recursos, augmentant la capacitat sense la necessitat de canviar tots els nodes de la WMN.

Una vegada s'han afegit les interfícies als nodes adequats, s'assignen els canals a les interfícies. Per fer-ho, s'utilitza una tècnica de colorejat de grafs [41] on cada canal es assignat de manera que dins el rang de cobertura del node hagi el major nombre de canals ortogonals seleccionats, minimitzant el nombre de col·lisions produïdes per la generació de tràfics, reduint al màxim les interferències degudes a la utilització del mateix canal a les interfícies dels diferents nodes que estan dins el mateix rang de cobertura.

## 4.5.2 Assignació dinàmica

### 4.5.2.1 Slotted Seeded Channel Hopping (SSCH)

La primera tècnica d'assignació dinàmica que presentarem és el **SSCH** [48]. La idea és similar i parteix, en certa mesura, d'una forma semblant a la que es proposa a altres solucions com **Multi-Channel Medium Access Control (MMAC)** [49]. L'objectiu dels autors és crear un protocol multicanal a nivell de capa d'enllaç amb capacitat per gestionar múltiples canals amb la major equitat possible, incrementant la capacitat total del sistema.

La idea principal de l'algorisme es fonamenta en aprofitar la canalització del medi assignant als paquets un canal d'una llista de canals ortogonals depenent de quin sigui el millor moment per aquesta transmissió. **SSCH** serà un protocol distribuït entre tots el nodes que coneixeran les planificacions de tots els seus nodes veïns, per tal de facilitar la **sincronització**. Per transmetre es divideix l'espai del temps en *slots* de 10ms on es visualitzarà en quin canal s'ha de fer la transmissió del paquet. Aquestes transmissions vendran determinades per una **planificació** dels canals a utilitzar i per una **assignació de prioritats** dels paquets.  Es realitza un *scheduling* dels paquets i un *scheduling* dels canals. Per fer això, **SSCH** es basa en IEEE 802.11 Distributed Coordination Function (DCF) [50] i, per tant, en l'arquitectura basada en els missatges RTS/CTS i a on, per transmetre s'ha de esperar un període DIFS en què el medi estigui buit.

Els requisits per suportar **SSCH** són els de tenir nodes amb una sola interfície ràdio, Network Interface Card (NIC) amb capacitat de funcionar a nivell de IEEE 802.11a i, per tant, de poder oferir fins a 13 canals ortogonals. Un altre requisit important és que el retard produït pel canvi de canal a la targeta *wireless* sigui baix (màxim de 80μs) ja que aquests canvis seran freqüents. Apart d'això, les NIC hauran de tenir la capacitat d'oferir cues de paquets de longitud considerable.

Com s'ha dit, el funcionament de **SSCH** es basa en dos factors principals: el ***scheduling de paquets*** i el ***scheduling de canals***. El de paquets es refereix al moment en què s'han de transmetre aquests paquets. Així, els paquets s'aniran emmagatzemant a cues FIFO generades pel node, i haurà una cua per cada veí del node en qüestió. Haurà un planificador que serà l'encarregat de donar permís per la transmissió dels paquets. En principi, el planificador actuarà com a *round-robin* però disminuirà la prioritat de les cues cada vegada que un dels paquets d'aquestes cues es perdi. D'aquesta forma **SSCH** ens assegura que no es malbaraten recursos.

Pel cas dels paquets *broadcast*, haurà d'arribar a totes les estacions i s'ha de tenir en compte que aquestes poden estar sintonitzades a canals diferents. En aquest cas s'opta per la retransmissió de paquets per tots els canals del sistema.

Un node amb **SSCH** pot descobrir que un veí pot estar a un altre canal quan no es rep cap CTS quan es transmet un RTS. En aquest cas el paquet es retingut a la cua del node fins que es rep el CTS corresponent. Aquest paquet serà descartat quan aquest veí no estigui en cap dels canals després d'haver completat un cicle del *scheduling* de canals.

L'altre aspecte bàsic de **SSCH** és el *scheduling* de canals que té per objectius bàsics el sincronisme de totes les estacions del sistema, que els nodes que no tinguin res a transmetre no interfereixin en els altres veïns i assegurar que tots els nodes estaran en contacte en algun moment de la comunicació. Per tant, es tracta d'idear un sistema que **reparteixi** els recursos d'una forma equitativa entre tots els nodes que conformen la xarxa. Per aconseguir això, s'utilitza una operació matemàtica molt simple (equació 18).

$$x_i \leftarrow (x_i + a_i) \bmod n$$

**Equació 18: Operació SSCH**

On $x_i$ és el canal en qüestió, $a_i$ és la llavor amb la que l'operació després determinarà el pròxim canal i $n$ serà el nombre de canals ortogonals. D'aquesta forma, s'aconsegueix que si tenim $n$ canals, en $n$ slots es podrà arribar a transmetre per tots els canals. A la figura 4.11 es té un exemple de com es fa servir aquesta idea. A l'exemple es tenen 3 canals ortogonals i depenent la dupla ($x_i$, $a_i$) anirem configurant el següent canal. A l'exemple es tenen dos *slots* per canal i es veuen dos casos (A i B) diferenciats amb diferents valors de llavor on es donen finalment els mateixos *slots* per a tots els canals però de forma diferent.

**Figura 4.11: Slots a SSCH**

Al final es té un *slot* de paritat que pot servir per sincronitzar les diferents estacions del sistema.

Els valors de llavor es poden canviar per obtenir un altre tipus de configuració. Això es fa per adaptar-se al medi transmissor de la millor manera possible.

#### 4.5.2.2  Hyacinth

**Hyacinth** [51] és una arquitectura WMN específica on es recullen una sèrie de tècniques noves que permeten millorar el rendiment de les xarxes multicanal. Al contrari del que Raniwala *et al.* proposaven a [43], on els algorismes proposats eren centralitzats des d'uns punts concrets, aquí parlem d'algorismes de naturalesa distribuïda. Els objectius clau de Hyacinth són:

- Un algorisme d'**assignació de canals distribuït** que permeten adaptar el tràfic d'una forma dinàmica.
- Un algorisme d'**encaminament basat en el balanceig de la càrrega** que sigui capaç d'adaptar-se als canvis de càrrega de tràfic i les fallades de la xarxa.
- La **implementació d'un prototipus** que demostri el guany que s'aconsegueix respecte al cas on s'utilitza un sol canal.

Per configurar l'arquitectura en qüestió es proposa un **algorisme d'encaminament** que també realitza l'assignació de canals. Aquest algorisme és **distribuït** i presenta una sèrie de particularitats.

Normalment les WMN s'adrecen cap a la xarxa cablada i, per tant, cadascuna de les WMN necessitarà un *gateway* per connectar-se a aquesta xarxa cablada. Aquest *gateway* serà l'origen de l'arbre expansiu que seguiran les rutes per arribar al medi cablat. Aquest element actuarà de pare de la resta de topologia i anirà anunciant als seus veïns sense fils els costos (dependrà de la mètrica usada) que suposarà arribar a la xarxa cablada amb un missatge ADVERTISE. Quan els nodes sense fils reben aquest ADVERTISE, afegeixen la ruta del pare si no la tenen i, si la tenen, actualitzen les mètriques. Tot seguit envien un missatge JOIN al node pare que afegirà la destinació a la llista dels fills. Al cas de què el node fill tingués com a *gateway* a un altre node amb pitjor cost, aquesta estació enviaria un missatge LEAVE cap al seu antic *gateway* i se sumaria al que té millor resultat amb un missatge JOIN. D'aquesta manera s'anirien configurant les taules de rutes de tots els nodes involucrats de la xarxa (veure figura 4.12).

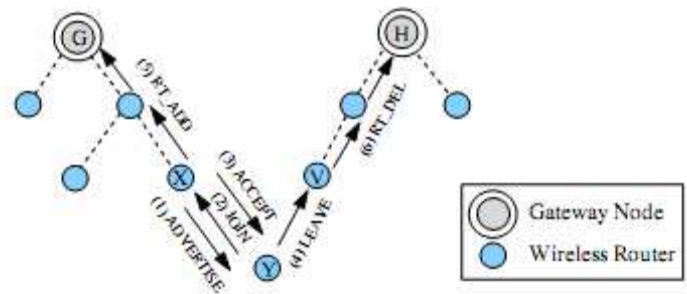

**Figura 4.12: Operacions de Hyacinth**

La **mètrica** utilitzada pel càlcul de costos pot ser variada. En aquest cas d'estudi s'han contemplat tres opcions. Una primera que seria una mètrica que compta els salts que s'han de fer per arribar a dites estacions. En canvi, la segona mètrica estudiada és la de conformar els costos depenent de la capacitat de l'enllaç que va del *gateway* a la xarxa cablada. La raó per elegir aquest camí per la mètrica és perquè suposem que aquest enllaç és el coll de botella de la xarxa. Finalment, es proposa una tercera mètrica que calcula els costos a partir de la capacitat del camí, és a dir, qualsevol constituent de la xarxa pot ocasionar els colls de botella.

En quant a la part d'**assignació de canals**, a Hyacinth s'utilitza una opció distribuïda. És un problema que es pot subdivir en dues parts: una primera que seria l'assignació d'interfície a un veí i una segona on se li fixaria un radiocanal a dita interfície.

El principal problema que tenim és que quan es comuniquen dos interfícies de diferents nodes han d'estar sintonitzats a una mateixa freqüència. Al ser un mètode distribuït, hi ha una gran dificultat per gestionar com s'han de canviar els canals depenent les condicions de l'entorn. Per fer front a aquest problema, els autors han ideat un sistema on s'utilitzi la **jerarquia** de pares-fills creada amb el protocol d'encaminament. D'aquesta manera, el que es faria seria assignar una sèrie d'interfícies com a UP-NIC i unes altres com a DOWN-NIC. Les UP-NIC es comunicarien amb les DOWN-NIC del node que està a damunt a la jerarquia, essent la responsabilitat d'assignar un canal del node de jerarquia superior. D'aquesta manera s'asseguren de què hagi el mateix canal entre aquests veïns.

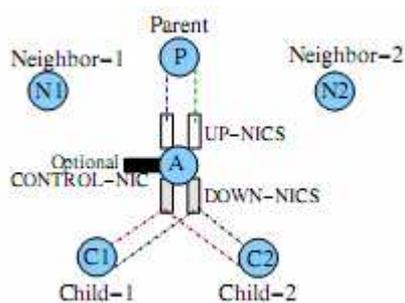

**Figura 4.13: Jerarquia de Hyacinth**

L'assignació de la freqüència a la interfície es fa controlant la utilització dels canals amb missatges CHNL_USAGE que porten un nombre on es reflecteix l'ús que es fa dels diferents canals. La freqüència serà assignada al canal que tingui un valor de CHNL_USAGE més baix.

Una altra particularitat de Hyacinth és que en cas de produir-se algun problema amb els nodes o enllaços que constitueixen la xarxa, l'arquitectura es reorganitza novament seguint els processos acabats d'esmentar.

En resum, estem davant d'una **tècnica multicanal distribuïda** entre tots els nodes, a diferència de la solució centralitzada proposada pels mateixos autors a Raniwala *et al.* [43] i que permet encaminar els paquets per múltiples canals i múltiples interfícies.

### 4.5.2.3   Encaminament basat en balanceig de càrrega

Aquesta és una tècnica d'encaminament proposada a So i Vaidya [52] i desenvolupada amb més detall a un altre article de So i Vaidya [53]. L'objectiu és el d'utilitzar un entorn multicanal, on tenim un cert nombre de AP connectats a la xarxa sense fils, que donen cobertura a un nombre determinat d'estacions, on la xarxa sigui capaç de balancejar la càrrega entre els diversos nodes i així, poder augmentar la capacitat de la xarxa ja que s'eliminarien interferències fruit de la utilització dels mateixos canals als diferents enllaços.

La tècnica està pensada en un inici per nodes amb una sola interfície, però que suporten multicanal i una velocitat de canvi de canal elevada.

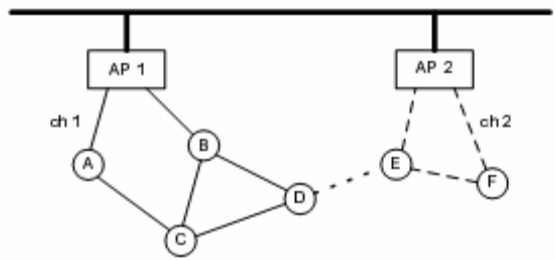

**Figura 4.14: Xarxa a balancejar**

La figura 4.14 ens exemplifica el tipus de sistema del què es pot treure profit balancejant càrrega. Utilitzant dos canals i, depenent de la càrrega dels enllaços es poden sintonitzar més estacions a un o altre canal. L'objectiu serà que la càrrega als dos canals estigui equilibrada aconseguint una capacitat similar per les dues freqüències.

L'algorisme que realitza el **balanceig de la càrrega** consta de passos similars als que realitzen els algorismes d'encaminament en MANET. Primerament, es realitza el procés d'establiment de rutes cap al punt d'accés. Per fer aquest descobriment de rutes, els diferents nodes fan un *broadcast* d'un missatge SCAN, que s'envia per tots els canals de la xarxa. D'aquesta forma ens assegurem d'arribar als Access Point (AP) del sistema. L'AP sintonitzat a un cert canal contestarà de forma *unicast* al node, realitzant-se l'associació del node amb el punt d'accés. Cada node mantindrà una taula d'encaminament composta per un conjunt de camps com: següent salt, salts a l'AP, tipus, canal, camí...

El sistema pot estar subjecte a modificacions de càrrega i, per tant, les taules dels nodes necessitaran actualitzar-se, podent-se donar el cas de què es requereixin canvis en la sintonització dels canals als diferents enllaços. Cada node haurà d'anar mesurant de forma continuada la **càrrega** del sistema. En aquest protocol s'empra una càrrega amb pesos, depenent dels salts als que s'està del AP. La informació sobre la càrrega a la subxarxa que comprèn l'AP, serà distribuïda mitjançant missatges HELLO als altres nodes, que agafaran aquesta informació calculant les mitjanes ponderades i actualitzant les seves taules de rutes.

Quan es tenen recopilades totes aquestes dades, els nodes estaran en disposició de prendre les seves decisions en quan a selecció del millor arbre de rutes per accedir a un dels AP. En cas de canvi per part d'aquest node, es comunica amb missatges SWITCH a tots els nodes fills del node el seu canvi de canal i, després s'enviaran els conseqüents missatges d'associació per ajuntar-se amb els nodes i l'AP que funcionaven a sobre de la segona subxarxa.

D'aquesta manera s'aconsegueix que, en cas de què hagi una gran quantitat de nodes sintonitzats a un dels canals, es pugui adreçar informació cap a un altre canal de la subxarxa. Així, a priori, s'arriba a tenir un major rendiment del sistema en global.

### 4.5.2.4   Multi Channel Extremely Opportunistic Routing (MCExOR)

Aquest protocol neix amb la idea d'adaptar a l'entorn multicanal l'innovador protocol Extremely Opportunistic Routing (ExOR) [54], que era un protocol d'encaminament per xarxes ad-hoc però que només actuava sobre un sol canal. És un protocol que no està basat en salts tal com ho estan la majoria d'altres protocols d'encaminamet (AODV, OLSR, DSR...), sinó que tracta d'aprofitar les qualitats del medi aeri. Els autors proposen que els algorismes esmentats utilitzen un mode de funcionament molt similar al que empren els protocols que actuen damunt medis cablats, és a dir, s'utilitza la mesura de salts a l'igual que amb una xarxa cablada. A ExOR tracten d'aprofitar que es pot accedir a nodes llunyans sense necessitat de realitzar salts per accedir a un determinat node. Així, mentre protocols com AODV té una ruta preestablerta amb les estacions per les quals s'ha de passar, ExOR té un conjunt de candidats als que es pot passar els paquets per tal d'arribar a la destinació. Es tracta d'aplicar un nou concepte de configurar les rutes i d'utilitzar-les.

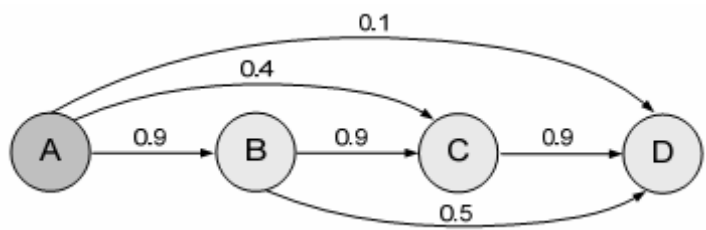

**Figura 4.15: Mètrica MCExOR**

Un exemple de les possibles rutes que es poden tenir, es pot veure a la topologia que mostra la figura 4.15. Amb ExOR es **contemplen totes les possibles rutes** per arribar a destí. Les quantitats que apareixen fan referència a la possibilitat de què una transmissió sigui reeixida.

Amb **MCExOR** [55] es tracta d'adaptar l'ExOR a un entorn multicanal amb intenció d'augmentar la seva capacitat. El protocol està dissenyat per aplicar-ho damunt estacions amb una sola interfície. A l'igual que al **Multi-channel OLSR** [56], i contràriament a **Multi Channel Routing Protocol (MCRP)** [57], a MCExOR s'opta per una tècnica d'**assignació de canal a nodes**. Per tant, per transmetre s'haurà de sintonitzar la targeta al canal de la destinació.

El primer dels passos que realitza **MCExOR** és el de l'assignació de canals als nodes. El canal que es sol assignar serà aleatori. També cap la possibilitat de què **MCExOR** assigni als nodes els canals menys utilitzats per tal de minimitzar la influència dels nodes veïns. Per aconseguir això, es fan mesures de la qualitat dels enllaços, seleccionant el canal d'acord amb aquestes mesures. Aquestes mesures es poden repetir en el temps ja que les condicions del sistema poden canviar donant resultats completament diferents i que poden requerir canvis de canal als nodes.

Quan es tenen tots els canals assignats, es passa a fer el descobriment de tots els canals que estan dins el rang del node. Aquest procés es realitza enviant un *broadcast* cap a tots els nodes en tots els canals dels què disposem i aquests contesten amb un paquet cap a l'origen. L'origen, al rebre aquest reconeixement, calcula la mètrica (molt semblant a ETX [38]) que es necessita per arribar a cadascun dels veïns dels que s'ha rebut resposta.

Amb el coneixement de veïns es passa a calcular les rutes. **MCExOR** dóna l'opció de realitzar aquesta tasca de dues maneres possibles: de forma proactiva o de forma reactiva. Cadascuna de les tècniques és semblant a com es fa en els seus protocols (proactius com l'OLSR, amb rutes més temps establertes, i reactius com AODV, creant rutes sota demanda) amb la diferència de què els *broadcast* es fan sobre múltiples canals, fent servir la mètrica calculada i possibilitant que els RREP tornin seguint la tècnica explicada a ExOR, és a dir, sense necessitat de seguir el mateix camí que ha seguit RREQ. Aquests missatges proporcionaran informació als veïns sobre el canal que estan emprant en aquells moments.

Finalment, quan es tenen totes les rutes configurades, es fa l'enviament de paquets. Per realitzar aquest enviament, es selecciona un conjunt de nodes que seran aptes per transmetre per cadascun dels diferents canals que tenim i, després es selecciona aquell conjunt de candidats més prometedor a l'hora de realitzar la transmissió pertinent. Els algorismes que s'empren per configurar primer els conjunts de candidats i, després, per seleccionar millor els candidats es poden veure a l'article de **MCExOR** [55].

## 4.5.3  Assignació híbrida

### 4.5.3.1  Multi-Channel Medium Access Control (MMAC)

**MMAC** [49] es basa en altres tècniques anteriors presentades per Wu *et al.* [58] i a on es descrivia una nova forma d'assignació de canals anomenada **Dynamic Channel Assignment (DCA)**. En aquesta tècnica es mantenia constantment un canal dedicat per transmetre informació de control, mentre que els altres canals estaven destinats a la transmissió de dades. El canal dedicat a control serà l'encarregat de realitzar l'elecció del canal pel qual es transmetran llavors les dades, cosa que es decidirà mitjançant un procés d'intercanvi entre el node emissor i el receptor i que correspondria a l'intercanvi de missatges RTS/CTS. A diferència de **MMAC** aquí sí que s'utilitzaven més d'un transceptor per node. L'objectiu de **MMAC** és adaptar les funcionalitats de **DCA** a un sistema amb una única interfície.

Com ja s'ha comentat en aquest document en alguna ocasió, un dels problemes principals que ens trobem a un entorn multicanal és la visibilitat que tenen els nodes veïns entre ells. Aquests nodes hauran de compartir el mateix canal per poder comunicar-se i, per això, abans de realitzar qualsevol comunicació, es posen d'acord en quin canal s'ha de fer la transmissión de dades. Un dels problemes que pot aparèixer és el de terminal multicanal ocult. Es produeix quan dos nodes es posen d'acord en la utilització d'un canal, mentre que altres dos estan trasmetent en aquest mateix canal, sense adonar-se'n que l'altre parell ho fa en la mateixa freqüència. Això provoca una col·lisió de dades, degradant la capacitat del sistema. El cas comentat es pot apreciar la figura adjuntada i a on es veu que es produeix una col·lisió deguda a que el node A no coneix el canal del node C, sintonitzat a una altra freqüència i que impossibilita la comunicació (figura 4.16). La superació d'aquest problema és un dels objectius principals de **MMAC**.

El **MMAC** realitzarà modificacions al funcionament normal de IEEE 802.11 DCF [50] i de IEEE 802.11 Power Saving Mechanism (PSM) [49] per resoldre aquestes dificultats.

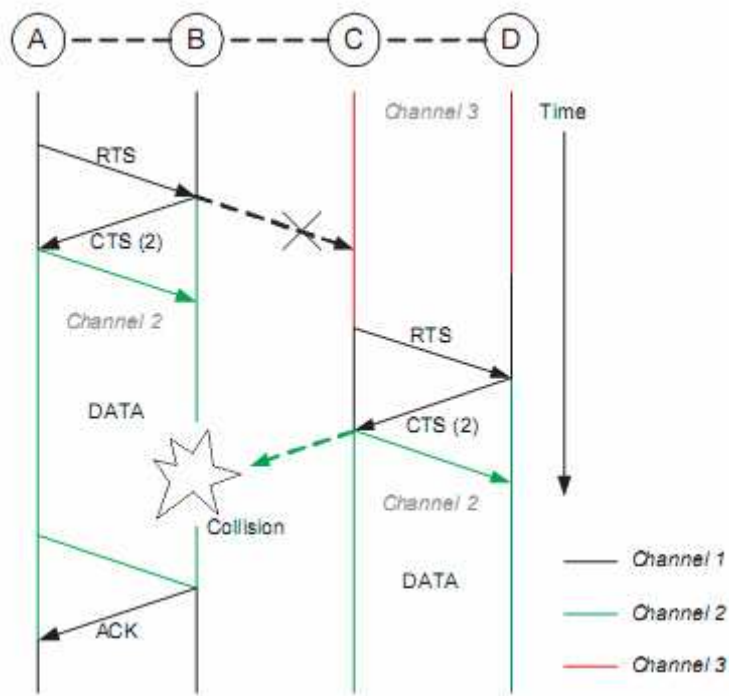

Figura 4.16: Diàleg entre estacions a MMAC

El funcionament de **MMAC** es basarà en la utilització de les finestres Ad-hoc Traffic Indication Messages (ATIM) que utilitza IEEE 802.11 PSM basant-se sempre en el funcionament de **DCA**. Així durant l'interval de temps que ocupa la finestra ATIM, totes les estacions han d'escoltar a un canal de control que serà el mateix per totes les estacions i que haurà estat establert amb la configuració del sistema. Les estacions que volen transmetre paquets ho fan en el canal de control definit per defecte i transmetran un paquet ATIM cap a la destinació desitjada incloent-hi al paquet la **llista de canals preferits** (PCL). Quan la destinació rep aquesta llista decideix quins és el millor dels canals possibles i respon cap a la destinació amb un ATIM-ACK que portarà quina ha estat la selecció de canal realitzada. Quan l'origen rep aquesta confirmació aquest envia un paquet ATIM-Reservation per confirmar que es pot utilitzar aquell canal definitivament. Aquest missatge també servirà per actualitzar l'estat dels canals als diferents Preferable Channel List (PCL). Després d'això, les estacions implicades en la comunicació, sintonitzen el canal acordat i s'inicia la transmissió de dades habitual que marca l'IEEE 802.11 DCF amb els cicles RTS/CTS emprats pel protocol. Els canals que apareixeran al

PCL són ortogonals i, per tant, no haurà cap interferència entre canals diferents a una altra banda de la xarxa. En cas de què no hagués hagut confirmació, les estacions haurien d'esperar al següent interval *beacon* per poder transmetre.

Per evitar **col·lisions** s'utilitza una tècnica molt similar al Carrier Sense Multiple Access – Collision Avoidance (CSMA-CA), esperant un interval aleatori per poder transmetre el paquet ATIM.

El criteri de **selecció de canal** es realitza a les dues estacions, que configuren la seva pròpia PCL. Cadascun dels canals apareixerà a aquesta llista amb un nivell de prioritats que poden ser de tres valors diferents: HIGH, MID o LOW. La decisió es prendrà tenint en compte aquestes prioritats, donant una major importància a la prioritat marcada pel receptor del missatge.

### 4.5.3.2   Multi Channel Routing Protocol (MCRP)

**MCRP** és un protocol d'encaminament per xarxes mallades sense fils que és multicanal i que suporta múltiples interfícies. El protocol ha estat desenvolupat per la University of Illinois dins del projecte Net-X [59], que estudia solucions per xarxes WMN. També es poden veure les particularitats del sistema Net-X a l'article de Kyasanur i Vaidya [40].

Inicialment **MCRP** era un protocol multicanal pensat per transmetre a sobre d'un sol transceptor i, per tant, utilitzant una sola interfície [53]. Aquí ja es varen proposar les línies a seguir per **MCRP** i que després han desembocat en el desenvolupament d'un **protocol multicanal i de múltiples interfícies** que es resumirà a l'apartat. Per arribar a la solució proposada pel grup, s'han fet diversos estudis per demostrar per què s'ha elegit la implementació d'aquest protocol i quins són els factors que han conduit al MCRP [57, 60].

El sistema tracta de xarxes amb estacions que contenen múltiples interfícies (m interfícies) i un nombre superior de canals (c canals). Per tant, els autors ideen una tècnica que permeti una eficient utilització dels recursos en un entorn on **m << c**.

Per fer-ho es tenen en compte varis aspectes. Primer de tot, un algorisme de **selecció d'interfícies**, després una **assignació de canals** damunt d'aquestes interfícies i finalment, la creació d'una nova **mètrica** que permeti configurar rutes en un entorn multicanal de forma més eficient.

Per a la selecció d'interfícies a **MCRP** s'utilitza un protocol de capa d'enllaç per facilitat la interacció amb aquestes estructures ja que permet que no s'hagin de sincronitzar el transmissor i el receptor (no com a **SSCH** [48]) i la utilització senzilla del conjunt de canals, podent-se escalar si aquest nombre augmenta.

El protocol fa una diferenciació entre dos tipus d'interfícies: les fixes i les configurables. Les interfícies fixes són aquelles que tenen un canal assignat durant un període de temps llarg, mentre que les altres van canviant ràpidament de canal depenent de les necessitats del sistema. Per a la interfície fixa s'elegeix el millor dels canals i això es fa consultat una llista anomenada *ChannelUsageList*, que retorna dades de qualitat dels diferents canals. A més, cada node tindrà un registre anomenat *NeighborTable* que contendrà els canals fixes utilitzats per cadascun dels veïns. Les interfícies fixes podran canviar de canal quan el *ChannelUsageList* supera un cert llindar. En aquest cas es procedeix al canvi de canal en la interfície fixa comunicant-ho a l'entorn.

L'actualització d'aquestes taules es pot fer mitjançant un *broadcast* cap a tots els veïns. En el cas d'aquesta tècnica s'ha configurat una component d'aquest tipus per tal de poder fer transmissions *broadcast* per tots els canals de la xarxa. A més la component *broadcast* serà necessària a l'hora de descobrir rutes en el protocol d'encaminament.

Les interfícies configurables seran les que ens permetran comunicar-nos amb altres interfícies fixes o interfícies configurables d'altres nodes veïns. D'aquesta manera, els nodes es poden connectar amb qualsevol canal sintonitzat.

Per a la transmissió *unicast* de paquets, es creen cues per cadascun dels canals que es poden manejar de forma adient amb el protocol que s'està usant. Aquestes cues es van buidant quan es serveixen per les interfícies. La interfície fixa va servint a mesura que li van arribant transmissions mentre que per servir les cues dels altres canals es fa un *scheduling* de tal manera que no es serveixi una cua que estigui buida i que una d'aquestes cues no es serveixi durant un temps major a *MaxSwitchTime*. La descripció respecte a les interfícies i les cues s'exemplifica amb la figura 4.17.

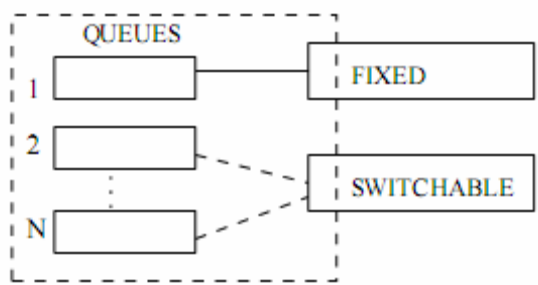

**Figura 4.17: Tipus de cues**

La següent passa de la solució integrada, proposada a l'article de **MCRP** [57], és la creació d'una eina que sigui capaç de trobar les millors rutes cap a les destinacions. La primera idea és la d'utilitzar un protocol d'encaminament **sota demanda** tipus Ad-hoc On-Demand Distance Vector (AODV) [22] o Dynamic Source Routing (DSR) [14], però que estigui adaptat a un entorn amb múltiples canals i múltiples interfícies i que, per tant, tingui en compte el cost que significa fer canvis de canals i d'interfícies. El que es fa és utilitzar la mètrica **Weighted Cumulative Expected Transmission Time (WCETT)** [39] ja explicada a l'apartat 4.3.2.2 i adaptar-la afegint-hi la variable dels cost de canviar el canal (*switching cost*). D'aquesta manera es calculen els paràmetres d'**Expected Transmission Count (ETX)** [38], el paràmetre de diversitat i finalment el *switching cost*. Ajuntant tota l'expressió es tindria que la mètrica Multi Channel Routing (MCR) seria la següent:

$$MCR = (1 - \beta) \cdot \sum_{i=1}^{n} (ETT_i + SC(c_i)) + \beta \cdot \max_{1 \le j \le c} X_j$$

**Equació 19: Mètrica MCR**

On SC vendria a representar el cost de canvi de canal i que al seu mateix vendria donat per l'expressió 19 que es descriu a continuació:

$$SC(j) = p_s(j) \cdot switchingDelay$$

**Equació 20: Switching Cost**

$$p_s(j) = \sum_{\forall i \ne j} InterfaceUsage(i)$$

**Equació 21: Ús de les interfícies**

El paràmetre *InterfaceUsage* marca el temps que la interfície està ocupada mentre que *switchingDelay* és la latència del canvi d'interfície estimada quan no està en execució el programa. D'aquesta manera ja es pot confeccionar la mètrica a emprar pels nodes de la WMN.

Una vegada es tenen les mètriques desenvolupades es pot passar a calcular les rutes òptimes per accedir als diferents nodes. La sistemàtica utilitzada seria molt semblant a la que es fa servir a DSR. Primer s'envia un RREQ damunt tots els canals que és enviat que aniran actualitzant es seu cost a mesura que van passant pels diferents nodes fins que arriben al destí. Quan la destinació rep el RREQ, contesta amb un RREP de forma *unicast* cap al node origen. El node destinatari pot rebre més d'un missatge RREQ i seleccionarà aquell que la seva utilització suposi un menor cost. El protocol també recull altres funcionalitats semblants a DSR, com el RERR que s'aplica en cas de fallada d'algun component de la xarxa.

Com s'ha vist la solució completa està constituïda per un element de capa d'enllaç i per un altre a nivell d'usuari. Per fer-ho els autors ens mostren el camí que han seguit per la seva **implementació** pràctica i posterior comprovació [61].

El primer que cal en aquests casos per poder emprar la solució presentada és tenir un sistema adient amb un kernel que suporti el funcionament multicanal que permeti fer canvis de canal i un *driver* que permeti aquests canvis. El sistema haurà de permetre que es pugui fer *broadcast* a tots els veïns de la xarxa i que es pugui fer canvis d'interfícies.

El problema de la utilització d'interfícies es resol amb una tècnica que permet seleccionar a cada moment la interfície per la qual es vol transmetre. Aquesta tècnica s'anomena *bonding* [62] i el que fa es crear una interfície virtual damunt de totes les interfícies reals. Un major detall del seu funcionament serà descrit més endavant. Així es pot crear un mòdul que separi tot el que està relacionat amb canals de la resta de l'algorisme. Es crearia el que ells anomenen ***Channel Abstraction Module***, i tot això s'englobaria com a un mòdul del kernel.

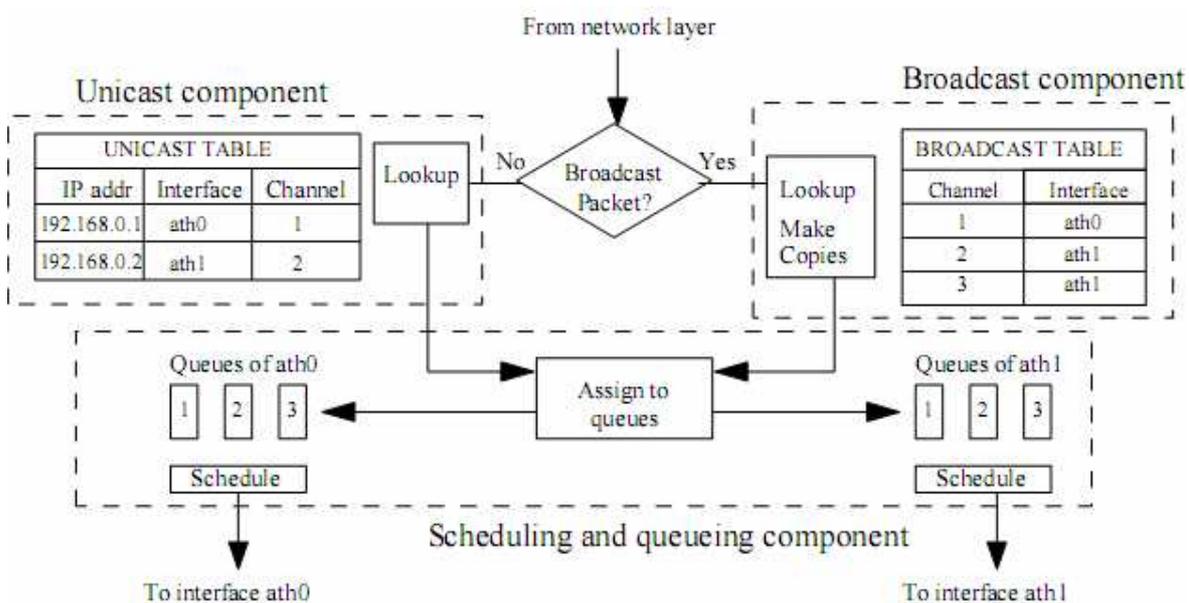

**Figura 4.18: Procés complet de MCRP**

A la figura 4.18 es poden veure tots els component d'aquest *Channel Abstraction Module*, així com el seu funcionament general i interrelacionat.

En canvi, la part de l'algorisme d'**encaminament** quedaria per damunt del mòdul de canal dissenyat i quedaria emplaçat a l'espai d'usuari podent-se utilitzar normalment les crides Input/Output Control (IOCTL) i crides al sistema habituals. El diagrama resumit d'aquesta implementació és el que es presenta a la figura 4.19.

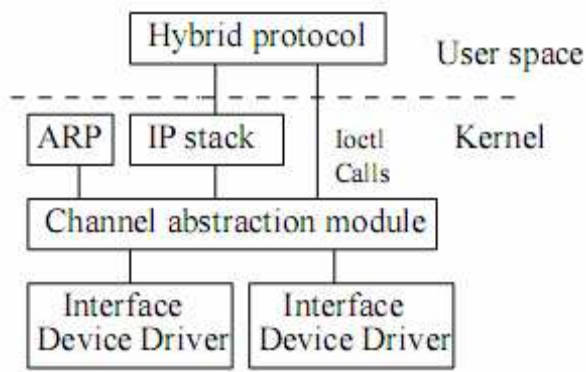

**Figura 4.19: Diagrama de blocs de MCRP**

A més, per configurar del tot la solució s'empra un *driver* anomenat *madwifi* [64] que permet modificar una sèrie de paràmetres relacionats amb el temps de **canvi de canal** i

d'**associació** d'aquests canals, així com d'altres canvis per suportar millor el *scheduling*. D'aquesta manera es pot guanyar en rendiment realitzant aquestes modificacions directament damunt el codi del *driver*.

### 4.5.3.3  Multi-channel Optimized Link State Routing

Una altra possibilitat per augmentar la capacitat dels entorns multicanal és utilitzant tècniques on es modifiquin protocols d'encaminament que tinguin en compte els canals que utilitzen els veïns del node que vol realitzar la comunicació. Un exemple d'aquest tipus de tècniques és el **Multi-channel OLSR** [56], on el que es fa bàsicament és afegir funcionalitats al comportament normal del protocol per tal de treure major profit de l'entorn multicanal.

Per suportar aquesta modificació d'OLSR es necessiten nodes amb més d'una interfície, una d'elles serà de control, mentre que les restants s'utilitzaran per transmetre dades. Aquesta interfície de control serà dedicada, a diferència de protocols com **Slotted Seeded Channel Hopping (SSCH)** [48] on les dades d'informació i control circulen per les mateixes interfícies, i per ella s'intercanviaran tots els missatges que permetran controlar els canals per on s'ha de transmetre. Aquesta interfície de control tindrà sintonitzat el mateix canal a totes les estacions per tal de facilitar l'enteniment entre elles.

En quant al tema d'**assignació de canals**, una gran part de les tècniques utilitzen una assignació per flux on emissor, receptor i nodes intermedis per on circula el paquet comparteixen el mateix canal com és el cas de **Multi Channel Routing Protocol (MCRP)** [57], però a **Multi-channel OLSR** s'utilitza un mètode d'assignació per node anomenat **Receiver-based Channel Assignment (RCA)**. Amb RCA s'assigna a tots els nodes dins el rang de la pròpia estació un canal diferent, és a dir, a una topologia en línia de tres enllaços s'assignarien tres canals diferents. Per realitzar la comunicació es realitza un canvi de canal cap a la destinació que és coneguda pel node gràcies a l'intercanvi de missatges de senyalització del protocol **OLSR** pel canal de senyalització dedicat.

Com s'ha dit, el protocol és una modificació d'**OLSR**. **OLSR** és necessari per crear l'arbre de rutes entre els nodes. S'utilitzarà aquesta interfície de control que porten les estacions per crear-lo i per intercanviar missatges HELLO entre ells. Per afegir les funcionalitats de multicanal i que, per tant, tots els nodes coneguin els canals dels veïns aquests missatges HELLO són modificats, afegint-li alguns camps nous. Les modificacions a un missatge HELLO es poden observar a la figura 4.20. Els nous camps estan marcats en negreta.

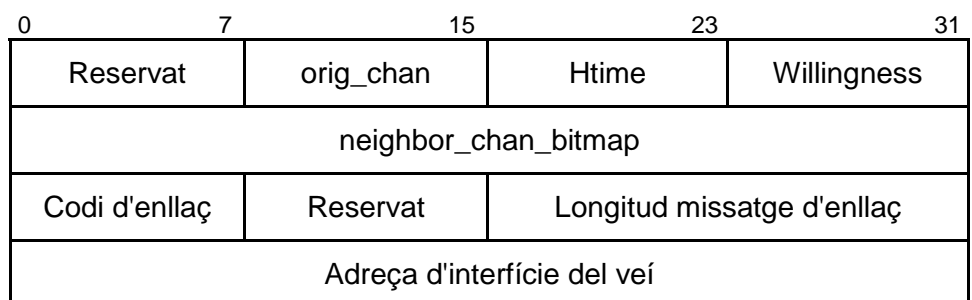

**Figura 4.20: OLSR Multicanal**

Les modificacions realitzades a un missatge HELLO estàndard són les següents.

- El camp **Reservat** veu reduïda la seva longitud a la meitat respecte a l'original
- Aquesta meitat és ocupada per un camp anomenat **orig_chan** que contendrà el valor del canal sintonitzat a aquell moment.
- S'afegeix el camp **neighbor_chan_bitmap** de 32 bits i que conté tota la informació sobre com estan ocupats els canals amb 2 bits d'informació per cada canal.

El significat d'aquests dos bits és pot veure a la taula 4.2.

| 2 bits de status | Funció |
|---|---|
| 00 | Canal lliure |
| 01 | Usat per un node veí |
| 11 | Usat per múltiples veïns |

**Taula 4.2: Bits de status**

També s'estén la **neighborTuple** d'**OLSR** on hi ha les entrades d'aquests veïns. S'afegeixen dos camps equivalents als que s'han creat al missatge HELLO: el **N_neighbor_channel** que conté el canal del veí i el **N_2hop_channel_bitmap** que conté els *status* dels seus respectius veïns.

Aplicant aquestes modificacions l'algorisme d'assignació de canal pot ser utilitzat ja que els intercanvis per la interfície de control arribaran a tots els nodes de la xarxa. El criteri de selecció serà el d'elegir un dels canals lliure. En cas de què no hagi cap, l'estació roman al mateix canal. Com al protocol es postula que hagin dues interfícies de dades (una de transmissió i una altra de recepció), el que es fa és canviar el canal de recepció segons les informacions transmeses pels HELLO, mentre que la transmissió sempre canviaria al pròxim canal de receptor veí i que estaria inserit en un dels nous camps de la **neighborTuple**. Aquest és un dels punts on poden aparèixer els majors problemes, ja que tot dependrà de la capacitat del sistema per canviar el canal d'una forma el suficientment ràpida perquè el sistema resulti profitós cosa que dependrà del *hardware* a emprar.

# 5 DISSENY DEL SISTEMA

A aquest apartat es detallarà quina solució s'ha dissenyat i quin són els passos que s'han seguit per arribar a la solució definitiva. Després d'haver estudiat quines propostes hi ha sobre els protocols multicanal i de veure una base tecnològica suficient, es passaran a relatar com alguns d'aquests conceptes s'han pogut aprofitar en el desenvolupament de del projecte, quines alternatives de projecte han sorgit en el procés de creació i quins factors han estat els que ens han conduit a arribar fins al nostre **protocol multicanal** escollit.

També es comentaran aspectes de la implementació com són les plataformes que s'han fet servir per realitzar el disseny del prototipus i les proves del sistema a més de problemes que han sorgit durant el temps que el projecte ha estat en procés de desenvolupament.

## 5.1 PLATAFORMA D'IMPLEMENTACIÓ

Abans de començar a explicar quin ha estat el disseny a seguir, es presentarà quina ha estat la plataforma que ens ha permès la creació del nostre prototipus i quines utilitats s'han fet servir per aconseguir realitzar el disseny complet.

### 5.1.1 4G AccessCube

L'eina fonamental utilitzada és el nomenat **4G AccessCube** o simplement **Meshcube** [65] i està desenvolupat per la companyia 4G Systems. És bàsicament un router *mesh* que permet configurar escenaris diferents de xarxes *mesh* i ens permetrà provar en temps real el treball implementat.

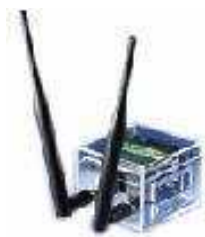

**Figura 5.1: 4G Access Cube**

El **4G AccessCube** porta una versió de Linux Debian encastada que s'anomena **Nylon**. Aquesta versió de Linux no té totes les funcionalitats aportades per una versió estàndard de Debian i equival a una versió del kernel de Linux 2.4 (al nostre cas, la 2.4.27). Per tant, estaríem davant d'una versió de Linux simplificada, ja que l'eina té una capacitat limitada que no pot suportar tot el processament que un ordenador normal sí que podria processar amb l'execució d'un sistema operatiu Debian corrent degut a que el seu processador és més limitat i que compta amb molta menys memòria que les màquines actuals.

Per tant, en principi, el sistema Nylon tindrà un mínim d'utilitats instal·lades al 4G AccessCube i sempre que es necessiti alguna aplicació a l'estació *mesh*, aquesta haurà de ser instal·lada externament. Per fer-ho, es pot consultar a la web[2] [65] per veure si apareixen els paquets desitjats, ja compilats i preparats per la instal·lació a l'AccessCube. Són paquets compatibles amb la versió Nylon utilitzada i hi ha comandes específiques per poder procedir a la instal·lació una vegada que aquests paquets han estat copiats dins l'eina.

De totes formes, serà molt freqüent que l'aplicació que vulguem emprar no aparegui entre els paquets preparats per la descàrrega. Per poder utilitzar aquest paquet en l'estació, s'haurà de utilitzar un procés anomenat **compilació creuada** (*cross-compiling*). Per poder emprar aquesta compilació, és necessari descarregar a l'ordenador personal una aplicació de desenvolupament anomenada *Development Build System* i que és el que portarà aquest compilador específic que serà el que permetrà obtenir executables compatibles amb l'arquitectura *MIPSEL* que és la que es fa servir a les estacions. S'ha de tenir en compte que l'aplicació, a l'hora de fer la compilació, utilitzarà els seus propis recursos, entenent per això, com les seves llibreries, variables, que poden ser diferents a les emprades per un compilador rutinari *gcc* que crea executables per una versió estàndard de Linux.

---

[2] Des del mes de novembre de 2007 la web ha deixat de ser operativa, amb la conseqüència de què ja no es pot accedir als repositoris i funcionalitats ofertes pels *meshcube*.

D'aquest tipus d'estacions hi han de dos tipus diferents: uns de longitud més petita (figura 5.2) i uns altres de més majors (figura 5.1). Encara que, en principi, els dos elements realitzen les mateixes funcions, hi ha certes diferències, ja que els AccessCube petits tenen una menor capacitat de processament i no són capaços de dissipar la mateixa calor generada internament que els grans, resultant en un retardament de les seves operacions. Aquest fet ha provocat que hagin alguns paquets específics per les estacions petites. En aquest projecte, a la majoria de situacions s'han emprat estacions grans.

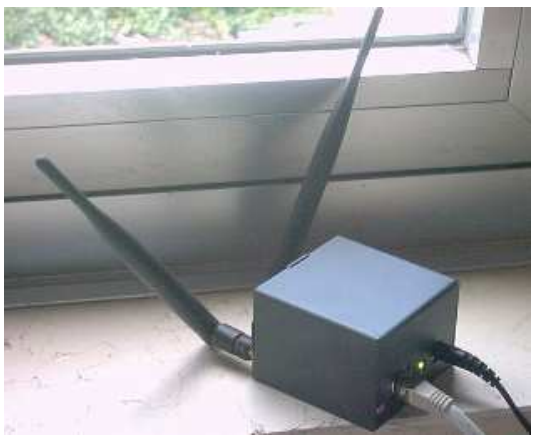

**Figura 5.2: 4G Access Cube petit**

La raó d'aquest ús no és només pel seu millor rendiment, sinó que, té molta importància la quantitat d'espai disponible per la inserció de noves targetes. Inicialment, les estacions duen dues targetes per xarxes sense fils amb dues antenes que actuen com a transmissor. Els AccessCube grans ens donen opció d'augmentar aquest nombre de targetes *wireless*, amb la possibilitat d'incorporar més transmissors. El tenir múltiples interfícies pot ser interessant al treball i d'aquí la raó principal per decantar-se per aquestes estacions. Les especificacions *hardware* i *software* més detallades del **4G AccessCube** es poden veure a l'annex A.1.

## 5.1.2  *Drivers wireless*

Els **4G AccessCube** habitualment compten amb dues interfícies ràdio que són les que permetran configurar les WMN desitjades i poder realitzar els experiments ideats. Com totes les interfícies ràdio que empren un *hardware* determinat per realitzar aquesta comunicació, es necessitarà un **driver** que ens permeti poder fer-lo utilitzable pel sistema operatiu de torn. En el nostre cas, els *drivers* que possibilitaran que les interfícies ràdio es puguin manejar de la millor manera possible seran el **madwifi** [64] i el **hostap** [66].

*Madwifi* és un projecte *open-source* en continu canvi i que presenta unes característiques bastant avançades. Les seves targetes suporten les versions d'IEEE 802.11a i d'IEEE 802.11b. Les interfícies sobre *Madwifi* s'anomenen com athX, on X és el número de interfície assignada.

*Hostap* és també un *driver* que aporta característiques privades interessants, tot i que les targetes emprades sobre *Hostap* en aquest projecte no suporten una altra versió que no sigui IEEE 802.11b. En aquest cas, les targetes seran anomenades al AccessCube com a wlanX. Més endavant s'explicarà més a fons quines són les causes d'emprar aquests tipus de *drivers.*

# 5.2  PROPOSTA INICIAL

La proposta inicial del projecte ha anat enfocada a superar els problemes que apareixen fruit de la utilització d'un entorn a un sol canal. Per això el disseny contempla un sistema amb el que siguem capaços d'aprofitar l'espectre multicanal, o com a mínim, que millori les mancances de l'entorn d'únic canal. Dit això i abans de començar la implementació, s'arriba a la conclusió de què per aconseguir una millora en les comunicacions de les xarxes *mesh*, farien faltes les següents tasques (figura 5.3).

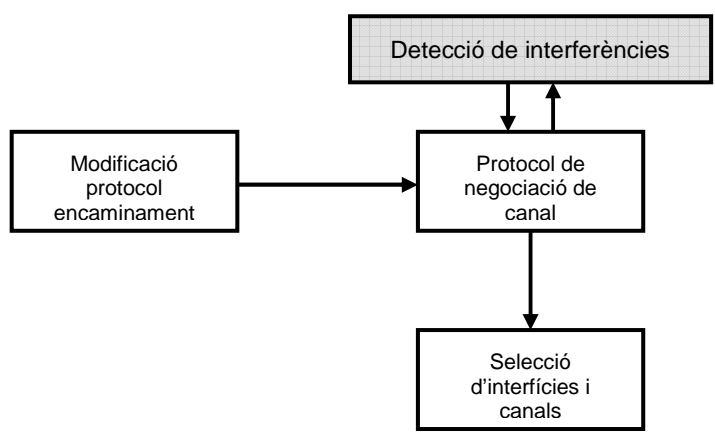

**Figura 5.3: Diagrama de blocs del sistema**

- Modificació del **protocol d'encaminament** per tal d'adaptar-lo a la utilització sobre múltiples interfícies i múltiples canals. El protocol seleccionat per realitzar aquesta adaptació és l'Optimized Link State Routing (OLSR), concretament en la seva adaptació d'*unik* [25], que és la versió suportada pels 4G AccessCube.

- Especificació d'un canal ràdio dedicat exclusivament a transportar la **senyalització** generada tant de l'algorisme d'encaminament com dels altres protocols que seran dissenyats a la solució definitiva.

- Algorisme de **detecció d'interferències** externes i que servirà per tenir localitzats quins són els canals més interferents i aquells que pateixen menys distorsió externa. Aquesta informació és la que pot ajudar a realitzar una selecció de canal millor a cada moment.

- Creació d'un **protocol de negociació de canal**, on primer els nodes realitzarien una assignació inicial dels canals segons la disposició que tenen i a on, després d'haver assignats aquests canals, els diferents nodes s'aniran relacionant entre sí intercanviant la informació obtinguda amb l'algorisme anterior, i realitzant càlculs i preses de decisions sobre quin és el millor canal a escollir en el moment en qüestió.

- Implementar un procediment on es puguin manejar de forma flexible les interfícies ràdio i els múltiples canals que disposem. Aquest procediment s'haurà de relacionar amb tots els protocols i solucions creades anteriorment, ja que aquí serà on es realitzarà la decisió física de **selecció de canal i interfície** calculades abans.

Per a la realització de tots aquests objectius, es pensa en equipar tots els nodes inicialment amb dues interfícies com a mínim, on tinguem una que es pugui dedicar a la senyalització esmentada als objectius de la proposta mentre que l'altre estigui preparada per a la transmissió de dades i que sigui aquella que ens permetrà la utilització de múltiples aspectes ràdio. Aquesta era la intenció inicial perquè, com s'anirà veient al desenvolupament de la solució, aquest plantejament inicial s'anirà modificant, ateses les característiques que presenten les eines utilitzades i del major partit que es poden treure d'altres solucions.

Representat gràficament es podria veure de la manera en què apareix a la figura 5.4.

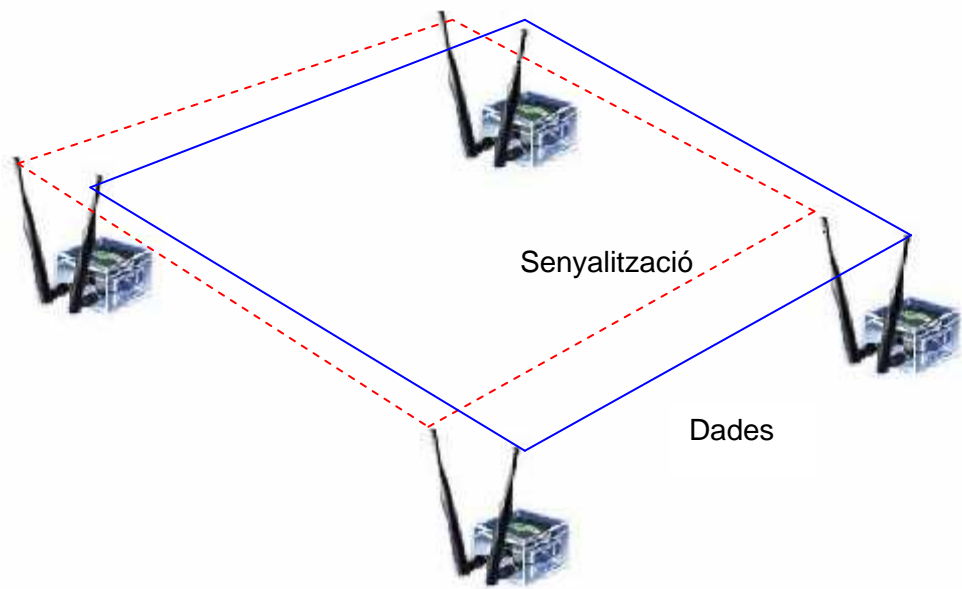

Senyalització

Dades

**Figura 5.4: Interfícies de senyalització i de dades**

Tindríem, per tant, una solució de les que en la classificació anterior (apartat 4.5) hem anomenat com a **híbrides**, amb una certa semblança a **Multi Channel Routing Protocol (MCRP)** [40], ja que aquí també teníem una interfície assignada a un canal de forma fixa, mentre que l'altra interfície seria variable podent ser diferent als canals utilitzats. Seria una solució on majoritàriament actuaríem a nivell d'aplicació o de xarxa, però on inevitablement, tenint en compte la natura dels temes dels que parlem, haurem de realitzar un bon grapat de funcions a nivell MAC.

També podríem trobar semblances amb altres solucions híbrides com **Multi-Channel Medium Access Control (MMAC)** [49] on s'utilitza l'intercanvi de missatges per aconseguir la taula de canals preferits amb una sola interfície o dinàmiques com **Slotted Seeded Channel Hopping (SSCH)** [48] on es reserva una part de l'*scheduling* per senyalització del sistema, encara que a aquest projecte es proposa una solució a més alt nivell.

## 5.3  ALTERNATIVES DE DISSENY

Per a la realització de la proposta inicial de disseny exposada a l'anterior apartat, s'han tractat de seguir diferents camins per tal de poder desenvolupar aquesta idea definitiva. En aquest apartat es relataran les diferents **alternatives de disseny** que s'han seguit i com s'ha anat modificant el que era la idea inicial, tenint en compte factors tècnics, d'implementació o de rendiment.

### 5.3.1  OLSR Multicanal

Una de les primeres opcions que es va sospesar va ser el d'implementar el protocol multicanal a dissenyar modificant directament el protocol d'encaminament **OLSR** . Alguna idea similar a aquesta ha estat utilitzada anteriorment a solucions com el **Multi-Channel OLSR** [56] on el que es fa és crear les rutes entre diferents nodes tenint en compte que estem davant d'un entorn multicanal. El que es feia a aquesta opció era modificar els camps dels missatges que s'intercanviava OLSR, i algunes estructures que manejava el protocol adaptant-ho al entorn multicanal, de tal manera que es recollissin camps de canal, necessaris per sintonitzar les interfícies. S'ha de recordar que a aquesta solució no s'assignava el canal per flux sinó que es feia per node.

El que es va tractar d'implementar era en certa manera una emulació d'aquesta solució però amb algunes diferències. Es va voler adaptar la **mètrica** que utilitza OLSR per configurar les rutes (l'ETX), a l'entorn multicanal. Per fer això una opció era la d'anar canviant de canal freqüentment per poder calcular la mètrica a les diferents freqüències. D'aquesta manera es podia configurar la millor ruta cap a una destinació amb el millor canal possible. Aquesta solució també implicaria la modificació de vàries **estructures** que apareixen al codi del protocol d'encaminament OLSR com són les estructures referides a les taules d'encaminament d'un sol *hop*, les taules d'encaminament de dos salts i altres taules gestionades pel protocol on es contenguin rutes cap a altres destinacions. També, tal com es plantejava a la solució esmentada com a referència, seria adient un canvi dels missatges HELLO que, al protocol OLSR, els nodes s'intercanvien constantment, introduint-li camps de canal. D'aquesta manera aquestes taules que abans s'han mencionat, es podrien anar actualitzant en el temps.

Amb aquesta solució queda patent la necessitat de dedicar recursos a la **senyalització** entre els nodes. Una de les interfícies haurà d'estar dedicada a suportar tota la senyalització per on circularan tots els missatges d'intercanvi del protocol d'enrutament més els càlculs de les mètriques dels diferents canals. Per tant, es podria seguir amb una de les premisses que s'havia plantejat inicialment, el utilitzar una interfície de senyalització dedicada per la comunicació.

Mentrestant, per realitzar les transmissions de dades, l'altra interfície s'anirà sintonitzant segons la destinació que tingui en cada moment i que li hagi indicat anteriorment el protocol d'encaminament.

L'eina OLSRD [25] és una implementació d'**OLSR** suportada per les estacions en les quals es faran les proves. Per tant, inicialment semblava que podria ser una opció factible i desenvolupable tècnicament ja que es tractaria de canviar directament el codi font d'OLSRD i després realitzar l'execució del protocol d'encaminament a l'entorn pràctic.

Però una vegada realitzades algunes aproximacions, aquesta opció va ser **descartada** com a tal, degut a incapacitats tècniques fruit de les limitacions del *hardware* que s'està utilitzant. Les incapacitats es donen a l'hora de realitzar canvis de canal ja que la tecnologia que disposem no els permet fer de forma immediata. Cada canvi de canal requereix un temps elevat, ja no sols per aquest fet, sinó perquè cada vegada que canviem de freqüència es crea de nou una nova cel·la amb el seu Base Station Subsystem Identification (BSSID), que serà diferent a cadascuna de les targetes on es canviï el canal ja que, habitualment, el BSSID pren com a model l'adreça MAC de la targeta. Per a què hagi comunicació entre dues estacions, apart d'estar sintonitzades a la mateixa freqüència ambdues, aquestes hauran de tenir assignat el mateix BSSID, cosa que s'aconsegueix amb un procés d'**associació** entre les estacions quan aquestes reben els *beacons* provinents de les altres estacions. Per tant, quan es realitza un canvi de canal, també es necessitarà un temps d'associació de cel·la que, en moltes ocasions, pot ser bastant elevat (fins i tot, segons).

A més, l'**OLSR** haurà de ser modificat en gran mesura i en gran quantitat d'estructures, tenint en compte multitud de factors a l'hora de gestionar correctament totes les taules. S'ha de fer notar que les actualitzacions haurien de ser constants ja que s'han de comunicar immediatament els canvis que s'han dut a terme a qualsevol moment. Aquest és un punt important ja que, recordem, per haver comunicació entre les estacions, les dues han de compartir el mateix canal i, per tant, aquí es veu la dificultat que suposarà aplicar qualsevol esquema d'assignació de canals per nodes.

Tot això, tenint en compte que estem parlant de canvis de canal, que es realitzarien a un nivell inferior al de xarxa que tracta el protocol d'encaminament i a on, per tant, es necessitarà afegir un suport que permeti interaccionar les complexitats del protocol amb les decisions respecte als canals que s'hagin de prendre a un nivell inferior que el que s'està tractant a l'OLSR.

Tots aquests factors (temps d'associació entre cel·les i problemes amb el protocol) provocaran una pèrdua de capacitat del sistema i un important retard degut a als temps d'associació. Degut a aquestes limitacions que ens imposa el propi sistema, es va optar per descartar aquesta possible solució.

## 5.3.2   Canvi de canal directe

Una altra forma per tractar d'arribar a la solució proposada inicialment és la de creació d'un protocol que prengués una sèrie de decisions i que gestionés els canvis de canal a les interfícies de forma **directa**.

La idea bàsica de la proposta es basa en la creació d'una aplicació que reculli dades per veure quin és el millor canal per dur a terme les comunicacions entre estacions, i a on després es realitza la selecció d'aquest canal que es sintonitzarà a la interfície corresponent. Com es veu, la intenció és similar a la que teníem a l'apartat anterior, però aplicada de manera diferent, tal com es veurà a continuació.

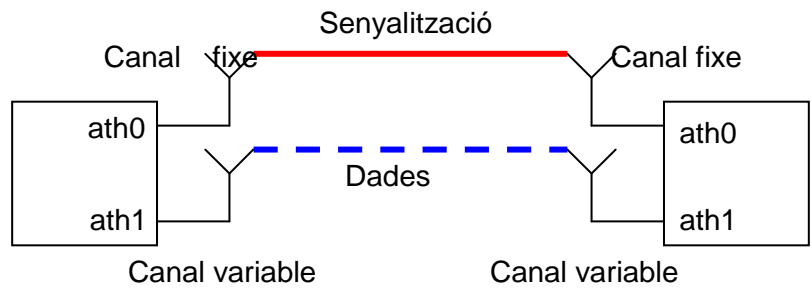

**Figura 5.5: Interfície per la comunicació**

La solució que es vol desenvolupar pren com a model, en certa mesura, la idea que s'havia proposat en MMAC [49] on els diferents nodes que conformen la xarxa es van intercanviant una **llista de canals preferits**. D'aquesta manera els nodes coneixen les llistes de canals que tenen els seus veïns.

A la implementació es vol aconseguir una situació similar. Es proposa que cadascuna de les estacions reculli dades sobre la qualitat dels canals. Aquestes dades s'enviaran a les estacions veïnes, que al mateix temps, realitzaran en paral·lel el mateix procés. Quan aquestes estacions reben informació de qualitat de les altres, comparen amb les dades pròpies, confeccionant la llista de qualitat global.

A diferència de l'apartat anterior, on utilitzàvem Expected Transmission Count (ETX) per mesurar la qualitat que ens proporcionava cada canal, aquí es fa un recull del **nivell de senyal interferent** rebut a les diferents freqüències. Aquells canals que presentin un menor nivell d'interferència seran els més adequats per a la transmissió de dades.

Com a MMAC, els nodes s'aniran intercanviant les dades per conèixer les de les estacions veïnes. D'aquesta forma s'evita l'haver de recórrer a solucions massa complicades en quan a l'assignació de canals com les centralitzades que es proposen a Load Aware Channel Assignment and Routing [43].

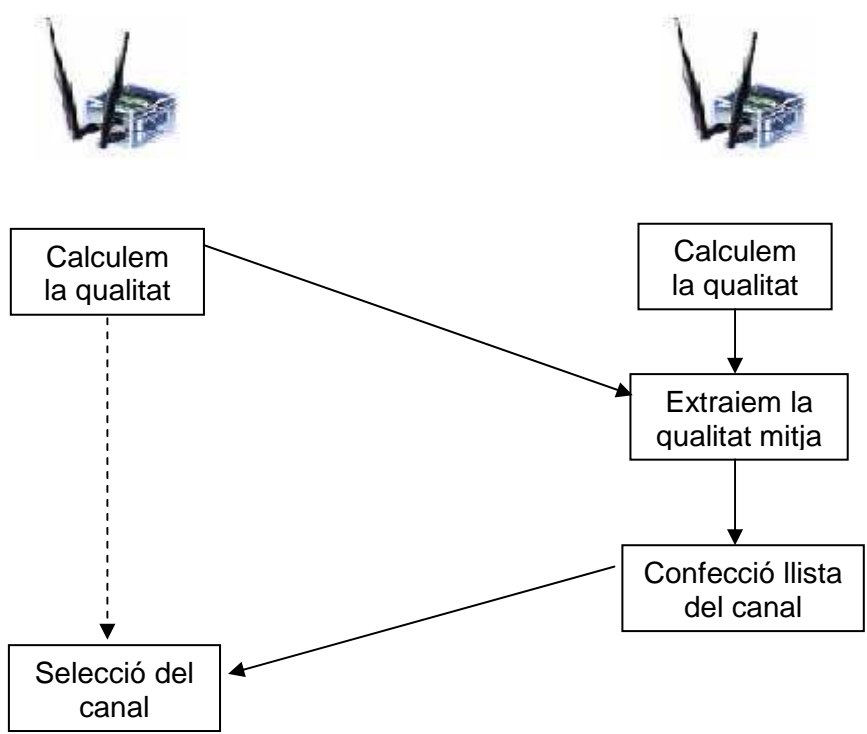

**Figura 5.6: Tasques de les estacions**

Tot això serà realitzat per una aplicació que, quan selecciona el canal òptim als dos extrems de la comunicació, realitza el **canvi de canal** escrivint la comanda necessària per realitzar aquesta operació. Com en el cas relatat abans, tota aquesta informació d'intercanvi circularà per l'enllaç que conformen les interfícies de senyalització de les estacions, mentre que l'altra interfície seria la que es reservaria per la transmissió de dades.

En aquesta solució, l'aplicació no crearà un arbre de rutes expansiu cap a les altres estacions. Per això, cal engegar algun mecanisme que ens proporcioni aquestes rutes per poder comunicar-nos amb els nodes que conformen la xarxa. Aquest mecanisme serà el protocol **OLSR**, que és suportat per la nostra arquitectura, i que configurarà les rutes òptimes en l'entorn. Com s'ha dit abans, es té una interfície de senyalització dedicada i aquesta serà aprofitada per transmetre els missatges generats pel protocol d'encaminament. Així ens asssegurem de què aquests missatges no molestaran a la transmissió de dades útils.

El problema residirà en que es crearan les rutes sobre les interfícies de senyalització, enlloc de fer-ho damunt les de dades. Per solucionar aquest entrebanc, cal realitzar una modificació al codi font d'OLSRD per tal de què les rutes creades a la interfície de senyalització es **mapegin** a sobre de la interfície de dades. D'aquesta manera les dades seran enviades per la interfície de dades mentre que els missatges de senyalització seguiran essent enviats per la interfície de senyalització.

De totes formes, encara que s'hagin solventat problemes que sorgien a la solució provada de 5.3.1 reduint-la en complexitat degut a la creació d'aquesta nova aplicació, es segueixen presentant dificultats en la visibilitat entre estacions degut als elevats temps d'associació de cel·la. Cada canvi de canal suposa una pèrdua de visibilitat entre les dues estacions que només recuperen quan comparteixen canal i quan ha passat el temps d'associació entre cel·les degut al procés d'intercanvi de *beacons*. Això comporta una pèrdua inacceptable de capacitat fent que l'opció en sí quedi descartada, però a l'hora amb una estructura de solució aprofitable posteriorment.

### 5.3.3  Supressió del temps d'associació

Per tant, la solució necessitava d'alguna tècnica per la qual es pogués reduir o suprimir el **temps d'associació de cel·la** que provoca que una opció multicanal amb canvis de canal a les interfícies sigui totalment inviable.

Com s'ha dit al subapartat 5.3.1, les estacions han d'estar associades a la mateixa cel·la per què pugui haver comunicació entre les estacions. El principal problema es produeix quan es generen noves cel·les independents fruit del canvi d'un canal a un altre necessitant-se un temps per a què les estacions es relacionin entre elles fins que queden agrupades a la mateixa cel·la. Aquestes cel·les seran identificades amb una adreça que, normalment, serà la mateixa que l'adreça física d'una de les targetes dels nodes.

Entre aquestes cel·les s'intercanvien uns petits paquets anomenats ***beacons***, que porten uns identificadors de la cel·la i que permeten la detecció d'aquesta pels nodes que estan al voltant. Aquestes transmissions de *beacons* ajuden en el procés d'associació entre cel·les realitzant la identificació dels nodes. Això comportarà un retard en aquesta associació provocant una degradació en la capacitat del sistema.

Però els *drivers wireless* de *madwifi* [64] i de *hostap* [66] que són els que es manegen, presenten una característica que pot resultar molt útil pel cas que tracta: el nomenat **pseudo-Base Station Subsystem (pseudo-BSS)**.

El **pseudo-BSS** s'activa mitjançant una comanda privada (*iwpriv*, veure annex A.8) que porten els *drivers* i consisteix en eliminar tot el procés *beaconing* eliminant tots els identificadors de cel·la adoptant com a identificador el 00:00:00:00:00:00 que serà el mateix per totes les interfícies que tinguin activada aquesta opció. D'aquesta manera es suprimeix tot el temps d'associació entre cel·les ja que, al adoptar aquest mateix BSSID, no es necessita intercanviar cap tipus de *beacon*. D'aquesta manera s'evita que al canviar de canal s'hagi d'activar un nou procés d'**associació** entre cel·les, fent-se immediatament aquesta associació.

Amb aquesta opció activada només es requerirà que les estacions que es volen comunicar comparteixin les mateixes freqüències i, per tant, es pugui adaptar la solució descrita al subapartat anterior amb aquesta opció de pseudo-BSS activada. L'únic retard que trobem seria el de **sincronitzar** les diferents estacions per tal de què coincideixin en la utilització del mateix canal.

El **pseudo-BSS** presenta altres tipus de problemes derivats de l'absència de la transmissió de *beacons* que impediran la mesura d'activitat de les estacions quan transmeten. Això dificultarà la detecció de senyals interferents que pertanyin a la mateixa xarxa associada i que, per tant, podria afectar directament a la resolució de l'aplicació que determina quin és el millor canal possible per la transmissió.

Amb aquesta opció activada tampoc es donen uns resultats massa satisfactoris, tot i el bon funcionament del **pseudo-BSS** a l'hora de realitzar aquest **canvi de canal**. Això és degut a que la utilització de només dues interfícies, tenint una dedicada a la senyalització no produeix un gran guany. El canviar de canal enmig d'una transmissió requerirà **sincronitzar** les estacions d'una comunicació reduint els guanys que es podrien obtenir si aquests canals ja estiguessin sintonitzats amb anterioritat. Això ens fa replantejar-nos el disseny del sistema amb dues interfícies tal com s'havia pensat fins a aquests moment.

### 5.3.4   Utilització de tres interfícies

Ja que el *hardware* que s'empra en la realització del projecte permet incorporar més targetes *wireless*, una opció que es té per no variar el plantejament en excés és la **d'afegir una interfície més**. Aquesta interfície ens aportaria majors possibilitats a l'hora d'aprofitar l'espectre multicanal sempre que el seu ús estigui correctament planificat i dissenyat.

Amb la **interfície extra** es pot enfocar el disseny d'una manera diferenciada a com s'havia vist fins a aquell moment. Així, mentre que amb dos interfícies s'estava dissenyant un sistema dinàmic, on els canvis de canal eren freqüents, ara es podrà reduir la **complexitat** del sistema a dissenyar ideant una solució basada en assignacions de canal comunes, és a dir, assignacions a un mateix canal durant un període llarg de temps. Aquesta interfície extra es destinarà a transmetre dades entre les estacions. Així es tindran **dues interfícies de dades** més **una altra de senyalització**, que conserva intacta la seva funció. Encara que en aquest cas s'ha optat per tenir tres interfíces, aquest cas és generalitzable a la utilització de més interfícies assignades a dades, sempre que el *hardware* emprat suporti el maneig d'aquestes targetes afegides.

Les interfícies de dades poden estar sintonitzades a canals ortogonals diferents, de tal manera que la seva utilització simultània no interfereixi mútuament les comunicacions. També la interfície de senyalització pot estar sintonitzada permanentment a un altre canal que sigui ortogonal del que es té a les interfícies de dades i a on, per tant, els missatges de control es puguin enviar independentment sense afectar negativament a la resta de les comunicacions. A la figura 5.7 es pot veure un exemple d'una estació amb tres interfícies.

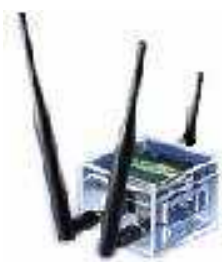

**Figura 5.7: 4G Access Cube amb 3 interfícies**

El tenir les interfícies assignades a un canal de forma permanent facilita la tasca d'emprar el multicanal. Ara l'objectiu principal passa a ser el de trobar un mecanisme que permeti realitzar la **selecció de les interfícies** en cada moment i depenent de la situació. Haurà de ser un mecanisme flexible i manejable ja que s'ha d'adaptar a l'aplicació que mesura el millor canal.

Aquest cas que es tracta podria ser similar en plantejament al que es proposava en la solució de MobiMESH [45], on havia elements amb dues interfícies i amb dues adreces que havien d'identificar que pertanyien a un mateix node i que, al mateix temps, gestionaven múltiples canals ja que cada interfície es sintonitzava a un canal diferent.

Així mateix, el fet d'afegir més interfícies per a la comunicació també crea noves dificultats. Una d'elles és la **duplicació** d'adreces a un mateix node. Cada interfície té la seva corresponent adreça IP i adreça MAC encara que comparteixin el mateix node. D'aquesta manera, es tindran tres adreces IP i MAC diferents. Una de les adreces correspondrà a la senyalització i, realment, no afectarà a la resta de la comunicació ja que per transmetre de dades no farà falta saber aquesta adreça. Però al cas de les dues interfícies de dades, qualsevol d'aquestes poden ser vàlides per arribar a qualsevol destinació, i això implicarà que s'ha d'identificar que les dues interfícies pertanyen al mateix node.

També s'hauran de realitzar més canvis per aconseguir un bon funcionament del sistema. L'**OLSR** haurà de canviar el seu mapeig i generalitzar aquell funcionament que tenia per les dues interfícies a casos amb més interfícies. D'aquesta manera s'haurà d'introduir a la taula de rutes, totes les possibles rutes cap a una determinada destinació ressenyant la interfície per la qual es pot arribar a aquests nodes. Per la seva part, l'aplicació haurà de canviar en seleccionar el millor canal i només tornarà aquells canals que es poden emprar amb la configuració dels *meshcube*, elegint el millor canal del grup de seleccionables.

En definitiva, el problema es reduirà a trobar un **mecanisme** que realitzi la funció de **selecció d'interfícies** de la manera més eficient possible. S'han seguit diverses línies per arribar a l'elecció definitiva del mecanisme de selecció d'interfícies i, a continuació, s'explicaran quines són les opcions que s'han tingut en compte abans d'aconseguir trobar la solució final que realitzés la commutació de les interfícies.

### 5.3.4.1  Netfilter

És la primera opció que es va manejar per realitzar la selecció d'interfícies, la utilització del **filtratge de paquets** ofert per un sistema Linux. L'objectiu de l'ús d'aquesta tècnica és la d'assignar la interfície de sortida que es desitgi a tots els paquets que s'hagin de passar cap a les seves destinacions.

**Netfilter** [67] realitza les mateixes funcions que les *iptables*, comanda utilitzada extensament a entorns GNU/Linux i que permet configurar regles amb les que tractar els paquets que circulen per la xarxa. Les *iptables* són una de les eines més emprades a l'hora de configurar *firewalls* ja que permet bloquejar els paquets depenent de la seva procedència, direcció o altre camp que sigui susceptible de ser filtrat. Un esquema bàsic de com es realitza la presa de decisions a *iptables* es pot veure a la figura 5.8.

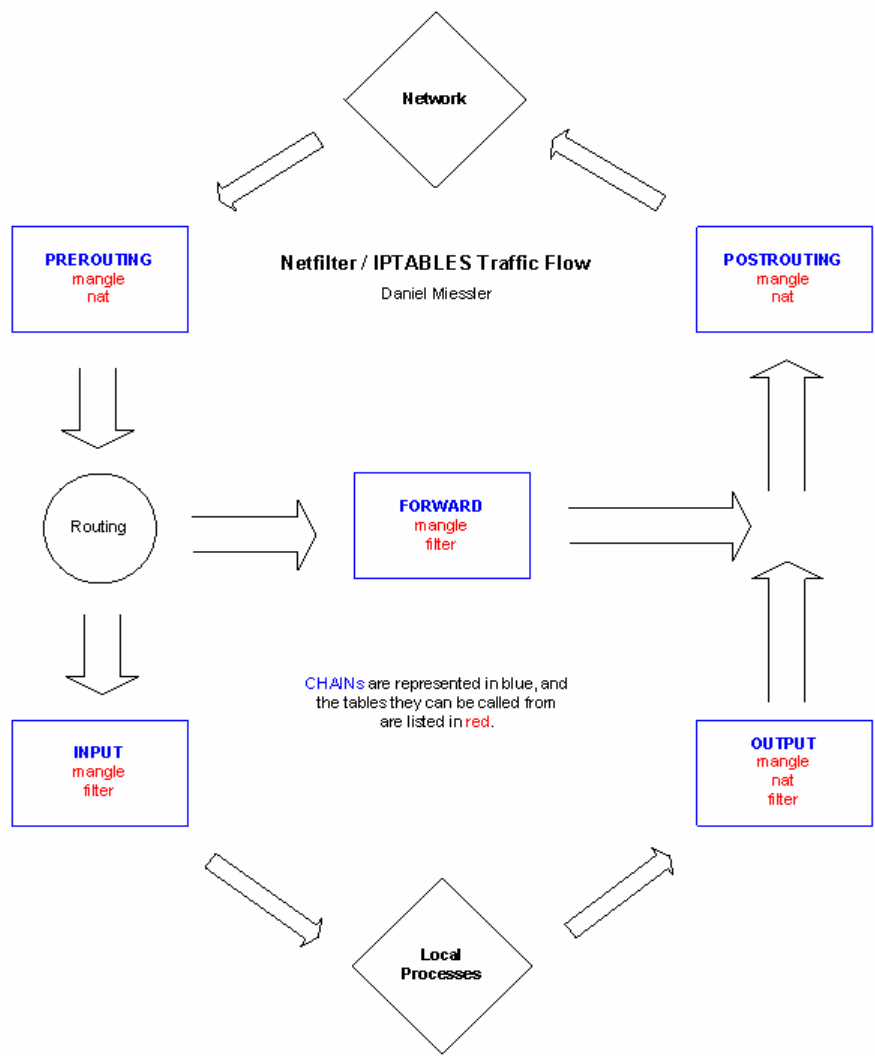

**Figura 5.8: Esquema de Netfilter**

**Netfilter** permet la utilització de les eines *iptables* des d'un programa de l'usuari. Per poder realitzar-ho cal implementar el programa en el *kernel space* o espai del *kernel*. El *kernel space* és on el *kernel* executa i proveeix les seves funcions que tenen un control complet de tot el que succeeix al sistema. A l'espai del *kernel* es poden prendre decisions directes respecte als *drivers* o a instruccions que afecten al funcionament del sistema. Amb la implementació d'un **mòdul**, programant-ho des del *kernel* es pot accedir a modificar el comportament intern d'aquestes aplicacions del *kernel*. Aquest nivell de programació s'ha de diferenciar del que solen emprar la majoria de la resta de programes i que estan situats al que s'anomena **user space**. La raó d'emprar programació en el **kernel space** vendrà donada per la necessitat de modificar operacions internes del *kernel* com són, en aquest cas, les operacions de *iptables*.

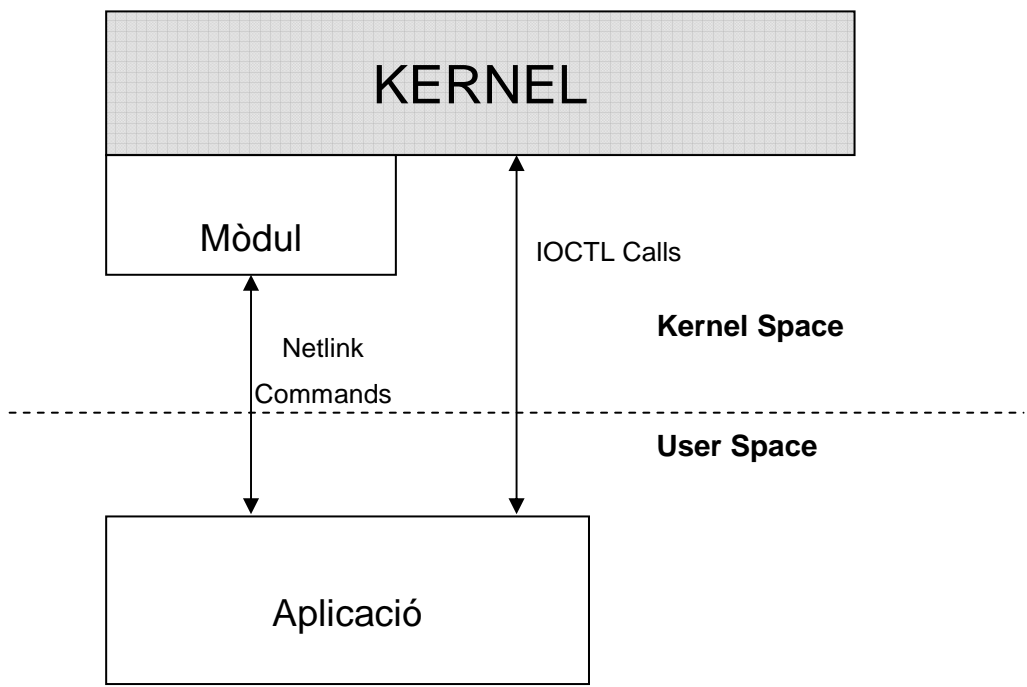

**Figura 5.9: User space vs. Kernel space**

La programació d'un **mòdul** del *kernel* [68] tindrà les seves particularitats enfront de les que s'ofereixen normalment. Al programar un mòdul directament al *kernel space* obtindrem un objecte *.o* que, per poder funcionar amb ell, s'haurà d'inserir a dins del sistema amb la comanda adient. Aquest mòdul compartirà la memòria comú que té el *kernel* per realitzar les seves operacions havent de tenir major cura amb la gestió de memòria que es fa a aquests programes degut a que es realitzen operacions més crítiques per al sistema global. Com s'ha dit, aquest mòdul tindrà accés a *drivers* inserits al sistema Linux així com a comandes internes tipus *iptables*, podent interactuar i, fins i tot, modificar el seu comportament estàndard.

Al ser el resultat de la compilació d'un **mòdul** un objecte *.o*, i no passar pel procés de *linkatge*, moltes de les funcionalitats aportades per llibreries externes del compilador no podran ser utilitzades com les llibreries d'escriptura de pantalla o algunes llibreries de comunicació.

L'objectiu d'utilitzar programació al *kernel* era la de poder seleccionar les interfícies de sortida quan es transmeten els paquets. D'aquesta manera es poden consultar els camps de la capçalera IP de tots els paquets i filtrar-los, amb una rutina *output_handler*, cap a la interfície de sortida desitjada depenent del paquet i el node en què estem tractant.

Per tractar de complir el nostre objectiu, el que es fa és implementar l'aplicació que teníem d'apartats anteriors i connectar-la amb el **mòdul** que s'ha dissenyat per **filtrar** els paquets. D'aquesta manera quan és necessari un **canvi de canal** (i conseqüentment d'interfície) es comunica al mòdul que és necessari dur a terme aquesta acció i el mòdul farà les operacions per realitzar aquest **canvi d'interfícies**.

La comunicació entre el *user space* i el *kernel space* es pot realitzar de diverses maneres. Una d'elles és utilitzar *netlink sockets* [69], definits al *user space* que poden enviar dades cap al *kernel space* esperant resposta de l'altra banda gràcies a unes MACROS definides per aquests tipus de *sockets*. Una altra forma de comunicació és la d'emprar els fitxers */proc* [70], utilitzats als sistemes de Linux. En aquest cas s'anirien modificant els fitxers depenent dels resultats obtinguts al programa. Al *kernel space* es poden crear i modificar fitxers dins */proc* que, al mateix temps, des de l'aplicació del *user space*, es podrà consultar els canvis aplicats des d'aquest *kernel space* i, finalment, poder actuar en conseqüència depenent de les dades guardades al fitxer.

De totes formes, encara que la programació al *kernel space* serà necessària per la realització del projecte, es va veure que **canviar d'interfícies** directament emprant ***netfilter*** era impossible ja que les interfícies tant d'entrada com de sortida es poden consultar per crear el conjunt de regles pel filtratge, però no es poden modificar tal com era la intenció. Per aquest motiu, l'opció de ***netfilter*** va ser **descartada** per la impossibilitat de ser implementada amb aquests medis.

### 5.3.4.2  Raw Sockets

Una altra opció diferent que es va manejar va ser la de provar de modificar les interfícies de sortida dels paquets sense haver de crear **mòduls** dins el *kernel space*. Teòricament, aquesta opció és més simple, ja que ens permet gestionar tot el programa de forma centralitzada sense haver de dependre de l'execució en diferents àmbits i evitant els problemes que sorgeixen dels mètodes de comunicació que hi ha entre el *kernel space* i el *user space*.

La clau està en trobar un mètode amb el que es pugui interceptar el tràfic que circula pels nodes per després poder canviar la informació continguda en aquests camps d'informació. Amb aquesta intenció, el que es vol realitzar és la intercepció del tràfic i modificar els camps de paquets de manera que es faci correctament la **selecció de les interfícies**.

Els **raw sockets** [71] són una classe de *sockets* que ens permeten realitzar la lectura de les dades que circulen per la xarxa. A més els **raw sockets** es poden definir de manera senzilla ja que la seva declaració és molt similar a la d'un *socket* convencional sols que durà les opcions específiques per crear-lo com a **raw socket**. Algunes funcions específiques d'aquests tipus de *sockets* seran les que ens facilitaran interceptar el tipus de tràfic que es vol filtrar (tràfic ICMP, UDP…). Serà la funció *setsockopt* la que permet fixar aquestes premisses. A més, els processos de lectura i escriptura que empren aquests tipus de mecanisme són de fàcil integració amb la resta del programa, ja que estem parlant de funcions semblants a les crides de sistema UNIX basades en la lectura i l'escriptura a sobre del descriptor que hagi retornat el *socket*.

El principal avantatge dels **raw sockets** consisteix en la facilitat d'integrar l'aplicació de negociació de canal dissenyada abans amb els mecanismes que permetran la selecció d'interfícies. Bastaria amb la simple introducció d'aquests processos dins del programa per tal d'aconseguir els objectius proposats.

Cal fer notar que l'adreçament de les interfícies no es realitza directament, ja que no hi ha cap camp a consultar que modifiqui directament aquesta informació, sinó que es fa indirectament, mapejant les IP que corresponen a la interfície primera a unes altres adreces que corresponen a la segona interfície del node. Això pot ser una dificultat perquè voldrà dir que s'hauran de tenir uns patrons d'adreces ben definits inicialment per tal de què es pugui realitzar correctament aquest mapeig entre les adreces de les diferents interfícies.

A més, a l'hora d'aplicar la idea presentada, apareixen més dificultats. Això és degut a les limitacions en el moment de realitzar la intercepció de grans volums de paquets d'informació, com previsiblement es tindran a la realitat, reduint el seu rendiment. Per raons com aquesta i l'anterior, ja comentada, dels problemes de coneixement de les adreces IP assignades, aquesta opció va ser **descartada**, al menys en la forma plantejada fins a aquell moment.

### 5.3.4.3   Netlink Firewall

Seguint amb les solucions basades en la intercepció del tràfic, es va trobar una possibilitat que combinava les dues opcions per la selecció d'interfícies que abans s'havien provat. Es tracta del *Netlink Firewall* [69].

El *Netlink Firewall* emula el comportament que segueixen les *iptables* dins del *user space* evitant l'haver de realitzar implementacions al *kernel space*. Per aconseguir això, es defineix el *Netlink Firewall* com una nova opció quan es crea un nou *raw socket*. Per tant, mitjançant un *raw socket* estem creant un mètode de filtratge de paquets, tal com es feia al *netfilter*. La definició del *Netlink Firewall* també es pot aconseguir utilitzant les funcions que ofereix la llibreria *libipq* a C i que no són més que un desenvolupament major de les funcions de creació, lectura i escriptura que empren els *raw sockets*.

Per tant, com està basat en un *raw socket* les funcions que s'utilitzaran per llegir i escriure damunt el descriptor del *socket* seran les mateixes que s'utilitzaven a l'apartat anterior. La novetat en aquest mètode la trobem a l'hora de definir les estructures que completen les funcions de creació i enviament d'informació utilitzant els *sockets*. Són unes estructures específiques de *netlink* que permeten lligar la informació que tenim a una altra interfície de les que hi ha disponibles. La tècnica que es va emprar va ser la de crear un *netlink socket* per cadascuna de les interfícies, de tal manera que després es pogués lligar la informació a enviar a qualsevol de les dues interfícies.

Per fer això, es va optar per marcar tots els paquets de forma idèntica amb informació de què esperessin a la cua del *netfilter*. Això s'aconsegueix marcant amb NF_QUEUE el camp de l'estructura corresponent. D'aquesta manera es tenen els paquets a l'espera d'una confirmació per ser enviats a la seva destinació correcta. Com que els missatges *netlink* permeten consultar els continguts d'un paquet com al cas del *output_handler* (apartat 5.3.4.1), podent-se consultar informació com la adreça IP de destinació, la d'origen, les interfícies d'entrada i de sortida, el que es va tractar de fer és mirar la informació respecte a quina era la interfície de sortida. Si aquesta interfície coincidia amb el resultat desitjat, es confirmava amb un missatge NF_ACCEPT que el paquet podia continuar amb el seu camí. Si el resultat no és coincident, el que es fa és lligar la informació a enviar al *socket* creat per l'altra interfície, validant després que es pot enviar el paquet. D'aquesta manera s'aconsegueix que, encara que l'adreça dels paquets sigui la mateixa, aquesta informació es pugui adreçar correctament per les interfícies que nosaltres desitgem.

Aquesta solució presentava els avantatges típics d'una solució que només empra el *user space*, simplicitat en la resolució del protocol. Bastaria combinar l'aplicació de mesura de qualitat amb les estructures de **netlink** i els corresponents **raw sockets** per aconseguir aquest mecanisme de **canvi d'interfícies**.

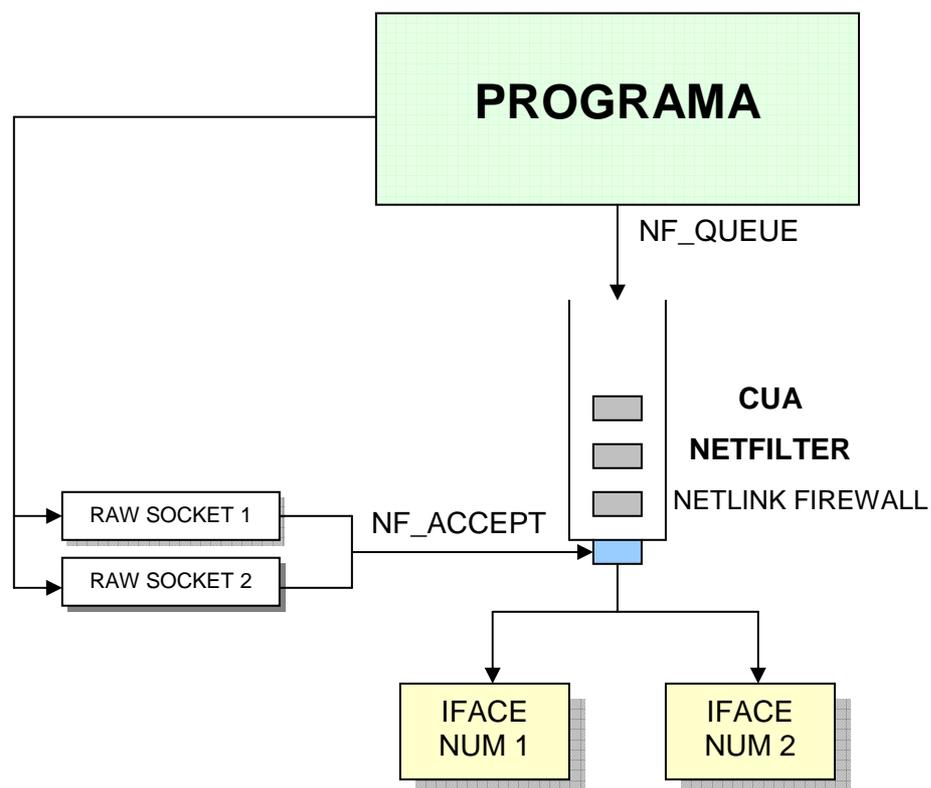

**Figura 5.10: Solució amb raw sockets**

Desafortunadament, aquesta proposta també presenta problemes que fan plantejar-nos la fiabilitat de la solució dissenyada. Tot i adreçar correctament per les interfícies, el rendiment es veurà ressentit pel fet d'haver d'encolar tota la informació. A més, en casos de transmissions grans, la solució pot presentar problemes d'estabilitat fruit de la sobrecàrrega dels *buffers* del *netfilter* proporcionats pel sistema. A tot això se li pot afegir un problema de conflicte d'adreces ja que s'envia per una interfície diferent una informació que inicialment havia d'anar destinada a una altra de les interfícies, provocant que es pugui donar qualque tipus de confusió en la recepció d'aquesta informació.

Per tots aquests factors negatius, també es va passar a **descartar** aquest com a mecanisme per a la **selecció d'interfícies**, tot i ser el primer que realment feia el canvi d'interfícies desitjat en el moment que pertocava per la informació que es manejava.

### 5.3.4.4  Internet Protocol Rules

Una altra modalitat de solució que es va plantejar va ser la de crear una sèrie de regles d'encaminament (**IP rules**) per tal d'assegurar el bon adreçament de la informació. És tracta d'aprofitar l'anomenada **Routing Policy DataBase (RPDB)** [72] suportada per Linux i que és la que s'encarrega de realitzar l'enrutament de forma correcta.

La **RPDB** s'encarrega de gestionar les diferents taules d'encaminament de les que consta el sistema. Per defecte es té una taula principal que és on es realitzen aquests encaminaments, però es poden crear unes altres amb unes regles d'encaminament diferenciades segons distints paràmetres com tipus de servei, origen i destinació del paquet o altres tipus de factors. Es poden crear fins a $2^8$ taules d'encaminament diferents atenent a criteris de classificació diversos.

 Els paquets utilitzaran una o altra taula d'encaminament depenent de com hagin estat marcats per les aplicacions que els envien. Depenent d'aquesta marca, es seleccionaran una o altra taula d'encaminament. Això sempre que abans el sistema no hagi trobat una semblança dins la *caché* de rutes disponible per fer més immediata la selecció de rutes i no haver d'emprar la RPDB.

Per adaptar aquesta tècnica de rutes al problema de selecció d'interfícies, es va pensar en confeccionar una **RPDB** on es generin tantes taules d'encaminament com interfícies diferents, ficant a cada taula de rutes les adreces corresponents a la interfície sobre la que s'està actuant.

Després, quan es realitza la transmissió d'informació, es van marcant tots els paquets depenent de la interfície que hagin d'utilitzar amb un paràmetre o un altre. Així es pot adreçar les informacions cap a la interfície desitjada. Per poder aconseguir això, s'hauran de mapejar les adreces IP depenent de la sortida que es tingui.

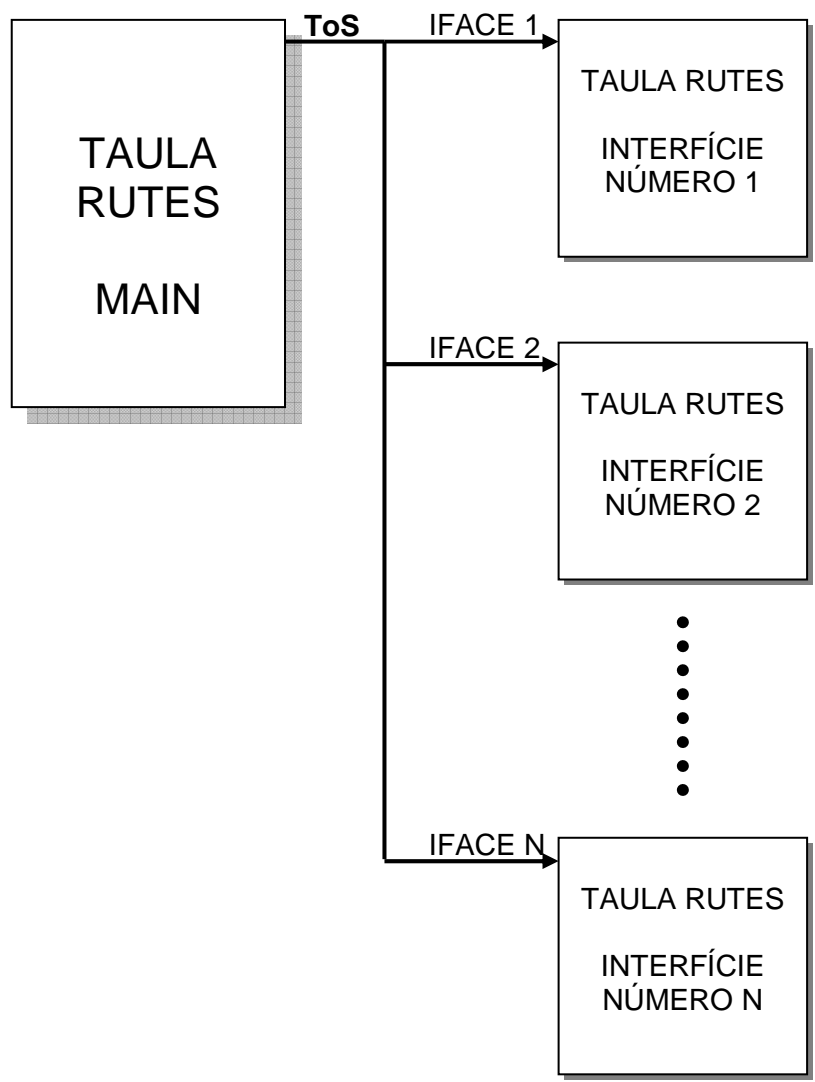

**Figura 5.11: IP rules**

Encara que pugui semblar una solució simple i efectiva, presenta nous problemes. Un d'ells està relacionat amb la *caché* de les rutes. Aquesta *caché* sempre serà la primera consulta que el sistema realitzarà abans d'utilitzar la **RPDB** creada i, serà un element que pot causar incoherències al sistema, ja que si no s'esborra adequadament poden aparèixer rutes no esperades. Un altre problema que ja havia sortit anteriorment és el relacionat amb el tema de les adreces IP. Aquest problema ve d'identificar clarament el mateix node amb diverses IP diferents, així com del mapeig que s'haurà de realitzar quan es vol accedir a una altra interfície. Aquest mapeig implicarà que les adreces IP hagin de ser preestablertes amb un patró que permeti fer els canvis de forma automatitzada des de l'adreça d'una interfície a l'adreça d'aquesta segona interfície.

Com s'ha vist, el tema dels identificadors i de les adreces IP és més complex del que pugui semblar en un principi, ja que limitarà quines adreces IP poden tenir les diferents interfícies i, per tant, qualsevol opció consistent en tenir un identificador per interfície serà una opció a descartar inicialment, preferint aquelles solucions que ens permetin tractar els diferents nodes amb $n$ interfícies com a una sola entitat.

Després d'haver sondejat aquesta possibilitat de solució al problema i rebutjar-la, es va passar a implementar la resolució definitiva gràcies a un **driver** virtual anomenat **bonding** i que es detallarà al següent apartat.

## 5.4   IMPLEMENTACIÓ DEL DISSENY

Una vegada presentada la plataforma *hardware* que es farà servir per a la realització del projecte, de proposar un plantejament inicial a desenvolupar i d'haver estudiat una sèrie de mètodes per arribar a la solució definitiva del problema plantejat, és el moment de descriure quina ha estat la **solució definitiva**, quins passos s'han seguit fins a aconseguir arribar a aquesta solució, com s'ha estructurat el disseny final i com s'han integrat finalment cadascuna de les diferents parts del sistema.

Recordar que amb el disseny el que es vol fer és **crear una WMN** amb diversos nodes, alineats amb diferents jerarquies i que ens serviran per justificar la necessitat d'incorporar un mecanisme per aprofitar millor l'entorn multicanal.

Com s'ha comentat a les alternatives del disseny, finalment es va optar per incorporar *meshcubes* amb **tres interfícies sense fils**. D'aquesta manera es pot conservar una interfície dedicada a senyalització afegint una interfície més per a la transmissió de dades. En resum, els nostres components estaran constituïts per dues interfícies de dades, que les anomenarem "ath0" i "ath1", més una interfície de senyalització, que anomenarem "wlan0". Tots els elements que intervinguin en la comunicació tindran els mateixos noms d'interfícies amb l'objectiu de facilitar la seva identificació i el seu maneig al funcionament del prototipus. Aquestes interfícies estaran configurades en **mode ad-hoc**, de tal forma que es puguin anar associant automàticament entre elles sense la necessitat de dependre d'un node central que actuï de *master*.

Cadascuna d'aquestes interfícies estarà sintonitzada a tres canals diferents, que seran ortogonals per no interferir en les seves transmissions (per exemple: canals de dades a 1 i a 11, canal de senyalització a 6) i que, per tant, seran les que ens donaran la capacitat multicanal que desitgem. Per reduir possibles incompatibilitats i contribuir a una major simplicitat del disseny, es configuraran totes les interfícies dels *meshcubes* amb la mateixa assignació de canals. En definitiva, estarem parlant d'un sistema d'assignació de canal comú, tal com eren opcions com **Multi-Radio Link Quality Source Routing (MR-LQSR)** [39] o **MobiMESH** [45].

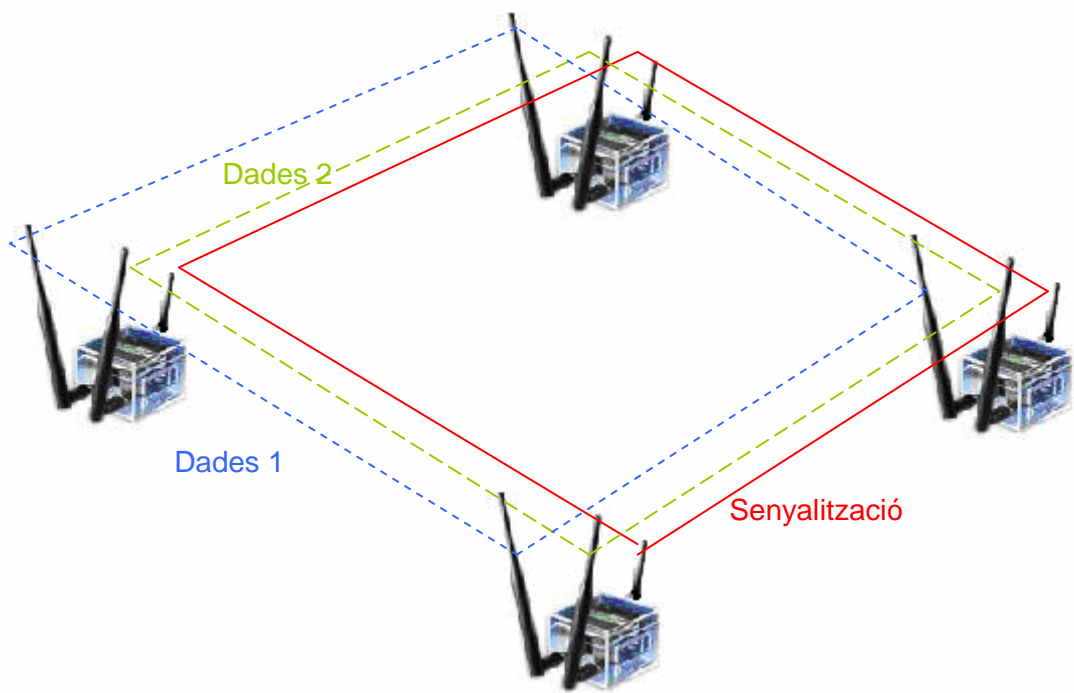

**Figura 5.12: Topologia amb 3 interfícies**

Aquestes interfícies implantades a dins dels *meshcubes* empraran *drivers* diferenciats. En un principi estava pensat que totes les targetes tinguessin instal·lat el **driver madwifi**, ja que és un *driver* amb versions actualitzades, que aporten característiques interessants i són, en general, versions amb un rendiment bastant satisfactori. Una de les característiques més interessants és la de suportar les freqüències de 2,4 GHz i de 5 GHz simultàniament, podent-se canviar de mode de funcionament sense cap tipus de problema.

Però les targetes *madwifi* presenten una mancança en el nostre sistema ja que no tenen la capacitat de realitzar un escaneig dels canals en mode ad-hoc. Amb *madwifi* només es tenen resultats aprofitables en un escaneig quan aquestes targetes estan en mode *managed*. Això implicaria haver de fer canvis constants a les targetes perdent-se la informació de senyalització que circula per aquestes targetes quan es realitzen aquests canvis cap a l'estat de *managed*.

Per tant, el que es va buscar va ser un *driver wireless* que permetés fer l'escaneig de l'entorn sense haver de canviar el seu estat en ad-hoc i, per tant, sense que es produeixin pèrdues derivades d'aquests canvis. El **driver hostap**, permet realitzar aquests escaneigs en mode ad-hoc, i per aquesta raó, va ser incorporat al nostre sistema com el *driver* destinat a la senyalització del sistema i als *scans* de l'entorn per mesurar les qualitats dels diferents canals. Tanmateix, aquests tipus de targetes presenten unes altres excepcions ja que no suporten el mode de funcionament en freqüències de 5 GHz, dificultant la detecció d'interferències en aquest espectre de freqüències.

També cal remarcar que finalment es va elegir un sistema on per cada node existís un sol identificador. Això és una sola adreça IP i una sola adreça MAC per node, evitant-nos les duplicitats que comportaven les múltiples adreces amb que comptaven al tenir vàries interfícies. El tenir un sol identificador per node s'aconsegueix amb el mecanisme de selecció d'interfícies **bonding** que s'explicarà més endavant.

Una vegada introduïts els aspectes bàsics en la configuració del material utilitzat pel banc de proves del disseny, passarem a definir quina és l'estructura del nostre disseny i de quins blocs bàsics està compost.

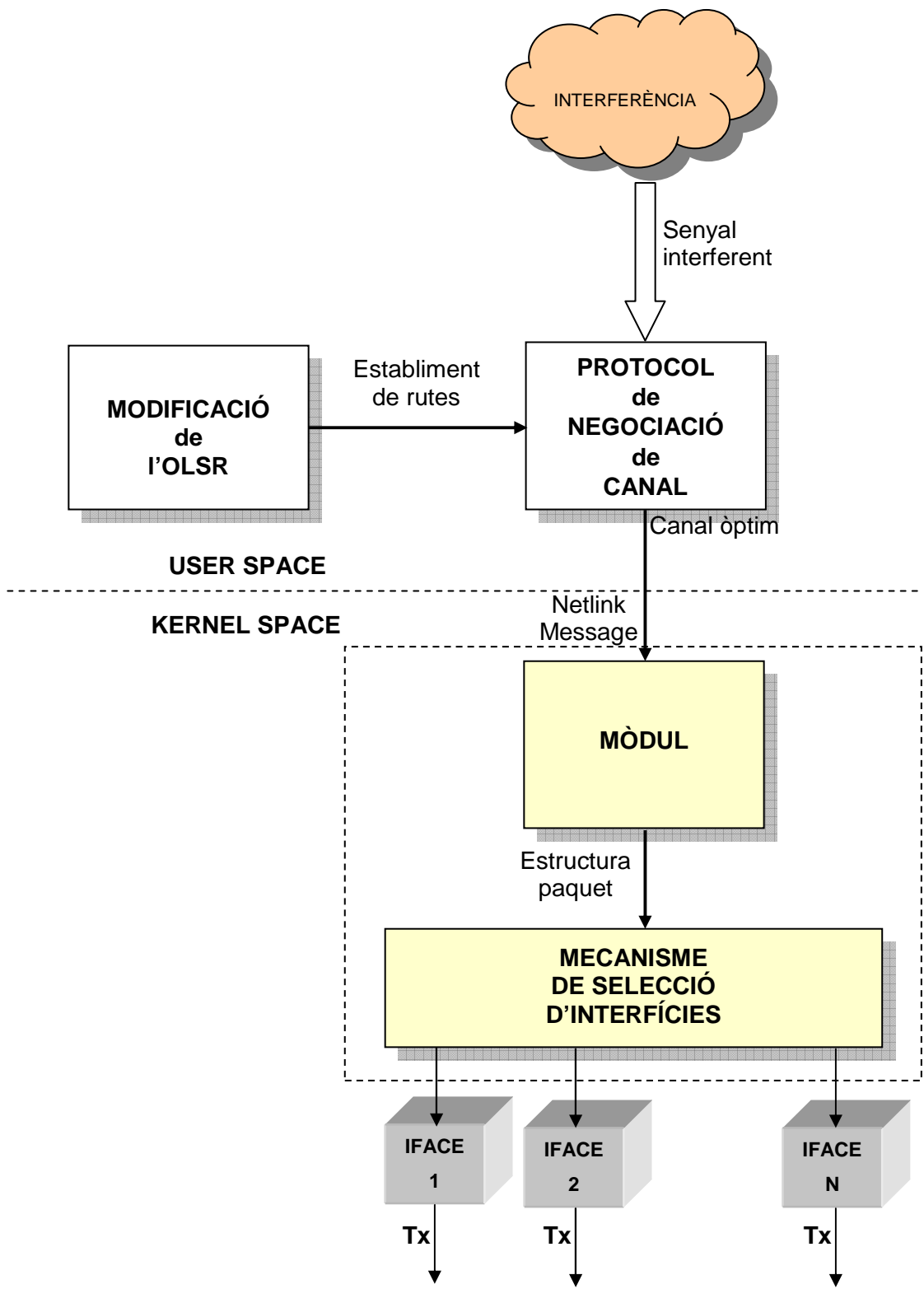

Figura 5.13: Diagrama de blocs del sistema

Els tres principals components que s'han desenvolupat són els següents:

- **OLSR modificat.** Per tenir configurades en tot moment les rutes, és necessari el concurs d'un protocol d'encaminament que ens permetrà l'accés a cadascun dels diferents nodes. Damunt d'aquest protocol es realitzaran una sèrie de modificacions per adaptar el funcionament per tal de què s'utilitzin de forma correcta les diferents interfícies del sistema i amb el propòsit amb el que es varen definir.
- **Protocol de negociació de canal.** Serà una aplicació de diàleg entre els diferents nodes on es realitzaran les funcions de mesura de qualitat de l'entorn i que, serà el que ens permetrà fer la selecció del millor canal segons les condicions que l'escaneig hagi detectat. A l'apartat corresponent es veurà tot el procés que segueix aquest protocol de diàleg entre les entitats del sistema.
- **Mecanisme de selecció d'interfícies.** És aquell que ens permetrà seleccionar entre les diferents interfícies quina és la que ens ofereix millors prestacions. Al realitzar aquesta selecció d'interfícies, serà la que aportarà els beneficis de l'entorn multicanal a la WMN creada. S'explicarà en detall com aquest mecanisme és aconseguit per un *driver* anomenat **bonding**.

## 5.4.1  OLSR modificat

Pel bon funcionament d'una WMN, cal tenir molt clar quines són les rutes que s'han de seguir per arribar a les destinacions desitjades. A més, tenint en compte que estem parlant de xarxes ad-hoc i que suporten mobilitat es requereixen protocols d'encaminament amb capacitat de trobar les millors rutes fins a cada destinació i que tinguin capacitat per suportar mobilitat en les estacions.

En el cas d'aquest projecte, també és necessari tenir aquest **protocol d'encaminament** que configuri les **rutes òptimes** en tot moment i ens permeti accedir als diferents nodes que composen la nostra xarxa amb la major eficiència possible. El protocol utilitzat per fer aquestes tasques serà l'**OLSR**. La raó d'escollir aquest protocol envers d'un altre és deguda a tres raons bàsiques: disponibilitat, fiabilitat i adequació al sistema.

- La **disponibilitat** és un fet bàsic ja que, com s'ha dit anteriorment, estem parlant de sistemes encastats que reduiran les opcions de suportar un o altre algorisme d'encaminament. Els creadors d'OLSRD van presentar una versió d'OLSR que encaixa a la perfecció amb els sistemes de Linux encastat dels *meshcube*, que ve inclosa dins els paquets proporcionats per *nylon* i que, a més, és fàcil d'aconseguir, actualitzar i distribuir.

- El protocol **OLSR** presenta uns resultats de funcionament molt **fiables**, essent bastant robust i consistent en el càlcul de les rutes, en les seves actualitzacions periòdics, així com en la inclusió i tractament d'altres elements externs de la xarxa.

- Afegint als dos factors abans anomenats, **l'adequació a l'entorn** que estem és també important, ja que a la WMN, encara que pugui haver mobilitat, les estacions romandran temps relativament llargs en una mateixa localització, essent un protocol proactiu com **OLSR** més adient que altres reactius com **AODV**.

Amb **AODV** també podem aconseguir bons resultats, ja que també és suportat en aquests sistemes encastats, donant resultats fiables, però la facilitat en el maneig de les interfícies d'OLSR, així com la millor adaptabilitat del sistema a un protocol proactiu va fer decantar-nos en l'ús de l'eina OLSRD ja inclosa com a paquet independent al sistema.

Encara que el protocol sigui capaç de suportar el funcionament sobre **múltiples interfícies**, proporcionant visibilitat a totes elles, no ens interessarà emprar la funcionalitat MID del algorisme. Això és degut a que les interfícies de dades posteriorment seran tractades com una tota sola, amb una única IP i una única adreça MAC. A més, la utilització de MID farà que els missatges de senyalització d'**OLSR** circulin per les interfícies que es volien dedicar a un ús exclusiu per dades.

Per tant, es va limitar el funcionament d'**OLSR** a una sola interfície, la "wlan0", és a dir, la interfície de senyalització. Però com el que ens interessa és que les rutes estiguin configurades sobre les interfícies de dades, es va haver de modificar el codi font d'OLSRD en certs aspectes.

La modificació consisteix en realitzar un **mapeig** d'adreces i d'interfícies a l'hora de realitzar l'operació de afegir les rutes al *kernel*. Així, es mapejaran les adreces IP de senyalització per les corresponents de l'entitat que engloba les interfícies de dades i, després, es seguirà el mateix procés amb les interfícies que s'han d'introduir a la taula de rutes, realitzant el conseqüent mapeig d'interfícies des de "wlan0" a la de l'entitat global aconseguida amb el mecanisme de selecció d'interfícies que sempre serà "bond0".

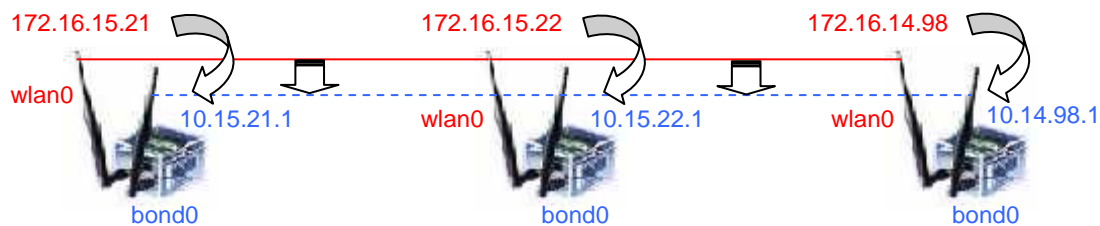

**Figura 5.14: Mapeig d'interfícies**

Amb aquestes operacions, s'aconsegueix que a la taula de rutes del *kernel* apareguin les rutes de les interfícies de dades sense que circulin per elles cap tipus de missatge de control del protocol. Aquests missatges de senyalització no desapareixen i seguiran transmetent-se per la **interfície dedicada** a la senyalització tal com era l'objectiu inicial del projecte.

Per a la realització d'aquestes modificacions s'han hagut de modificar els arxius destinats a la inserció de noves rutes, el de la gestió de les interfícies, així com els que tenen en compte les rutes cap a elements externs que no suporten OLSR (HNA). Alguns detalls modificats es poden veure a l'annex A.7.

## 5.4.2 Protocol de negociació de canal

Un dels objectius bàsics de tot el treball desenvolupat és el de poder **seleccionar** el millor canal depenent de les condicions que ens envolten i de les interferències que els propis nodes detecten.

Aquest protocol actuarà bàsicament intercanviant la informació entre els diferents nodes i, a partir d'aquesta informació, conformarà la seva pròpia que serà amb la que es traurà les informacions respecte les idoneïtats dels diferents canals. Llavors, es realitzarà la **selecció** del millor canal amb la utilització d'aquestes dades que, finalment seran enviades al mecanisme que gestiona l'elecció de la interfície corresponent. Després d'un període de temps detingut, el procés anterior es torna a repetir de la mateixa manera que s'ha fet anteriorment. El protocol en sí, no prendrà cap tipus de decisió directa respecte als possibles canvis d'interfície ja que això es gestionarà amb la informació generada per aquest mateix protocol, però al nivell del ***driver bonding*** que més tard es detallarà.

Cal remarcar alguns aspectes bàsics de la implementació d'aquest protocol per entendre les etapes en què constituïm el protocol. Cadascun dels nodes executarà l'aplicació basant-se en una **arquitectura client-servidor** que serà la que després ens facilitarà l'intercanvi de dades entre les estacions. Aquesta arquitectura s'activarà amb la utilització de *pthreads* [73] amb el que s'activen dos fils simultanis dins cadascun dels nodes: el fil que correspon al client i el fil del servidor. Amb una configuració com aquesta caldrà tenir molt en compte la sincronia a l'hora de realitzar els processos procurant que les peticions coincideixin en el temps als clients i servidors de diferents nodes. Un dels mètodes amb els que s'aconsegueix aquest **sincronisme** és amb la introducció de semàfors abans d'avançar cap a les següents etapes de la implementació.

Per reduir una mica la complexitat del que pot ser tenir un important nombre de nodes en tota la xarxa, s'opta per reduir totes les relacions que poden tenir els nodes a aquells veïns contigus. Això és possible ja que les interferències que a nosaltres ens poden afectar seran aquelles que provinguin d'altres veïns i no de les comunicacions del propi node amb les estacions a un sol *hop*. A més, així s'aconsegueix tenir un **sistema distribuït** entre totes les estacions, molt més manejable que un sistema centralitzat des d'un dels nodes o des d'altres entitats externes a la WMN.

Dins el programa es definiran vàries **estructures** que serviran per gestionar les dades generades pel protocol, però haurà una en especial que serà l'estructura central del protocol i que resultarà clau en el desenvolupament final de la solució, serà l'estructura que nosaltres hem anomenat "**paquet**" i que al mateix temps està composta per tres camps: un string on es recull l'adreça IP del veí en qüestió i que té una longitud igual a la longitud màxima que pot assolir una adreça IP, un string on es guarda l'adreça MAC d'aquest mateix veí i amb una longitud igual a la mesura de les adreces MAC afegint-li els punts separadors entre les parelles de nombres i, finalment, un camp tipus *int* on apareixerà quin és el **canal òptim** per arribar al veí. La definició de l'estructura es pot veure a continuació.

```
struct paquet
{
        struct iph dir_ip;
        char mac[LONGMAC];
        int channel;
        int channel_meu;
};
```

Per omplir aquesta estructura i després poder-la utilitza per poder complir el nostre objectiu final s'hauran de seguir les etapes mostrades al següent diagrama (figura 5.15).

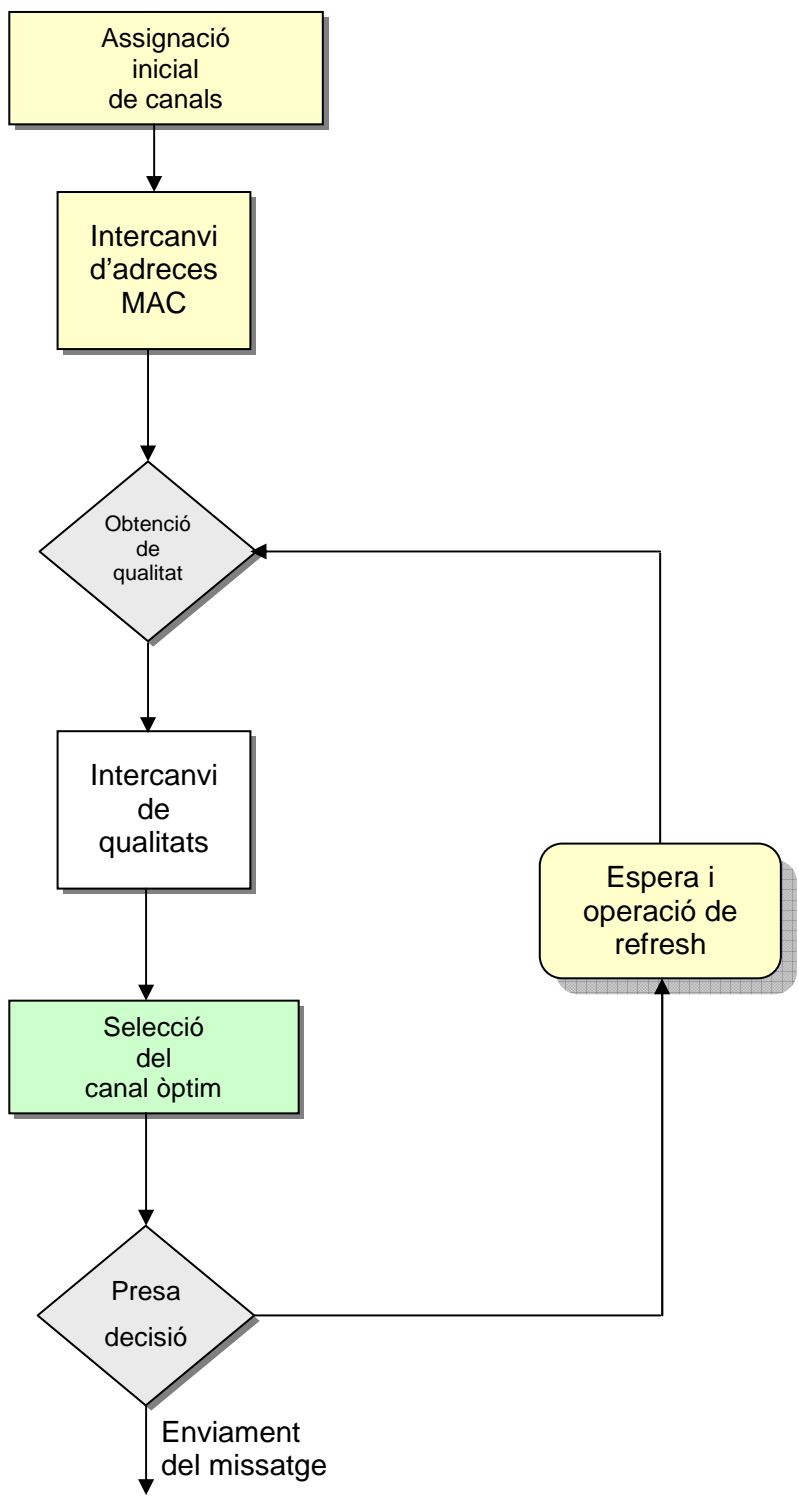

**Figura 5.15: Protocol de negociació de canal**

Les etapes que segueix el **protocol de diàleg** i que es poden observar gràficament a la figura 5.15 són les següents:

- **Assignació inicial de canals.** Abans de començar a realitzar cap tipus de mesura de qualitat, s'**assignen** els canals inicials pels que han de circular els enllaços. Aquests enllaços seran assignats tenint en compte els enllaços que han assignat els veïns, evitant així interferències pròpies generades pel mateix sistema ja a la configuració inicial. Per exemplificar la situació dir que si a un dels enllaços s'assigna al canal 1 (que és el de "ath0"), a l'enllaç contigu s'assignarà el canal 11 (corresponent a "ath1"). Per realitzar aquesta etapa s'aniran assignant els canals a mesura que es té coneixement dels que han estat assignats pels nodes veïns. Tot aquest procés s'anirà materialitzant amb l'intercanvi de dades entre els clients i servidors dels diferents nodes. Aquesta etapa d'inicialització només es realitzarà en la primera execució del programa, en les altres execucions, aquesta etapa serà evitada, tal com es pot veure al diagrama de la figura 5.16.

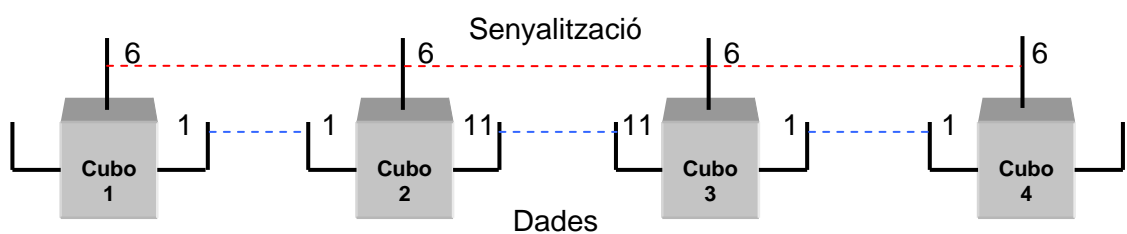

**Figura 5.16: Assignació inicial de canals**

- **Intercanvi d'adreces MAC.** La informació de l'adreça MAC ens servirà per completar un dels camps de l'estructura bàsica del programa, però sobretot, ens servirà per realitzar les funcions de escaneig correctament, així com de consulta de qualitat dels canals. Per aconseguir conèixer les MAC s'executen les instruccions pertinents per extreure la MAC del node i es van enviant entre els nodes veïns aquesta informació. Amb aquest intercanvi es determinarà l'adreça MAC del veí, havent enviat la pròpia al node contigu que, igualment, confeccionarà les seves pròpies taules.

- **Obtenció de les qualitats.** A partir d'aquesta etapa ja entrem al cicle **recursiu** del programa, repetint-se aquest pas tantes vegades com sigui necessari. Aquesta etapa és realitzada pels clients de tots els nodes. El que es fa és realitzar l'escaneig de l'entorn obtenint a canvi unes mesures de qualitat. Aquest escaneig és realitzat amb un *script* de Perl (veure annex A.9), i el que fa és realitzar una instrucció Linux de *scan* a sobre de la **interfície de senyalització** que hem designat abans (aquí seria la "wlan0"). Aquesta operació retornarà una informació amb totes les interferents que el *meshcube* detecta i que, d'alguna manera, estan interferint en el nostre sistema. La informació que ens interessarà de la què ens retorni la instrucció, serà aquella que es referirà a quina freqüència està actuant el senyal interferent en qüestió i el nivell de senyal mesurat en dBm que apareixerà en el llistat. Quan es tenen les dades, aquestes es recopilen en una llista de canals on es mostraran el nivell de senyal interferent en dBm màxim dels dos veïns que rebem de cada canal.

- **Intercanvi de qualitat i càlcul definitiu de qualitats.** Aquesta operació es realitzarà íntegrament al servidor del node. Aquest rebrà la llista de **senyal interferent** confeccionada pel client del node veí i farà la mateixa operació d'escaneig que s'havia realitzat al client. Quan es tenen les pròpies dades, calculades al **servidor**, es comparen amb les dades rebudes del client del veí i s'elabora una sola llista que recollirà el pitjor dels valors per confeccionar-la. El pitjor valor serà el que tingui l'interferent més elevat en dBm. Aquest criteri d'elecció és fa així ja que el que estem fent realment és assignar un canal a un enllaç que dependrà de les interferències rebudes als dos extrems de la comunicació. I aquesta és també la raó per la qual es repeteix la mateixa operació als dos *meshcube* que, a més, incrementa la fiabilitat del resultat final.

```
for (i = 0; i < num_canals; i++)
{
    if (qual_client[i] < qual_server[i])
          qual = qual_client[i];
    else
          qual = qual_server[i];
}
```

- **Selecció del canal òptim.** També és una operació realitzada al servidor del node. Quan es tenen les dades definitives respecte al senyal interferent, s'executa l'algorisme de **selecció del canal**. El criteri per fer l'elecció és el de fer la selecció d'entre el ventall de possibles canals suportats pel sistema. Als nodes que s'han implementat les opcions eren només dues, les que oferien els canals sintonitzats per "ath0" i per "ath1". De totes formes s'ha dissenyat un sistema que elabora una llista ordenada d'aquells canals que es poden utilitzar de forma ordenada de millor qualitat a menor. Per realitzar aquesta mesura s'ha tingut en compte com afectaven els canals contigus i, per tant, no ortogonals als canals que es podien elegir. Per exemple, s'han comptat les interferències detectades al canal 1, al 2, al 3 i al 4 per determinar la qualitat del canal 1. El nivell de senyal interferent que s'ha seleccionat en aquests casos sempre ha estat el més elevat. Quan s'ha fet la selecció de canal aquest s'envia cap al client del node veí.

```
for (i = 0; i < num_canals; i = i + SEPARACIO)
{
    if (i-4 < 0) j = 0; //per si no són ortogonals
    else j = i-4;

    if (i+4 > num_canals-1) k = num_canals-1;
    else k = i+4; //per si no són ortogonals

    for (j; j <= k; j++)
    {
        if (num2 < qual_mitja[j])
        {
            num2 = qual_mitja[j]; //anem agafant el millor
        }
    }

    if (num2 <= num)
    {
        num = num2;
        canal = i+1; //anem agafant el millor
    }

    num2 = -100;
}
```

- **Presa de decisions i enviament de l'estructura al mòdul.** Quan el client rep la informació sobre el **canal òptim**, aquest decidirà si el sistema necessita un canvi de canal. Amb això es vol dir que es vol deixar un marge a l'hora de fer un canvi, ja que depenent de si la diferència de resultats entre qualitats no és massa àmplia, pot no convenir realitzar el canvi. Al nostre cas s'ha solucionat amb un **llindar** en dBm que superat, realitza el canvi indicat pel servidor. Aquí també entra la possibilitat d'afegir memòria al sistema, tenint en compte les dades obtingudes a passades anteriors. Això podria ser una bona opció per no realitzar canvis massa immediats però presenta problemes a l'hora del càlcul matemàtic a realitzar. Una operació de *refresh* s'haurà de fer tenint en compte que les dades que es tenen estan preses en dBm, i que necessitaran ser operades amb inverses de log, cosa que podria donar problemes al compilador creuat dels *meshcubes*. Per tant, es va optar per no incloure aquestes operacions en aquest protocol, quedant marcat que poden ser utilitzades en un futur. Després, quan s'hagi pres la decisió i fet tots els càlculs, s'enviarà tota l'estructura "paquet" cap al mòdul (estem al *user space*) que és on es prendran les decisions directes de canvi d'interfície.

```
qual_ponderada = qual1*param_refresh + qual2*(1-param_refresh);
```

- **Espera i repetició de les operacions.** Quan s'han acabat de realitzar totes les operacions descrites, s'espera un cert temps, modificable per l'usuari i, després es torna a executar tot el procés des de l'etapa de **obtenció de qualitat**. La repetició es realitza per detectar canvis en les fonts de les interferències de l'entorn i que, per tant, requereixin d'un canvi de canal. D'aquesta manera sempre es mantindrà **actualitzada** la configuració dels canals del sistema.

Convé realitzar un apunt sobre el mètode d'escaneig de qualitat dels canals. El sistema utilitzat no és d'una fiabilitat total, i aquesta és una de les causes per duplicar les proves. A més amb el *scan* no tenim en compte les interferències produïdes pel node en qüestió que realitza l'operació ni per les produïdes per les que estan a la mateixa xarxa. Això, com es veurà al següent apartat, pot causar problemes si l'extensió de la xarxa és prou gran, abraçant varis salts.

Per fer-se una idea real de com es configuren en el temps aquestes operacions descrites i com es realitza la relació entre els diferents fils dels nodes, es mostrarà un exemple (figura 5.17) entre dues estacions que executen el protocol de negociació de canal.

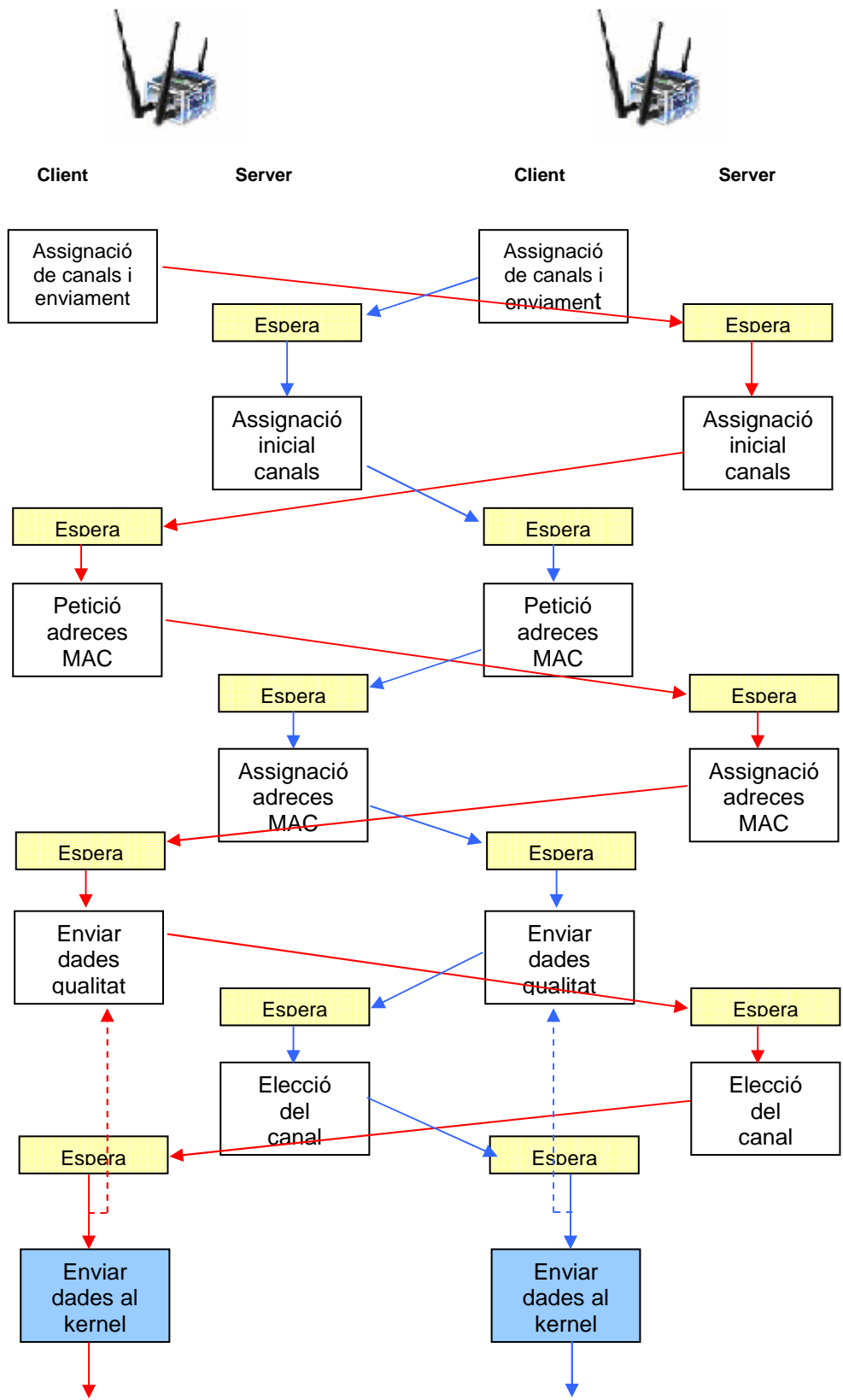

**Figura 5.17: Diàleg entre dos estacions**

### 5.4.3 Mecanisme de selecció d'interfícies

Aquest ha estat el punt que més costós ha resultat a l'hora d'implementar el disseny plantejat, com s'ha comentat a l'apartat d'alternatives d'implementació. Finalment, el mètode que ha resultat més convincent i que ha donat resultats positius en la seva aplicació ha estat el del **bonding driver**.

El **bonding driver** [62] és una eina que en principi va ser dissenyat per ser emprat en escenaris Ethernet a nodes on es tenien vàries targetes d'aquest tipus baix un mateix node. L'objectiu que es tenia amb aquest *driver* virtual era el d'aconseguir balancejar de la millor manera possible les transmissions de dades entre totes les targetes Ethernet integrades a dit node. Per fer-ho, **bonding** desenvolupava una sèrie de rutines on es permetien gestionar canvis d'interfície seguint distintes polítiques d'utilització, moltes d'elles bastant populars al món de la tecnologia. Algunes d'aquestes polítiques són el *round-robin*, on s'alternen les transmissions canviant d'interfície després d'haver transmès un paquet o el balanceig de càrrega, on s'envia un nombre similar de paquets cap a una interfície o una altra, tot això entre d'altres opcions que han estat implementades per diferents autors, complementant el codi font del **bonding driver**.

Però autors com Kim i Ko, van aconseguir utilitzar el mateix *driver* virtual per interfícies sense fils [63] i alguns altres com Chereddi *et al.* el varen utilitzar per aconseguir tenir un protocol multicanal [61] i van demostrar que, convenientment modificat, és una opció molt interessant que permet escollir les interfícies segons el disseny de **selecció d'interfícies** que es realitzi independentment. Per aquesta raó, es va decidir la utilització del **bonding driver** en aquest projecte.

El mode de funcionament del **bonding driver** és, a més, molt convenient i encaixa perfectament al nostre disseny. A l'aplicar-ho sobre un dels nodes, el que es fa és crear una estructura d'interfícies on una d'elles és la *master* i les altres seran les *slave* d'aquesta primera. La interfície *master* serà una estructura virtual que es crearà per damunt de les altres i que s'anomenarà "bond0". Per davall estaran aquelles interfícies que, mitjançat la instrucció *ifenslave* (veure més detalls a annex A.5), fan de *slave* de la interfície "bond0". Des d'aquesta interfície primària es podran controlar les altres secundàries i serà la que elegirà sobre quina secundària (que són les interfícies reals) s'haurà de utilitzar.

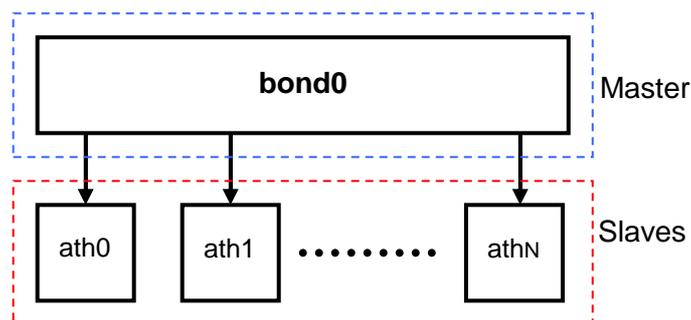

**Figura 5.18: Bonding driver**

Al nostre cas, el ***bonding*** es realitza sobre les dues interfícies de dades ("ath0" i "ath1") que funcionen a diferents freqüències. Una avantatge del *bonding* és que uniformitza totes les interfícies a la mateixa adreça IP i a la mateixa adreça física. D'aquesta manera s'aconsegueix identificar un únic node com a una entitat depenent de la interfície virtual creada, la "bond0". Les adreces imposades poden ser introduïdes externament per l'usuari o simplement s'agafen les dades d'una de les interfícies "esclavitzades".

Dir que, per a què el ***bonding driver*** funcionés i no es perdessin paquets en els canvis d'interfície, i tampoc apareguessin paquets duplicats, es varen **uniformitzar** les adreces físiques de totes les interfícies esclavitzades a una mateixa i es va posar el mateix identificador de xarxa a totes les estacions del sistema (mateix Extended Service Set Identifier (ESSID)), i a totes les interfícies de dades d'aquests nodes. D'aquesta manera ens assegurem de què totes les estacions s'associïn entre elles i, per tant, formin part d'una mateixa xarxa, evitant possibles duplicats dels paquets que es confonien en el moment d'enviar paquets per xarxes diferents.

Una vegada que es va trobar l'eina que més s'adaptava a la solució que s'estava conformant, va tocar modificar algunes parts del ***bonding driver***, creant un mòdul propi que aprofités la informació provinent del protocol de diàleg i prengués les decisions pertinents respecte al **canvi d'interfícies**. Les decisions es prendran a nivell de paquet, tal com es feia amb la solució del ***Netfilter***, i serà amb cada paquet que es realitzaran totes les operacions implementades al mòdul. L'única funció del mòdul serà la guardar les estructures de paquets que va rebent periòdicament del protocol de diàleg, on després aquesta informació és aprofitada pel ***bonding driver***, a la que se li afegeix una rutina anomenada *bond_xmit_protocol*, que el que es fa és veure quin és el **canal òptim** guardat a les estructures i canviar d'interfície si la situació així ho requereix. Serà, per tant, realitzar una petita operació de **comparació** i de **canvi d'interfícies** per cada paquet, tal com es pot veure al següent pseudocodi.

```
if ((pack.channel != channel_meu) && (pack.channel != 0))
{
        name = treure_interficie(pack.channel);
        while (strcmp(current_slave, name) != 0)
        {
                bond->current_slave = slave->next;
        }
        channel_meu = pack.channel;
        }
}
```

Com hem vist, finalment, s'ha requerit del concurs de la programació al *kernel*, degut a que és aquesta la manera més simple per fer comparacions a nivell de paquet, a més de ser la manera que ens serveix per relacionar-nos amb el ***bonding driver***, que s'emprarà al sistema com un **mòdul** més.

El problema d'aquest sistema és que les operacions es realitzaran sobre tots els paquets que s'hagin de transmetre per les interfícies de dades. Si el **mòdul** és molt complex el rendiment de les transmissions es veurà afectat. Per aquesta raó, s'ha procurat crear unes rutines que haguessin de fer pocs recorreguts i poques comparacions per conservar al màxim els avantatges que pot proporcionar el mecanisme de selecció d'interfícies. Així i tot, a les proves es veurà que totes les operacions que realitza ***bonding driver*** influeixen una mica en el rendiment final del muntatge.

Per finalitzar amb l'apartat d'implementació, comentar que, finalment, va haver un altre tipus de problemes tecnològics produïts internament pels mateixos *meshcube*. Robinson *et al.* [34] o Zhu *et al.* [37] ja havien descrit successos similars a proves específiques realitzades sobre aparells semblants als que nosaltres utilitzem en la realització del projecte. Es tracta d'interferències produïdes per l'acoblament que es produeix a les antenes al canviar d'interfícies en un mateix node degudes a la **proximitat** entre els propis transceptors. L'energia transmesa per l'antena transmissora és suficientment gran per distorsionar els filtres interns i els amplificadors del receptor, degradant el rendiment del sistema. Als *meshcube* la distància a la que venien les antenes era d'aproximadament uns 5 cm, distància massa petita per què no es produeixi acoblament. Aquest factor serà estudiat a les proves i es veurà una **evolució** de com afecta la distància entre transceptors a les transferències d'informació.

Una altra **limitació** detectada damunt el prototipus dissenyat es dóna a les proves físiques damunt freqüències que corresponen a la IEEE 802.11a, és a dir, les de 5 GHz. Com hem vist teòricament a la web [27], a IEEE 802.11a hi ha fins a 13 canals ortogonals repartits en tot l'espectre disponible cada 20 MHz. Però una vegada duta a terme la realització pràctica del **prototipus**, es veu que realment, hi ha molts menys canals ortogonals, havent de ser la separació entre les freqüències de treball molt majors que 20 MHz. De fet, s'han comptat **4 canals absolutament ortogonals** entre ells que és una xifra semblant al nombre de canals ortogonals a IEEE 802.11b. La raó per la que pot succeir aquesta limitació és per què al ser una versió basada en OFDM, requereix un nivell de senyal-soroll major que a la versió IEEE 802.11b. Les estacions utilitzades al disseny no poden transmetre amb una major potència de 18 dBm i en OFDM, al compartir les bandes més espectre, si no es transmet en major potència, aquestes són més difícils de discriminar. Això provoca que hagi interferències laterals que afecten als canals adjacents, essent l'efecte més gran quan més propers són els canals [75]. A l'apartat següent veurem unes il·lustracions de com evoluciona la capacitat a mesura que es va separant la freqüència i de quina és la separació que necessitem per tenir canals ortogonals.

# 6 AVALUACIÓ DELS RESULTATS

Una vegada s'ha realitzat la implementació del sistema tal com s'ha descrit al capítol anterior, toca provar quin rendiment assoleix el nostre prototipus amb les eines implementades i avaluar els avantatges d'emprar el disseny presentat al projecte, així com de valorar els resultats obtingut a les proves realitzades.

Per a la mesura de la qualitat dels enllaços s'ha utilitzat una eina que Linux i *Nylon* ens permet utilitzar als seus sistemes: el **iperf** [74]. Aquest *software* ens permet mesurar la capacitat total de l'enllaç i ho fa saturant la capacitat de l'enllaç. És a dir, s'envien el màxim de paquets cap a la destinació indicada omplint l'enllaç al màxim de la seva capacitat. El mateix *software* retorna els resultats de nombre de paquets enviats i de la capacitat de l'enllaç fins la destinació seleccionada. A més, **iperf** és una aplicació que ofereix un ampli ventall d'opcions que ens facilitarà la lectura de les dades en diferents situacions. Així, ens permet realitzar les transmissions de paquets emprant el protocol Transmission Control Protocol (TCP) o el protocol User Datagram Protocol (UDP), es pot fer la selecció d'ample de banda a transmetre, el temps en el que es transmet el flux de dades, l'interval de temps amb el que es vol que es rebin resultats... (més informació, veure annex A.6)

A aquest treball els diferents escenaris que es proposen i sobre els que s'obtindran els resultats estan extrets amb l'aplicació **iperf**. Cadascun d'aquests escenaris ha estat avaluat en 10 proves de 30 segons cadascuna d'elles i realitzades amb els dos protocols de transports disponibles: **TCP** i **UDP**.

Els escenaris estaran compostos de vàries estacions a una distància d'aproximadament uns dos metres entre elles. Per tant, tots els nodes tindran visibilitat entre ells, fent-se necessaris mecanismes de filtratge per poder configurar les topologies desitjades. El filtratge es realitza amb *iptables* (veure annex A.4) descartant paquets depenent de la seva adreça MAC. Això permet que el protocol d'encaminament configuri la topologia corresponent, encara que les estacions es vegin entre elles.

La longitud dels paquets que s'enviaran amb **iperf** seran els enviats per defecte per l'aplicació, és a dir, de 8 Kbytes, i la potència amb la que es faran les transmissions és la que ve donada per defecte pels *meshcube* quan s'inicien que, a més, és la màxima en què poden transmetre, 18 dBm. La velocitat en què es transmetran les dades dependrà de la versió IEEE que s'estigui utilitzant. Si s'està utilitzant la **IEEE 802.11b**, la modulació que tindran seleccionada les estacions serà la de 11Mbps, en canvi, per **IEEE 802.11a**, la modulació correspondrà a la de 12Mbps. Finalment, remarcar que la separació entre les antenes de les interfícies estan separades uns 30 cm (excepte en el cas específic de 6.4).

Les proves s'han realitzat sobre els diferents escenaris amb el protocol d'encaminament funcionant a la interfície de senyalització, amb el protocol de negociació de canal engegat i amb el *bonding driver* aplicat sobre les interfícies de dades que composen cadascuna de les estacions. Aquestes execucions comportaran un cert retard extra en cada paquet transmès. A continuació es fa un recull de dades a la taula 6.1 amb diferents casos a sobre d'una arquitectura amb tres estacions. El cas 1 indica transmissions sense engegar cap mòdul de la solució, el cas 2 té executats el protocol d'encaminament modificat i el *bonding driver* i el cas 3 té totes les funcionalitats en marxa. Els resultats de la taula són en milisegons (ms).

| Cas 1 | Cas 2 | Cas 3 |
|-------|-------|-------|
| 1,8 | 2,1 | 2,9 |
| 1,8 | 2,2 | 3,4 |
| 1,8 | 1,9 | 3,6 |
| 1,8 | 2 | 3,2 |
| 1,9 | 2,2 | 2,3 |
| 1,8 | 2 | 3,3 |
| 1,8 | 2,2 | 2,9 |
| 1,8 | 2 | 3,1 |
| 1,7 | 2,1 | 2,1 |
| 1,8 | 2,1 | 2,7 |
| **1,8** | **2,08** | **2,95** |

Taula 6.1: Retard dels paquets

Com es veu, la diferència és de més o menys 1 ms, retard tolerable tenint en compte les grans millores que aquesta solució ens presenta.

A continuació es mostraran els diferents reculls de dades amb les seves gràfiques comparatives associades dels diferents casos. Les figures dels diferents casos indiquen el nombre d'estacions de l'escenari i el canal que es sintonitza a l'enllaç entre els nodes. El nombre en vermell que apareix correspon al canal utilitzat a les proves.

# 6.1   ESCENARIS AMB UN CANAL

Per  començar a avaluar resultats pràctics que apareixen a la nostra implementació, primer es mesurarà l'evolució que pateixen els escenaris on s'utilitza el mateix canal a tots els enllaços entre les estacions a l'afegir nous nodes que interactuen en la comunicació de la xarxa. Les proves físiques es realitzaran en escenaris de dues, tres, quatre i cinc estacions, amb el mateix canal enllaçant entre elles. A continuació s'elaborarà un llistat dels resultats obtinguts pels diferents casos.

## 6.1.1   Proves amb IEEE 802.11a

Primerament es faran aquestes proves amb freqüències de **5 GHz** que, en general, pateixen menys interferències que les de **2,4 GHz** degut a que són menys utilitzades tant pels punts d'accés com per altres aparells sense fils.

Les proves amb IEEE 802.11a s'han mesurat amb les estacions a una velocitat de transmissió a les estacions de 12 Mbps, pròxim als 11 Mbps que pot arribar a atorgar IEEE 802.11b, i que ens servirà per tenir valors semblants a les altres proves que ens permetran extreure més conclusions.

**Cas 1: Dues estacions**

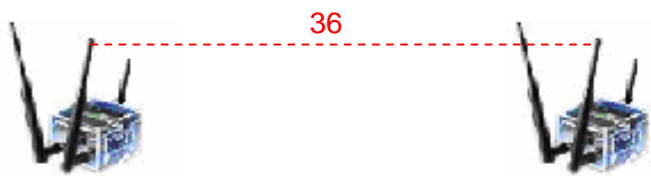

**Figura 6.1: Dos estacions, un canal**

| TCP (Mbps) | UDP (Mbps) |
|:---:|:---:|
| 8,86 | 9,93 |
| 8,83 | 9,93 |
| 8,83 | 9,93 |
| 8,79 | 9,93 |
| 8,81 | 9,93 |
| 8,8 | 9,93 |
| 8,82 | 9,93 |
| 8,81 | 9,93 |
| 8,83 | 9,92 |
| 8,83 | 9,92 |
| **8,82** | **9,93** |

Taula 6.2: Dos estacions, un canal

Com es veu a la taula 6.2, la transmissió de dades és la màxima assolible, voltant els **9 Mbps** pel cas de **TCP** i els **10** per **UDP**. No s'arriba a 12 Mbps (la velocitat assignada a les estacions) per culpa de reconeixements a diferents velocitats i capçaleres inherents als protocols [28].

**Cas 2: Tres estacions**

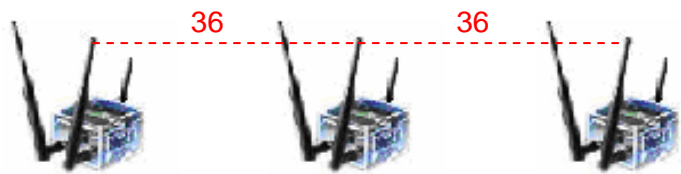

Figura 6.2: Tres estacions, un canal

| TCP (Mbps) | UDP (Mbps) |
|:---:|:---:|
| 4,37 | 5,01 |
| 4,37 | 5 |
| 4,38 | 5,01 |
| 4,39 | 5,02 |
| 4,38 | 5 |
| 4,39 | 5,01 |
| 4,39 | 5,01 |
| 4,41 | 5,02 |
| 4,41 | 5,01 |
| 4,39 | 5,01 |
| **4,39** | **5,01** |

Taula 6.3: Tres estacions, un canal

Aquí es confirma el que s'ha comentat a l'apartat 4.3.2, on s'assegurava que per cada *hop* afegit a una jerarquia en línia, la capacitat total de la xarxa s'aniria reduint en $\frac{1}{N}$, on N és el nombre total d'enllaços entre les estacions. A aquest cas, a l'haver dos enllaços sintonitzats al mateix canal, la capacitat de la xarxa es veurà reduïda aproximadament a la meitat de la capacitat original calculada al primer cas.

**Cas 3: Quatre estacions**

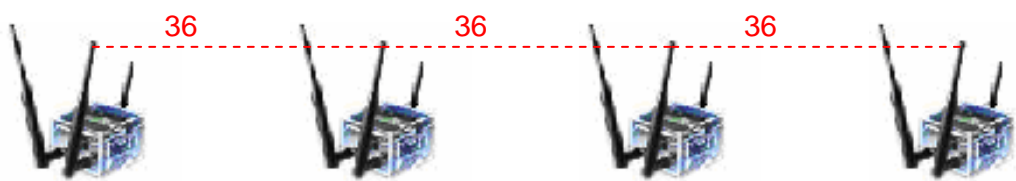

**Figura 6.3: Quatre estacions, un canal**

| TCP (Mbps) | UDP (Mbps) |
|:---:|:---:|
| 2,92 | 3,25 |
| 2,9 | 3,26 |
| 2,89 | 3,23 |
| 2,91 | 3,25 |
| 2,9 | 3,26 |
| 2,9 | 3,26 |
| 2,92 | 3,24 |
| 2,88 | 3,24 |
| 2,9 | 3,23 |
| 2,89 | 3,25 |
| **2,90** | **3,25** |

**Taula 6.4: Quatre estacions, un canal**

En aquest cas hi ha tres enllaços iguals que connecten quatre estacions en línia, confirmant-se segons els resultats de la taula 6.3 que es va dividint la capacitat inicial pel nombre d'enllaços amb el mateix canal sintonitzat.

**Cas 4: Cinc estacions**

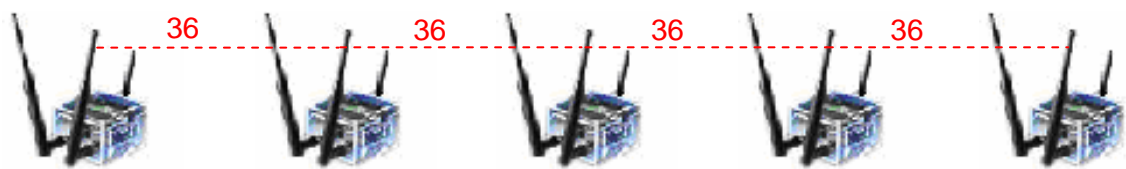

**Figura 6.4: Cinc estacions, un canal**

| TCP (Mbps) | UDP (Mbps) |
|:---:|:---:|
| 2,19 | 2,43 |
| 2,17 | 2,43 |
| 2,18 | 2,43 |
| 2,17 | 2,43 |
| 2,18 | 2,43 |
| 2,19 | 2,43 |
| 2,18 | 2,43 |
| 2,18 | 2,43 |
| 2,19 | 2,43 |
| 2,19 | 2,43 |
| **2,18** | **2,43** |

**Taula 6.5: Cinc estacions, un canal**

El cas 4 confirma una vegada més les teories sobre disminució de capacitat exposades anteriorment, ja que amb cinc estacions i quatre enllaços, la capacitat original es veu reduïda a una quarta part de l'original.

## 6.1.2 Proves amb IEEE 802.11b

Els mateixos casos avaluats amb freqüències de 5 GHz, també han estat avaluats per freqüències de **2,4 GHz**. Les freqüències d'IEEE 802.11b en el nostre entorn de treball són més utilitzades que les d'IEEE 802.11a, ja que els punts d'accés sense fils per a la connexió a la xarxa Internet utilitzen aquestes freqüències provocant en molts de casos interferències en les nostres proves.

Els resultats han estat obtinguts en moments de baixa interferència, però al contrari que amb IEEE 802.11a els resultats, en general, són més irregulars. A més, s'ha de tenir en compte que les estacions ofereixen velocitats diferents que la versió de 5 GHz. Per obtenir resultats semblants s'han configurats els nodes a una velocitat de 11 Mbps i així poder comparar des d'uns valors similars als que teníem abans.

**Cas 1: Dues estacions**

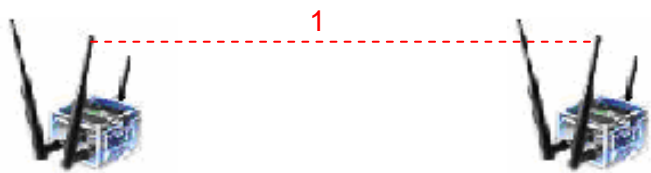

**Figura 6.5: Tres estacions, un canal 2,4 GHz**

| TCP (Mbps) | UDP (Mbps) |
|:---:|:---:|
| 6,39 | 7,34 |
| 6,3 | 7,4 |
| 6,41 | 7,37 |
| 6,36 | 7,41 |
| 6,34 | 7,35 |
| 6,28 | 7,37 |
| 6,41 | 7,35 |
| 6,37 | 7,39 |
| 6,36 | 7,32 |
| 6,25 | 7,44 |
| **6,35** | **7,37** |

**Taula 6.6: Dos estacions, un canal 2,4 GHz**

Aquests resultats mostren el màxim de capacitat que es pot obtenir amb una comunicació directa entre dues estacions a 11 Mbps compartint el mateix canal. Recordem que la pèrdua de capacitat es deu a les capçaleres implícites de les comunicacions sense fils i de la pila de protocols.

**Cas 2: Tres estacions**

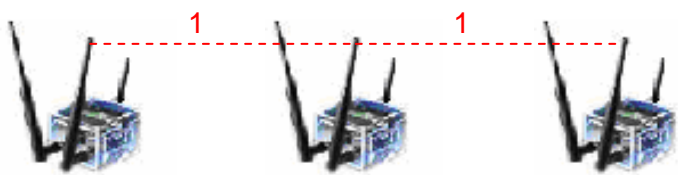

**Figura 6.6: Tres estacions, un canal 2,4 GHz**

| TCP (Mbps) | UDP (Mbps) |
|:---:|:---:|
| 3,28 | 3,71 |
| 3,28 | 3,72 |
| 3,28 | 3,69 |
| 3,22 | 3,72 |
| 3,22 | 3,71 |
| 3,19 | 3,73 |
| 3,23 | 3,62 |
| 3,18 | 3,76 |
| 3,15 | 3,72 |
| 3,18 | 3,75 |
| **3,22** | **3,71** |

**Taula 6.7: Tres estacions, un canal 2,4 GHz**

Com es veu, a l'afegir una estació en línia més a sobre del mateix canal, la capacitat que els enllaços ens ofereixen es veu reduïda a, aproximadament, la **meitat** de la capacitat que teníem al cas de comunicació directa entre estacions adjacents.

**Cas 3: Quatre estacions**

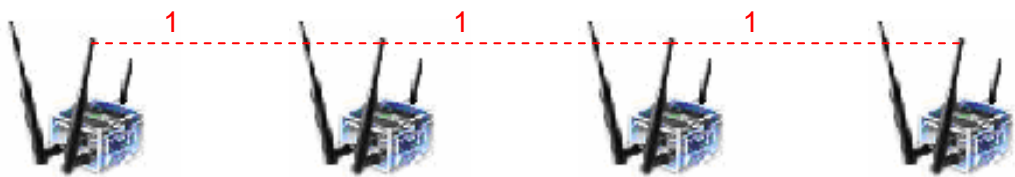

**Figura 6.7: Quatre estacions, un canal 2,4 GHz**

| TCP (Mbps) | UDP (Mbps) |
|:---:|:---:|
| 2,11 | 2,55 |
| 2,11 | 2,54 |
| 2,09 | 2,55 |
| 2,11 | 2,58 |
| 2,12 | 2,56 |
| 2,13 | 2,57 |
| 2,08 | 2,54 |
| 2,11 | 2,56 |
| 2,06 | 2,55 |
| 2,12 | 2,56 |
| **2,10** | **2,56** |

**Taula 6.8: Quatre estacions, un canal**

A aquesta configuració li afegim una estació en línia més que al cas anterior i com veurem al resultats posteriors recollits a la taula 6.8, veiem que la capacitat és una **tercera part** de la que es tenia al primer cas, de comunicació directa entre els nodes.

<u>**Cas 4: Cinc estacions**</u>

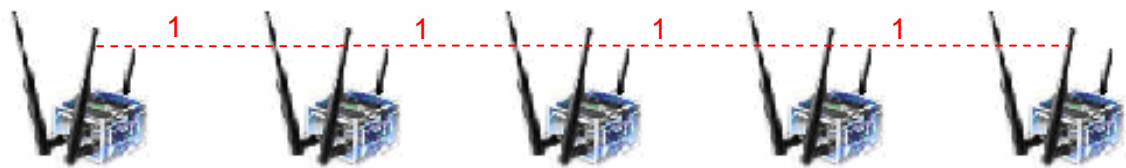

**Figura 6.8: Cinc estacions, un canal 2,4 GHz**

| TCP (Mbps) | UDP (Mbps) |
|:---:|:---:|
| 1,63 | 1,93 |
| 1,6 | 1,92 |
| 1,6 | 1,93 |
| 1,63 | 1,92 |
| 1,62 | 1,92 |
| 1,6 | 1,93 |
| 1,6 | 1,93 |
| 1,62 | 1,92 |
| 1,62 | 1,93 |
| 1,61 | 1,93 |
| **1,61** | **1,93** |

**Taula 6.9: Cinc estacions, un canal 2,4 GHz**

Per a cinc estacions es repeteixen els patrons ja comentats anteriorment. Al tenir 4 enllaços dins un mateix rang de cobertura, la capacitat total es **dividirà entre els 4 enllaços** per igual, essent una quarta part de l'original.

## 6.1.3  Valoracions

Amb aquestes proves queda demostrat el que s'afirmava als apartats teòrics presentats anteriorment, a on s'assegurava que l'augment de estacions connectades al mateix canal comportava una important disminució en les prestacions de la xarxa sense fils. En les següents gràfiques es pot veure com va evolucionant la capacitat de les dues versions IEEE 802.11 anteriorment provades i pels dos protocols emprats.

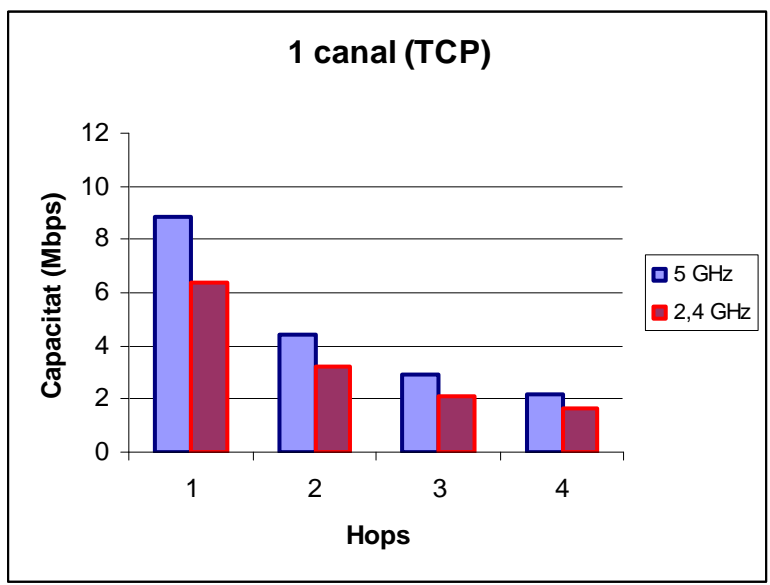

**Figura 6.9: Evolució capacitat un canal per TCP**

Una cosa similar es té pel cas UDP:

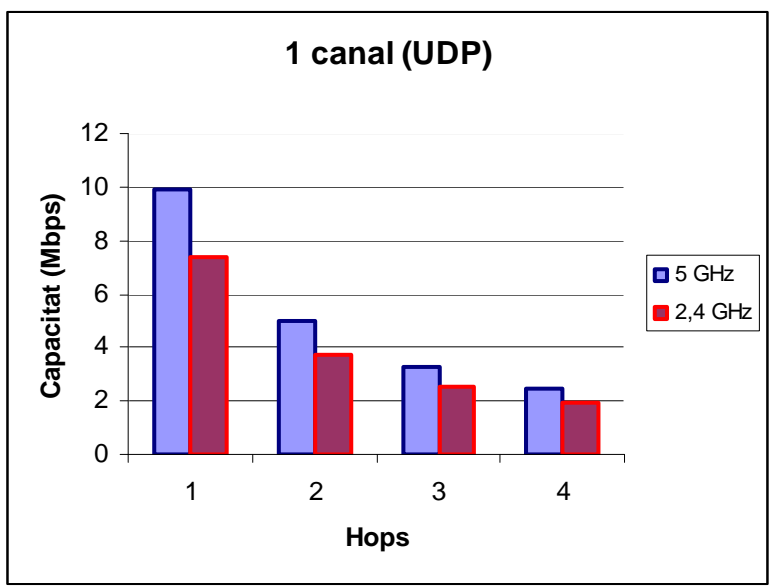

**Figura 6.10: Evolució capacitat un canal per UDP**

Visualitzant a les gràfiques es veu que en tots els casos es segueix el mateix patró de disminució de capacitat, una reducció que es produeix de forma exponencial invertida $\frac{1}{N}$ on N és el nombre de salts del sistema.

Dels resultats obtinguts, també es pot ressaltar la diferència de capacitat transmesa entre les proves a 5 GHz i les de 2,4 GHz. S'ha dit anteriorment que a les primeres la velocitat de les estacions era de 12 Mbps, mentre que a la segona era de 11 Mbps. A les gràfiques es veu que la diferència és major que la que pertocaria en un entorn sense interferències i amb els mateixos funcionaments. S'ha introduït com una de les causes l'aparició de més senyals funcionant en la versió IEEE 802.11b ja que les seves freqüències són més utilitzades per les connexions a les xarxes sense fils.

Un altre factor a tenir en compte és el de les velocitats dels ACK propis del funcionament dels *meshcube*. En aquests casos, encara que la velocitat sintonitzada sigui de 11 Mbps, la velocitat de l'ACK és la mínima que es pot utilitzar a IEEE 802.11b, que és 1 Mbps. En canvi, la mínima velocitat a IEEE 802.11a és de 6 Mbps. Tot això implica una menor degradació en les prestacions que s'obtenen al sistema i que queden reflectides a les gràfiques anteriors.

## 6.2  ESCENARIS AMB 2 CANALS

Aquí aprofitarem l'estructura que ens ofereixen els *meshcube* que hem estat utilitzant per tal de poder oferir **dos canals** diferents en les comunicacions. Teòricament, el fet d'oferir 2 canals diferents sobre els que transmetre hauria de millorar la capacitat dels escenaris amb 1 canal. De fet, hauria de proporcionar la màxima capacitat pel cas d'una configuració en línia de 3 estacions o 2 salts.

A continuació, s'avaluaran tots els escenaris utilitzats en el cas anterior i es mesuraran els nous resultats fruit de les modificacions aplicades a les estacions. Aquests resultats seran obtinguts en les dues versions suportades d'IEEE pels *meshcube* i pels dos protocols de transport suportats per l'aplicació *iperf* (TCP i UDP).

A més, una vegada obtinguts aquests resultats, es compararan les dades amb les que s'havien calculat amb els escenaris anteriors de topologies en línia. També s'afegiran nous casos per avaluar els **avantatges** de tenir escenaris amb dos canals, a més de realitzar-se un estudi sobre l'ortogonalitat dels canals amb les estacions que estem manejant i amb les diferents versions d'IEEE.

Finalment, es mostraran els resultats mesurats amb diferents separacions entre antenes i com influeix aquest factor en les comunicacions sense fils, tal com s'avançava a Robinson *et al.* [34].

## 6.2.1  Proves amb IEEE 802.11a

Les proves realitzades a freqüències de 5 GHz són les mateixes que les realitzades a l'apartat 6.1.1, és a dir, proves sobre topologies en línia formades per vàries estacions. També s'ha afegit alguna configuració diferent a la configuració en línia per comprovar aquest rendiment.

Per realitzar les proves, els canals que s'han seleccionat per a les transmissions són totalment **ortogonals**. En el cas d'aquestes proves de 5 GHz, s'ha optat per elegir els canals 36 i 64 (140 MHz de diferència), que estan suficientment separats i asseguren que no hagi cap tipus d'interferència amb l'ús simultani dels dos canals.

Com abans, la velocitat que tindran sintonitzada les estacions serà la de 12 Mbps a cadascuna d'elles.

**Cas 1: Tres estacions**

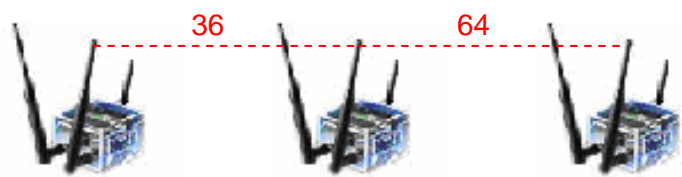

**Figura 6.11: Tres estacions, dos canals**

| TCP (Mbps) | UDP (Mbps) |
|:---:|:---:|
| 8,3 | 9,88 |
| 8,32 | 9,88 |
| 8,38 | 9,88 |
| 8,36 | 9,88 |
| 8,35 | 9,88 |
| 8,37 | 9,88 |
| 8,27 | 9,88 |
| 8,28 | 9,88 |
| 8,3 | 9,88 |
| 8,31 | 9,86 |
| **8,32** | **9,88** |

**Taula 6.10: Tres estacions, dos canals**

En aquest escenari amb dos salts veiem que, al canviar de canal en l'estació intermèdia, podem tenir plena capacitat gràcies a l'ortogonalitat dels canal, que possibiliten que les estacions no interfereixin. La capacitat és pràcticament igual a la que s'aconsegueix amb un canal i les petites pèrdues que es produeixen són degudes als canvis d'interfície realitzats en el node intermedi. Cal destacar també que les pèrdues a TCP són superiors a les que es donen a UDP, conseqüència dels reconeixements propis d'un protocol de transport fiable com TCP. De totes formes, es pot observar que amb aquest sistema estem duplicant la capacitat que teníem en contrast amb el que teníem amb un únic canal.

**Cas 2a: Quatre estacions**

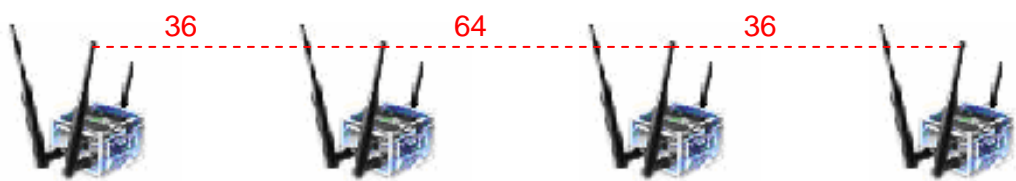

**Figura 6.12: Quatre estacions, dos canals alternants**

| TCP (Mbps) | UDP (Mbps) |
|:---:|:---:|
| 4,35 | 4,86 |
| 4,37 | 4,81 |
| 4,38 | 4,89 |
| 4,44 | 4,83 |
| 4,28 | 4,86 |
| 4,36 | 4,88 |
| 4,37 | 4,9 |
| 4,33 | 4,9 |
| 4,4 | 4,9 |
| 4,4 | 4,9 |
| **4,37** | **4,87** |

**Taula 6.11: Quatre estacions, dos canals (cas I)**

També apliquem el cas de dos canals per a quatre estacions i es pot observar que les dades que s'obtenen són semblants a les obtingudes amb dos estacions amb un sol canal de funcionament. La millora que es produeix enfront al cas d'un canal és de

$$\frac{1/N_{2canals}}{1/N_{1canal}} = \frac{1/2}{1/3} = \frac{3}{2}$$, que és una millora bastant significativa. Aquest cas és, teòricament,

el millor dels que es pot tenir amb quatre estacions i dos canals ja que hi ha menys possibilitats d'ocupació de l'espai del canal amb enllaços alternants en els seus canals que no al contrari. Al nostre cas, com que les distàncies entre estacions són molt petites, les diferències amb altres casos de quatre estacions i dos canals seran quasi inexistents.

**Cas 2b: Quatre estacions**

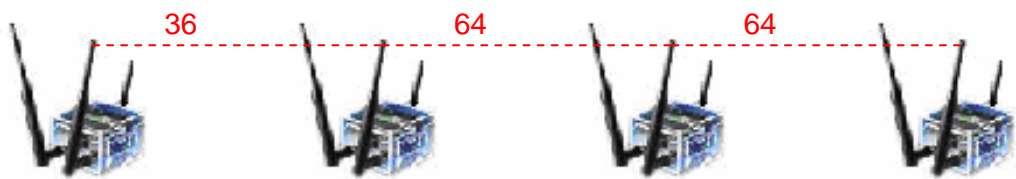

**Figura 6.13: Quatre estacions, dos canals (cas II)**

També s'ha provat el cas de quatre estacions canviant la configuració dels canals entre els enllaços.

| TCP (Mbps) | UDP (Mbps) |
|:---:|:---:|
| 4,52 | 4,85 |
| 4,51 | 4,89 |
| 4,51 | 4,89 |
| 4,46 | 4,66 |
| 4,5 | 4,69 |
| 4,36 | 4,67 |
| 4,37 | 4,93 |
| 4,33 | 4,62 |
| 4,4 | 4,93 |
| 4,4 | 4,93 |
| **4,44** | **4,81** |

**Taula 6.12: Quatre estacions, dos canals (cas II)**

Com es veu als resultats, aquests són molt similars als calculats al cas 2a. Això és degut a que les estacions estan bastant properes i, els extrems de la comunicació es veuen, compartint els canals el mateix espectre de freqüències, provocant la compartició de la capacitat entre les diferents estacions.

**Cas 3: Cinc estacions**

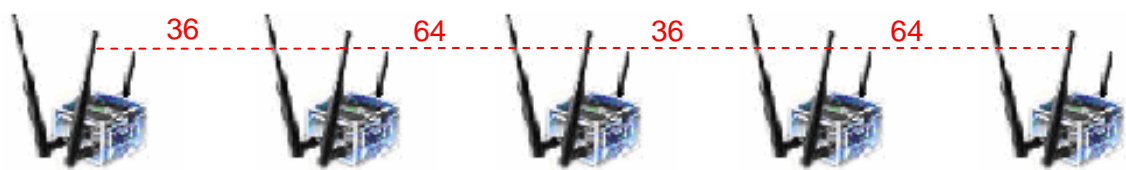

**Figura 6.14: Cinc estacions, dos canals**

| TCP (Mbps) | UDP (Mbps) |
|:---:|:---:|
| 4,37 | 5,02 |
| 4,4 | 5,02 |
| 4,37 | 5,02 |
| 4,37 | 5,02 |
| 4,44 | 5,02 |
| 4,46 | 5,02 |
| 4,45 | 5,02 |
| 4,47 | 5,01 |
| 4,47 | 5,02 |
| 4,44 | 5,02 |
| **4,42** | **5,02** |

**Taula 6.13: Cinc estacions, dos canals**

Pel cas de cinc estacions, si es distribueixen els enllaços de tal manera que es vagi canviant de canal després de cada hop, s'arriba a duplicar la taxa que teníem pel cas d'un sol canal. Per tant, el prototipus ens proporciona una millora bastant destacada.

**Cas 4a: Quadrat (camí 1)**

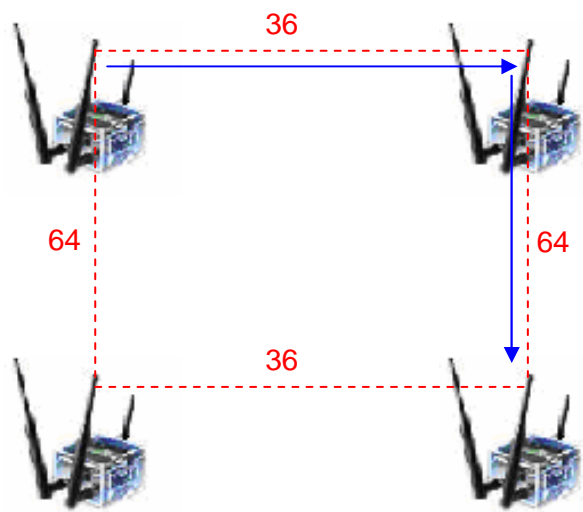

**Figura 6.15: Quadrat, dos canals, camí 1**

| TCP (Mbps) | UDP (Mbps) |
|:----------:|:----------:|
| 8,02 | 9,69 |
| 8,03 | 9,69 |
| 8,13 | 9,68 |
| 8,21 | 9,69 |
| 8,17 | 9,68 |
| 8,19 | 9,68 |
| 8,23 | 9,67 |
| 8,39 | 9,68 |
| 8,15 | 9,68 |
| 8,2 | 9,68 |
| **8,17** | **9,68** |

**Taula 6.14: Quadrat, dos canals, camí 1**

S'ha fet també una prova amb una configuració diferent a la de la topologia en línia, adoptant una topologia quadrada on cadascun dels nodes està comunicat amb el seus nodes contigus, excepte el que es troba en diagonal, que serà amb aquell amb el que ens voldrem comunicar. Les distàncies entre els costats de la configuració és de dos metres.

Per arribar al destí, sempre s'hauran de realitzar dos salts i el prototipus distribuirà els canals de tal manera que els dos enllaços pels que passa tinguin dos canals ortogonals. Una vegada distribuïts, el rendiment mesurat és pràcticament igual que l'obtingut al cas òptim.

**Cas 4b: Quadrat (camí 2)**

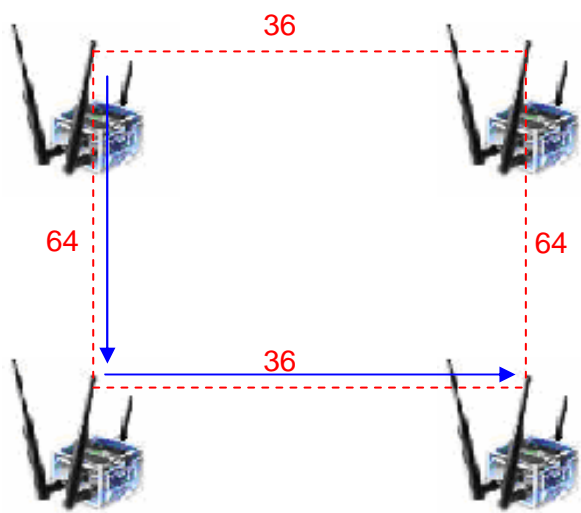

**Figura 6.16: Quadrat, dos canals, camí 2**

| TCP (Mbps) | UDP (Mbps) |
|:---:|:---:|
| 8 | 9,9 |
| 7,5 | 9,9 |
| 7,93 | 9,9 |
| 8,01 | 9,9 |
| 8,17 | 9,9 |
| 8,12 | 9,9 |
| 8,12 | 9,9 |
| 8,37 | 9,9 |
| 8,13 | 9,9 |
| 8,41 | 9,9 |
| **8,08** | **9,90** |

**Taula 6.15: Quadrat, dos canals, camí 2**

Si al mateix cas, fem que els paquets segueixin l'altre camí possible per arribar a la destinació, veiem que els resultats obtinguts són molt similars, influint mínimament el camí agafat, sempre que realment existeixi l'alternança dels canals entre els enllaços.

## 6.2.2  Proves amb IEEE 802.11b

Com s'ha esmentat abans, la versió IEEE 802.11b pateix moltes més **interferències** que la versió IEEE 802.11a. A les següents proves, aquestes interferències estaran presents fent que els resultats assolits siguin inferiors als que es podrien obtenir en absència d'aquestes.

La selecció dels canals es realitzarà tenint en compte que la norma de l'ortogonalitat entre ells es compleixi. Teòricament, amb una separació de 20 MHz hauria de ser suficient per assegurar l'ortogonalitat entre les freqüències i, d'aquesta manera, a l'espectre de 2,4 GHz es poden tenir tres canals ortogonals (1, 6 i 11). Recordem que amb els *drivers* que s'utilitzen només es tenen 11 canals i no els 13 que podria arribar a assolir l'espectre. Això limita el ventall d'opcions de canals que es poden utilitzar, així com el de separar encara més les freqüències perquè es produeixin unes interferències encara menors.

Per assegurar l'ortogonalitat, els dos canals seleccionats per realitzar les proves han estat **el 1 i el 11** que tenen fins 50 MHz de separació entre ells. Després, a l'estudi d'ortogonalitat s'utilitzaran altres canals per veure el rendiment que aquests proporcionen.

**Cas 1: Tres estacions**

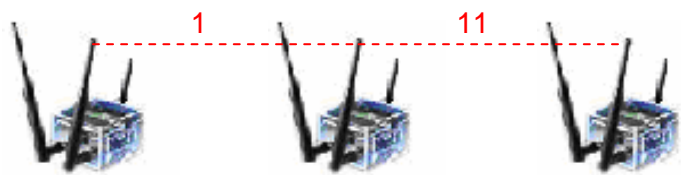

**Figura 6.17: Tres estacions, dos canals 2,4 GHz**

| TCP (Mbps) | UDP (Mbps) |
|:---:|:---:|
| 5,28 | 6,67 |
| 5,21 | 6,56 |
| 5,28 | 6,65 |
| 5,3 | 6,6 |
| 5,25 | 6,6 |
| 5,12 | 6,53 |
| 5,08 | 6,76 |
| 5,23 | 6,77 |
| 5,02 | 6,54 |
| 4,97 | 6,63 |
| **5,17** | **6,63** |

**Taula 6.16: Tres estacions, dos canals 2,4 GHz**

Veiem que el guany amb aquestes freqüències també és considerable, però a diferència de la versió IEEE 802.11a, presenta majors irregularitats en la distribució dels resultats sortints, perdent una part de la seva capacitat en comparació del rendiment que podria assolir. Tanmateix, la millora que es produeix és evident, gairebé duplicant el guany respecte al cas d'un sol canal.

**Cas 2a: Quatre estacions**

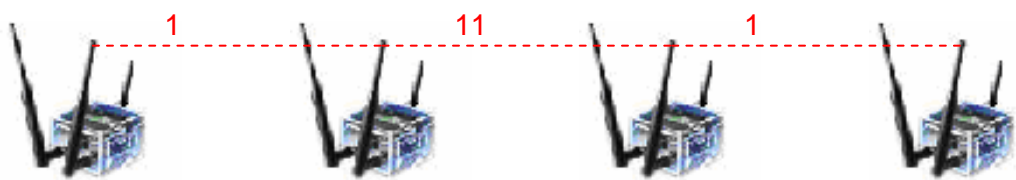

**Figura 6.18: Quatre estacions, dos canals (cas I)**

| TCP (Mbps) | UDP (Mbps) |
|:---:|:---:|
| 2,59 | 3,8 |
| 3,09 | 3,48 |
| 2,96 | 3,81 |
| 3,04 | 3,86 |
| 3,08 | 3,87 |
| 3,06 | 3,86 |
| 3,1 | 3,79 |
| 3,03 | 3,8 |
| 3,13 | 3,86 |
| 2,89 | 3,87 |
| **3,00** | **3,80** |

**Taula 6.17: Quatre estacions, dos canals (cas I)**

Pel cas de quatre estacions tenim uns resultats semblants als que teníem amb tres estacions i un sol canal. El guany que tenim és igual al que aconseguim amb la versió IEEE 802.11a, i el factor de guany que tenim amb el cas d'un canal és de $\frac{3}{2}$, tal com es deia abans.

**Cas 2b: Quatre estacions**

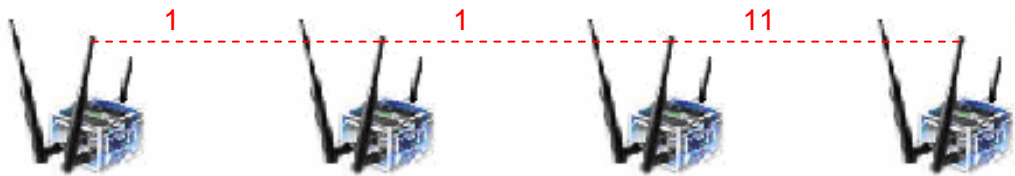

**Figura 6.19: Quatre estacions, dos canals (cas II)**

| TCP (Mbps) | UDP (Mbps) |
|:---:|:---:|
| 3,11 | 3,35 |
| 2,56 | 3,77 |
| 3,15 | 3,76 |
| 3,26 | 3,79 |
| 2,89 | 3,8 |
| 2,77 | 3,81 |
| 2,81 | 3,79 |
| 3,23 | 3,83 |
| 3,37 | 3,81 |
| 3,3 | 3,8 |
| **3,05** | **3,75** |

**Taula 6.18: Quatre estacions, dos canals (cas II)**

Un cas molt similar al d'abans és el que es presenta, on només canviem la distribució dels canals que comuniquen les diferents estacions. Com es podrà observar a la taula 6.18, els resultats que ens donen les proves són gairebé iguals als recollits en la taula 6.17 del cas anterior.

Aquí es pot arribar a la mateixa conclusió que a l'apartat 6.2.1 on sempre que hagi casos de proximitat entre les estacions, sempre haurà compartició de recursos quan es sintonitzi el mateix canal. Com s'ha dit amb IEEE 802.11a, sempre serà millor tenir el canals alternants per si no es veuen els extrems de la comunicació.

**Cas 3: Cinc estacions**

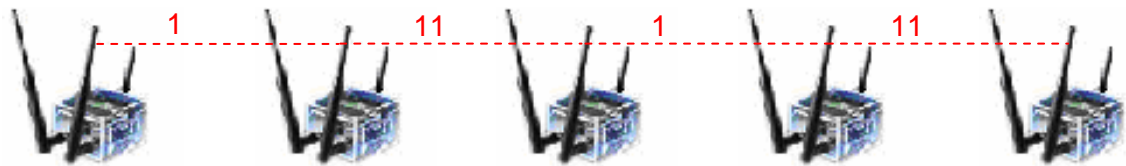

**Figura 6.20: Cinc estacions, dos canals**

| TCP (Mbps) | UDP (Mbps) |
|:----------:|:----------:|
| 2,86 | 3,22 |
| 2,81 | 3,33 |
| 2,8 | 3,3 |
| 2,82 | 3,1 |
| 2,59 | 3,69 |
| 2,84 | 3,53 |
| 2,87 | 3,53 |
| 2,59 | 3,4 |
| 2,81 | 3,26 |
| 2,71 | 3,32 |
| **2,77** | **3,37** |

**Taula 6.19: Cinc estacions, dos canals**

Tal com succeïa a l'apartat anterior, amb dos canals podem arribar a tenir configuracions de cinc estacions en línia amb el mateix rendiment que amb tres estacions en línia amb un únic canal. Cal fer notar, un petit descens dels casos de quatre estacions que poden ser degudes a l'aparició de petites interferències o per pèrdues de capacitat produïdes pels canvis de canal als diferents nodes.

**Cas 4a: Quadrat (camí 1)**

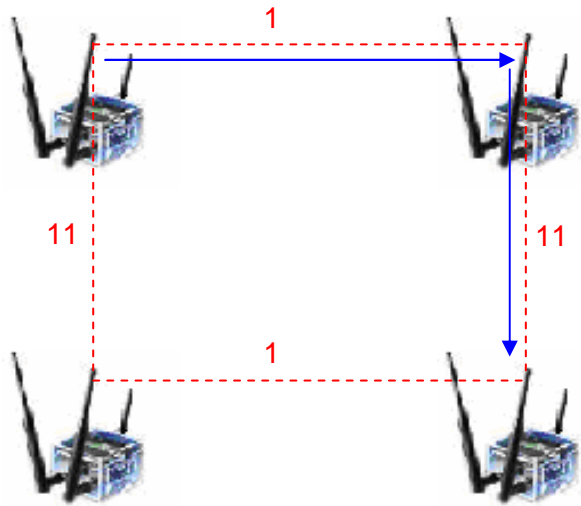

**Figura 6.21: Quadrat, dos canals, camí 1**

| TCP (Mbps) | UDP (Mbps) |
|:---:|:---:|
| 3,57 | 5,45 |
| 3,85 | 5,14 |
| 3,13 | 5,28 |
| 3,1 | 5,15 |
| 3,05 | 5,27 |
| 3,08 | 4,79 |
| 3,19 | 4,68 |
| 3,3 | 5,14 |
| 3,16 | 5,03 |
| 3,21 | 5,14 |
| **3,26** | **5,11** |

**Taula 6.20: Quadrat, dos canals, camí 1**

Les proves amb aquesta configuració quadrada, com havíem referenciat abans, dóna resultats més aviat pobres en rendiment. Realment, el resultat esperat per aquest experiment era el d'obtenir un resultat una mica inferior que el cas 1, cosa que no es compleix per aquest cas. De totes formes, el resultat que dóna el protocol UDP fa pensar que el resultat que dóna TCP estigui afectat per interferències externes a la prova realitzada, i que la configuració i distribució realitzada sí que sigui realment correcta.

**Cas 4b: Quadrat (camí 2)**

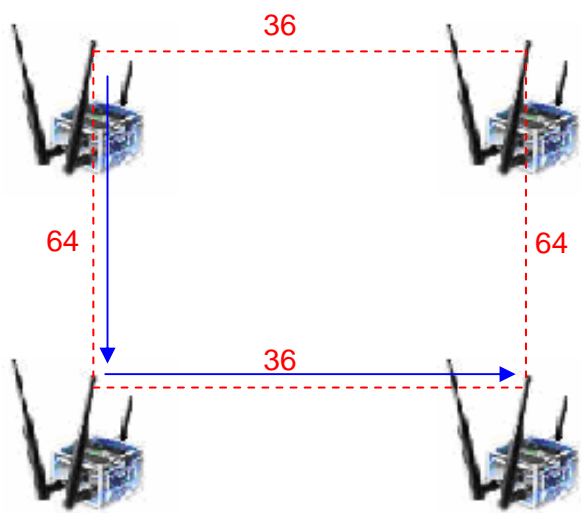

Figura 6.22: Quadrat, dos canals, camí 2

| TCP (Mbps) | UDP (Mbps) |
|:---:|:---:|
| 4,04 | 5,45 |
| 3,88 | 5,14 |
| 3,8 | 5,28 |
| 4,33 | 5,15 |
| 3,7 | 5,27 |
| 4,42 | 4,79 |
| 4,03 | 4,68 |
| 4,18 | 5,14 |
| 4,03 | 5,03 |
| 3,83 | 5,14 |
| **4,02** | **5,11** |

Taula 6.21: Quadrat, dos canals, camí 2

Fent servir l'altre camí possible, veiem que el resultat del protocol TCP canvia, essent un nombre més lògic pel cas que ens ocupa. La reducció respecte al cas, es pot explicar per la possible interferència produïda pels altres enllaços al propi enllaç útil.

## 6.2.3  Valoracions

El prototipus dissenyat ha permès la utilització, a qualsevol jerarquia, de dos canals diferenciats per transmetre qualsevol tipus d'informació. Si aquests dos canals són ortogonals, els resultats ens indiquen que hi ha una **millora**.

Gràficament, podem veure quin és el rendiment que assoleixen els diferents casos, amb els diferents protocols de transport que es poden utilitzar a l'aplicació *iperf*.

Pel cas TCP podem veure a la figura 6.23 com evoluciona la capacitat amb l'augment de salts a partir dels resultats calculats.

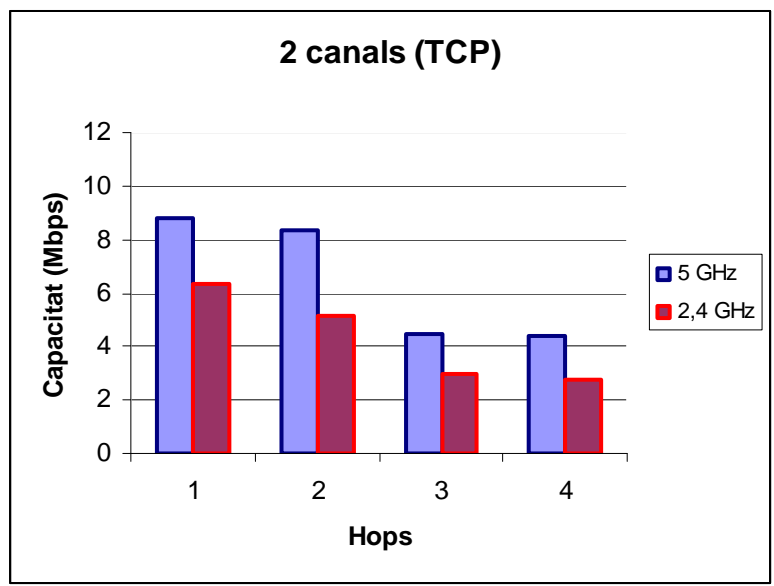

**Figura 6.23: Evolució capacitat amb dos canals per TCP**

Aquí es veu el que s'ha comentat amb les taules de resultats, és a dir, com es manté la capacitat pel cas de 2 salts en relació al cas de només 1 *hop* i com es manté la capacitat als casos de 3 i 4 salts.

Pel cas d'UDP (figura 6.24) es poden treure valoracions molt semblants, però amb una major capacitat degut a la particularitat pròpia d'UDP, que no requereix cap tipus de mecanisme d'autenticació ni de reconeixement.

Fer notar que per 2,4 GHz, quan es perd més capacitat de l'original que amb la banda de freqüències de 5 GHz.

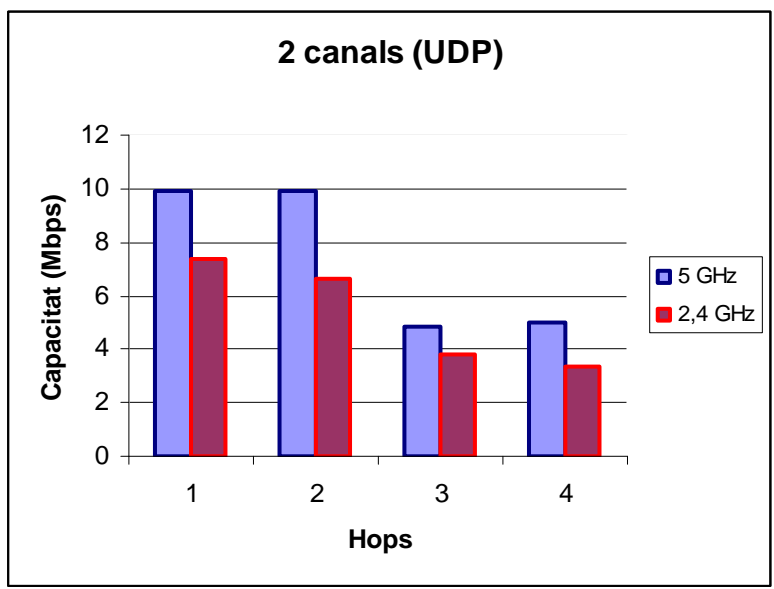

**Figura 6.24: Evolució capacitat amb 2 canals per UDP**

Per tant, es pot observar una certa millora a tots els salts, quantificada a les següents gràfiques i que ens donen una idea dels avantatges d'emprar sistemes de 2 canals enfront del d'únic canal utilitzat habitualment. A les gràfiques surt indicat quin percentatge respecte al màxim s'assoleix per cadascun dels casos. El màxim és el cas d'una transmissió entre dos estacions adjacents. Per altra banda, surt un factor de millora del guany d'emprar dos canals respecte al cas d'utilitzar un sol canal.

La figura 6.25 correspon al cas d'una transmissió d'un flux al protocol TCP amb una freqüència de 5 GHz.

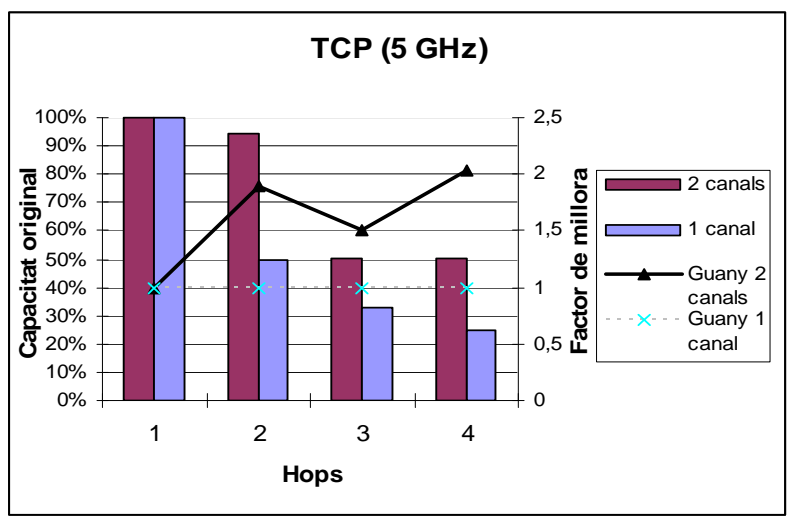

**Figura 6.25: Comparativa entre dos canals i un canal per TCP a 5 GHz**

A la següent, el mateix cas però per UDP.

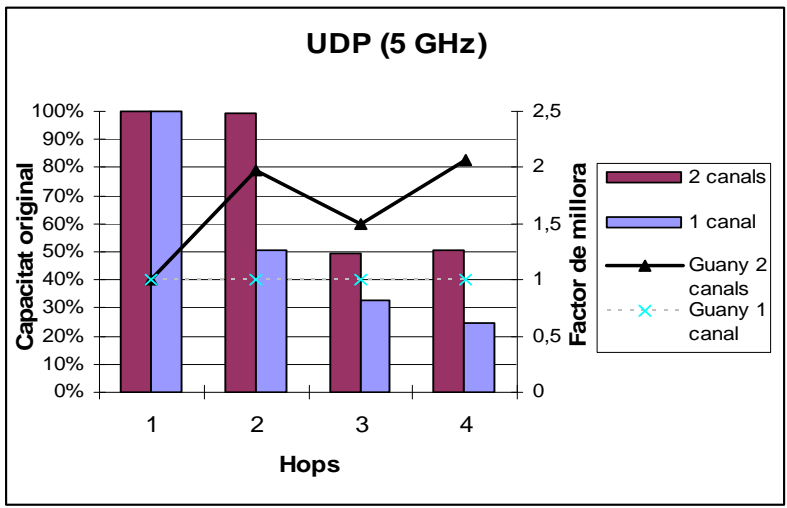

**Figura 6.26: Comparativa entre dos canals i un canal per UDP a 5 GHz**

Després podem veure la comparació de les dades a una freqüència de 2,4 GHz. Per TCP ho podem veure a la figura 6.27.

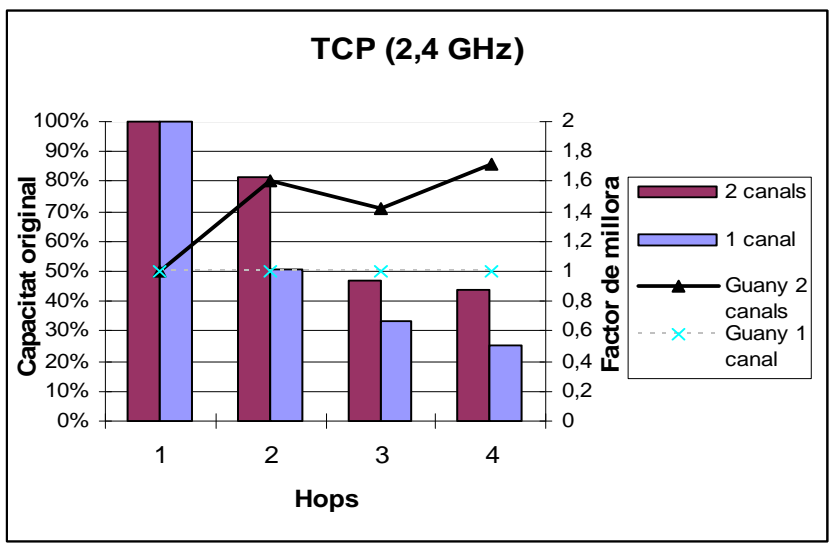

**Figura 6.27: Comparativa entre dos canals i un canal per TCP a 2,4 GHz**

I per UDP, la següent figura.

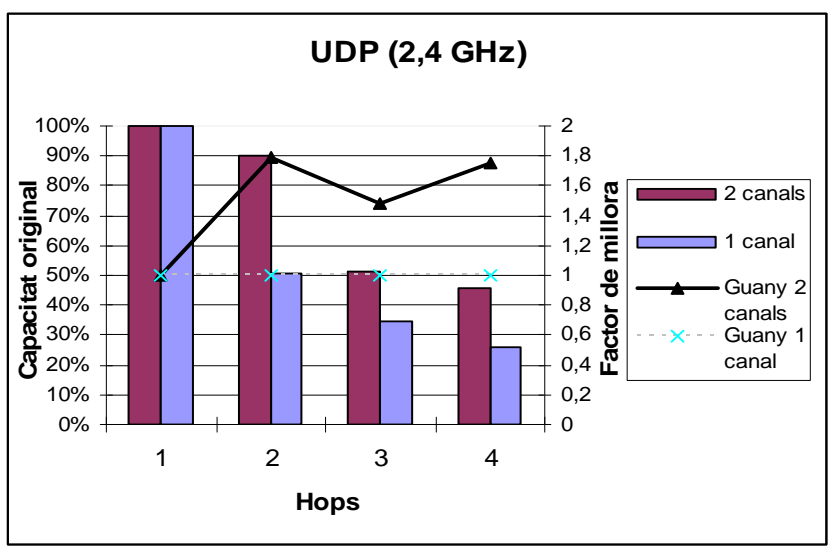

**Figura 6.28: Comparativa entre dos canals i un canal per UDP a 2,4 GHz**

És comú per tots els casos que la millora en dos canals sigui evident i que el factor de millora respecte al cas d'un canal sigui sempre **superior a 1,5** essent, per tant, una millora més que considerable. Si a aquestes jerarquies s'afegissin més estacions, es el resultat obtingut per aquests casos es podria generalitzar per més topologies oscil·lant el factor de millora entre 1,5 i 2. Pels casos de configuracions amb nombre de salts parell, el resultat sempre estarà més prop del factor 2 (es duplica la capacitat que es té amb 1 canal), mentre que amb configuracions amb nombre de salts senar, el factor serà més proper a 1,5 augmentant una mica quan més gran és la topologia. Comentar que en el cas de 2,4 GHz, sobretot amb el protocol TCP, hi ha una disminució de la capacitat per les raons que ja havíem comentat abans: interferències externes i reconeixements a velocitats molt més baixes que a 5 GHz (1 Mbps enfront a 6 Mbps).

## 6.3   ORTOGONALITAT DELS CANALS

A aquest apartat, s'ha fet un estudi pràctic de l'**ortogonalitat dels canals**. Teòricament, a apartats anteriors, s'ha presentat un llistat d'aquells canals que són totalment ortogonals entre sí, però les particularitats que té el *hardware* fa que el que es compleix teòricament no sempre es doni als casos pràctics. Això pot ser degut a imperfeccions d'alguns components del *hardware* o alguna mancança en el procés de fabricació d'aquests components.

Al projecte que ens ocupa s'han realitzat proves amb diferents casos i que, finalment, han servit per extreure quins són realment els canals que es poden utilitzar com a totalment ortogonals. Recordem que amb una **separació** de 20 MHz els canals ja no haurien d'interferir encara que això, com es veurà a continuació, no sempre es compleix. Els estudis sobre ortogonalitat es faran per les dues versions de IEEE: la 802.11a (5 GHz) i la 802.11b (2,4 GHz).

## 6.3.1 Proves amb IEEE 802.11a

Teòricament hi han 13 canals ortogonals a la freqüència de **5 GHz** essent canals d'aquest tipus aquells que tenen diferència de 4 entre ells. Així, el 36 és ortogonal del 40, el 40 del 44, el 44 del 48… i així successivament.

Per determinar si això és cert, s'han fet diverses proves amb escenaris de dos canals a una topologia de tres estacions en línia, on un dels canal és sintonitzat al 36 i un altre és canviant, permetent-nos mesurar aquesta ortogonalitat. Les proves ens demostraran que, a la pràctica, a aquest espectre tindrem molts menys canals no interferents disponibles.

**Cas 1: Canals 36 i 40**

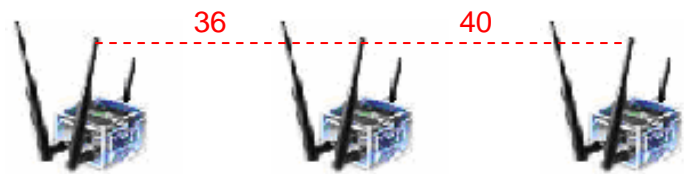

**Figura 6.29: Canals 36 i 40**

| TCP (Mbps) | UDP (Mbps) |
|:---:|:---:|
| 4,38 | 4,98 |
| 4,48 | 4,98 |
| 4,44 | 4,95 |
| 4,44 | 5 |
| 4,3 | 4,97 |
| 4,37 | 4,98 |
| 4,22 | 4,95 |
| 4,14 | 4,98 |
| 4,37 | 4,93 |
| 4,34 | 4,97 |
| **4,35** | **4,97** |

**Taula 6.22: Canals 36 i 40**

Els resultats que ens dóna aquest cas són iguals que els que ens donen pel cas en què els dos enllaços es sintonitzen a 36. Per tant, encara que teòricament no haurien de presentar cap tipus d'interferència, a la pràctica és com utilitzar el mateix canal als dos nodes.

**Cas 2: Canals 36 i 44**

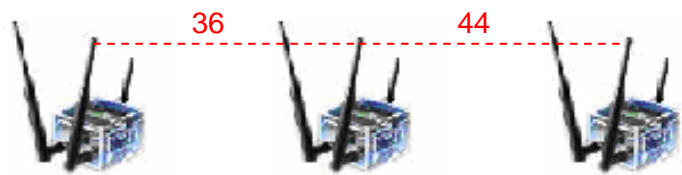

**Figura 6.30: Canals 36 i 44**

| TCP (Mbps) | UDP (Mbps) |
|:---:|:---:|
| 5,51 | 6,56 |
| 5,65 | 6,55 |
| 5,7 | 6,55 |
| 5,71 | 6,51 |
| 5,64 | 6,59 |
| 5,66 | 6,61 |
| 5,71 | 6,59 |
| 5,68 | 6,63 |
| 5,73 | 6,64 |
| 5,67 | 6,64 |
| **5,67** | **6,59** |

**Taula 6.23: Canals 36 i 44**

Les dades indiquen una millora de rendiment respecte a l'últim canal, però encara estem enfora del resultat desitjat obtingut a les proves de l'apartat 6.2. Per tant, amb una separació de 40 MHz no tenim suficient per tenir independència als canals existint encara una important interferència en la comunicació.

**Cas 3: Canals 36 i 48**

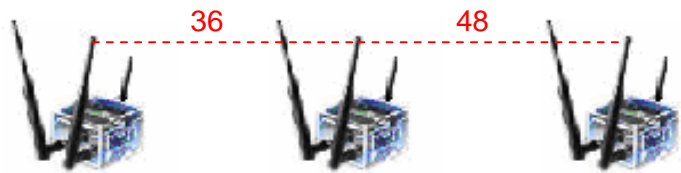

**Figura 6.31: Canals 36 i 48**

| TCP (Mbps) | UDP (Mbps) |
|:----------:|:----------:|
| 6,08 | 6,77 |
| 6,16 | 6,61 |
| 6,15 | 6,64 |
| 6,04 | 6,66 |
| 6,08 | 7,02 |
| 6,2 | 6,85 |
| 6,12 | 7,15 |
| 6,16 | 6,84 |
| 6,16 | 6,74 |
| 6,08 | 6,96 |
| **6,12** | **6,82** |

**Taula 6.24: Canals 36 i 48**

Separant encara més els nodes en freqüència, veiem que s'incrementa la capacitat que tenim, però seguim estant enfora de les fites que es poden assolir per un rendiment més òptim.

**Cas 4: Canals 36 i 52**

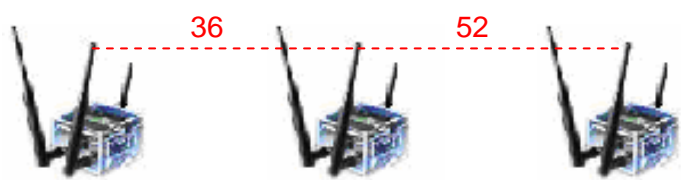

**Figura 6.32: Canals 36 i 52**

| TCP (Mbps) | UDP (Mbps) |
|:---:|:---:|
| 6,52 | 7,01 |
| 6,31 | 7,15 |
| 6,41 | 8,02 |
| 6,73 | 7,77 |
| 6,37 | 8,79 |
| 6,54 | 9,07 |
| 6,86 | 8,74 |
| 6,85 | 8,26 |
| 6,93 | 8,38 |
| 6,79 | 8,51 |
| **6,63** | **8,17** |

**Taula 6.25: Canals 36 i 52**

A la taula de dades observem com la capacitat augmenta respecte als casos anteriors i, cada vegada ens aproximem més a les dades òptimes dels canals ortogonals.

**Cas 5: Canals 36 i 56**

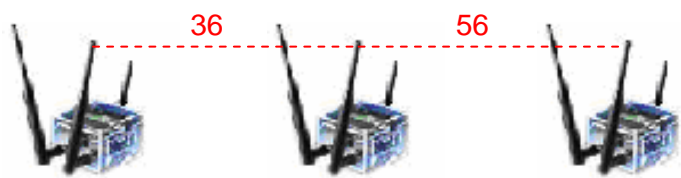

**Figura 6.33: Canals 36 i 56**

| TCP (Mbps) | UDP (Mbps) |
|:---:|:---:|
| 7,58 | 8,41 |
| 7,27 | 9,35 |
| 6,83 | 9,65 |
| 7,23 | 9,65 |
| 7,44 | 9,59 |
| 7,18 | 9,32 |
| 7,06 | 9,7 |
| 7,33 | 9,67 |
| 7,26 | 9,6 |
| 7,22 | 9,58 |
| **7,24** | **9,45** |

**Taula 6.26: Canals 36 i 56**

Amb una separació de 100 MHz entre els canals seleccionats ja veiem que pràcticament tenim els resultats màxims que es poden assolir, especialment en el cas UDP, que al no requerir retransmissions ni reconeixements fan que es minimitzi la interferència d'un canal provocat en un altre.

**Cas 6: Canals 36 i 60**

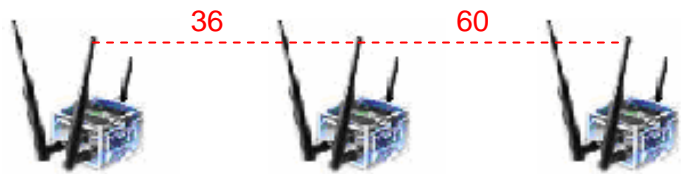

**Figura 6.34: Canals 36 i 60**

| TCP (Mbps) | UDP (Mbps) |
|:---:|:---:|
| 7,83 | 9,88 |
| 8,12 | 9,68 |
| 8,14 | 9,78 |
| 8,22 | 9,81 |
| 8,25 | 9,63 |
| 8,24 | 9,73 |
| 8,16 | 9,75 |
| 8,22 | 9,76 |
| 8,29 | 9,84 |
| 8,27 | 9,78 |
| **8,17** | **9,76** |

**Taula 6.27: Canals 36 i 64**

En aquest cas, els resultats obtinguts amb aquests canals, tal com es veuen reflectits a la taula 6.27, són gairebé iguals que els que s'han mostrat als escenaris de dos canals al subapartat anterior. Per tant, ja es pot dir que l'ortogonalitat a IEEE 802.11a en aquests prototipus arriba a partir d'una separació freqüencial de 120 MHz.

**Conclusió**

Com veiem als diferents casos, a major separació de canals, major rendiment es té. Però s'està molt enfora de complir-se allò que s'exposava teòricament i que deia que els canals 36 i 40 eren **independents** i no interferien entre ells. L'evolució de TCP i d'UDP vista als casos anteriors es pot veure a la següents gràfica.

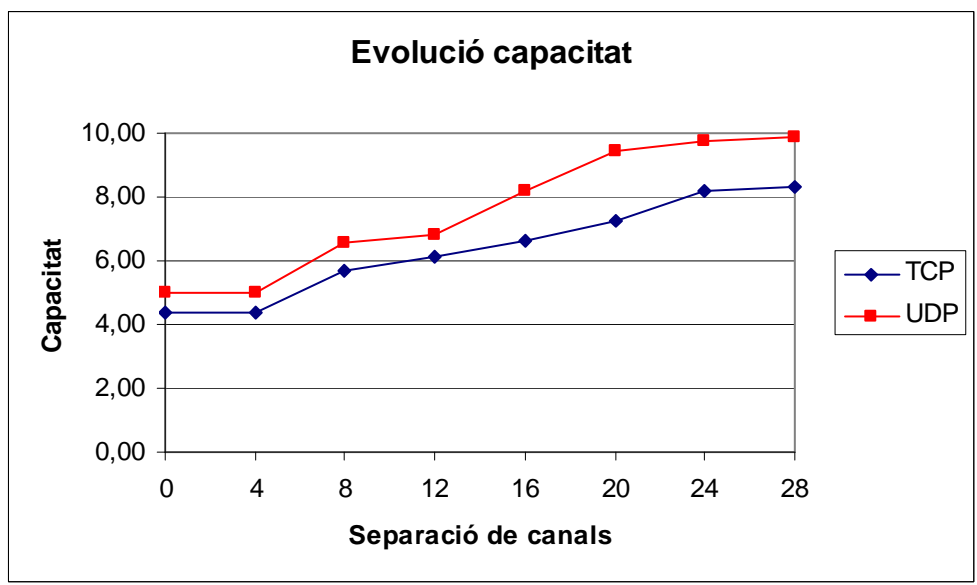

**Figura 6.35: Evolució ortogonalitat a 5 GHz**

L'explicació que se li pot donar a aquestes interferències podria ser el d'una pobra implementació dels filtres del sintonitzador de freqüències, que no elimina les interferències secundàries i que afecta directament al rendiment que tenim en altres freqüències. En resum, vist el problema dels canals ortogonals a la IEEE 802.11a, resultarà que només es tindran 4 canals totalment ortogonals i que es poden utilitzar simultàniament.

## 6.3.2  Proves amb IEEE 802.11b

En aquesta versió d'IEEE, segons el que s'ha estudiat, tenim tres canals completament ortogonals: el 1, el 6 i el 11, essent la separació freqüencial entre ells de 25 MHz.

A aquest subapartat es mostrarà com evoluciona la capacitat a mesura que es va canviant la freqüència del segon enllaç. Les proves d'ortogonalitat d'aquest punt es centraran en fixar un dels canals a 1 mentre que l'altre canal s'anirà canviant de 1 fins a 6, comprovant si realment és certa aquesta ortogonalitat pel cas pràctic.

**Cas 1: Canals 1 i 2**

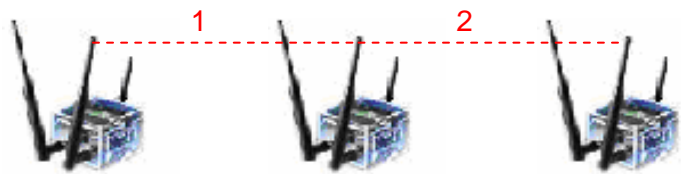

**Figura 6.36: Canals 1 i 2**

| TCP (Mbps) | UDP (Mbps) |
|:---:|:---:|
| 3,05 | 3,25 |
| 3,01 | 3,25 |
| 3,02 | 3,73 |
| 3,04 | 3,73 |
| 3,03 | 3,61 |
| 3,02 | 3,45 |
| 3,01 | 3,66 |
| 2,99 | 3,62 |
| 3,05 | 3,58 |
| 3,07 | 3,65 |
| **3,03** | **3,55** |

**Taula 6.28: Canals 1 i 2**

Sintonitzant els canals 1 i 2 veiem que el resultat donat és l'esperat, essent molt similar al que es té quan ambdós enllaços tenen sintonitzat el mateix canal. Això coincideix amb el que s'havia comentat, ja que amb separacions menors de 20 MHz no existeix ortogonalitat.

**Cas 2: Canals 1 i 3**

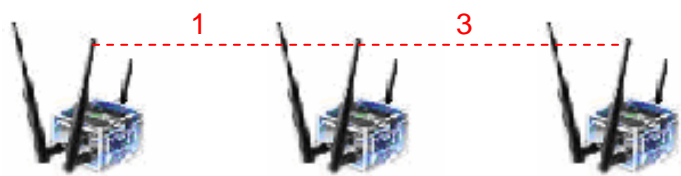

**Figura 6.37: Canals 1 i 3**

| TCP (Mbps) | UDP (Mbps) |
|:---:|:---:|
| 2,22 | 3,59 |
| 2,74 | 3,58 |
| 2,59 | 3,45 |
| 2,6 | 3,32 |
| 2,23 | 3,57 |
| 3,22 | 3,6 |
| 2,48 | 3,26 |
| 2,41 | 3,48 |
| 2,48 | 3,44 |
| 2,82 | 2,94 |
| **2,58** | **3,42** |

Taula 6.29: Canals 1 i 3

Canviant el canal 2 pel 3, es pot fer notar que el rendiment obtingut pel sistema és, fins i tot, inferior que aplicant el mateix canal i que el cas anterior. La interferència que es pateix és superior que els casos anteriors i que emprant el mateix canal per tota la jerarquia.

**Cas 3: Canals 1 i 4**

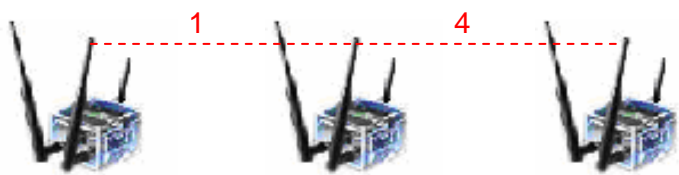

Figura 6.38: Canals 1 i 4

| TCP (Mbps) | UDP (Mbps) |
|:---:|:---:|
| 2,67 | 3,37 |
| 3,15 | 3,42 |
| 2,58 | 3,38 |
| 2,8 | 3,21 |
| 2,72 | 3,16 |
| 3,16 | 3,9 |
| 2,64 | 3,48 |
| 2,68 | 4 |
| 2,76 | 3,4 |
| 2,16 | 3,37 |
| **2,73** | **3,47** |

Taula 6.30: Canals 1 i 4

Els resultats calculats d'aquest tipus de configuració són molt similars al cas 2, on es sintonitzaven els canals 1 i 3 i amb resultats pitjors en mitjana que els que teníem pel cas on teníem fixat els canals 1 i 2 i pels casos d'únic canal.

**Cas 4: Canals 1 i 5**

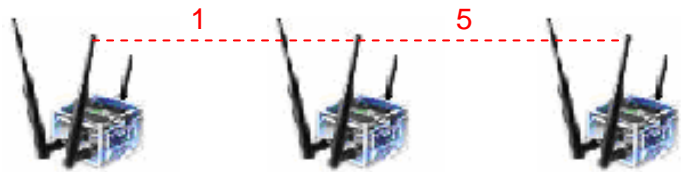

**Figura 6.39: Canals 1 i 5**

| TCP (Mbps) | UDP (Mbps) |
|:----------:|:----------:|
| 3,88 | 4,67 |
| 4,35 | 5,42 |
| 3,76 | 5,42 |
| 3,51 | 5,58 |
| 3,6 | 4,76 |
| 3,11 | 4,97 |
| 2,93 | 4,97 |
| 3,79 | 5,35 |
| 2,41 | 5,35 |
| 2,75 | 5,35 |
| **3,41** | **5,18** |

**Taula 6.31: Canals 1 i 5**

Per aquest cas, es pot observar que la capacitat a la que arribem és superior a tots els casos que s'han introduït anteriorment, però segueix estant enfora del cas òptim vist als escenaris de 2 canals completament ortogonals.

**Cas 5: Canals 1 i 6**

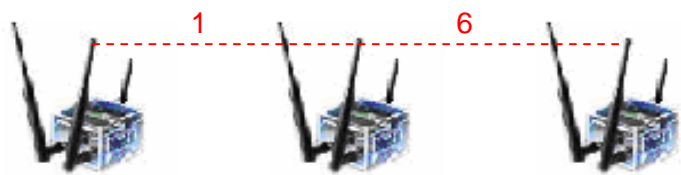

**Figura 6.40: Canals 1 i 6**

| TCP (Mbps) | UDP (Mbps) |
|:---:|:---:|
| 5,43 | 6,82 |
| 5,5 | 6,79 |
| 5,53 | 6,93 |
| 5,55 | 6,86 |
| 5,48 | 6,89 |
| 5,58 | 6,81 |
| 5,59 | 6,88 |
| 5,67 | 6,87 |
| 5,39 | 6,87 |
| 5,62 | 6,76 |
| **5,53** | **6,85** |

**Taula 6.32: Canals 1 i 6**

Provant amb els canals 1 i 6, teòricament s'havien d'obtenir uns resultats semblants als escenaris de 2 canals abans esmentats. Els resultats que reflecteix la taula són, fins i tot, millors que els que s'havien calculat pel cas dels canals 1 i 11, també ortogonals entre sí.

**Conclusió**

A diferència de l'anterior versió d'IEEE, l'ortogonalitat dels canals a 2,4 GHz és tal com s'esperava, essent el canal 1, el 6 i el 11 absolutament no interferents, podent-se utilitzar per realitzar configuracions multicanal a topologies més extenses.

Un fet que crida l'atenció d'aquestes proves, és la **disminució** de capacitat que es produeix als canals 3 i 4 respecte al cas d'únic canal. És pitjor utilitzar els canals 1 i 3, que generalitzar l'ús d'un dels dos canals a tota la topologia. A partir de tenir una diferència de 20 MHz, l'augment de capacitat comença a ser considerable aconseguint independència entre els tres canals teòricament **ortogonals** (1, 6 i 11). La raó de què sigui pitjor és deguda a un problema en la detecció de les col·lisions. Si es tenen dos canals iguals a les dues interfícies, la capa MAC detectarà que el medi està ocupat, compartint el medi. Però si estan a freqüències diferents, no es produeix aquesta detecció encara que siguin freqüències interferents i, per tant, encara es degradarà més les prestacions del sistema.

L'evolució de la capacitat en aquests canals és visible a la següent gràfica adjunta (figura 6.41).

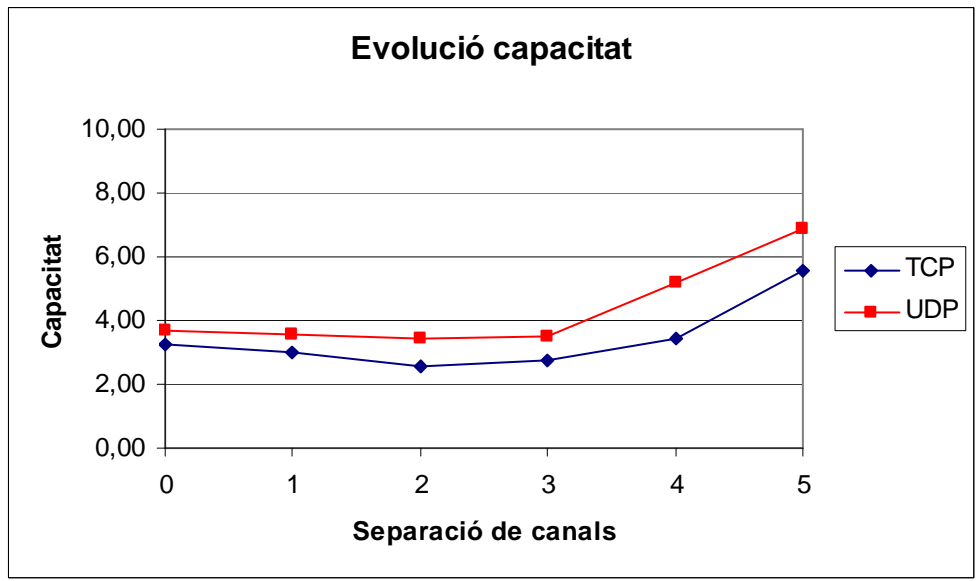

**Figura 6.41: Evolució ortogonalitat a 2,4 GHz**

## 6.4  SEPARACIÓ D'ANTENES

Abans, s'ha parlat extensament de les limitacions tècniques que podien tenir els components a l'hora de dissenyar un sistema. Hi ha una pila d'articles que parlen damunt d'aquestes limitacions ([34], [35], [36] i [37]), sofertes al realitzar una implementació pràctica amb sistemes multicanal.

El nostre prototipus no escapa d'aquestes limitacions i també les pateix provocant una disminució de les prestacions del sistema dissenyat. Concretament, és la **separació entre les antenes** que conformen el *meshcube* la que afecta negativament al nostre prototipus.

Fent proves en escenaris de dos canals, amb dos canals independents seleccionats a cadascun dels enllaços, es va veure que el rendiment que s'obtenia com a resultat era molt pobre envers dels resultats esperats. Després d'una sèrie de proves, es va comprovar que la separació entre les antenes del *meshcube* intermedi (el que canvia el canal) afectava directament al rendiment de les comunicacions.

Per això, a aquest subapartat es fa un estudi de com afecta la **distància** entre aquestes antenes en la capacitat total. L'estudi s'ha realitzat sobre escenaris amb tres estacions i dos canals ortogonals seleccionats entre ells, partint d'una distància inicial nul·la i augmentant per cada cas la distància en 5 cm. Les proves s'han fet en IEEE 802.11b i les mesures s'han realitzat sintonitzant en un dels casos els canals 1 i 6, mentre que a l'altre cas es sintonitzava els canals 1 i 11. Pels dos casos s'han fet proves pel protocol TCP i pel protocol UDP. Per cadascun dels casos i de les distàncies s'han fet cinc proves de les què s'ha extret un resultat mitjà. Totes les dades resultants són en Mbps.

A la il·lustració 6.42 es pot veure a quina distància ens referim quan parlem de distància entre les antenes del *meshcube*.

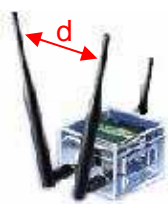

**Figura 6.42: Distància entre antenes**

A continuació es mostrarà el recull de dades fruit d'aquestes proves i unes gràfiques que mostren l'evolució de capacitat, a mesura que es va separant la distància *d* entre les antenes.

**Cas 1: Separació entre antenes amb canal 1 i 11**

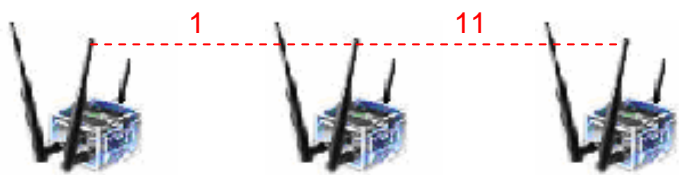

**Figura 6.43: Separació entre antenes, canals 1 i 11**

Pel cas de TCP i per diferents distàncies mesurades al *meshcube* intermedi tindrem els següents valors.

| TCP | | | | | | |
|------|------|-------|-------|-------|-------|-------|
| **0 cm** | **5 cm** | **10 cm** | **15 cm** | **20 cm** | **25 cm** | **30 cm** |
| 2,31 | 2,75 | 3,36 | 3,56 | 4,57 | 5,32 | 5,74 |
| 2,31 | 2,79 | 3,29 | 3,49 | 4,25 | 5,53 | 5,52 |
| 2,31 | 3,04 | 3,35 | 3,45 | 4,55 | 5,54 | 5,61 |
| 2,29 | 3,18 | 3,4 | 3,41 | 4,71 | 5,42 | 5,52 |
| 2,35 | 2,97 | 3,38 | 3,46 | 4,9 | 5,38 | 5,54 |
| **2,31** | **2,95** | **3,36** | **3,47** | **4,60** | **5,44** | **5,59** |

**Taula 6.33: Capacitat-separació d'antenes als canals 1 i 11 per TCP**

El mateix cas per UDP ens dóna els valors de la taula 6.33.

| UDP | | | | | | |
|------|------|-------|-------|-------|-------|-------|
| **0 cm** | **5 cm** | **10 cm** | **15 cm** | **20 cm** | **25 cm** | **30 cm** |
| 2,88 | 3,69 | 4,09 | 4,01 | 5,74 | 6,39 | 6,87 |
| 2,87 | 3,69 | 4,09 | 4,51 | 5,82 | 6,6 | 6,95 |
| 2,91 | 3,51 | 3,97 | 4,13 | 5,92 | 6,43 | 6,74 |
| 2,85 | 3,59 | 3,97 | 4,13 | 5,77 | 6,39 | 6,59 |
| 2,83 | 3,21 | 3,92 | 4,23 | 5,44 | 6,3 | 6,79 |
| **2,87** | **3,54** | **4,01** | **4,20** | **5,74** | **6,42** | **6,79** |

**Taula 6.34: Capacitat-separació d'antenes als canals 1 i 11 per UDP**

Per UDP, lògicament, la capacitat és major en totes les distàncies.

Amb les dades recollides a les diferents taules, podem concloure que a partir d'una distància de **25 cm** ja tenim uns resultats que es poden considerar absolutament **vàlids**. A la figura 6.44 podem veure l'evolució de la capacitat depenent de la distància on apareix el llindar que abans s'havia marcat com a resultat vàlid als anteriors apartats.

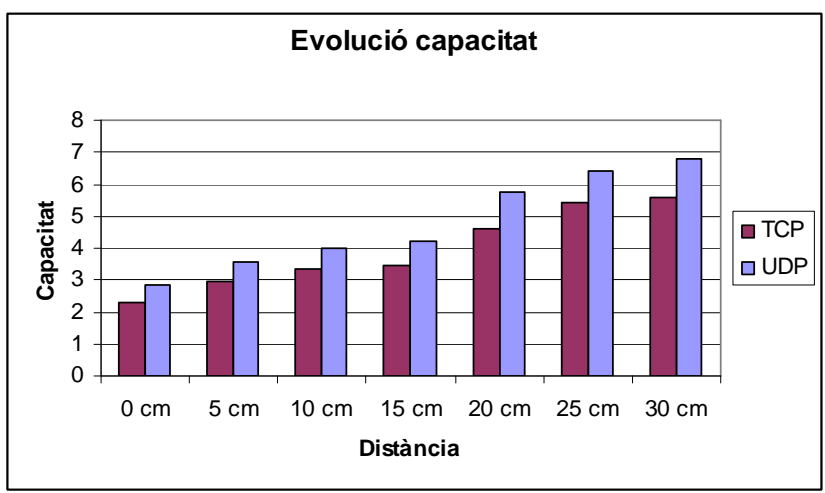

**Figura 6.44: Evolució capacitat-distància, canals 1 i 11**

## Cas 2: Separació entre antenes amb canal 1 i 6

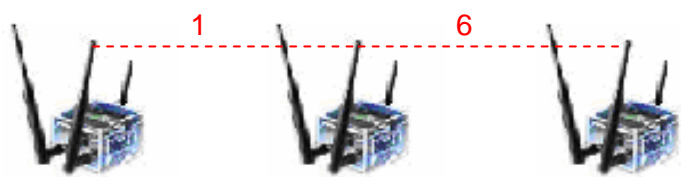

**Figura 6.45: Separació entre antenes, canals 1 i 6**

Per contrastar les dades anteriors, també s'han fet els mateixos experiments per un altre cas de canals no interferents com són el 1 i el 6.

Els resultats que ens ha donat pel protocol de transport TCP són els següents:

| TCP | | | | | | |
|---|---|---|---|---|---|---|
| **0 cm** | **5 cm** | **10 cm** | **15 cm** | **20 cm** | **25 cm** | **30 cm** |
| 2,64 | 2,66 | 3,67 | 3,8 | 3,95 | 5,15 | 5,19 |
| 2,54 | 2,65 | 3,6 | 3,68 | 3,85 | 5,06 | 5,22 |
| 2,56 | 2,62 | 3,61 | 3,63 | 3,79 | 4,88 | 5,26 |
| 2,61 | 2,62 | 3,5 | 3,52 | 3,79 | 5,06 | 5,27 |
| 2,55 | 2,62 | 3,5 | 3,55 | 3,84 | 5,03 | 5,31 |
| **2,58** | **2,63** | **3,58** | **3,64** | **3,84** | **5,04** | **5,25** |

**Taula 6.35: Capacitat-separació d'antenes canals 1 i 6 per TCP**

Mentre que per UDP, l'execució dels programes donava els següents resultats:

| UDP | | | | | | |
|---|---|---|---|---|---|---|
| 0 cm | 5 cm | 10 cm | 15 cm | 20 cm | 25 cm | 30 cm |
| 3,09 | 3,09 | 4,24 | 4,19 | 4,37 | 6,94 | 7,1 |
| 3,08 | 3,1 | 4,07 | 4,19 | 4,33 | 6,96 | 7,13 |
| 3,09 | 3,14 | 4,07 | 4 | 4,44 | 6,88 | 6,55 |
| 3,14 | 3,12 | 3,83 | 4,21 | 4,45 | 6,85 | 6,94 |
| 3,13 | 3,17 | 4,08 | 4,21 | 4,45 | 6,8 | 7,13 |
| **3,11** | **3,12** | **4,06** | **4,16** | **4,41** | **6,89** | **6,97** |

**Taula 6.36: Capacitat-separació d'antenes canals 1 i 6 per UDP**

Tant en un cas com en l'altre, es veu que la capacitat augmenta amb la distància entre les antenes arribant a tenir valors vàlids, semblants als obtinguts en apartats anteriors, a partir dels 25 cm, coincidint aquesta observació amb la que s'havia realitzat al primer cas avaluat. Per tant, es pot concloure d'aquest apartat que aquells nodes que canvien de canal han de tenir les antenes separades com a mínim uns 25 cm per poder tenir resultats fiables.

Gràficament es pot visualitzar a la figura 6.46.

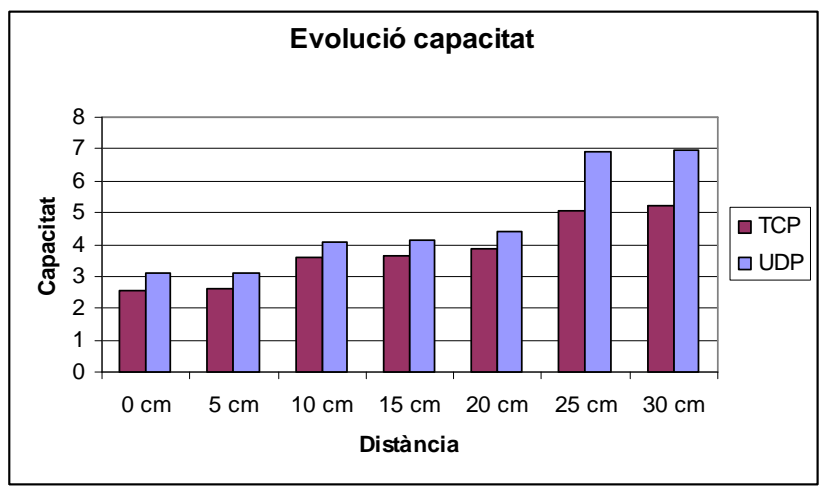

**Figura 6.46: Evolució capacitat-distància, canals 1 i 6**

# 7 CONCLUSIONS I LÍNIES FUTURES

Les **xarxes mallades sense fils** (les WMN) són ideals per adoptar sistemes de comunicació sense fils de major abast i de major flexibilitat que els sistemes sense fils que s'havien utilitzat fins ara. Les WMN, gràcies a la seva configuració, possibiliten la interconnexió de diferents xarxes aïllades de tipus wi-fi, permetent l'accés a recursos fora de la xarxa immediata de la que es forma part.

A més, s'ha vist que les WMN suporten **mobilitat** en els seus nodes sempre que hagi punts d'accés convenientment posicionats, i suposen una bona opció per flotes de vehicles (flotes privades de camions, vehicles d'emergències...), així com per medis de transport massius (metro, tren, tramvia...).

Però aquests tipus de xarxes encara formen part d'una idea bastant recent que encara es pot desenvolupar degut a les particularitats que poden presentar aquestes tecnologies. S'hauran d'assumir nous reptes per aconseguir un bon funcionament de la WMN en qüestió, ja que seran necessàries millores en els protocols d'encaminament, per optimitzar l'adreçament dels paquets, millores en la seguretat de les comunicacions, introducció de sistemes multicanal per tal de poder incrementar la capacitat total de la xarxa...

En concret, en aquest projecte s'elabora una **solució multicanal** amb el propòsit d'incrementar les **prestacions** d'una xarxa d'únic canal normalment utilitzades. Per l'elaboració del projecte s'ha fet la realització pràctica a un prototipus on s'ha experimentat una sèrie de problemes fruit de la dificultat en la seva aplicació pràctica a sobre d'una plataforma *hardware*.

A continuació, es remarcaran quines han estat les principals conclusions que s'han extret de la implementació d'aquest prototipus i dels resultats obtinguts degut a l'execució del prototipus, així com les possibles futures línies d'investigació, línies que podrien servir per perfeccionar el que s'ha dissenyat.

## 7.1  CONCLUSIONS

De les proves realitzades, la primera conclusió que es pot extreure és que quan més gran és la xarxa que estem manejant, major serà la interferència que tindrem si els enllaços que connecten els diferents nodes utilitzen el mateix canal. Per tant, el primer que es veu és que serà necessària alguna solució que permeti millorar les prestacions que presenta una xarxa d'únic canal. Una possible solució podria ser el **protocol multicanal** dissenyat a aquest treball.

A més, s'ha pogut comprovar com les interferències externes al nostre prototipus també afectaven directament a les comunicacions realitzades, reduint de manera destacada la capacitat teòrica que el sistema ens hauria d'oferir. En resum, tant les interferències internes com externes que provenen de freqüències no ortogonals a la que es té sintonitzat a l'enllaç en aquell moment, són un dels principals causants de l'empitjorament del rendiment de les comunicacions.

Amb el procés elaborat en l'aplicació pràctica del sistema, s'ha pogut veure que, al provar la implementació teòrica dissenyada damunt les plataformes *hardware* emprades, sorgeixen importants limitacions degudes a la forma en què està fabricada dita plataforma *hardware* que, al nostre cas, es tracten dels *meshcube*. Aquestes limitacions **condicionen** el nostre disseny i les aspiracions inicials es veuen limitades per aquest fet.

Amb això el que es vol dir és que, després d'haver realitzat el recull d'informació que servirà com a *background* i d'haver ideat una proposta de solució pel problema que ocupa el projecte, el **disseny pràctic** que finalment s'implementa té en compte els condicionants que imposa la infraestructura *hardware* que disposem.

En un principi, es tenia pensat el desenvolupament d'un protocol multicanal d'assignació dinàmica amb possibilitats de canvis freqüents de canal en el moment de les comunicacions. En el procés d'elaboració de la solució, es va veure que algunes proposicions plantejades eren inassolibles, com la d'aquests canvis continus de canal ja que la seva implementació significava obtenir unes prestacions molt pobres que empitjoraven els resultats d'únic canal. Per això es va haver de modificar alguns plantejaments inicials fins que es va arribar a la solució dissenyada i explicada en aquest projecte, donant els resultats mostrats a l'apartat anterior.

Aquesta solució és la d'un **prototipus multicanal**, amb un sistema d'assignació fix de canals a cadascuna de les interfícies que constitueixen els canals. D'aquesta manera s'aconsegueix duplicar els recursos assignats a la xarxa general essent una millora sensible en les prestacions, tal com s'ha descrit al capítol 6.

El prototipus dissenyat està compost de tres **interfícies** diferents: dues interfícies de dades i una interfície de senyalització. Aquesta actuació permet separar les dades útils d'aquelles de control que podrien interferir en les transmissions de dades realitzades entre els diferents *meshcube.* Les dades de control són aquelles que permeten fer la selecció dels canals i que, a més, porten les informacions de control del protocol d'encaminament.

Per poder utilitzar aquestes interfícies s'ha modificat el protocol d'encaminament utilitzat per la configuració de les rutes òptimes, l'**OLSR**. Amb aquesta modificació, s'ha aconseguit dirigir la informació de control cap a una interfície desitjada, assignant les rutes per transmetre dades a les interfícies que han quedat lliures amb l'objectiu d'enviar només informació útil. D'aquesta manera, s'aconsegueix tenir un protocol d'encaminament suportat pel *hardware* emprat pel disseny del prototipus convenientment modificat pel cas particular que ens ocupa.

Per la implementació de les solució multicanal, s'ha implementat un **protocol de negociació del canal**. Aquest protocol assigna inicialment els canals en tota la topologia fent una distribució depenent de la posició que ocupi el node dins la jerarquia conformada per realitzar les proves. Després, intercanviant dades entre els diferents nodes sobre la qualitat dels diferents canals seleccionables, s'ha construït un algorisme que selecciona el millor canal possible tenint en compte les interferències que pateixen els enllaços degut a transmissions a l'entorn. Els canals seleccionables són aquells que estan sintonitzats de manera fixa a les interfícies de dades dels *meshcube*.

Aquest algorisme segueix una estructura **client-servidor**, és **iteratiu** i conté opcions per què sigui **adaptable** a les condicions de l'entorn. S'estarà executant de forma continuada a cadascuna de les estacions i anirà confeccionant les seves seleccions depenent de les condicions que l'envolten.

Una part important del treball ha estat l'elaboració d'un **mecanisme de selecció d'interfícies**. A aquest projecte, aquest mecanisme és el que aporta la particularitat multicanal al sistema ja que, a l'estar sintonitzades a diferents canals ambdues interfícies, al canviar d'interfície per on transmetre dades estem canviant al mateix temps el canal.

El *bonding driver* és el que ha permès realitzar aquesta selecció d'interfícies i el que possibilita fer aquesta funció física de transmissió per un diferent canal. Amb el *bonding* es crea una interfície lògica a sobre de les interfícies físiques aconseguint prendre decisions del seu funcionament manejant la interfície lògica.

El mecanisme de selecció d'interfícies s'alimenta de la informació que li proporciona el protocol de negociació de canal, passant-li quin és el millor canal en aquell moment concret per la transmissió de dades. El mecanisme **tradueix** la informació relacionant el canal amb la interfície que pertoca i duent a terme tots els processos que es necessiten per fer els canvis d'interfície pertinents.

Ajuntant els distints components lògics s'arriba a tenir aquesta solució global multicanal amb canals d'assignació fixa a les diferents interfícies i amb una interfície de senyalització que aporta més funcionalitats al sistema global. El protocol d'encaminament OLSR, modificat amb el propòsit d'adaptar-lo a aquest cas, serà el que permetrà crear les rutes i fer les deteccions de noves situacions amb l'alta o baixa de noves estacions.

Els resultats han demostrat que l'aplicació multicanal permet obtenir uns rendiments a la xarxa francament millors. De totes formes, abans d'obtenir aquests resultats, es van haver de superar uns cert problemes de tipus pràctic derivats de la utilització dels *meshcube* i que hem vist com afectaven al comportament de les comunicacions. Són interferències degudes a la **proximitat d'antenes** i els problemes sorgits amb els canals de l'espectre IEEE 802.11a. Tots ells, encara que teòricament no estan tipificats, suposen una important **pèrdua de capacitat** que també s'ha de tenir en compte en qualsevol tipus d'implementació pràctica en una temàtica similar.

En resum, doncs, estem davant d'un **prototipus** que incrementa en un important factor el rendiment global de la xarxa en un factor com a mínim de 1,5 fins a 2 respecte als casos d'únic canal i que, a més, és fàcilment aplicable a les eines de les què disposem amb les seves característiques particulars.

## 7.2   FUTURES LÍNIES D'INVESTIGACIÓ

El tema de les solucions multicanal és una de les branques amb més recorregut i amb més possibilitats de les xarxes *mesh* en particular i de les xarxes sense fils en general. La possibilitat d'augmentar les prestacions d'aquest tipus de xarxes és una opció molt interessant, sobretot si es té en compte la gran quantitat de factors que limiten la capacitat de les comunicacions entre els nodes.

Aquest projecte s'emmarca dins d'aquestes opcions per obtenir importants millores de rendiment. Tot i això, aquest subapartat es centrarà en les **possibles millores** que se li poden aplicar a la nostra solució, ja siguin canviant punts concrets del disseny o aplicant les noves solucions a noves plataformes *hardware*.

Una primera possible millora per tal de tenir un major nombre de canals seria el d'**afegir més targetes *wireless*** dins el node emprat. Fent això, aconseguim tenir un major nombre d'interfícies i, per tant, un major nombre de canals que impliquen, amb el sistema que tenim dissenyat, un major rendiment en la xarxa.

Amb el nostre prototipus, l'afegir més interfícies pot suposar problemes d'escalfament, i que els *meshcube* quedin bloquejats per les elevades temperatures que assoleix l'estació.

Una altra possible millora que es podria realitzar seria la d'implementar un protocol multicanal amb **assignació dinàmica** de canals. D'aquesta manera es podrien utilitzar més canals amb un nombre més reduït d'interfícies. Aquesta millora implicaria també canvis en els algorismes, possibilitant la sintonització i els canvis de canal a qualsevol moment, sempre que sigui necessari aquest canvi.

El que es proposa al paràgraf anterior es podria fer amb estacions similars a les que tenim, amb la mateixa estructura i el mateix nombre d'interfícies. Per fer-ho es podria emprar, fins i tot, el mateix mecanisme de selecció d'interfícies, el *bonding* driver, ampliant la seva funcionalitat de multi-interfície a multicanal.

Comentar que el material del que es disposa per a les realitzacions presenta algunes limitacions a l'hora d'implementar una solució com l'anterior degut a l'elevat temps de canvi dels canals. Per tant, possiblement, s'hauria de provar amb un altre plataforma *hardware* en el que el canvi de canal i les associacions derivades d'aquests canvis fossin gairebé imperceptibles.

A més, si realment es vol accedir a molts de canals en el sistema, hem de tenir estacions que realment suportin molts de canals ortogonals. Hem vist que, a la pràctica, només podíem arribar a tenir fins a quatre canals independents per IEEE 802.11a, quan, teòricament, es poden tenir fins a 13 ortogonals.

Una de les altres possibles futures línies d'investigació a tenir en compte, serien aquelles que tenen que veure amb les **modificacions** que es poden aplicar al protocol de negociació del canal. Des d'aquest treball es proposen diferents possibles modificacions a fer a aquest protocol. Una d'elles ve en el mètode de detecció d'interferències. El mètode emprat detecta amb un *scan* les possibles fonts interferents que, després, serveix per confeccionar la llista de qualitats dels canals. Una opció que es proposa és la de detectar aquestes interferències fent servir el mètode del *ping quiet*, on es detecta major interferència a major retard dels paquets. Aquest mètode podria ser aplicat amb la interfície de senyalització.

Llavors, el protocol està preparat per introduir possibles balancejos entre client i servidor o operacions de *refresh* introduint memòria al sistema, afectant el que s'havia calculat a iteracions del protocol anteriors.

També hi ha altres idees que es recullen a altres articles i que podrien ser vàlides per protocols multicanal. Les més destacades són les que tenen que veure amb l'assignació de canals a topologies compostes de diferents nodes com són les **tècniques de grafs**, i que han estat introduïdes a la memòria, en l'apartat teòric dels protocols multicanal, així com protocols proposats que realitzen l'assignació dels canals d'una forma **centralitzada**, tenint el control de tots els nodes que conformen la xarxa i distribuint de la millor manera possible els canals dels que es disposen entre tots els enllaços que uneixen aquests nodes.

El problema que hi ha amb aquestes idees és que, encara que sembli que siguin les més idònies per implementar una solució multicanal, és la seva aplicació pràctica, ja que habitualment, no hi ha components pràctics que suportin de forma òptima el comportament teòric d'aquestes estratègies, especialment degut a que proposen constants canvis de canal, que duen temps i problemes d'associació entre cel·les. A més, les contínues modificacions que es donen en una xarxa *mesh* dels nodes que conformen una xarxa dificulta l'aplicació de solucions on es controlin tots els enllaços de la xarxa.

# 8 REFERÈNCIES

# 9 ACRÒNIMS

| | |
|---|---|
| **ACK** | Acknoledge Message |
| **AES** | Advanced Encryption Standard |
| **AMRIS** | Ad-hoc Multicast Routing protocol using Increasing id-numberS |
| **AMRoute** | Ad-hoc Multicast Route |
| **ANSN** | Advertised Neighbor Sequence Number |
| **AODV** | Ad-hoc On-Demand Distance Vector |
| **AP** | Access Point |
| **AR** | Access Router |
| **ARP** | Address Resolution Protocol |
| **ATIM** | Ad-hoc Traffic Indication Messages |
| **BSS** | Base Station Subsystem |
| **BSSID** | Base Station Subsystem Identification |
| **CNAF** | Cuadro Nacional de Atribución de Frecuencias |
| **CS** | Carrier Sense |
| **CSMA-CA** | Carrier Sense Multiple Access – Collision Avoidance |
| **CTS** | Clear To Send |
| **DCA** | Dynamic Channel Assignment |
| **DCF** | Distributed Coordination Function |
| **DIFS** | Distributed Inter-Frame Space |
| **DHCP** | Dynamic Host Configuration Protocol |
| **DLL** | Dynamic Linking Library |
| **DSDV** | Destination-Sequenced Distance Vector |
| **DSR** | Dynamic Source Routing |
| **DYMO** | Dynamic MANET On-Demand |
| **EAP** | Extensible Authentication Protocol |
| **ESSID** | Extended Service Set Identifier |
| **ETSI** | European Telecommunication Standard Institute |
| **ETT** | Expected Transmission Time |
| **ETX** | Expected Transmission Count |
| **ExOR** | Extremely Opportunistic Routing |
| **GCRP** | Geographic Circuit Routing Protocol |
| **GPRS** | General Packet Radio Service |
| **GSM** | Global System for Mobile Communications |
| **HNA** | Host and Network Association |
| **ICMP** | Internet Control Message Protocol |
| **IEEE** | Institute of Electrical and Electronics Engineers |
| **IETF** | Internet Engineering Task Force |
| **IOCTL** | Input/Output Control |
| **IP** | Internet Protocol |
| **ISM** | Industrial, Scientific and Medical |
| **LAN** | Local Area Network |
| **LQSR** | Link Quality Source Routing |
| **MAC** | Medium Access Control |
| **MANET** | Mobile Ad-hoc Network |
| **MCExOR** | Multi Channel Extremely Opportunistic Routing |
| **MCR** | Multi Channel Routing |
| **MCRP** | Multi Channel Routing Protocol |
| **MID** | Multiple Interface Declaration |
| **MIPS** | Microprocessor without Interlocked Pipeline Stages |
| **MMAC** | Multi-Channel Medium Access Control |
| **MPR** | Multi Point Relay |
| **MR-LQSR** | Multi-Radio Link Quality Source Routing |
| **MUP** | Multi-Radio Unification Protocol |
| **NIC** | Network Interface Card |
| **NAT** | Network Address Translation |
| **ODMRP** | On Demand Multicast Routing Protocol |
| **OFDM** | Orthogonal Frequency Division Multiplexing |
| **OLSR** | Optimized Link State Routing |
| **OLSRD** | Optimized Link State Routing Daemon |
| **PCL** | Preferable Channel List |
| **PSM** | Power Saving Mechanism |
| **QoS** | Quality of Service |
| **QoSRBonBERP** | Quality of Service aware Routing Based on Bandwidth Estimation Routing Protocol |
| **QoSSIRP** | Quality of Service aware Source-Initiated Routing Protocol |
| **QRREP** | Quality Route Reply |
| **QRREQ** | Quality Route Request |
| **RADIUS** | Remote Authenticated Dial-In User Service |
| **RCA** | Receiver-based Channel Assignment |
| **RERR** | Route Error |
| **RFC** | Request For Comment |
| **RPDB** | Routing Policy DataBase |
| **RREP** | Route Reply |
| **RREQ** | Route Request |
| **RTS** | Request To Send |
| **RTT** | Round Trip Time |
| **SSCH** | Slotted Seeded Channel Hopping |
| **SRTT** | Smoothed Round Trip Time |
| **TBRPF** | Topology Dissemination Based on Reverse Path Forwarding |
| **TC** | Topology Control |
| **TCP** | Transmission Control Protocol |
| **TMT** | Theoretical Maximum Throughput |
| **TTL** | Time To Live |
| **UCSB** | University of California, Santa Barbara |
| **UDP** | User Datagram Protocol |
| **UMTS** | Universal Mobile Telephone System |
| **VoIP** | Voice over Internet Protocol |
| **WCETT** | Weighted Cumulative Expected Transmission Time |
| **WiFi** | Wireless Fidelity |
| **WiMax** | Worldwide Interoperability for Microwave Access |
| **WLAN** | Wireless Local Area Network |
| **WMN** | Wireless Mesh Network |
| **WRP** | Wireless Routing Protocol |
| **ZRP** | Zone Routing Protocol |

# ANNEXES

## A.1. DISTRIBUCIÓ DE FREQÜÈNCIES

Aquí veiem una petita mostra del CNAF del que s'ha parlat al capítol 2 i com s'estructuren les diferents freqüències de l'espectre a nivell nacional.

| Freqüència | Assignació |
|---|---|
| 830 - 862 MHz | Televisió digital |
| 870 - 880 MHz | GSM-R |
| 880 - 960 MHz | GSM i E-GSM |
| 1,35 - 1,71 GHz | Serveis de radiolocalització i fix |
| 1,71 - 1,88 GHz | DCS-1800 |
| 1,88 - 1,9 GHz | DECT |
| 1,9 - 2,025 GHz | UMTS/IMT-2000 |
| 2,025 - 2,11 GHz | Servei fix |
| 2,11 - 2,2 GHz | UMTS/IMT-2000 |
| 2,2 - 2,29 GHz | Servei fix |
| 2,3 - 2,483 GHz | Radioenllaços mòbil per TV |
| **2,4 - 2,483 GHz** | **Banda ISM** |
| 2,5 - 2,7 GHz | UMTS/IMT-2000 |
| 2,7 - 3,6 GHz | Radars militars |
| 3,6 - 4,2 GHz | Radioenllaços per TV i telefonia analògica |
| 4,2 - 5 GHz | Usos militars |
| 5 – 5,15 GHz | Radionavegació |
| **5,15 - 5,35 GHz** | **Mòbil** |
| 5,35 - 5,725 GHz | Radiolocalització |
| **5,725 - 5,875 GHz** | **Banda ISM** |
| 5,9 - 7,1 GHz | Radioenllaços telefònics i analògics |

**Taula A.1: Distribució de freqüències**

Marcades en negreta estan el rang de freqüències que s'empren en la realització global d'aquest muntatge. Les de 2,4 GHz corresponen a les freqüències d'IEEE 802.11b, mentre que els altres dos intervals corresponen a les freqüències d'IEEE 802.11a. En les següents taules es pot veure els canals que es poden utilitzar als *meshcube* tant a una versió com a l'altra.

Per IEEE 802.11b tenim la següent taula:

| Canal | Freqüència |
|-------|-----------|
| 1 | 2,412 GHz |
| 2 | 2,417 GHz |
| 3 | 2,422 GHz |
| 4 | 2,427 GHz |
| 5 | 2,432 GHz |
| 6 | 2,437 GHz |
| 7 | 2,442 GHz |
| 8 | 2,447 GHz |
| 9 | 2,452 GHz |
| 10 | 2,457 GHz |
| 11 | 2,462 GHz |

**Taula A.2: Freqüències d'IEEE 802.11b**

Per IEEE 802.11a la taula de freqüències corresponent és la següent:

| Canal | Freqüència |
|-------|-----------|
| 36 | 5,18 GHz |
| 40 | 5,2 GHz |
| 44 | 5,22 GHz |
| 48 | 5,24 GHz |
| 52 | 5,26 GHz |
| 56 | 5,28 GHz |
| 60 | 5,3 GHz |
| 64 | 5,32 GHz |
| ... | ....... |
| 149 | 5,75 GHz |
| 153 | 5,77 GHz |
| 157 | 5,79 GHz |
| 161 | 5,81 GHz |
| 165 | 5,83 GHz |

**Taula A.3: Freqüències d'IEEE 802.11a**

# A.2. ESPECIFICACIONS MESHCUBE

A continuació es presenta una taula on es recullen les especificacions tant de tipus *hardware* com *software* dels *meshcube*. La informació ha estat extreta de la pàgina del 4G Access Cube [65].

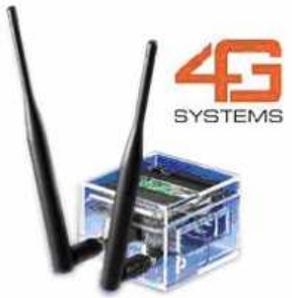

| | Hardware | Software |
|---|---|---|
| | • Procesador MIPS 400MHz.<br>• Memoria RAM de 64MB.<br>• Memoria Flash de 32MB.<br>• Puerto Ethernet 100Mbps.<br>• 2 Interfaces miniPCI multibanda IEEE 802.11 con antenas exteriores omnidireccionales (extensible hasta 8 interfaces).<br>• Dimensiones: 7 x 5 x 7cm.<br>• Otros: puerto usb, PoE, consumo de potencia 4-6W (depende del número y tipo de tarjetas miniPCI). | • Sistema operativo Nylon (versión embebida basada en Debian Linux, **kernel 2.4.x**).<br>• Herramientas Unix implementadas en un binario simple (BusyBox), precompiladas para la arquitectura MIPS.<br>• Herramientas y drivers adicionales disponibles como módulos del kernel en la comunidad de soporte meshcube.org. |

**Taula A.4: Especificacions Meshcube**

En quan a les especificacions *hardware* afegir que hi han dos mides diferents de *meshcube*, existint un més major que el que s'indica al quadre amb unes dimensions de 8,5x8,5x8,5 cm. També indicar que es poden afegir més interfícies que les 2 que surten indicades al quadre d'especificacions. De fet, teòricament hi ha espai per introduir fins a 8 interfícies, encara que realment el sistema suporta fins a 4 interfícies, ja que un nombre major provoca problemes d'estabilitat de l'estació.

Comentar també, que els paquets que porten aplicacions per ser utilitzats en la versió encastada de Linux, són paquets específics pel sistema operatiu Nylon i que, per tant, són diferents als paquets de Debian. La pàgina oficial de *meshcube* conté, disponibles per a la descàrrega, els paquets més significatius i que es poden fer servir més sovint.

Tots aquells paquets que no apareguin al llistat de descàrrega de la pàgina poden ser adaptats al sistema operatiu Nylon compilant el paquet Debian convencional amb el *cross-compiler* MIPSEL que tradueix aquest paquet a informació comprensible per Nylon. Per a la realització d'aquest procés cal tenir instal·lat el *cross-compiler* a un ordinador personal on després es passin les dades cap als *meshcube*.

# A.3. TRAMES DELS PROTOCOLS D'ENCAMINAMENT

En aquest annex es mostraran quines són les diferents trames que utilitzen els dos protocols d'encaminament descrits en aquest projecte i que són les dues millors opcions per la detecció de nous veïns. Es farà també una breu descripció del significat dels camps que apareixen a les diferents trames que surten plasmades al treball

**Ad-hoc On-Demand Distance Vector (AODV)**

Els missatges d'AODV són aquells que es transmeten entre els diferents usuaris i corresponen als ja comentats missatges de RREQ, per realitzar peticions als veïns i poder configurar les rutes, missatges de RREP, que són els missatges de resposta als RREQ i els missatges RERR, que són missatges que es generen fruit d'alguna irregularitat succeïda a l'entorn de la xarxa.

**Missatge RREQ:**

| 0 | 7 | 15 | 23 | 31 |
|---|---|---|---|---|
| Tipus | Flags | Reservat | | Compte de salts |
| RREQ ID | | | | |
| Adreça de la destinació | | | | |
| Número de seqüència de la destinació | | | | |
| Adreça de l'origen | | | | |
| Número de seqüència de l'origen | | | | |

**Figura A.1: RREQ**

**Missatge RREP:**

| 0 | 7 | 15 | 23 | 31 |
|---|---|---|---|---|
| Tipus | Flags | Reservat | Long. Prefix | Compte de salts |
| Adreça de la destinació | | | | |
| Número de seqüència de la destinació | | | | |
| Adreça de l'origen | | | | |
| Temps de vida | | | | |

**Figura A.2: RREP**

**Missatge RERR:**

| Tipus | N | Reservat | Compte destí |
|---|---|---|---|
| Adreça de la destinació no assolible | | | |
| Número de seqüència de la destinació no assolible | | | |
| Adreça de la destinació no assolible addicional (si és necessari) | | | |
| Número de seqüència de la destinació no assolible addicional (si és necessari) | | | |

**Figura A.3: RERR**

El significat dels camps de les trames vistes són les següents:

- **Tipus.** És el camp que indica quin tipus de missatge s'està enviant, és a dir, si és RREQ (tipus = 1), RREP (tipus = 2) o RERR (tipus = 3).

- **Flags.** Indiquen particularitats damunt del paquet que s'envia. RREQ té 5 flags diferents: el flag J, reservat per *multicast*, el flag R, de reparació també reservat per *multicast*, el flag G, que marca si el RREP ha d'anar de forma *unicast* cap a la destinació, el flag D, que si està activat indica que només la destinació pot respondre i el flag U, que s'activarà en cas de què el nombre de seqüència sigui desconegut. RREP té només dos flags: el R i el A, que requereix un reconeixement a la seva resposta. RERR en té un anomenat N i que si està activat indica que s'ha reparat un enllaç i que no s'ha d'esborrar la ruta en qüestió.

- **Reservat.** Són una pila de bits reservats per un futur ús.

- **Longitud del prefix.** És propi de RREP i està compost per 5 bits que si no són 0 pot ser utilitzat per algun altre node amb el mateix prefix.

- **Compte de salts.** És un nombre que compta els salts que hi ha entre l'adreça d'origen i l'adreça destí. Apareix a RREQ i a RREP.

- **Compte destí.** És un camp propi de RERR i conté el nombre de destinacions a les que no es pot arribar. Aquest valor com a mínim sempre haurà de valer 1.

- **RREQ ID.** És un identificador únic pel missatge RREQ enviat.

- **Adreça de destinació.** Conté l'adreça a la qual volem arribar per generar una ruta des de l'origen del qual hem enviat la petició.

- **Adreça d'origen.** Contendrà l'adreça del node que generi la petició per configurar la ruta cap a la destinació.

- **Número de seqüència de la destinació.** És el darrer número de seqüència rebut per la mateixa adreça d'origen. Aquests tres darrers camps surten al RREQ i al RREP.

- **Número de seqüència de l'origen.** L'actual número de seqüència a utilitzar quan apuntem cap a l'adreça originària. És propi només de RREQ.
- **Temps de vida.** És el temps en milisegons durant el qual els nodes que reben el RREP poden considerar vàlid el missatge.
- **Adreça de la destinació no assolible.** Surt la primera adreça de la destinació a la que no es pot arribar. Si hi ha més destinacions a les que no es pot arribar, aquestes surten indicades als posteriors camps addicionals, que tindran la mateixa estructura que aquest.
- **Número de seqüència de la destinació no assolible.** Reflectirà el nombre de seqüència d'aquesta destinació a la que no es pot arribar.

**Optimized Link State Routing (OLSR)**

El missatge estàndard d'OLSR conté aquests camps. Després, a continuació, veurem casos de missatges particulars que prenen aquests camps i els adapten a les necessitats per les quals els missatges van ser generats. El que fan aquests missatges particular és omplir part de l'espai que està destinat al missatge, afegint les particularitats pròpies d'aquests missatges específics.

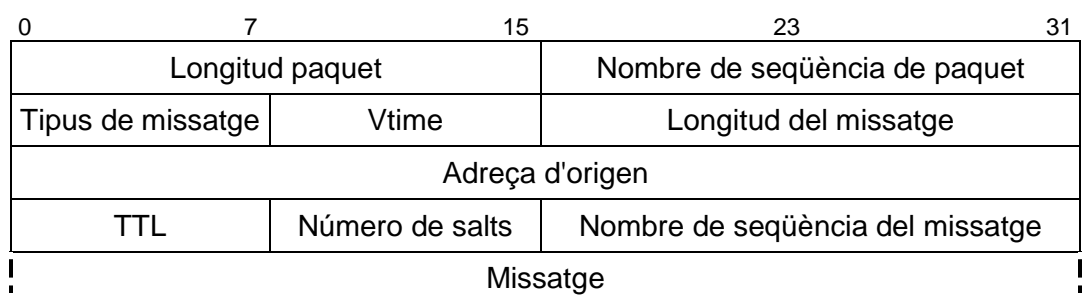

| 0 | 7 | 15 | 23 | 31 |
|---|---|---|---|---|
| Longitud paquet | | | Nombre de seqüència de paquet | |
| Tipus de missatge | Vtime | | Longitud del missatge | |
| Adreça d'origen | | | | |
| TTL | Número de salts | | Nombre de seqüència del missatge | |
| Missatge | | | | |

**Figura A.4: Missatge genèric d'OLSR**

- **Longitud del paquet.** La longitud total en bytes del paquet, amb la capçalera inclosa.
- **Nombre de seqüència del paquet.** Cada vegada que un paquet es transmet pel host, aquest nombre s'incrementa en una unitat. Les diferents interfícies tenen nombres de seqüència diferents i allunyats en valor.
- **Tipus de missatge.** Ens indica quin tipus d'OLSR missatge és el paquet en qüestió.
- **Vtime.** Camp que representa el temps durant el qual el missatge que es transmet serà vàlid.

- **Longitud del missatge.** Ens diu la longitud del missatge continguda a l'estructura general, amb la capçalera del missatge inclosa.
- **Adreça origen.** Conté l'adreça IP del node que ha generat el missatge.
- **TTL.** El nombre màxim de salts que el missatge pot realitzar. És el camp que determina el radi màxim que pot arribar a tenir la xarxa.
- **Número de salts.** Indica la quantitat de vegades que el missatge s'ha reenviat.
- **Nombre de seqüència del missatge.** Un nombre de seqüència que s'incrementa cada vegada que es transmet un nou paquet OLSR per aquell host.

A continuació mostrarem els detalls dels quatre missatges d'OLSR que podem tenir: el missatge HELLO, el TC, el MID i el HNA.

El missatge **HELLO**:

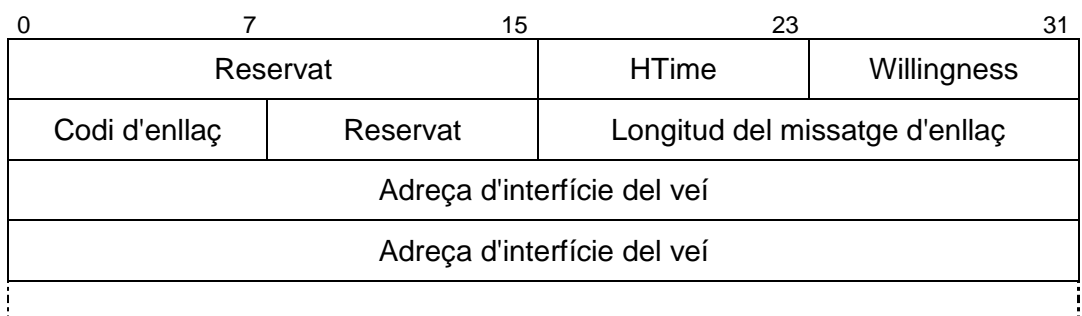

**Figura A.5: Missatge HELLO**

- **Reservat.** Bits reservat per un futur ús.
- **HTime.** Representa el temps d'emissió de cada missatge HELLO per aquell node per una determinada interfície.
- **Willingness.** És un camp que fa referència a com serà utilitzat per transmetre tràfic cap a altres nodes. Els nodes amb aquest camp elevat seran seleccionats dins l'arbre MPR, en canvi, si tenen un valor baix o nul, no seran seleccionats com a integrants del MPR.
- **Codi d'enllaç.** Té informació respecte a l'enllaç entre la interfície del transmissor i la llista d'interfícies veïnes destinatàries. Especifica el *status* del veí marcant si estem en un enllaç simètric, si ha hagut errors…
- **Longitud del missatge d'enllaç.** És la mida del missatge de l'enllaç mesurat en bytes.
- **Adreça d'interfície del veí.** L'adreça d'una interfície d'un node veí.

El missatge **TC**:

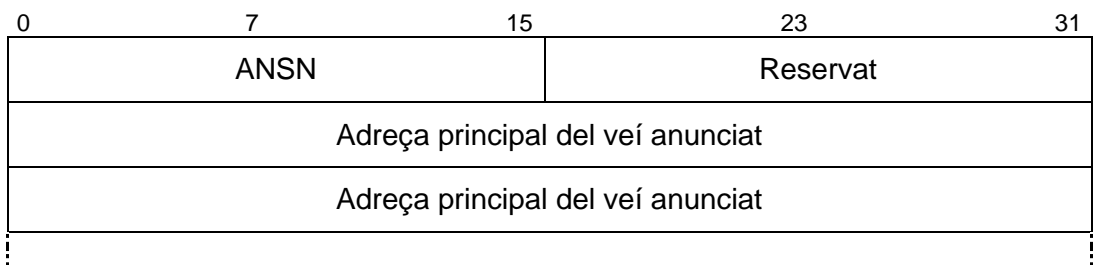

**Figura A.6: Missatge TC**

- **ANSN.** És un nombre de seqüència associat al conjunt de veïns coneguts pel node. Cada vegada que hi ha un canvi aquest nombre es va actualitzant sumant un nombre. Amb aquest identificador es pot tenir controlat si la informació rebuda amb aquest missatge TC és més recent que la que tenia el node.
- **Reservat.** Bits reservats per un ús futur, estan fixats tots a 0.
- **Adreça principal del veí anunciat.** Aquest camp conté l'adreça principal del veí. Totes les adreces dels nodes veïns són recollides al missatge TC en múltiples camps.

El missatge **MID:**

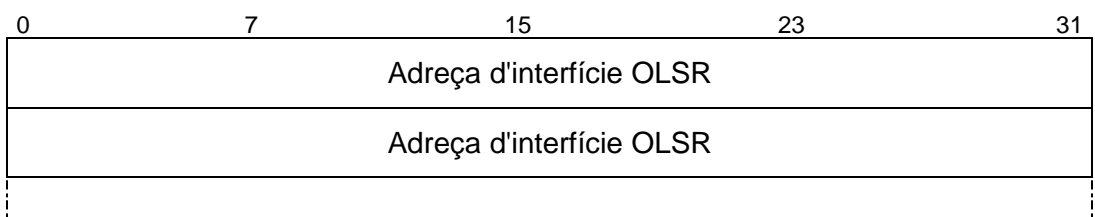

**Figura A.7: Missatge MID**

- **Adreça d'interfície OLSR.** Conté l'adreça d'una altra interfície del node, diferent a la principal d'aquest mateix. OLSR suporta vàries interfícies.

El missatge **HNA:**

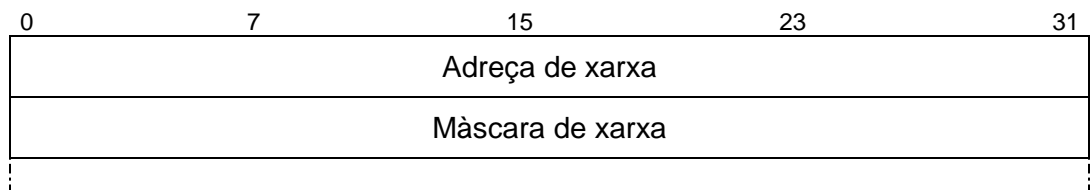

**Figura A.8: Missatge HNA**

- **Adreça de xarxa.** Camp que contendrà l'adreça de xarxa d'aquesta xarxa externa que no suporta OLSR.
- **Màscara de xarxa.** Serà la màscara de xarxa corresponent a l'adreça de xarxa abans calculada. Aquests dos camps es poden repetir sempre que hagi més xarxes externes a l'entorn en què es mesuren les rutes.

# A.4. SCRIPTS DEL *MESHCUBE*

Per la configuració del *meshcube* a cada cas i poder obtenir els resultats mesurats al capítol d'avaluació de resultats, ha estat necessària la creació i modificació de diversos *scripts*. D'aquesta manera s'aconseguia crear les condicions necessàries per les proves realitzades. Al següent llistat apareixeran aquells *scripts* modificats o creats amb una breu explicació de la seva utilitat en l'execució dels processos.

**olsrd.conf**

És el paràmetre de configuració del protocol d'encaminament d'OLSR. En aquest cas, s'han configurat els paràmetres de tal manera que no hagi xarxes HNA, que s'utilitzi el paràmetre *Link Quality* complet i que el protocol circuli només per una interfície, la wlan0, que és la de senyalització, que després serà mapejada a la taula de rutes per la bond0. El *script* és el següent.

```
#
# olsr.org configuration file
#

#
# Periodically print the internal state, including information on
# route calculation
#

DebugLevel          2

#
# Clear the screen each time the internal state changes
#

ClearScreen         yes

#
# IP version to use
#

IpVersion           4

#
# IPv4 HNAs - syntax: netaddr netmask
#

Hna4
```

```
{
#      0.0.0.0 0.0.0.0
#192.168.21.0 255.255.255.0
}

#
# Do not use hysteresis
#

UseHysteresis              no

#
# If using hysteresis, use "smooth" parameters
#

HystScaling        0.10
HystThrHigh        0.80
HystThrLow         0.30

#
# Which neighbours should be advertised via TC messages?
#
# 0 - only advertise our MPR selectors
# 1 - advertise our MPR selectors and our MPRs
# 2 - advertise all neighbors
#

TcRedundancy       2

#
#  Specifies  by  how  many  MPRs  each  two-hop  neighbour  should  be
covered
#

MprCoverage        3

#
# Enable the link quality extensions?
#
# 0 - no
# 1 - yes, use link quality for MPR selection
# 2 - yes, use link quality for MPR selection and routing
#

LinkQualityLevel   2

#
# Windows size for packet loss calculation
#

LinkQualityWinSize 20
```

```
#
# Load plugins
#

#LoadPlugin "olsrd_httpinfo.so.0.1"
#{
#       PlParam "Port" "8080"
#       PlParam "Net" "0.0.0.0 0.0.0.0"
#}

#LoadPlugin "olsrd_dot_draw.so.0.3"
#{
        # accept connection from IP:
        # default 127.0.0.1 (localhost)
        #PlParam      "accept" "192.168.0.5"
#}

#
# Interfaces -
# change to the name of your WLAN interface
#

Interface "wlan0"
{

        #
        # HELLO interval in seconds (float)
        #

        HelloInterval           2.0

        #
        # HELLO validity time
        #

        HelloValidityTime  40.0

        #
        # TC interval in seconds (float)
        #

        TcInterval         3.0

        #
        # TC validity time
        #

        TcValidityTime          15.0

        #
        # MID interval in seconds (float)
        #
```

```
MidInterval        5.0

#
# MID validity time
#

MidValidityTime        15.0

#
# HNA interval in seconds (float)
#

HnaInterval        5.0

#
# HNA validity time
#

HnaValidityTime    15.0
}
```

**startCubexxxx.sh**

Totes les estacions executen, per defecte, una sèrie de paràmetres que poden ser utilitzats. Al nostre cas no ens interessa executar tots aquests paràmetres, i possiblement voldrem engegar altres aplicacions més propícies per les proves que es volen fer.

L'arxiu està a dins del directori /etc/init.d/ que és on es configuren les aplicacions o paràmetres que volem executar cada vegada que l'estació és inicialitzada. Per tant, tot el que aparegui a aquest *script* d'inici serà executat al sistema operatiu de l'estació en qüestió. L'estructura de tots aquests programes és igual i el que es plasmarà aquí correspondrà al node 1521, amb les úniques diferències de les adreces entre ells.

```
#!/bin/bash

#stop olsrd and cron
/etc/init.d/olsrd stop
/etc/init.d/firewall stop
/etc/init.d/busybox-cron stop
/etc/init.d/ipsec stop
iwconfig ath0 rate 11M
iwconfig ath1 rate 11M
```

**iptables.sh**

Per configurar les topologies seran necessàries algunes regles de filtratge entre les diferents estacions. En un cas ideal, les estacions que composen una topologia en línia només veurien als seus veïns adjacents, però en un entorn limitat com és el que s'ha provat totes les estacions es detecten i, per tant, sorgeix la necessitat de filtrar tots els paquets que provinguin d'aquelles estacions que no són adjacents.

En aquest annex mostrarem un exemple de l'estructura d'un d'aquests *scripts*. En concret, es tractarà de l'estació 1521 que és un dels extrems de la comunicació i que, per tant, haurà de filtrar totes les dades que no provinguin del seu veí adjacent. Per les altres estacions només varia en les adreces a filtrar. El filtratge dels diferents nodes és realitza descartant els paquets amb adreça MAC igual al d'un d'aquests nodes.

```
#Flush de Regles
iptables -F
iptables -X
iptables -Z
iptables -t nat -F
iptables -t mangle -F
iptables -t filter -F

#Politica per defecte (ACCEPT)
iptables -P INPUT ACCEPT
iptables -P OUTPUT ACCEPT
iptables -P FORWARD ACCEPT
iptables -t nat -P PREROUTING ACCEPT
iptables -t nat -P POSTROUTING ACCEPT

#Filtrem el trafic que vengui de 1498
iptables -t filter -A INPUT -m mac --mac-source 00:02:6F:35:52:2B
-j DROP

#Filtrem el trafic que vengui de 1533
iptables -t filter -A INPUT -m mac --mac-source 02:02:61:1D:52:AE
-j DROP

#Filtrem el trafic que vengui de 1535
iptables -t filter -A INPUT -m mac --mac-source 00:02:6F:35:52:AE
-j DROP
```

**if_bonding.sh**

El darrer dels *scripts* que quedaran reflectits a aquesta memòria és un de nova creació, anomenat if_bonding.sh i que té totes les instruccions a seguir per, primerament, posar les mateixes adreces MAC a totes les interfícies transmissores de dades útils i després aplicar el *bonding driver* a aquestes mateixes interfícies.

```
#!/bin/bash

ifconfig ath1 down
ifconfig ath1 hw ether 00:02:6F:2F:04:88
ifconfig ath1 up

iwconfig ath0 essid xarxa
iwconfig ath1 essid xarxa

insmod bonding.o
ifconfig    bond0    10.15.21.1    netmask    255.0.0.0    broadcast
10.255.255.255
ifenslave bond0 ath0 ath1
```

## A.5. BONDING DRIVER

Al projecte el *bonding driver* ha estat una eina fonamental ja que ha estat gràcies al *driver* que s'ha pogut implementar un mecanisme fiable de selecció d'interfícies, modificant el codi font del mateix *driver*.

En un principi, el *bonding driver* estava destinat per entorn de xarxa cablada com Ethernet que permetien crear una interfície lògica damunt de les interfícies físiques que convivien a un mateix node. D'aquesta manera cadascun dels nodes és molt manejable ja que totes les interfícies, que sempre estan identificades per la seva pròpia adreça IP i la seva pròpia adreça MAC, compartirien una mateixa adreça de xarxa i una mateixa adreça física. Aquest concepte també es pot reutilitzar per xarxes amb interfícies *wireless*, fent exactament la mateixa funció. Així, seguint un procés determinat, la interfície lògica seleccionaria per quina de les interfícies físiques es volen transmetre aquestes dades.

Per poder utilitzar correctament el *driver* abans s'han de realitzar una sèrie d'operacions des de la línia de comandes. Es tracta de carregar primer el mòdul *bonding* amb la instrucció següent.

```
insmod bonding.o
```

Després es realitzar el procés d'esclavització de les interfícies a on, amb la instrucció *ifenslave* s'indica quines interfícies s'esclavitzen (seran les *slave*) i quina serà la interfície lògica que actuarà de mestre (serà la *master*) anomenant-li de la manera més adient possible (bond0 és l'opció per defecte). Un cas on es vol fer *bonding* damunt de tres interfícies diferents (ath0, ath1 i ath2) es faria introduint la següent instrucció en la línia de comandes.

```
ifenslave bond0 ath0 ath1 ath2
```

Per aplicar aquesta darrera instrucció als *meshcube* s'ha de fer passar el seu codi original pel *cross-compiler* específic que empra Nylon.

El criteri de quina interfície física seleccionar dependrà del *bonding driver* que atenent al mètode que es vol seleccionar, elegeix la interfície física corresponent transmetent les dades per la mateixa.

El *driver* lògic va ser creat a l'any 1999 per Thomas Davis com una eina *dummy*. Posteriorment, nombrosos autors han anat afegint noves funcionalitats al codi original i han anat corregint algunes mancances del disseny original. Entre les noves característiques que es van incorporar apareixen les de nous criteris de selecció d'interfícies.

El procediment que el *bonding driver* segueix és similar al que segueix el *netfilter*, ja que com aquest, agafa les dades del *buffer skb*, que porta tota la informació damunt del paquet que es vol transmetre, ja estructurada en les capçaleres que du el paquet (la de transport, la de xarxa i la MAC). Després hi ha una funció de transmissió al *bonding* anomenada *bond_dev_queue_xmit*, que envia les dades per la interfície esclava que ha estat seleccionada pel programa.

Alguns d'aquests criteris s'enumeren a la següent llista:

- **Round Robin.** En aquest cas es transmet un paquet per cadascuna de les interfícies canviant la interfície física a la que apuntar després d'haver dut a terme la transmissió. D'aquesta manera s'aconsegueix un sistema equitatiu entre totes les targetes *wireless*.
- **Active-Backup.** Aquí es transmetrà la informació sempre que la interfície esclava corresponent sigui vàlida. En cas contrari, es canviaria d'interfície física.
- **Mode XOR.** És un mode on es determina la interfície destí mitjançant l'operació d'una XOR entre les adreces físiques de l'origen i de la destinació. En el cas de què aquesta interfície no estigui activada, la interfície elegida seria la següent disponible.
- **Broadcast.** S'envien les dades pendents per totes les interfícies disponibles.
- **802.3ad.** Es tracta d'un sistema d'agregació de grups que comparteixen les mateixes estacions i funcionalitats dúplex. Totes les interfícies esclaves són utilitzades per a la transmissió de dades cap aquest grup agregat.
- **Load-Balancing.** Es balanceja la càrrega enviada entre totes les interfícies esclaves. Aquest balanceig és calculat respecte a la velocitat de transmissió. Tot el tràfic entrant és rebut pel actual esclau seleccionat. En el cas de produir-se algun problema amb la interfície actual, es realitzaria un canvi d'interfície a la següent disponible.

- **Adaptive-Balancing.** És un cas evolucionat del sistema anterior. Proveeix tant el *load-balancing* com el *load-balancing* rebut des de la negociació amb ARP des d'IPv4. Requereix l'habilitat de canviar l'adreça MAC d'una eina mentre aquesta estigui oberta.

Aquests són alguns dels exemples que es podrien fer servir, però es podrien generar tantes opcions com es volgués, sempre que es modifiquessin alguns paràmetres al programa principal i s'afegissin noves rutines *bond_xmit_...*

En aquest projecte, es va introduir nou codi, modificant alguns paràmetres per a què es seleccionés la nostra solució, afegint a l'arxiu principal del *bonding* la rutina *bond_xmit-_protocol.*

# A.6. SOFTWARE DE LES PROVES

A aquest apartat es relatarà quines han estat les eines utilitzades per l'avaluació dels resultat i s'entrarà més en detall en les diferents funcionalitats que poden arribar a tenir cadascuna d'aquestes eines. Els dos *software* més emprats en la realització del projecte han estat l'aplicació *iperf* i l'aplicació *ping.*

**Iperf**

*Iperf* ha estat l'eina més important per realitzar les proves ja que és la que ens dóna resultats més fiables per la mesura de les capacitats del sistema que s'està provant.

La metodologia que utilitza *Iperf* és la de saturar els canals per on s'està transmetent, forma en la que es pot mesurar quina és la capacitat total que pot arribar a assolir aquell enllaç.

Per a l'execució d'*Iperf* es requereixen com a mínim dues estacions controlades per l'usuari, ja que l'aplicació segueix el model client-servidor. D'aquesta manera, l'extrem que vol iniciar la comunicació serà el que farà de client, indicant el destí, protocols i altres modificadors, mentre que el node destinació s'executarà com a servidor esperant a rebre el flux de dades enviat per l'altre extrem.

Una vegada acabada la connexió entre ambdós extrems, cadascuna de les estacions retornarà una sèrie de valors, com són la capacitat enviada, el nombre de paquets que s'han perdut en la transmissió i el *jitter* que ha aparegut en la comunicació.

A més, amb *Iperf* es poden configurar una gran quantitat d'opcions ja que l'aplicació ofereix un ampli ventall de diferents possibilitat que poden ser accedits emprant adequadament els diferents modificadors a l'hora de l'execució de l'aplicació.

Les opcions més interessants i que han estat àmpliament utilitzades en aquest treball, són les que fan referència als diferents protocols de transport amb els que transmetre la informació. Així, si no es marca cap modificador, per defecte, transmetrem informació a sobre del protocol TCP, mentre que si a la instrucció li afegim un modificador –u, la informació serà transmesa a sobre del protocol UDP. Pel cas de TCP per defecte, el sistema satura l'enllaç de connexió fins al màxim de la seva capacitat, mentre que per UDP sempre se li haurà d'indicar la longitud del flux de dades que es vol enviar.

Per tant, existiran opcions que permetran modificar la mida del flux de dades que s'envia per la configuració de xarxa que es té, així com altres que permetran canviar la durada d'aquesta transmissió.

A la següent taula es mostraran quins són els modificadors més interessants d'aquesta aplicació.

| Paràmetre | Funció |
|-----------|--------|
| -c | Client |
| -s | Servidor |
| -f | Format (KB, MB, GB…) |
| -u | Mode UDP |
| -a | Ample de banda a transmetre (TCP) |
| -b | Ample de banda a transmetre (UDP) |
| -t | Durada de la transmissió |
| -i | Interval sobre el que retorna info |
| -p | Port de la transmissió |
| -l | Longitud del buffer per llegir/escriure |
| -w | Longitud de la finestra TCP |
| -D | Corre el servidor en mode *daemon* |
| -n | Nombre de bytes a transmetre |
| -S | Tipus de servei |
| -T | Canviar el TTL |

**Taula A.5: Paràmetres de iperf**

Un exemple d'ús d'aquesta aplicació es pot veure en els dos exemples que es presenten a continuació.

En el primer exemple es vol enviar un flux de dades amb el protocol TCP des d'una estació cap a altra. S'afegirà la particularitat d'aquesta transmissió duri 120s i que volem que vagi mostrant dades en intervals cada 30s. Dir que les estacions estan fixades a una velocitat de 12 Mbps i que s'està emprant l'espectre de freqüències corresponent a IEEE 802.11a. Les instruccions que s'hauran de seguir es recullen a continuació.

Al servidor:

```
iperf -s
```

Al client:

```
iperf -c 10.15.21.1 -t 120 -i 30
```

L'execució d'ambdues comandes té com a resultat pel servidor:

```
------------------------------------------------------------
Server listening on TCP port 5001
TCP window size: 85.3 KByte (default)
------------------------------------------------------------
[  6] local 10.15.21.1 port 5001 connected with 10.15.33.1 port 1027
[ ID] Interval        Transfer      Bandwidth
[  6]  0.0-120.1 sec   123 MBytes  8.60 Mbits/sec
```

**Figura A.9: Resultat iperf al servidor amb TCP**

Mentre que pel client veiem la següent pantalla:

```
root@cube-1533:~# iperf -c 10.15.21.1 -t 120 -i 30
------------------------------------------------------------
Client connecting to 10.15.21.1, TCP port 5001
TCP window size: 16.0 KByte (default)
------------------------------------------------------------
[  5] local 10.15.33.1 port 1027 connected with 10.15.21.1 port 5001
[ ID] Interval        Transfer      Bandwidth
[  5]  0.0-30.0 sec   30.6 MBytes  8.55 Mbits/sec
[  5] 30.0-60.0 sec   30.8 MBytes  8.62 Mbits/sec
[  5] 60.0-90.0 sec   30.7 MBytes  8.57 Mbits/sec
[  5] 90.0-120.0 sec  31.1 MBytes  8.68 Mbits/sec
[  5]  0.0-120.1 sec   123 MBytes  8.60 Mbits/sec
```

**Figura A.10: Resultat iperf al client amb TCP**

El segon exemple que es presenta és el mateix que el primer però canviant el protocol de transport a UDP. Això implicarà que s'hagin d'introduir diferents modificadors pel protocol utilitzat, a més d'un altre que indiqui l'ample de banda que es vol transmetre. Per aquest cas les instruccions als dos extrems de la comunicació són les mostrades a continuació.

Pel servidor:

```
iperf -s -u
```

Pel client:

```
iperf -c 10.15.21.1 -t 120 -i 30 -u -b 12M
```

L'execució ens mostrarà pantalles similar als casos anteriors però amb diferents resultats degut a la utilització del protocol de transport UDP.

Pel servidor veuríem el següent:

```
root@cube-1521:~# iperf -s -u
------------------------------------------------------------
Server listening on UDP port 5001
Receiving 1470 byte datagrams
UDP buffer size:  105 KByte (default)
------------------------------------------------------------
[  3] local 10.15.21.1 port 5001 connected with 10.15.33.1 port 1025
[ ID] Interval        Transfer     Bandwidth        Jitter   Lost/Total Datagrams
[  3]  0.0-120.3 sec   142 MBytes  9.89 Mbits/sec  3.658 ms    0/101168 (0%)
```

**Figura A.11: Resultat iperf al servidor amb UDP**

Pel client, en canvi, veiem la següent pantalla:

```
------------------------------------------------------------
Client connecting to 10.15.21.1, UDP port 5001
Sending 1470 byte datagrams
UDP buffer size:  105 KByte (default)
------------------------------------------------------------
[  5] local 10.15.33.1 port 1025 connected with 10.15.21.1 port 5001
[ ID] Interval        Transfer     Bandwidth
[  5] -0.0-30.0 sec  35.7 MBytes  9.98 Mbits/sec
[  5] 30.0-60.0 sec  35.3 MBytes  9.88 Mbits/sec
[  5] 60.0-90.0 sec  35.4 MBytes  9.89 Mbits/sec
[  5] 90.0-120.0 sec  35.4 MBytes  9.90 Mbits/sec
[  5]  0.0-120.0 sec   142 MBytes  9.91 Mbits/sec
[  5] Server Report:
[  5]  0.0-120.3 sec   142 MBytes  9.89 Mbits/sec  3.658 ms    0/101168 (0%)
[  5] Sent 101168 datagrams
```

**Figura A.12: Resultat iperf al client amb UDP**

Tant per un resultat com per l'altre observem que les dades que es retornen són de capacitat de transmissió de l'enllaç, del nombre de paquets perduts en la transmissió i del *jitter* dels paquets.

**Ping**

*Ping* és una aplicació que realitza *echo* cap a les destinacions indicades i que, al trobar la destinació desitjada, respon amb un altre *echo* cap a l'origen del missatge rebent el mateix missatge original transmès. *Ping* retornarà una sèrie de valors de temps. De cada *echo* retornarà el seu RTT i quan acaba l'execució apareix imprès en pantalla els RTT mínims, mitjos i màxims de tota la sèrie d'*echos* que s'han generat.

És una de les aplicacions que es pot utilitzar en l'àmbit d'aquest projecte ja que serveix per detectar el veïns als quals es pot transmetre i per mesurar en mitjana la qualitat d'un enllaç. La qualitat es pot mesurar depenent del RTT mig que donin les transmissions. A major temps pitjor qualitat considerem que té el muntatge.

Al sistema operatiu encastat de Nylon la versió de *Ping* que ve incorporada és una versió limitada de l'aplicació original amb només les tres opcions més genèriques. Els tres modificadors a utilitzar en aquesta versió limitada de *Ping* queden recollits a la taula A.6.

| Paràmetre | Funció |
|-----------|--------|
| -c | Nombre d'*echo* a enviar |
| -s | Longitud dels paquets |
| -q | Execució en mode silenciós |

**Taula A.6: Paràmetres de ping**

Un exemple d'ús podria ser aquest:

```
ping –c 5 10.15.21.1
```

On la resposta a la instrucció concreta seria:

```
root@cube-1522:~# ping –c 5 10.15.21.1
PING 10.15.21.1 (10.15.21.1): 56 data bytes
64 bytes from 10.15.21.1: icmp_seq=0 ttl=64 time=3.6 ms
64 bytes from 10.15.21.1: icmp_seq=1 ttl=64 time=1.4 ms
64 bytes from 10.15.21.1: icmp_seq=2 ttl=64 time=4.3 ms
64 bytes from 10.15.21.1: icmp_seq=3 ttl=64 time=1.3 ms
64 bytes from 10.15.21.1: icmp_seq=4 ttl=64 time=1.3 ms

--- 10.15.21.1 ping statistics ---
5 packets transmitted, 5 packets received, 0% packet loss
round-trip min/avg/max = 1.3/2.3/4.3 ms
```

**Figura A.13: Resultat d'un ping**

Al resultat es poden veure clarament les dades de RTT que retorna l'aplicació i que abans s'havien comentat.

# A.7. NETFILTER

L'eina de *Netfilter* ha estat present durant aquest treball en unes quantes opcions de desenvolupament plantejades, a més de ser contínuament utilitzat per totes les estacions a l'hora de configurar les topologies desitjades per a la realització de diferents proves.

Les principals funcions de *Netfilter* són les següents:

- **Filtratge.** Mitjançant l'eina *iptables* present a Linux, es poden crear grans quantitats de regles diverses que permeten realitzar un filtratge de dades atenent a les regles creades.
- **Mangle.** La seva funció bàsica és la d'alteració del paquet del paquet, ja sigui modificant algun dels camps del paquet o realitzant alguna marca específica a aquest.
- **NAT.** Realitza una traducció d'adreces de xarxa i pot alterar tant l'adreça d'origen com de destinació.

La funcionalitat bàsica per la que s'ha fet servir a aquest projecte el *Netfilter*, ha estat la de filtratge. Aquest filtratge ha servit, especialment, per forçar diferents topologies en línia. Aquestes topologies, que han estat majoritàriament les que s'han fet servir per provar l'evolució del sistema, necessiten de regles amb les que descartar possibles paquets que arribin d'estacions adjacents. D'aquesta forma, les regles només deixen passar aquells paquets originats per veïns d'un sol *hop*.

Els filtratges es poden realitzar de moltes maneres diferents, però la manera més raonable pel cas que ens ocupa és la de realitzar-ho per adreces MAC. Encara que també es pot filtrar per IP d'origen o per qualsevol altre camp. Altres opcions de filtratge englobarien a qualsevol dels camps de les capçaleres dels protocols. Així, es pot filtrar per protocol, per tipus de servei, per nombre de salts o per qualsevol paràmetre susceptible de ser consultat.

Una altra de les característiques destacades del *Netfilter* és determinar a quin moment s'ha d'aplicar la regla. Amb *iptables* podem distingir tres modalitats diferents d'aplicar les regles:

- **INPUT.** La regla s'aplica abans de realitzar cap tipus de decisió, és a dir, s'aplica sobre el tràfic entrant al node.
- **FORWARD.** En aquest cas, la regla es fa servir pel tràfic generat pel propi node.
- **OUTPUT.** La regla implementada s'aplica per tot el tràfic sortint del node, és a dir, una vegada que els paquets han d'abandonar els nodes.

Existeixen multitud d'opcions per complementar una regla, que es poden consultar a una àmplia bibliografia a la pàgina web [67]. Una de les decisions que sempre haurà d'incorporar la regla és què es farà amb el paquet capturat pel filtre degut a la regla implementada. Encara que existeixen més opcions, les tres principals i més importants són les següents:

- **NF_ACCEPT.** El paquet en qüestió és acceptat i, per tant, es passa cap a la destinació correponent.
- **NF_QUEUE.** En aquest cas, el paquet es envia a una cua a l'espera d'una presa de decisions sobre quin és el comportament que ha de seguir el paquet. Si no hi ha cua definida, el paquet és descartat de manera immediata.
- **NF_DROP.** El paquet és descartat pel node en qüestió.

Tant **ACCEPT** com **DROP** poden ser les polítiques per defecte de les *iptables*. El que es fa, bàsicament, és acceptar tots els paquets per defecte en un cas, mentre en l'altre aquests serien rebutjats.

Un exemple d'utilització d'una regla on es filtren els paquets d'una determinada targeta, fent el filtratge per l'adreça MAC d'aquell node, filtrant per tot el tràfic entrant al node i rebutjant amb el DROP tot el tràfic que coincideixi amb dita cadena.

```
iptables -t filter -A INPUT -m mac --mac-source 00:02:6F:35:52:AE
-j DROP
```

# A.8. MODES DE LES TARGETES *WIRELESS*

A aquest annex farem un repàs dels diferents modes que poden assolir les diferents targetes inalàmbriques repassant quina utilitat tenen cadascun d'aquests modes i per a què es poden fer servir.

Els diferents modes en què poden funcionar les nostres targetes són els següents:

- **Ad-hoc.** És el mode més emprat a aquest projecte ja que permet que les diferents estacions es relacionin entre elles sense el concurs de cap punt d'accés. Les targetes que estan en mode ad-hoc s'associaran entre elles fins a conformar una única cel·la. Amb *madwifi*, les targetes en ad-hoc no realitzen correctament la funció de llistar punts d'accés interferents i, per això, les interfícies de senyalització van ser substituïdes per targetes amb *driver hostap*.

- **Managed.** El node es connecta a un entorn on hi ha molts de punts d'accés gràcies al procés de *roaming*. Amb aquest mode *madwifi* sí que obté resultats a l'hora de realitzar l'escaneig de l'entorn.

- **Master.** En aquest cas, la targeta actua com a punt d'accés per a altres nodes en ad-hoc. Serveix per a crear xarxes sense fils amb infraestructura, ja que el propi node tindrà el control de les operacions que es realitzaran dins l'àmbit de la xarxa.

- **Repeater.** El node actua com a un simple repetidor, sense realitzar cap tipus de funció que requereixi un mínim processament. Tots els paquets arribats al node són retransmesos cap als altres nodes sense fils.

- **Secondary.** El node funciona com un recanvi d'un node en mode *master* o en mode *repeater*. Només es posa en marxa quan el "principal" deixa de funcionar.

- **Monitor.** En aquest cas el node entra en estat passiu, associant-se a una de les cel·les existents. Quan està associat, monitoritza tots els paquets d'aquella freqüència en aquella cel·la concreta.

- **Auto.** És el mode automàtic del sistema en el que serà aquest el que decidirà quin mode assolirà la targeta *wireless*.

- **Pseudo_ibss.** És un mode privat de les targetes utilitzades a la realització del treball, però que té unes característiques interessants per a la realització de les proves. Amb pseudo_ibss, les estacions no necessiten cap tipus d'associació ja que no generaran cap cel·la (adreça 00:00:00:00:00:00). Per tant, és un mode molt útil per realitzar associacions entre una gran quantitat d'estacions, ja que aquestes es conformaran dins la mateixa xarxa de forma immediata. Al no requerir cap tipus d'informació respecte a la cel·la associada, amb *pseudo_ibss* no es requereix cap procés de *beaconing*, factor que impedeix que les targetes amb aquest mode seleccionat siguin detectats a l'hora de realitzar un escaneig a l'entorn.

Tots els modes, menys l'últim, són comuns a gairebé totes les targetes *wireless*, mentre que la darrera de totes és d'àmbit privat i pot haver un grapat de *drivers* que no suportin dit mode. Els *drivers* utilitzats a aquest projecte (*madwifi* i *hostap*) suporten de manera privada aquest mode de funcionament.

Per accedir als diferents modes cal introduir una comanda que indiqui quin d'ells és el que volem utilitzar. Per qualsevol dels modes, exceptuant el darrer de tots, la comanda que s'utilitza per l'accés és:

```
iwconfig nom_iface mode ad-hoc
iwconfig nom_iface mode managed
iwconfig nom_iface mode master
iwconfig nom_iface mode repeater
iwconfig nom_iface mode secondary
iwconfig nom_iface mode monitor
iwconfig nom_iface mode auto
```

Mentre que per tenir les interfícies en mode *pseudo_ibss* s'ha d'introduir una comanda que faci referència als paràmetres privats.

Per *madwifi*:
```
iwpriv nom_iface ibss 0
```

Per *hostap*:
```
iwpriv nom_iface ibss 1
```

# A.9. ARXIUS DEL CODI FONT

El codi font que s'ha utilitzat per a la confecció definitiva del projecte no serà inclòs a la memòria degut a la seva extensió. El codi complet podrà ser accedit a la versió digital entregada del projecte.

Tot i això, a l'annex es farà un petit resum dels arxius modificats i creats durant tot el procés amb una breu explicació de les funcions que realitzen les rutines incloses a tots els punts. Els arxius s'agruparan seguint els tres blocs diferenciats que, finalment, han donat lloc a la solució final. Aquests tres blocs són: l'OLSR modificat, el protocol de negociació de canal i el *bonding driver*. Tret d'alguna característica molt concreta, el codi s'ha elaborat en el llenguatge de programació C.

## OLSR modificat

Per a la realització d'aquestes modificacions, es va agafar una de les darreres versions d'OLSR i es va anar modificant fins a obtenir el resultat que s'esperava. Degut al gran nombre de fitxers que composen el protocol d'encaminament, en aquest annex només es recolliran aquells que han estat modificats, afegint-li rutines o realitzant alguns canvis en les rutines ja creades.

- **interfaces.c**

És on es realitzen gairebé totes les modificacions al protocol ja que és on es defineixen les funcions de mapeig de les adreces que es volen modificar per substituir les adreces de la interfície de senyalització per la de *bond0* i les funcions de mapeig d'interfície i de conversió d'aquestes, que requereixen consultes *ioctl*, per la consulta de paràmetres relacionats amb les interfícies i que s'aporten des del *kernel*.

- **interfaces.h**

A més de definir les noves funcions de mapeig, també es crea una nova estructura que ajudarà a la conversió entre la interfície de senyalització i la que s'utilitzarà com a interfície de dades posteriorment.

- **lq_route.c**

És l'arxiu que conté funcions que s'executen quan s'estan configurant les rutes calculades per OLSR. Aquí OLSR accedeix sempre que el *LinkQuality* al fitxer de configuració sigui igual a 2. A les funcions d'afegir rutes, es modificaran aquestes rutes fent crides a les funcions definides a **interfaces.c**.

- **routing_table.c**

Es fan les mateixes modificacions que al fitxer anterior. Pensem que les funcions recollides a aquest fitxer són equivalents a les anteriors però que s'accedeixen quan el fitxer de configuració té seleccionat un *Link Quality* igual a 1.

**Protocol de negociació del canal**

A diferència del bloc anterior, tots els fitxers d'aquest punt han estat creats de bell nou, des de zero. L'objectiu d'aquest protocol és el de seleccionar el millor dels canals depenent de les condicions que l'envolten. Els fitxer que composen aquest protocol són les següents:

- **principal.h**

Són una sèrie de definicions globals que serveixen per tot el protocol com la longitud de les adreces IP o MAC.

- **principal.c**

És el main del protocol i només presenta les rutines necessàries per llegir la interfície de dades en la que es realitzaran les operacions i per la generació dels *pthreads* que, a cada node, actuaran com a client i com a servidor simultàniament.

- **fils.h**

Arxiu on es defineixen una sèrie de constants a manejar en tot el protocol, unes estructures bàsiques per recopilar les dades que es consultaran a altres programes, entre elles l'estructura *paquet* i la definició d'unes funcions que s'empraran a altres fitxers.

- **client.c**

És el fitxer on estan plasmades totes les funcions que realitza el fil del client generat al programa principal. Dins d'aquest arxiu es veu com s'aniran omplint tots els camps de l'estructura definida a **fils.h** gràcies a l'intercanvi que tindrà lloc entre el client i el servidor. Al final del client s'hauran obtingut les adreces IP i MAC del veïns i el canal òptim per comunicar-se amb aquests veïns. Finalment, aquesta estructura s'envia amb una funció definida a un altre arxiu cap a l'espai del *kernel*. El client serà iteratiu i, per tant, el procés es repetirà indefinidament fins que s'interrompi l'execució del protocol.

- **server.c**

Correspon al fitxer que recull totes les rutines del fil del servidor del node. Aquest servidor proporciona totes les dades sol·licitades pel client i aquestes són passades mitjançant l'intercanvi esmentat abans. Dins aquest arxiu s'executen les operacions de qualitat, les funcions de la qual es troben descrites a un altre fitxer. A l'igual que el client, el servidor és iteratiu i sempre estarà a l'espera de rebre algun tipus d'informació per part del client.

- **canal.h**

Aquí es defineixen totes les funcions que tindran que veure amb manipulació de canals i obtenció de qualitats. A més, també es declara l'estructura *qualitat* que servirà per guardar els valors de qualitat dels escanejos damunt dels diferents canals.

- **canal.c**

Al fitxer es pot veure el desenvolupament de les tres funcions relacionades amb els canals que havien estat declarats a la seva llibreria. La primera funció és la que agafa el canal sintonitzat actualment per la interfície que es consulta, la segona obtén tots els valors de qualitat de tots els canals i la tercera sintonitza la targeta *wireless* al canal indicat.

- **funcions.c**

És un arxiu on es recopilen aquelles funcions auxiliars necessàries pel funcionament del client i del servidor i que estan separades per mantenir la bona estructuració del programa general. Les funcions definides són molt diverses ja que es té englobat en el mateix fitxer la funció d'agafar un número d'un fitxer, la funció de llegir una adreça IP, una operació de ponderació de la qualitat que es rep i, finalment, l'elecció del canal òptim segons les dades de qualitat obtingudes anteriorment.

- **d_netlink.h**

Només declara la funció que s'emprarà en el .c d'aquesta llibreria.

- **d_netlink.c**

Desenvolupa la funció declarada abans i el que tracta de fer és enviar tota l'estructura *paquet* plena de dades recopilades cap a l'espai del *kernel*, que serà on, llavors, es seguiran els mecanismes necessaris per fer la selecció d'interfícies.

- **scan_new_2.pl**

És un script en Perl, referenciat per una de les funcions establertes a **canal.c** i que té la funció d'executar les instruccions necessàries pel càlcul d'interferències. Simplement, executa una instrucció Linux, guardant el seu resultat en un fitxer manipulable per les funcions de canal.

**Bonding driver**

El codi del *bonding driver* es de lliure distribució i pot ser descarregat sense problemes des de la xarxa. En el nostre cas, el codi s'ha agafat de la pàgina del projecte Net-X [59], adaptant-lo posteriorment a les nostres necessitats. De fet, s'han creat dos fitxers nous al *bonding* i s'han modificat els principals. A diferència dels dos blocs anteriors, que havien estat desenvolupats en l'àmbit convencional de l'espai d'usuari, tota la programació que es fa del *bonding driver*, es realitza dins l'espai del *kernel* directament. Això implica que algunes funcions no es podran emprar i el resultat a l'hora de compilar les rutines desitjades és la de la creació d'un objecte .o que, després, s'haurà de carregar com a mòdul dins el sistema Linux encastat dels *meshcube*.

- **if_bonding.h**

S'afegeix la definició del mode de *bonding* específic dissenyat per aquest projecte. Simplement es tracta de afegir el nom del nostre procés relacionant-lo amb un nombre.

- **bond_main.c**

Aquí, es modifiquen una sèrie de paràmetres de tot el programa perquè el mètode de *bonding* apuntat sigui el que s'ha dissenyat. A més, s'afegeix la rutina *bond_xmit_protocol* on, agafant les dades dels canals òptims que es tenen, es fa el canvi d'interfície cap a la solució desitjada.

- **pr_netlink4.h**

Es declaren totes les funcions que s'utilitzaran al fitxer de desenvolupament, a més de l'estructura *paquet* equivalent per a l'espai del *kernel.*

- **pr_netlink4.c**

És el fitxer on es defineixen una sèrie de funcions que fan de pont al protocol de diàleg abans comentat amb el *bonding driver*. Tot i pertànyer a l'espai del *kernel*, fa de nexe d'unió entre els dos extrems. Així, les funcions escolten la informació que prové de l'espai d'usuari i la recopila en l'estructura equivalent que s'ha declarat a la llibreria. D'aquesta manera s'ha recollit la mateixa informació que es manejava en el protocol. Apart d'aquestes funcions, també té definides una sèrie de funcions extra com són, les *inet_ntoa* i les *inet_aton* especials pel *kernel space* i funcions de traducció entre els canals que s'han rebut i les interfícies que pertocaven per aquell canal.